\documentclass[11pt,letterpaper]{article}

\usepackage[margin=1in]{geometry}
\usepackage[utf8]{inputenc}

\usepackage{amsmath}
\usepackage{mathtools}
\usepackage{amsthm,amsfonts,amstext,amssymb}
\usepackage{bm}
\usepackage[dvipsnames]{xcolor}
\usepackage{comment}

\usepackage[hidelinks]{hyperref}
\usepackage[nameinlink,capitalize]{cleveref}
\Crefname{algocf}{Algorithm}{Algorithms}
\crefname{algocfline}{line}{lines}
\Crefname{claim}{Claim}{Claims}
\Crefname{remark}{Remark}{Remarks}

\usepackage{thmtools,thm-restate}
\usepackage[ruled,vlined]{algorithm2e}

 \usepackage{xfrac}
 \usepackage{stmaryrd}

 \usepackage{graphicx}

\usepackage[inline]{enumitem}

\SetAlgoSkip{nospace}
\SetKwFor{Upon}{upon}{do}{end}

\SetCommentSty{mycommfont}
\SetAlCapHSkip{0pt} 
\SetAlgoHangIndent{1em}

\newtheorem{theorem}{Theorem}[section]
\newtheorem{lemma}[theorem]{Lemma}
\newtheorem{claim}{Claim}[section]

\newtheorem{corollary}[theorem]{Corollary}
\newtheorem{observation}[theorem]{Observation}

\newcommand{\OPT}{\textsc{Opt}}
\newcommand{\opt}{\textsc{Opt}}
\newcommand{\eps}{\varepsilon}

\newcommand{\items}{\mathcal{I}}
\newcommand{\itemsT}[1]{\items^{#1}}
\newcommand{\iopt}{\mathcal{I}^{\sigma^\ast}}
\newcommand{\itemsoverlapt}[2]{\items_{#1(i) < #2 < #1(i) + w(i)}}
\newcommand{\itemsatt}[2]{\items^{#1}(#2)}
\newcommand{\mountainininterval}[3]{\items_{#1(i) \leq #2,\,  #1(i) + w(i) \geq #3}}

\newcommand{\diffitemsatt}[3]{{\color{black}#1^{#2}(#3)}}

\newcommand{\itemsM}{\mathcal{M}}
\newcommand{\iadd}{i_{\lambda}}

\newcommand{\area}{\mathrm{area}}
\newcommand{\start}{\sigma}
\newcommand{\packing}{\sigma}
\newcommand{\sopt}{\sigma^{\ast}}
\newcommand{\D}{D}

\usepackage{ tipa }
\newcommand{\leftl}{\ell\hspace{-.7ex}\raisebox{0.8ex}{\textlptr}}
\newcommand{\rightl}{\ell\hspace{-.7ex}\raisebox{0.8ex}{\textrptr}}
\newcommand{\leftr}{r\hspace{-.9ex}\raisebox{0.2ex}{\textlptr}}
\newcommand{\rightr}{r\hspace{-.9ex}\raisebox{0.2ex}{\textrptr}}

\newcommand{\leftlm}{\ell\hspace{-.9ex}\raisebox{0.8ex}{\textlptr}}
\newcommand{\rightlm}{\ell\hspace{-.9ex}\raisebox{0.8ex}{\textrptr}}
\newcommand{\leftrm}{r\hspace{-1.1ex}\raisebox{0.2ex}{\textlptr}}
\newcommand{\rightrm}{r\hspace{-1.1ex}\raisebox{0.2ex}{\textrptr}}

\newcommand{\srightl}{\ell\hspace{-.6ex}\raisebox{0.5ex}{\scriptsize\textrptr}}
\newcommand{\sleftr}{r\hspace{-.8ex}\raisebox{0.1ex}{\scriptsize\textlptr}}
\newcommand{\srightr}{r\hspace{-.8ex}\raisebox{0.1ex}{\scriptsize\textrptr}}

\newcommand{\insim}{\mathrel{\raisebox{.5ex}{$\in$}\hspace{-1.6ex}\raisebox{-.8ex}{$\sim$}}}

\newcommand{\cT}{\ensuremath{\mathcal{T}}}

\newcommand{\cH}{\ensuremath{\mathcal{H}}}
\newcommand{\cL}{\ensuremath{\mathcal{L}}}
\newcommand{\cS}{\ensuremath{\mathcal{S}}}
\newcommand{\cW}{\ensuremath{\mathcal{W}}}
\newcommand{\cF}{\ensuremath{\mathcal{F}}}

\newcommand{\complexityP}{\mathrm{P}}
\newcommand{\NP}{\mathrm{NP}}
\newcommand{\bigO}{\mathcal{O}}

\newcommand{\Ga}{I_r} 
\newcommand{\Gal}{I_\ell}
\newcommand{\Gb}{I_{\ell,r}} 
\newcommand{\Gd}{I_{\leq \ell}} 
\newcommand{\Ge}{I_{\geq r}} 
\newcommand{\Gaa}{I_{r,1}} 
\newcommand{\Gab}{I_{r,2}} 
\newcommand{\Gac}{I_{r,3}} 

\usepackage[textsize=tiny,textwidth=2cm,color=Red]{todonotes}

\newcommand{\alertBound}[1]{{#1}}

\usepackage{tikz}
\usetikzlibrary{shapes.geometric, arrows.meta, graphs, matrix,decorations.pathmorphing, arrows, automata, positioning, calc, decorations.pathreplacing} 
\usetikzlibrary{patterns}
\newboolean{BlackAndWhite}
\setboolean{BlackAndWhite}{true}

\definecolor{li}{HTML}{00677C} %
\definecolor{wi}{HTML}{A4C73B} %
\definecolor{ti}{HTML}{F29400} %
\definecolor{si}{HTML}{E43117} %
\definecolor{vi}{HTML}{39842E} %
\definecolor{hi}{HTML}{9B0a7d} %
\definecolor{di}{HTML}{B062D3} 
\definecolor{ki}{HTML}{9EDDE3}

\definecolor{myblue1}{RGB}{166,206,227}
\definecolor{myblue2}{RGB}{31,120,180}
\definecolor{mygreen1}{RGB}{178,223,138}
\definecolor{mygreen2}{RGB}{51,160,44}

\colorlet{colorb}{myblue1} 
\colorlet{colora}{mygreen1}
\colorlet{coloraa}{myblue2!90!white}
\colorlet{colorab}{mygreen2}
\colorlet{colorac}{mygreen1}
\colorlet{colord}{li!60!white}
\colorlet{colore}{vi!60!white}
\colorlet{coloriadd}{ti!70!white}
\colorlet{colorm}{mygreen1}
\colorlet{tallItemColor}{gray}
\colorlet{mediumItemColor}{gray!30}

\newcommand{\steinberg}{\textcolor{black}{Steinberg}}

\title{A Tight ($3/2 + \eps$)-Approximation Algorithm for~Demand~Strip~Packing}
\author{
    Franziska Eberle \\ Technische Universität Berlin \\ \texttt{f.eberle@tu-berlin.de}
    \\[2ex]
    Felix Hommelsheim \\ Universität Bremen \\ \texttt{fhommels@uni-bremen.de}
    \\[2ex]
    Malin Rau \\ Chalmers University of Technology \\ \texttt{malin.rau@chalmers.se}
    \\[2ex]
    Stefan Walzer \\ Karlsruher Institut für Technologie \\ \texttt{stefan.walzer@kit.edu}
}
\date{}

\begin{document}
\pagenumbering{gobble} 
\maketitle
\begin{abstract}
We consider the Demand Strip Packing problem (DSP), in which we are given a set of jobs, each specified by a processing time and a demand. 
The task is to schedule all jobs such that they are finished before some deadline $D$ while minimizing the peak demand, i.e., the maximum total demand of tasks executed at any point in time.
DSP is closely related to the Strip Packing problem (SP), in which we are given a set of axis-aligned rectangles that must be packed into a strip of fixed width while minimizing the maximum height.
DSP and SP are known to be NP-hard to approximate to within a factor below $\frac{3}{2}$.

To achieve the essentially best possible approximation guarantee, we prove a structural result. Any instance admits a solution with peak demand at most $\big(\frac32+\varepsilon\big)\textsc{Opt}$ satisfying one of two properties. Either (i) the solution leaves a gap for a job with demand $\textsc{Opt}$ and processing time $\mathcal O(\varepsilon D)$ or (ii) all jobs with demand greater than $\frac{\textsc{Opt}}2$ appear sorted by demand in immediate succession. 
We then provide two efficient algorithms 
that find a solution with maximum demand at most $\big(\frac32+\varepsilon\big)\textsc{Opt}$ in the respective case. 
A central observation, which sets our approach apart from previous ones for DSP, is that the properties (i) and (ii) need not be efficiently decidable: We can simply run both algorithms and use whichever solution is the better one.
\end{abstract}

\pagenumbering{arabic}

\section{Introduction}

Packing and scheduling problems are among the most fundamental problems in combinatorial optimization \cite{GalvezGIHKW21,DBLP:conf/soda/GalvezKMMPW22, DBLP:conf/soda/GrandoniMW22,DBLP:conf/stoc/0001MW22,GMW018}. 
In this paper, we consider the \textsc{Demand Strip Packing} problem (DSP), in which we are given a set of $n$ items (or \emph{jobs}) $\items = \{1, ..., n \}$ and a width $\D \in \mathbb{N}$ (or \emph{deadline}). 
Each item has a \emph{width} (or \emph{processing time}) $w(i) \in \mathbb{N}_{>0, \leq  \D}$ and a \emph{height} (or \emph{demand}) $h(i) \in \mathbb{N}_{>0}$.
A $\D$-\emph{feasible} packing (or \emph{schedule}) is an assignment $\start: \items \rightarrow [0, D]$ such that $0 \leq \start(i) \leq \D - w(i)$, i.e., no item starts before~$0$ and each item ends before~$\D$. 
For a packing $\start$ and $t \in [0, \D]$, let $\items^\start (t) \coloneqq \{ i \in \items \mid \start (i) \leq  t < \start(i) + w(i) \}$ be the items packed at point $t$. 
The objective in DSP is to compute a $\D$-feasible packing $\start$ that minimizes the \emph{height}
$h(\start) := \max_{t \in [0, \D] } \sum_{i \in \items^\start (t)} h(i)$.
An \emph{optimal} packing $\sopt$ of $\items$ with deadline $\D$ is a $\D$-feasible packing with minimum height.

This model captures a resource allocation problem, e.g., encountered in smart electricity grids \cite{alamdari2013smart,karbasioun2018asymptotically,siano2014demand}.
While the historical strategy was to provide an energy infrastructure that 
is able to supply whatever demand arises, 
we should expect a shift towards a world where, conversely, energy demand is adjusted according to energy supply.
With smart homes, smart factories, and the internet of things on the rise, such a shift is \emph{possible}, i.e.\ energy requirements for linked smart appliances such as charging stations of electric vehicles or washing machines can be met by flexibly planning their respective usage. Moreover, such a shift is \emph{necessary} due to the high fluctuation in electricity generated by renewable energy sources such as wind turbines and photovoltaic systems~\cite{masters2013renewable}.

A problem closely related to DSP is the \textsc{Strip Packing} problem (SP), in which we are given an axis-aligned half-strip of width $\D$ and infinite height and a set of axis-aligned open rectangles, where each rectangle $i$ has width $w(i) \in (0, \D]$ and height $h(i) > 0$.
The objective in SP is to compute an axis-aligned non-overlapping packing of all the rectangles within the strip while minimizing the maximum height spanned by the rectangles.
Clearly, feasible solutions to SP are feasible for DSP as well, while the converse is not necessarily true~\cite{BladekDGS15,Galvez0AK21}. However, for e.g. smart grid scheduling, DSP already 
captures the essence of the problem and might allow for simpler and better algorithms.

Since DSP and SP are~$\NP$-hard, we do not expect that a polynomial-time algorithm exists that solves either of these problems optimally. Thus, research mostly focused on designing efficient
algorithms that find near-optimal solutions: A polynomial-time algorithm is an $\alpha$-\emph{approximation algorithm} if 
\(
    h(\start) \leq \alpha \opt
\) 
for all instances $(\items,\D)$, 
where~$\start$ is the packing obtained by the algorithm and $\opt$ is the value $h(\sopt)$ for an optimal packing $\sopt$; $\alpha$ is called the \emph{approximation ratio}.

A simple reduction from the {\sc partition} problem shows that neither SP nor DSP~\cite{yaw2014peak} can be approximated within a factor $\frac{3}{2} - \eps$ for any $\eps > 0$, unless $\complexityP = \NP$. 
For both problems, SP~\cite{DBLP:journals/comgeo/HarrenJPS14} and DSP~\cite{DBLP:journals/algorithmica/DeppertJKRT23,Galvez0AK21}, the state-of-the-art are algorithms with approximation ratio $\big(\frac{5}{3}+\eps\big)$.

\subsection{Our main result and technical contribution}

Our main contribution is a $\big(\frac32 + \eps\big)$-approximation algorithm for \textsc{Demand Strip Packing}.  

\begin{restatable}{theorem}{thmthreehalfapproxalg}
    \label{theo:approxalg}
    For any $\eps > 0$, there exists a $\big(\frac32+\eps\big)$-approximation algorithm for \textsc{Demand Strip Packing}. 
\end{restatable}

\begin{figure}
    \centering
    \includegraphics[width=0.32\textwidth,page=3]{figures/neat-and-forgiving.pdf}~
    \includegraphics[width=0.32\textwidth,page=1]{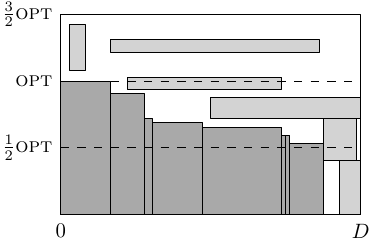}~
    \includegraphics[width=0.32\textwidth,page=2]{figures/neat-and-forgiving.pdf}
    \caption{An optimal packing (left), a $(0,\opt)$-neat packing of the same items (middle), and a $\lambda$-forgiving packing (right) of the same items and the extra item $i_\lambda$ (orange).}
    \label{fig:neat-and-forgiving}
\end{figure}

The key ingredient for proving the theorem is a strong structural result that guarantees that any optimal packing~$\sopt$ can be transformed into a packing~$\start$ of height at most~$\big(\frac32+\eps\big)\opt$ with much more structure than~$\sopt$: 
The new packing~$\start$ either packs all items of height at least $\frac\opt2$ sorted by non-increasing height starting at time~$0$ or can accommodate an extra item of width~$\lambda \D$ for some~$\lambda < \eps$. 
Concretely, for any $H$ and $\eps$, we define an \emph{$(\eps,H)$-neat packing} as a $\D$-feasible packing of the items $\items$ with height at most $\big(\frac{3}{2}+\eps\big)H$ where all items of height greater than $\frac H2$ are packed in immediate succession in non-increasing order of height starting at~$0$.
Additionally, we define a \emph{$\lambda$-forgiving packing} as a $\D$-feasible packing of height at most $\frac{3}{2}\opt$ that can accommodate an extra item~$\iadd$ of width~$\lambda \D$ for some~$\lambda \in \Theta(\eps)$.
An example of such packings is depicted in \cref{fig:neat-and-forgiving}.
Using these two definitions, we prove the following theorem. 

\begin{restatable}{theorem}{thmstructuralresult}
\label{thm:restructuring}
    For any $\eps> 0$, any \alertBound{$\lambda \leq \min\big\{\frac{\eps}{3(5+4\eps)}, \frac{1}{60}\big\}$}, and any instance $(\items, \D)$, any optimal packing for $(\items,\D)$ of height $\opt$ can be transformed into a packing $\start$ that is 
    \[\text{{\upshape(i)} a $\lambda$-forgiving packing \qquad or \qquad {\upshape(ii)} an $(\eps,\opt)$-neat packing.}\]
\end{restatable}

If the packing~$\start$ is $\lambda$-forgiving, we can employ a packing algorithm by \cite{DBLP:journals/algorithmica/DeppertJKRT23} to create a partial solution to a larger instance that additionally packs the extra item. 
The algorithm is LP-based and rounds a fractional solution to an integral solution that packs all items except some items that can be packed in an area of width~$\lambda \D$ and height $\opt$.
These items are then placed ``inside'' the extra item using standard packing strategies.
If the packing~$\start$ is $(\eps,\opt)$-neat, we design an algorithm that places items with height at most $\frac{\opt}{2}$ around the sorted items with height larger than $\frac{\opt}{2}$. 
This algorithm packs all items except for items with width at most $\frac{\eps}{1+\eps}\D$ (called \emph{squeezable items}; formally defined in \cref{sec:squeezing}). 
We design and analyze a packing procedure that \emph{squeezes in} those unpacked items into the packing relying on the fact that $\start$ is $(\eps,\opt)$-neat.

Restructuring optimal solutions to create good approximations is a common practice used in many packing algorithms that achieve an approximation ratio below 2 (see, e.g., \cite{DBLP:journals/algorithmica/DeppertJKRT23,Galvez0AK21}).
However, most of these structural results describe one of two structural properties: they can either fit an extra item with a small width and height $\opt$ (see, e.g., \cite{DBLP:journals/algorithmica/DeppertJKRT23}) or the restructured packing sorts the items with a height larger than $c \cdot \opt$ for some constant $c \geq \frac{1}{2}$ (see, e.g., \cite{Galvez0AK21}).
There are results, e.g., \cite{DBLP:journals/algorithmica/DeppertJKRT23}, that, in addition to proving a structural lemma, also reorder tall items in some cases. 
However, these cases depend on the instance's properties and not on the structure of an optimal solution.
One of our contributions is to rigorously combine these two restructuring methods into one structural theorem.
This allows us to design an essentially tight approximation algorithm that does not rely on case distinction anymore: We simply create two packings, one assuming (i) and one assuming (ii), and return the better of the two. 

\paragraph*{Key repacking procedures.}
Key ingredients to proving \cref{thm:restructuring} are two novel repacking strategies: the \emph{Stretching} and the \emph{Squeezing Algorithm}, formalized in \cref{sec:key_repacking_procedures}. 
The Stretching Algorithm is employed when restructuring an optimal solution. 
More precisely, when some fixed optimal packing~$\sopt$ has a segment of width~$w$ which densely packs items with height greater than $\frac\opt2$, 
i.e., subsegments of total width at most $w' \ll w$ do not contain such tall items,
we can repack the items of height at most $\frac\opt2$ originating from this segment into two areas of height $\frac\opt2$ and width $w + w'$ and $4 w'$, respectively: 
$\sopt$ restricted to these items \emph{already} has a height of at most $\frac\opt2$ over a width of $w - w'$. 
Hence, we only need to ``repair'' a relatively small width. 
This is achieved by removing items not overlapping with a tall item (creating the area of width $4w'$) and quite literally stretching the remaining items of height at most $\frac\opt2$ by $w'$. 

Compared to known packing procedures, the stretched packing is much denser and, most importantly for our purposes, the packing with width $w+w'$ still resembles the original packing of its items in $\sopt$. 
Employing, e.g., Steinberg's repacking procedure~\cite{Steinberg97} would result in a (literal) black-box packing of width $4w$ and height $\frac\opt2$. 
The strong similarity between the stretched packing and the original packing allows us to argue about the interplay between the stretched items and the non-stretched tall items. 
This fact is crucial in many of our subcases when proving \cref{thm:restructuring}.

The Squeezing Algorithm can be used for any $(\eps,\opt)$-neat packing.  
It allows us to add squeezable items to an $(\eps,\opt)$-neat packing
\emph{without} increasing the packing height beyond $\big(\frac{3}{2} +\eps\big)\opt$.
By shifting all items with height at most $\frac{\opt}{2}$ to the left as far as possible without increasing the packing height above $\big(\frac{3}{2} +\eps\big)\opt$, we create an empty area of width at least $\frac{\eps}{1+\eps}\D$ and height at least $\frac{\opt}{2}$ at the right end of our packing
for a squeezable item. For the sake of completeness, we include this algorithm here even though a similar technique has been developed by~\cite{Galvez0AK21}.

Having the Squeezing Algorithm at hand, we can focus on items of height greater than $\frac\opt2$ or width greater than $\frac{\eps}{1+\eps}\D$ when proving the existence of $(\eps,\opt)$-neat packings. 
This simplifies the exposition of our proofs as we require fewer case distinctions. 
Finally, when designing the $\big(\frac32+\eps\big)$-approximation algorithm, we do not need to create so-called medium items (as in \cite{DBLP:journals/algorithmica/DeppertJKRT23,Galvez0AK21}). 
Those items do not have to be rounded and the remaining items have more structure; combining these three observations allows us to design an approximation algorithm whose running time has single-exponential dependency on $\frac1\eps$ if the instance admits an $(\eps,\opt)$-neat packing
in contrast to the algorithms designed in \cite{DBLP:journals/algorithmica/DeppertJKRT23,Galvez0AK21}, which are at least double-exponential in $\frac1\eps$.

\paragraph*{Extension to other packing problems.} 
While some of our techniques, such as the Squeezing Algorithm, crucially exploit that, in DSP, we are not interested in a geometric packing, it is conceivable that other techniques such as the Stretching Algorithm can also be analyzed for packing problems where items cannot be sliced, e.g., \textsc{Strip Packing}.
Further, as for DSP, showing similar structural results for related packing problems such as \textsc{Geometric Knapsack}, \textsc{Storage Allocation} or \textsc{Unsplittable Flow on a Path} led to breakthrough results~\cite{bonsma2014constant,GalvezGIHKW21,DBLP:conf/icalp/MomkeW20}. Such a structural result might pave the way for also matching the lower bound of $\frac32$ for \textsc{Strip Packing}.

\subsection{Related work}

It has been known for almost a decade that DSP is $\NP$-hard to approximate within a factor better than $\frac32$~\cite{yaw2014peak}. 
We essentially match this lower bound of $\frac32$ by providing a $\big(\frac32+\eps\big)$-approximation algorithm.
The previously best approximation ratio of $\frac{5}{3} + \eps$ is achieved by~\cite{DBLP:journals/algorithmica/DeppertJKRT23} and~\cite{Galvez0AK21}. 
If all items are squares, i.e., their width is the same as their height, there is a  $\frac32$-approximation algorithm~\cite{Galvez0AK21}. 
To overcome the lower bound of~$\frac{3}{2}$, also pseudo-polynomial-time algorithms\footnote{An algorithm is pseudo-polynomial-time if its running time is $\mathrm{poly}(n,\D)$, where $\D \in \mathbb{N}$ is the width of the strip.} are studied.
Very recently, \cite{JansenRT24} gave such an algorithm with approximation ratio $\frac54+\eps$ and a matching lower bound.

As mentioned above, closely related to DSP is Strip Packing (SP), where the items have to be packed as rectangles.
The first approximation algorithm for SP dates back to 1980, when Baker et al.~\cite{BakerCR80} gave a 3-approximation algorithm. 
The current best approximation ratio is $\big(\frac53+\eps\big)$~\cite{DBLP:journals/comgeo/HarrenJPS14}.
If all items are relatively small in one dimension (measured by $\delta$), then there exists a $\big(\frac{3}{2}+\eps\big)$-approximation algorithm~\cite{DBLP:journals/algorithmica/GalvezGAJKR23} for $\eps > 0$ and $\delta \leq \eps^{1/\eps^{\mathcal{O}(1)}}$.
There is a pseudo-polynomial-time algorithm with approximation guarantee $\big(\frac{5}{4}+\eps\big)$~\cite{JansenR19}, essentially matching the lower bound by~\cite{HenningJRS20}.

In another closely related problem, \textsc{Parallel Task Scheduling} (PTS), a set of jobs $\mathcal{J}$ and a number of machines $m$ are given. 
For job $j \in \mathcal J$ to complete, it needs to be processed contiguously for $p(j)$ units of time on $q(j)$ many machines, while any machine can process at most one job at a time. 
The aim is to minimize the makespan, i.e., the latest job completion time, while guaranteeing that the total number of machines required by jobs is at most $m$ at any point in time. 
PTS and DSP are closely related by observing that the height of an item can be interpreted as demand for machines and the width of an item as processing time. 
However, the ``fixed'' sides of the packings in PTS and DSP are orthogonal: PTS requires the number of machines (or maximal total demand) to be bounded by $m$ while DSP requires the width (or makespan) to be bounded by~$\D$. 
Therefore, algorithms approximating PTS cannot be directly used for approximating DSP instances and vice versa. 
In terms of hardness of approximation, these two problems are identical: Reducing again from \textsc{Partition}, there cannot be a $\big(\frac32-\eps\big)$-approximation algorithm for any $\eps > 0$ for PTS, unless $\text{P}=\text{NP}$. This lower bound is essentially achieved by the $\big(\frac32 + \eps\big)$-approximation algorithm by~\cite{DBLP:conf/spaa/Jansen12}. In contrast to DSP, PTS admits a pseudo-polynomial-time $(1+\eps)$-approximation algorithm~\cite{DBLP:journals/siamcomp/JansenT10}.

\subsection{Preliminaries}

\paragraph*{Notation.} 
Recall that $h(i)$ and $w(i)$ refer to the height and width of an item $i \in \items$, respectively. Let $\area(i) := h(i) \cdot w(i)$. We overload these functions to also refer to the height, width, and area of a set of items $\items' \subseteq \items$, i.e.,~$h(\items') := \sum_{i \in \items'} h(i)$, $w(\items') := \sum_{i \in \items'} w(i)$, and $\area(\items') := \sum_{ i \in \items'} \area(i)$.

For any property $P$, we write $\items_P$ to refer to the subset of items of $\items$ that have property $P$, e.g., we use $\items_{h > a}$ for the set of items of height greater than~$a$, i.e., for $\{i \in \items \mid h(i) > a\}$. 

When packing items, we typically refer to a contiguous subset of~$[0,\D]$ as a \emph{segment} in order to distinguish it from arbitrary intervals. 
For a packing~$\start$, let~$\items_{[a,b]}^{\start}$ denote the set of items completely packed in the segment~$[a,b)$, i.e., items with $a \leq \start(i) < \start(i) + w(i) \leq b$. 
Recall that~$\itemsT{\start}(t):= \{i \in \items: \start(i) \leq t < \start(i) + w(i) \}$.
We refer to the height of the set $\itemsT{\start}(t)$ as \emph{packing height (at $t$)}. 
For conciseness, we borrow some notation from the scheduling view of DSP: We refer to~$t \in [0,\D]$ as a \emph{(time) point} and say that an item \emph{starts} at~$t$ if~$\start(i) = t$ and it \emph{ends} at~$t$ if~$\start(i) + w(i) = t$.

Recall that, for an optimal packing~$\sopt$, we used~$\opt(\items) := h(\sopt) = \max_{t \in [0,\D]} h\big(\itemsT{\sopt}(t)\big)$ to denote its maximal height. 
The above definition allows us to bound~$\area(\items) \leq \D \cdot \opt(\items)$.

For some $H$ (typically, $H = \opt$), we refer to~$\items_{h > \frac H2}$ and~$\items_{h \leq \frac H2}$ as $H$-\emph{tall} and \emph{non-$H$-tall} items, respectively. 
Since we mostly talk about \opt-tall items when proving the structural theorem, we drop the dependency on $\opt$ and use \emph{tall} and \emph{non-tall}.
In a packing $\start$ with height~$H$ and width~$W$, at most one $H$-tall item can overlap any time~$t \in [0,W)$ and, for each $H$-tall item~$T$, the height of the non-$H$-tall items overlapping~$t \in [\start(T), \start(T) + w(T)]$ is bounded; such items are said to be \emph{packed in parallel} to~$T$.

\begin{observation}\label{obs_height_G1G2} 
    Let $\start$ be a packing with height~$H$ and consider some $H$-tall item~$T$. Then
    $h(\itemsoverlapt{\start}{t}\setminus \{T\}) \leq H - h(T) \leq \frac H2$ for every $t \in [\start(T), \start(T) + w(T)]$.
\end{observation}

For an optimal packing $\sopt$, consider the set of maximal, non-empty, right-open segments where no tall item is packed. Its elements $[\ell,r) \subseteq [0,\D]$ are called \emph{gaps}. (Note that $\ell = 0$ and $r = \D$ are permitted.) The \emph{width} of a gap is $r - \ell > 0$.

\paragraph*{Some known repacking procedures.}

We frequently sort items by height and place them next to each other (formalized in \Cref{alg_items_next_to_each_other} in \Cref{app:packing_algorithms}). We also \emph{mirror} a packing~$\start$, i.e., create a packing where each~$i \in \items$ starts at $\D - \start(i) - w(i)$ and ends at~$\D - \start(i)$ (\Cref{alg_mirroring}).

We sometimes pack small (w.r.t.\ area) sets of non-tall items with the well-known packing algorithm by Steinberg~\cite{Steinberg97}.
This algorithm places items into a rectangular area with height $H$ and width $W$ if 
$H$ and $W$ satisfy certain properties with respect to the item sizes.
We refer to this process as \steinberg$(\items',H)$, where $\items'$ is the set of items to be repacked and $H$ is the target height, implicitly setting $W = 2\cdot \max\big\{ \frac{\area(\items')}{H}, \max\big\{  w(i)| i \in \items'\big\}\big\}$.
We restate the result 
and refer to~\cite[Theorem 1.1]{Steinberg97} for a proof. 
(As usual, $x^+ = \max\{x, 0\}$ for some $x \in \mathbb R$.) 

\begin{lemma}[Steinberg Lemma]\label{lem:steinberg} 
    There is a polynomial-time algorithm that, when given as input a set $\items$ of items, a width $W \in \mathbb{Q}$, and a height $H \in \mathbb{Q}$, computes a packing of $\items$ into a rectangle of width~$W$ and height~$H$ under the condition that
    \begin{align*}
        \max_{i \in \items} w(i) \leq W,    \quad \quad   \max_{i \in \items} h(i)  \leq H, \text{ and}\\
         2 \, \area(\items)  \leq H\cdot W - \big(2 \, \max_{i \in \items} w(i) - W \big)^+  \big(2 \, \max_{i \in \items} h(i) - H \big)^+.
    \end{align*}
\end{lemma}

If~$2 \max_{i \in \items} w(i) \leq W$ or~$2 \max_{i \in \items} h(i) \leq H$, then the last condition becomes $2 \, \area(\items) \leq H\cdot W$.

\section{Key repacking procedures}
\label{sec:key_repacking_procedures}

In this section, we design and analyze some of our novel repacking algorithms, which allow us to obtain a tight approximation guarantee for DSP: the Stretching and the Squeezing Algorithm. 

\subsection{Stretching Lemma}

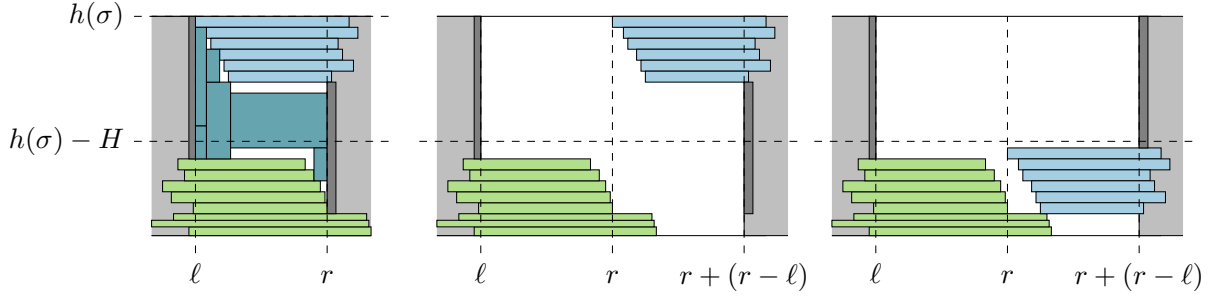
\begin{figure}
    \centering
    \resizebox{\textwidth}{!}{
    \begin{tikzpicture}
\pgfmathsetmacro{\w}{3}
\pgfmathsetmacro{\h}{3}
\pgfmathsetmacro{\os}{0.24}
\pgfmathsetmacro{\l}{0.2}
\pgfmathsetmacro{\r}{0.8}
\pgfmathsetmacro{\lamb}{0.04}

\draw (0,0) -- (\w,0);
\draw (0,\h) -- (\w,\h);


\begin{scope}[yscale =-1, yshift = -1*\h cm]
\draw[fill = lightgray] (0,0) -- (\l*\w,0) -- (\l*\w,\h) --(0,\h);
\draw[fill = lightgray] (\w,0) -- (\r*\w,0) -- (\r*\w,\h) --(\w,\h);

\draw[fill = colora] (\l*\w-0.03*\w,0.96*\h) rectangle (\r*\w+0.2*\w,\h);
\draw[fill = colora] (\l*\w-0.2*\w,0.93*\h) rectangle (\r*\w+0.19*\w,0.96*\h);
\draw[fill = colora] (\l*\w-0.1*\w,0.9*\h) rectangle (\r*\w+0.18*\w,0.93*\h);
\draw[fill = colora] (\l*\w-0.01*\w,0.85*\h) rectangle (\r*\w,0.9*\h);
\draw[fill = colora] (\l*\w-0.11*\w,0.8*\h) rectangle (\r*\w-0.01*\w,0.85*\h);
\draw[fill = colora] (\l*\w-0.15*\w,0.75*\h) rectangle (\r*\w-0.031*\w,0.8*\h);
\draw[fill = colora] (\l*\w-0.05*\w,0.7*\h) rectangle (\r*\w-0.06*\w,0.75*\h);
\draw[fill = colora] (\l*\w-0.08*\w,0.65*\h) rectangle (\r*\w-0.1*\w,0.7*\h);

\draw[fill = colorb] (\l*\w,0*\h) rectangle (\r*\w+0.1*\w,0.05*\h);
\draw[fill = colorb] (\l*\w+0.05*\w,0.05*\h) rectangle (\r*\w+0.14*\w,0.1*\h);
\draw[fill = colorb] (\l*\w+0.07*\w,0.1*\h) rectangle (\r*\w+0.05*\w,0.15*\h);
\draw[fill = colorb] (\l*\w+0.11*\w,0.15*\h) rectangle (\r*\w+0.07*\w,0.2*\h);
\draw[fill = colorb] (\l*\w+0.13*\w,0.2*\h) rectangle (\r*\w+0.12*\w,0.25*\h);
\draw[fill = colorb] (\l*\w+0.15*\w,0.25*\h) rectangle (\r*\w+0.02*\w,0.3*\h);

\draw[fill = colord] (\l*\w,0.05*\h) rectangle (\l*\w+0.05*\w,0.5*\h);
\draw[fill = colord] (\l*\w,0.5*\h) rectangle (\l*\w+0.05*\w,0.65*\h);
\draw[fill = colord] (\l*\w+0.05*\w,0.3*\h) rectangle (\l*\w+0.16*\w,0.65*\h);
\draw[fill = colord] (\l*\w+0.05*\w,0.15*\h) rectangle (\l*\w+0.11*\w,0.3*\h);
\draw[fill = colord] (\l*\w+0.16*\w,0.35*\h) rectangle (\r*\w,0.6*\h);
\draw[fill = colord] (\r*\w-0.06*\w,0.6*\h) rectangle (\r*\w,0.75*\h);

\draw[fill = tallItemColor] (\l*\w-0.03*\w,0) rectangle (\l*\w,0.65*\h);
\draw[fill = tallItemColor] (\r*\w,0.9*\h) rectangle (\r*\w+0.04*\w,0.3*\h);

\end{scope}

\draw[dashed] (\l*\w,-\os) node[below]{\vphantom{$r$}$\ell$} -- (\l*\w,\h);
\draw[dashed] (\r*\w,-\os) node[below]{\vphantom{$\ell$}$r$} -- (\r*\w,\h);

\draw[dashed] (-\os,0.43*\h) node[left]{$h(\sigma)-H$} -- (\w+\os,0.43*\h) ;
\draw[dashed] (-\os,1*\h) node[left]{$h(\sigma)$} -- (\w+\os,1*\h) ;


\begin{scope}[xshift = 1.3*\w cm]

\draw (0,0) -- (1.5*\w,0);
\draw (0,\h) -- (1.5*\w,\h);

\begin{scope}[yscale =-1, yshift = -1*\h cm]

\draw[fill = lightgray] (0,0) -- (\l*\w,0) -- (\l*\w,\h) --(0,\h);

\draw[fill = colora] (\l*\w-0.03*\w,0.96*\h) rectangle (\r*\w+0.2*\w,\h);
\draw[fill = colora] (\l*\w-0.2*\w,0.93*\h) rectangle (\r*\w+0.19*\w,0.96*\h);
\draw[fill = colora] (\l*\w-0.1*\w,0.9*\h) rectangle (\r*\w+0.18*\w,0.93*\h);
\draw[fill = colora] (\l*\w-0.01*\w,0.85*\h) rectangle (\r*\w,0.9*\h);
\draw[fill = colora] (\l*\w-0.11*\w,0.8*\h) rectangle (\r*\w-0.01*\w,0.85*\h);
\draw[fill = colora] (\l*\w-0.15*\w,0.75*\h) rectangle (\r*\w-0.031*\w,0.8*\h);
\draw[fill = colora] (\l*\w-0.05*\w,0.7*\h) rectangle (\r*\w-0.06*\w,0.75*\h);
\draw[fill = colora] (\l*\w-0.08*\w,0.65*\h) rectangle (\r*\w-0.1*\w,0.7*\h);

\begin{scope}[xshift = \r*\w cm -\l*\w cm]
\draw[fill = lightgray] (\w,0) -- (\r*\w,0) -- (\r*\w,\h) --(\w,\h);

\draw[fill = colorb] (\l*\w,0*\h) rectangle (\r*\w+0.1*\w,0.05*\h);
\draw[fill = colorb] (\l*\w+0.05*\w,0.05*\h) rectangle (\r*\w+0.14*\w,0.1*\h);
\draw[fill = colorb] (\l*\w+0.07*\w,0.1*\h) rectangle (\r*\w+0.05*\w,0.15*\h);
\draw[fill = colorb] (\l*\w+0.11*\w,0.15*\h) rectangle (\r*\w+0.07*\w,0.2*\h);
\draw[fill = colorb] (\l*\w+0.13*\w,0.2*\h) rectangle (\r*\w+0.12*\w,0.25*\h);
\draw[fill = colorb] (\l*\w+0.15*\w,0.25*\h) rectangle (\r*\w+0.02*\w,0.3*\h);

\draw[fill = tallItemColor] (\r*\w,0.9*\h) rectangle (\r*\w+0.04*\w,0.3*\h);
\end{scope}


\draw[fill = tallItemColor] (\l*\w-0.03*\w,0) rectangle (\l*\w,0.65*\h);
\end{scope}

\draw[dashed] (\l*\w,-\os) node[below]{\vphantom{$r$}$\ell$} -- (\l*\w,\h);
\draw[dashed] (\r*\w,-\os) node[below]{\vphantom{$\ell$}$r$} -- (\r*\w,\h);
\draw[dashed] (2*\r*\w -\l*\w,-\os) node[below]{$r + (r-\ell)$} -- (2*\r*\w -\l*\w,\h);

\draw[dashed] (-\os,0.43*\h) -- (1.59*\w+\os,0.43*\h) ;

\end{scope}

\begin{scope}[xshift = 3.1*\w cm]

\draw (0,0) -- (1.5*\w,0);
\draw (0,\h) -- (1.5*\w,\h);

\begin{scope}[yscale =-1, yshift = -1*\h cm]

\draw[fill = lightgray] (0,0) -- (\l*\w,0) -- (\l*\w,\h) --(0,\h);

\draw[fill = colora] (\l*\w-0.03*\w,0.96*\h) rectangle (\r*\w+0.2*\w,\h);
\draw[fill = colora] (\l*\w-0.2*\w,0.93*\h) rectangle (\r*\w+0.19*\w,0.96*\h);
\draw[fill = colora] (\l*\w-0.1*\w,0.9*\h) rectangle (\r*\w+0.18*\w,0.93*\h);
\draw[fill = colora] (\l*\w-0.01*\w,0.85*\h) rectangle (\r*\w,0.9*\h);
\draw[fill = colora] (\l*\w-0.11*\w,0.8*\h) rectangle (\r*\w-0.01*\w,0.85*\h);
\draw[fill = colora] (\l*\w-0.15*\w,0.75*\h) rectangle (\r*\w-0.031*\w,0.8*\h);
\draw[fill = colora] (\l*\w-0.05*\w,0.7*\h) rectangle (\r*\w-0.06*\w,0.75*\h);
\draw[fill = colora] (\l*\w-0.08*\w,0.65*\h) rectangle (\r*\w-0.1*\w,0.7*\h);

\begin{scope}[xshift = \r*\w cm -\l*\w cm]
\draw[fill = lightgray] (\w,0) -- (\r*\w,0) -- (\r*\w,\h) --(\w,\h);
\begin{scope}[yshift = 0.6*\h cm]
\draw[fill = colorb] (\l*\w,0*\h) rectangle (\r*\w+0.1*\w,0.05*\h);
\draw[fill = colorb] (\l*\w+0.05*\w,0.05*\h) rectangle (\r*\w+0.14*\w,0.1*\h);
\draw[fill = colorb] (\l*\w+0.07*\w,0.1*\h) rectangle (\r*\w+0.05*\w,0.15*\h);
\draw[fill = colorb] (\l*\w+0.11*\w,0.15*\h) rectangle (\r*\w+0.07*\w,0.2*\h);
\draw[fill = colorb] (\l*\w+0.13*\w,0.2*\h) rectangle (\r*\w+0.12*\w,0.25*\h);
\draw[fill = colorb] (\l*\w+0.15*\w,0.25*\h) rectangle (\r*\w+0.02*\w,0.3*\h);
\end{scope}

\draw[fill = tallItemColor] (\r*\w,0*\h) rectangle (\r*\w+0.04*\w,0.6*\h);
\end{scope}

\draw[fill = tallItemColor] (\l*\w-0.03*\w,0) rectangle (\l*\w,0.65*\h);
\end{scope}

\draw[dashed] (\l*\w,-\os) node[below]{\vphantom{$r$}$\ell$} -- (\l*\w,\h);
\draw[dashed] (\r*\w,-\os) node[below]{\vphantom{$\ell$}$r$} -- (\r*\w,\h);
\draw[dashed] (2*\r*\w -\l*\w,-\os) node[below]{$r + (r-\ell)$} -- (2*\r*\w -\l*\w,\h);

\draw[dashed] (-\os,0.43*\h) -- (1.59*\w+\os,0.43*\h) ;

\end{scope}

\end{tikzpicture}
    }
    \caption{One \textbf{while}-loop iteration of the Stretching Algorithm~\ref{alg_stretching} with the original (left) and the stretched packing (center), and the height bound (right). Removed items (green) are added to~$\items_S$. For visualization we also shift the tall item $T$ even though Algorithm~\ref{alg_stretching} does not pack them.}
    \label{fig:stretching}
\end{figure}

The Stretching Algorithm is best understood as repacking the items with height at most $H$ (for some given $H \geq \frac{h(\sigma)}2$) within some segment of a packing $\start$. 
More concretely, assume $[\tau_{\min},\tau_{\max}]$ is a segment of a packing $\sigma$ and $\items' \subseteq \items$ is a subset of the items that overlap with this segment, i.e., $[\start(i), \start(i) + w(i)) \cap [\tau_{\min}, \tau_{\max}] \neq \emptyset$ for each $i \in \items'$. 
Let $\mathcal T$ be the set of items in $\items'$ with $h(i) > H$. 
When restricted to $\items' \setminus \mathcal T$, i.e., to the items with height at most $H$, the packing $\start$ has a height of at most $h(\sigma)-H$, \emph{except} in places where no item from $\mathcal{T}$ was packed. 
Suppose that the total width in $[\tau_{\min},\tau_{\max}]$ where no item from $\mathcal T$ is packed is given by $d$. 
By removing some set of items $\items_S$ with small total area and by \emph{stretching} the remaining packing, a packing of the items $\items' \setminus \mathcal{T}$ with a slightly larger width (extended by $d$) is obtained that has a height of at most $h(\sigma)-H$. 

Informally, the algorithm inductively considers segments (\emph{gaps}) between two consecutively packed items with height greater than~$H$.
When it encounters such a gap $[\ell, r)$, the algorithm removes all items contained in the segment $[\ell, r)$ and increases the starting time of all items that start after~$\ell$ by the width of the gap. 
Hence, the algorithm quite figuratively ``stretches'' the packing to the right as can be seen in \cref{fig:stretching}.

\begin{algorithm}
\DontPrintSemicolon
\caption{Right-Stretching Algorithm}
\label{alg_stretching}\label{alg_right_stretching}
\textbf{Input:}  $(\start,H,\tau_{\min},\tau_{\max})$\;
\smallskip
$\mathcal{T} \leftarrow \items_{h > H}$ \tcp*[r]{tall items}
$G \leftarrow$ set of maximal segments $[\ell,r) \subseteq [\tau_{\min},\tau_{\max}]$ where $\start$ packs no item from $\mathcal{T}$\;
$\items_S \leftarrow \{i \in \items \mid [\sigma(i),\sigma(i)+w(i)) \text{ is a subsegment of an element in $G$}\}$ \tcp*[r]{removed items}
$\items' \leftarrow \items^{\start}_{h \leq H, [\start(i), \start(i) + w(i)) \cap [\tau_{\min}, \tau_{\max}] \neq \emptyset}$
\tcp*[r]{items overlapping $[\tau_{\min}, \tau_{\max}]$}
\For{$i \in\items' \setminus \items_S$}{
  $\sigma'(i) \leftarrow \sigma(i)$\;
}
\For{$(\ell,r) \in G$}{
  \For{$i \in \items'_{\sigma(i) \geq \ell} \setminus \items_S$}{
    $\sigma'(i) \leftarrow \sigma'(i) + (r-\ell)$\tcp*[r]{shift items starting after $\ell$ by the gap width $(r-\ell)$}
  }
}
\textbf{return} $(\sigma',\items_S)$ \;
\end{algorithm}

\begin{lemma}[Right-Stretching Lemma]
     \label{lem:stretching}\label{lem:right_stretching}
     Let $\sigma$ a packing of a set $\items$ of items and $H$ a number with $\frac{h(\sigma)}2 \leq H \leq h(\sigma)$. 
     Let $\mathcal{T} = \items_{h > H}$, and let $[\tau_{\min},\tau_{\max}]$ be a segment of $\sigma$ where an item from $\mathcal{T}$ ends at $\tau_{\min}$ and an item from $\mathcal{T}$ starts at $\tau_{\max}$. 
     Let $\items' = \items^{\start}_{h \leq H, [\start(i), \start(i) + w(i)) \cap [\tau_{\min}, \tau_{\max}] \neq \emptyset} $
     and let $d$ be the total width in $[\tau_{\min},\tau_{\max}]$ where $\sigma$ packs no item from $\mathcal{T}$.

    Then, there exists a set $\items_S \subseteq \items'$ with $\area(\items_S) \leq d\cdot h(\sigma)$ such that there exists a packing $\sigma'$ of $\items' \setminus \items_S$ with $h(\sigma') \leq h(\sigma) - H$ and $\sigma(i) \leq \sigma'(i) \leq \sigma(i)+d$ for each $i \in \items' \setminus \items_S$. 
\end{lemma}

\begin{proof}
   \Cref{alg_right_stretching} defines $\items_S$ as the items contained within gaps between items in $\mathcal{T}$. 
   By definition of $d$, we have $\area(\items_S) \leq d\cdot h(\sigma)$.
   Since $\sigma'$ arises from $\sigma$ by removing items and shifting items to the right by at most $d$, i.e., the sum of all gaps, it holds that $\sigma(i) \leq \sigma'(i) \leq \sigma(i)+d$ for $i \in \items'\setminus \items_S$. 

   Now consider the height of $\sigma'$. 
   If the items $\mathcal{T}$ are removed from the packing $\sigma$, we already obtain a packing $\sigma''$ that has height at most $h(\sigma) -H$ at all time points where an item from $\mathcal{T}$ is packed in $\sigma$. 
   However, the height of $\sigma''$ may exceed $h(\sigma) -H$ within a gap, either because the combined height of the items overlapping one of the end points is greater than $h(\sigma) - H$ or because of the items completely contained in the gap. 
   We obtain $\sigma'$ by fixing $\sigma''$ one gap at a time. 

   Consider a gap $[\ell,r)$ between two items $T_1, T_2 \in \mathcal T$. 
   Refer to \cref{fig:stretching} for an illustration. 
   First, remove all items packed completely in the gap (shown in green); \Cref{alg_right_stretching} adds such items to the set $\items_S$. 
   Then, shift all items starting at $\ell$ or later to the right by $r-\ell$. 
   Hence, no item starts in $[\ell,r)$ anymore. 
   In particular, the segment $[\ell,r)$ is only used by items starting earlier than $\ell$ and ending after $\ell$, which are the (blue) items that $\start$ packs in parallel to $T_1$. 
   Their height is at most $h(\sigma)-h(T_1) \leq h(\sigma) - H$ by definition of $\mathcal T$. 
   Similarly, the segment $[r,r+(r-\ell))$ is only used by items that were packed in parallel with $T_2$ (which includes all red items and potentially some of the blue items). 
   Thus, the height is at most $h(\sigma) - H$ in $[r,r+(r-\ell))$ as well.

   \Cref{alg_right_stretching} is obtained by applying such a fix to each of the gaps, summing up the effects. 
   It remains to argue that these fixes do not interact in a problematic way, i.e., applying two or more fixes does not increase the height. 
   To this end, we observe that when fixing a gap $[\ell,r)$, all affected items, i.e., those originally (in $\start$) starting at or after~$\ell$ will move to the right by $r-\ell$ in unison, which cannot cause the height to increase if the height of the non-moving items is non-increasing after $\ell$.
   To this end, we observe that the only items not moving are those that start earlier than $\ell$ in~$\start$ (pictured in blue): 
   Hence, after $\ell$, their combined height can only decrease. 
\end{proof}

By mirroring the packing~$\start$ before stretching the packing and then mirroring the output again, we obtain the Left-Stretching Algorithm, given as \Cref{alg_left_stretching} in \Cref{app:packing_algorithms}.
We conclude with two observations about the set~$\items_S$ of not-packed items returned by \Cref{alg_stretching} or \Cref{alg_left_stretching}.

\begin{corollary}\label{cor:stretching_without_narrow_items}
    If~$\items_{w(i) \leq d } = \emptyset$, then~$\items_S = \emptyset$.
\end{corollary}

Using the bounds on~$\items_S$ by \Cref{lem:stretching} and applying \steinberg$(\items_S, H)$, the next result follows.

\begin{corollary}\label{cor:steinberg_for_stretching}
    There is a packing of~$\items_S$ with height at most~$H$ and width at most $2 \frac {d h(\start)}{H}$.
\end{corollary}

\subsection{Squeezing Lemma}
\label{sec:squeezing}

In this section, we introduce the Squeezing Algorithm that allows us to \emph{squeeze in} items with height at most~$\frac H2$ and width at most~$\frac{\eps}{1+\eps}\D$ into a packing~$\start$ of height at most~$\big(\frac32 + \eps\big)H$ without exceeding the bound of~$\big(\frac32 + \eps\big)H$ on the height anywhere. 
We note that Lemma 12 in \cite{Galvez0AK21} provides a similar result. 
For the sake of completeness and since their upper bound on the width of items that can be added is slightly worse than ours, we still give the Squeezing Algorithm here.

Informally, the Squeezing Algorithm shifts non-$H$-tall items, i.e., those of height at most~$\frac H2$, to the left as long as such a shift does not increase the packing height beyond $\big(\frac 32 + \eps\big)H$. 
Using a volume bound, we then argue that after those shifts the height in~$\big[\big(1-\frac{\eps}{1+\eps}\big)\D, \D\big]$ is at most~$(1+\eps)H$, which allows us to \emph{squeeze in} items of width at most~$\frac{\eps}{1+\eps}\D$ and height at most~$\frac H2$; such items are called \emph{$(\eps,H)$-squeezable}.
Let $\items_{\mathrm{sq},\eps,H} = \items_{w\leq\frac{\eps}{1+\eps}\D,h<\frac H2}$.
If $\eps$ and $H$ are clear from the context, we call those items squeezable and denote them by $\items_{\mathrm{sq}}$.

\begin{figure}
\begin{minipage}[t]{0.49\textwidth}
\begin{algorithm}[H]
\DontPrintSemicolon
\caption{Squeezing Algorithm}
\label{alg:squeezing}
\textbf{input:} $(\start, H, \eps)$ \tcp*[r]{$\start$ is $(\eps,H)$-neat} 
\smallskip 
$\tau \leftarrow 0$ \; 
\While{true}{
    $\tau \leftarrow \min\{t \geq \tau:  h(\itemsT{\start}(t)) \leq (1+\eps)H \} $\;
    $i \leftarrow \arg\min\{\start(i) : i \in \items_{h \leq \frac H2}\cap\itemsT{\start}_{(\tau,\D]}\}$ \; 
    \If{$i$ undefined}{
        \textbf{break}\;
    }
    $\start(i) \leftarrow \tau$ \tcp*[r]{start $i$ at $\tau$}
}
\textbf{return} $\start, \tau$ \;
\end{algorithm}    
\end{minipage}~
\begin{minipage}[t]{0.48\textwidth}
\begin{algorithm}[H]
\DontPrintSemicolon
\caption{Iterated Squeezing}
\label{alg:iterated-squeezing}
\textbf{input:} $(\start, H, \eps, \mathcal{S})$\;
\tcp{$\start$ is $(\eps,H)$-neat packing of $\items \setminus \mathcal{S}$} 
\smallskip 
\For{$i \in \mathcal{S}$}{
    $(\start,\tau) \leftarrow $ SqueezingAlgorithm$(\start,H,\eps)$\;
    $\start(i) \leftarrow \tau$\;
}
\textbf{return} $\start$ \;
\end{algorithm}    
\end{minipage}
\end{figure}

\begin{lemma}[Squeezing Lemma]
\label{lem:squeezing}
Let $H \geq \max\big\{ \frac{\area(\items)}{D}, \opt(\items,\D) \big\}$ and let $\start$ be an $(\eps,H)$-neat packing of $\items \setminus \items_{\mathrm{sq},\eps,H}$. 
\Cref{alg:iterated-squeezing} on input $(\start,H,\eps, \items_{\mathrm{sq},\eps,H})$ computes an $(\eps,H)$-neat \mbox{packing $\sigma'$ of~$\items$.}
\end{lemma}

\begin{proof}
We first show by induction that \cref{alg:squeezing} never increases the height of the packing over $\big(\frac32 +\eps\big)H$: 
Consider a \textbf{while}-loop iteration beginning with a packing $\start$ and value $\tau$ where the start point of item $i$ is reduced from $\start(i)$ to $\tau$. The packing height can only increase within $[\tau,\start(i))$.
By definition of~$i$, the only items starting in $[\tau, \start(i))$ are $H$-tall items, whose height is non-increasing in~$t$.
Hence, before shifting~$i$,~$\start$ had a height of at most $(1+\eps)H$ for $t \in [\tau,\sigma(i))$ by definition of $\tau$. 
Further, $h(i) \leq \frac H2$. 
Hence, after repacking~$i$, the height of~$\start$ is still at most $\big(\frac32+\eps\big)H$.

Next we show that we can insert an item from $\items_{\mathrm{sq},\eps,H}$ at $\tau$, where $(\start,\tau)$ is returned by \cref{alg:squeezing}.
Note that $\tau$ is the first point in time where the packing height is at most $(1+\eps)H$.
Since the total area of the items is bounded by $\area(\items)\leq \D\cdot H$, it has to hold that $\tau \leq \frac{\area(\items)}{(1+\eps)H} \leq \frac{1}{1+\eps} \D = \big(1-\frac{\eps}{1+\eps}\big)\D$.
By construction, in~$\start$, no non-$H$-tall item starts in~$[\tau,\D]$ as otherwise \cref{alg:squeezing} would have decreased the starting time of such an item to~$\tau$. 
As the height of the $H$-tall items is non-increasing in~$t$, this implies that the height of~$\start$ is non-increasing in~$[\tau,\D]$. 
Hence, we can indeed place an extra item of height $\frac H2$ and width $\frac{\eps}{1+\eps}\D$ in the segment $\big[\big(1-\frac{\eps}{1+\eps}\big)\D, \D\big]$.
Employing binary search trees, this algorithm runs in time $\mathcal{O} (n \log n)$.
\end{proof}

\section{Paper overview}

In this section, we give a brief overview of the paper, our results, algorithms, and proof strategy.
Key to our $\big(\frac{3}{2} + \eps\big)$-approximation algorithm for DSP is the structural result, \Cref{thm:restructuring}.
Before giving more details on its proof, we briefly explain how to use it to obtain the polynomial-time $\big(\frac{3}{2} + \eps\big)$-approximation algorithm, which is given in full detail in \Cref{sec:algorithm}.

\subsection{Algorithm outline}
\Cref{thm:restructuring} suggests to design two algorithms: one algorithm for the case that there is a 
$\lambda$-forgiving packing and another algorithm for the case that there is an
$(\eps,\opt)$-neat packing.

If the given instance admits a $\lambda$-forgiving packing, we use an algorithm by \cite{DBLP:journals/algorithmica/DeppertJKRT23}, which essentially packs \emph{almost all} items into a packing of width $\D$ and height at most $(1 + c \eps') \OPT$ and the remaining items that are not packed fit into a box of width $\eps' \D$ and height $(1 + c \eps') \OPT$
for some constant~$c > 0$.
Using that the instance can accommodate an extra item of height $\OPT$ and width $\lambda \D$ without exceeding the height of $\big(\frac{3}{2} + \eps\big) \OPT$, we can employ the above algorithm, where the items that are packed into the box can be viewed as the extra item $i_\lambda$ (\Cref{lem:algorithm_if_extra_item}).
This allows us to find the desired packing of all items without exceeding the height of
$\big(\frac 32 + \eps\big) \OPT$.

If the given instance admits an $(\eps,\opt)$-neat packing
we design a new algorithm, which re-uses some ideas from~\cite{Galvez0AK21}.
Let $\bar{\sigma}$ be the $(\eps,\opt)$-neat packing.
We partition the set of items into certain subsets based on their width and height and show that we can restructure $\bar{\sigma}$ which allows us to efficiently guess the position of most items.
The remaining items are $(\eps,\opt)$-squeezable. 
This is summarized in \Cref{lem:algorithm_if_tall_sorted}, and its proof is deferred to \Cref{sec:proof-of-alg-if-tall-sorted}.

Of course, we do neither know which of the two cases applies to the given instance nor do we know the value of $\opt$. 
By simply running both algorithms and embedding these algorithms into a binary search on the value of $\OPT$, we obtain the desired result. 
Here, we use that any $(\eps,H)$-neat packing is also $(\eps,H')$-neat for $H'\geq H$.
The full algorithm is given in \Cref{alg_three_half} in \cref{sec:algorithm}.

\subsection{Outline of the proof of Theorem \ref{thm:restructuring}}
The main part of the paper is dedicated to proving \Cref{thm:restructuring}.
Throughout the paper, we consider some optimal packing $\sopt$ of height $\OPT$ and consider the position of tall items, i.e., items of height greater than $\frac\opt2$. 
Since $\eps$, $\lambda$, and $\opt$ are fixed during the proof of \Cref{thm:restructuring}, we use neat and forgiving to refer to $(\eps,\opt)$-neat and $\lambda$-forgiving, respectively.
The proof is a case distinction based on the widths of different gaps in $\sopt$. Hence, as a corner case, we have to deal with instances where the total width of tall items is \emph{close to} $\D$:
\begin{enumerate}
    \item The total width of tall items is at least $\big(1 - \frac{\eps}{5 + 4 \eps}\big) \D \approx \D$. We show that $(\items,\D)$ admits a neat packing in \Cref{sec:case_a}. 
\end{enumerate}
In the other extreme, i.e., for instances without tall items, every optimal solution~$\sopt$ is neat. Hence, for every remaining instance, every optimal solution admits significant gaps (w.r.t. total width).
\begin{enumerate}
    \setcounter{enumi}{1}
   \item There is a sequence of consecutive gaps with a combined width between $\lambda \D$ and $2 \lambda \D$. We show that 
    $(\items,\D)$ admits a forgiving packing in \Cref{sec:case_b} by ``fusing'' the gaps.
\end{enumerate}
This case lays the foundation for the remaining cases: 
From now on, we can divide the segment $[0,\D]$ into sub-segments where either the tall items are packed very densely (creating gaps of total width at most $\bigO(\lambda\D)$ and somewhat making it easier to sort the tall items and obtain a neat packing) or where there are already non-narrow gaps of width at least $\lambda\D$ (implying a forgiving packing if we can clear one such gap). 
The remaining cases distinguish the width of those gaps and whether the corresponding gap is cleared (leading to a forgiving packing) or the tall items around this gap are sorted (leading to a neat packing):
\begin{enumerate}
    \setcounter{enumi}{2}
    \item There is a gap of width at least $\lambda \D$ and at most $\big(\frac 12 - 3 \lambda\big) \D$. We show that $(\items,\D)$ admits a forgiving packing in \Cref{subsec:between_lambda_and_1/2}.
    \item There is exactly one \emph{wide} gap of width at least $\big(\frac 12 - 3 \lambda\big) \D$ and the total width of the other gaps is \emph{almost $0$}. 
    We show that $(\items,\D)$ admits a neat packing in \Cref{sec:one-wide-gap}.
    \item There are exactly two wide gaps of width at least $\big(\frac 12 - 3 \lambda\big) \D$ each. 
    We show that $(\items,\D)$ admits a neat packing in \Cref{sec:two-large-gaps}.
\end{enumerate}

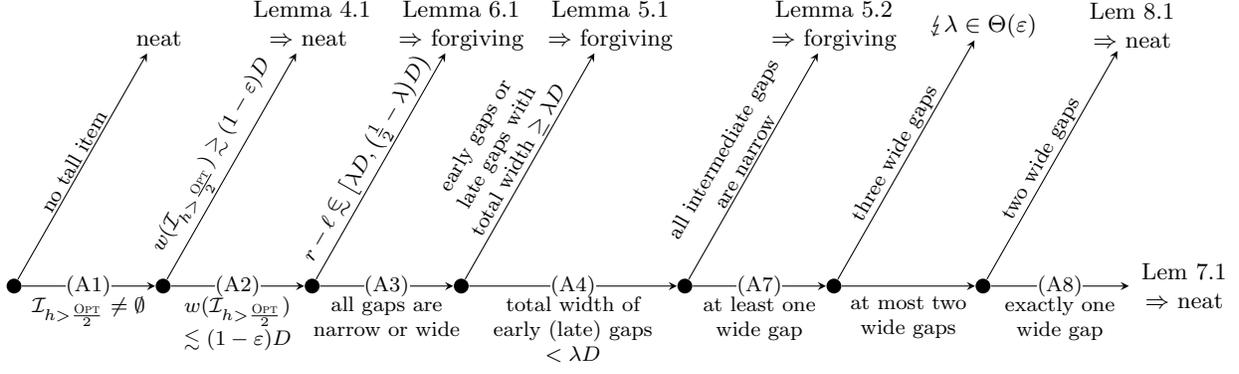
\begin{figure}
\centering
\resizebox{\textwidth}{!}{
\begin{tikzpicture}
\tikzstyle{alongedge}=[midway, sloped, anchor=south, inner sep = 1pt, font = \scriptsize]
\tikzstyle{addassumption} = [midway, sloped, anchor=center, fill=white, font = \scriptsize, inner sep = 0pt]
\tikzstyle{assumption} = [midway, sloped, anchor=north, font = \scriptsize]
\tikzstyle{instance}=[font = \footnotesize, inner sep = 1pt, align = center]
\tikzstyle{dot}=[circle, fill = black, inner sep = 2pt]

\def \hi {3.5}
\node[instance] at (2,\hi) (i1) {\vphantom{lem 4.1} \\ neat \vphantom{g}}; 
\node[instance] at (4,\hi) (i2) {\cref{lem:tall_items_are_wide} \\ $\Rightarrow$ neat \vphantom{g}};
\node[instance] at (6,\hi) (i5) {\cref{lem:full-case-C} \\ $\Rightarrow$ forgiving}; 
\node[instance] at (8,\hi) (i6top) {\cref{lem:small-gap-at-border} \\ $\Rightarrow$ forgiving};
\node[instance] at (11,\hi) (i7) {\cref{lem:small-gap-center} \\ $\Rightarrow$ forgiving};
\node[instance] at (15,\hi) (itwogaps) {Lem \ref{lem:two_large_gaps_G_3_small} \\ $\Rightarrow$ neat};
\node[instance] at (13,\hi) (ithreegaps) { $\lightning \lambda \in \Theta(\eps)$};
\node[instance] at (15.7,0) (ionegap2) {Lem \ref{lem:one-wide-gap} \\ $\Rightarrow$ neat};

\foreach \x/\a in {0/0, 2/1, 4/2, 6/5, 9/55, 11/8, 13/9}{
    \node[dot] at (\x,0) (a\a) {};
}
\node[instance] at (15,0) (aonegap) {};

\draw[-stealth] (a0) to node[alongedge] {no tall item} (i1); 
\draw[-stealth]  (a1) to node[alongedge] {$w(\items_{h > \frac\opt2}) \gtrsim (1-\eps) \D$} (i2); 
\draw[-stealth]  (a2) to node[alongedge, align = center] {$ r-\ell \insim \big[\lambda \D, \big(\frac12-\lambda\big)\D \big)$} (i5); 
\draw[-stealth]  (a5) to node[alongedge, align = center] {early gaps or \\ late gaps with \\ total width $\geq \lambda\D$}
(i6top); 
\draw[-stealth]  (a55) to node[alongedge, align = center] {all intermediate gaps \\ are narrow}
(i7); 
\draw[-stealth]  (a8) to node[alongedge, align = center] {three wide gaps} (ithreegaps); 
\draw[-stealth]  (a9) to node[alongedge, align = center] {two wide gaps} (itwogaps) ;

\draw[-stealth] (a0) to 
    node[addassumption] {\eqref{eq:at-least-one-tall}} 
    node[assumption] {$\items_{h > \frac\opt2} \neq \emptyset$} 
(a1);
\draw[-stealth] (a1) to 
    node[addassumption] {\eqref{eq:width-tall-upper-bounded}} 
    node[assumption, align = center] {$w(\items_{h > \frac\opt2})$ \\ $  \lesssim (1-\eps) \D$} 
(a2);
\draw[-stealth] (a2) to 
    node[addassumption] {\eqref{eq:distance-two-tall}}
    node[assumption, align = center] {all gaps are \\ narrow or wide }
(a5);
\draw[-stealth] (a5) to 
    node[addassumption] {\eqref{eq:early-and-late-gaps}}
    node[assumption, align = center] {total width of \\ early (late) gaps \\ $< \lambda \D$}
(a55);
\draw[-stealth] (a55) to 
    node[addassumption] {\eqref{eq:gap_at_least_one_wide_gap}}
    node[assumption, align = center] {at least one \\ wide gap }
(a8);
\draw[-stealth] (a8) to 
    node[assumption, align = center] {at most two \\ wide gaps}
(a9);
\draw[-stealth] (a9) to
    node[addassumption] {\eqref{eq:exactly_one_wide_gap}}
    node[assumption, align = center] {exactly one \\ wide gap} 
(aonegap);

\end{tikzpicture}
}
\caption{Proof sketch for \cref{thm:restructuring}. The time points $\ell,r \in [0,\D]$ refer to the gap $[\ell,r)$ of interest. Further, the symbols $\lesssim, \gtrsim,$ and $\insim$ are used to simplify notation; they imply that the right-hand side is only approximately the necessary condition.}
\label{fig:structural-lemma-decision-tree}
\end{figure}

Finally, in \Cref{sec:proof-thm1} we give the proof of \Cref{thm:restructuring}, i.e., we show that
any optimal solution $\sopt$ must satisfy one of the above five cases, for which we have shown the existence of either a neat or a forgiving packing. 
\Cref{fig:structural-lemma-decision-tree} in \Cref{sec:proof-thm1} gives an overview of this proof.

We remark that in all sections we provide figures that illustrate the algorithms and give an intuition for the proof.
Typically, we divide the interval $[0, \D]$ into smaller segments for which we then separately bound the height.
Therein, we usually consider certain subsets of items and show that some of these subsets have a combined height of, e.g., $\OPT$ or $\frac \OPT 2$.
This is also highlighted in the respective figures, where item sets that have a combined height of, say at most $\OPT$ in some segment, are also packed between $0$ and $\OPT$.

The next five paragraphs give an overview of these algorithms used for restructuring the solution and give some intuition on the correctness.
We remark that in those three cases in which we guarantee the existence of a neat packing, we do not have to consider squeezable items as we can add these afterwards using the Squeezing Lemma~\ref{lem:squeezing}.

\paragraph*{(i) The width of the tall items is almost $\D$.}
In this case, we create a neat packing, starting by placing the tall items as required. 
Further, we define \emph{medium} items, which are non-tall items of height at least $\frac{1}{4} \OPT$ and \emph{wide} items, which are items which are neither tall nor medium and have a width of at least $\big(\frac 12 + 2 \cdot \frac{\eps}{5 + 4 \eps}\big) \D$.
We sort the medium items by non-increasing height and start those at $0$ (with one exception) and let the wide items end at $\D$, creating two intricately aligned \emph{L-packings} of height at most $\frac32\opt$.
After pushing the wide items iteratively (in decreasing order of width) as far to the left as possible without exceeding a height of $\big(\frac 32 + \eps\big) \OPT$, we iteratively take not yet packed items and pack them as early as possible without exceeding the desired height.

By construction, the maximum height is at most $\big(\frac 32 + \eps\big) \OPT$, and we show that the resulting packing is $\D$-feasible arguing about the total area (\Cref{lem:caseA-main}).
This case is slightly different than the other cases as we do not restructure an optimal solution but construct a neat packing of height at most $\big(\frac 32  + \eps\big)$ from scratch.

\paragraph*{(ii) Fusing gaps.} 
We treat the cases where the sequence of gaps occurs on the very left, the very right or \emph{close} to the center $\frac 12 \D$.
Here, we only describe the first case, where a tall item $T$ begins at time point $\ell$ and the gaps to the left of $\ell$ have a combined width in $[\lambda D,2\lambda D]$.
We define $I_{\leq \ell}$ to be the set of non-tall items that finish before $\ell$. 
Intuitively, we first shift all tall items that start before $\ell$ in the optimal packing $\sopt$ to the left, such that the first tall item starts at $0$, there is now a gap (ending at $\ell$) and the order of the tall items w.r.t.\ the starting times is the same as in $\sopt$.
We then apply the Stretching Lemma~\ref{lem:stretching} to $I_{\leq \ell}$ and pack the resulting two sets on top of the packing, one starting at $0$ and the other finishing at $\D$.
The remaining items ($I_{\ell, \D}$) are packed as in~$\sopt$.

Observe that the stretched packings (packed on top) have a height of at most $\frac12 \opt$.
Recall that up to time $\ell$ the total width of the gaps is in $[\lambda D,2\lambda D]$.
Hence, we can show that we can fit an extra item $i_\lambda$ of height $\OPT$ and width $\lambda \D$ into the gap ending at~$\ell$, such that the height bound of $\big(\frac 32 + \eps\big) \OPT$ is never exceeded (Lemmas~\ref{lem:small-gap-at-border} and~\ref{lem:small-gap-center}).

\paragraph*{(iii) A gap of width at least $\lambda \D$ and at most $\big(\frac 12 - 3 \lambda\big) \D$.}
Let~$[\ell,r)$ be the gap considered. 
Our goal is to remove some items that are completely packed within the gap and that have a combined height of at least $\frac \opt2$ in a segment of width at least $\lambda \D$ and pack them on top of the packing to the left or to the right.
To this end, we distinguish several cases (based on the position of the segment we want to clear) in which we remove different item sets.
Again depending on the case, these removed items are packed greedily and / or using Steinberg's algorithm while ensuring that this repacking does not increase the height beyond $\frac32\opt$.
By removing items of total height at least $\frac \opt2$ in a segment of width at least $\lambda \D$, we can pack the extra item $\iadd$ of height $\OPT$ and width~$\lambda \D$ in this segment without exceeding the maximum height of $\frac32 \OPT$ anywhere (\Cref{lem:medium-wide-gap}).

\paragraph*{(iv) Exactly one gap of width at least $\big(\frac 12 - 3 \lambda\big) \D$.}
Let~$[\ell,r)$ be the gap considered. 
We develop three different algorithms depending on the values of $\ell$ and $r$.
Here, we describe the algorithm and the proof idea for the most general case (see \Cref{lem:repacking_large-gap_ell-large}), in which we additionally require $\frac{\eps}{1+ \eps} \leq \ell \leq D - r$. Note that $\ell \leq \D - r$ holds w.l.o.g.\ as we can simply mirror the packing.

We consider some packing $\sopt$ of height at most $\OPT$ and give an algorithm that restructures this packing to a neat packing $\start$.
To this end, we define four different sets of non-tall items according to their positions in $\sopt$: $I_{\leq \ell}$ is the set of non-tall items ending before~$\ell$,~$I_{\geq r}$ is the set of non-tall items starting after~$r$, $I_{\ell, r}$ is the set of non-tall items not in $I_{\leq \ell}$ that end before $r$ or shortly after, and~$I_r$ is the set of items overlapping time point~$r$ and some time point shortly after~$r$.
The packing~$\start$ packs the tall items as desired and shifts the items in $I_{\ell, r}$ such that they end at $\D$.
The algorithm then shifts the items $I_r$ to the right such that they essentially end at $\D$. 
Finally, we stretch $I_{\leq \ell}$ and $I_{\geq r}$ to the right and left, respectively, and place them similarly to their previous positions in~$\sopt$.

To prove the desired bound on the height, we highly rely on the Stretching Lemma~\ref{lem:stretching}: It allows us to essentially use the structure of the original packing (i.e., in $\sopt$) of $I_{\leq \ell}$ and $I_{\geq r}$, which is crucial to bounding the height for all time points overlapped by items from these sets. 
Furthermore, the width of the stretched packings of $I_{\leq \ell}$ and $I_{\ell, r}$ is not much larger than the width of the non-stretched packings of $I_{\leq \ell}$ and $I_{\ell, r}$ by the second assumption of this case, which we crucially exploit in the height analysis for the segments where the stretched sets are packed, but not the original ones. 
Here, we additionally use that we do not consider squeezable items which allows us to argue that the height of the stretched packings is monotone in those segments. 

\paragraph*{(v) Exactly two gaps of width at least $\big(\frac 12 - 3\lambda\big) \D$.}
Let $[\leftl, \leftr)$ and $[\rightl, \rightr)$ be the two gaps considered. 
When choosing $\lambda \in \Theta(\eps)$ small enough, we can ignore the non-tall items packed in $[0, \leftl]$, $[\leftr, \rightl]$, and $[\rightr, \D]$ as these items are squeezable. 
We divide the other non-tall items into four different item sets: $G_L$ is the set of items ending before~$\rightl$, $G_{1, R}$ is the set of items overlapping~$\rightr$, $G_{2, R}$ is the set of items not in $G_{1, R}$ and overlapping~$\rightl$, and~$G_{3, R}$ is the set of items completely packed in the interval $[\rightl, \rightr]$ in~$\sopt$.
Our restructuring algorithm for this case packs the tall items as desired, shifts $G_{3, R}$ in unison to the right such that its last items ends at $\D$ and shifts $G_L$ slightly to the right such that the tall items and $G_L$ do not exceed the height of $\OPT$.
Furthermore, we pack all items in $G_{1, R}$ such that they end at $\D$ and pack all items in $G_{2, R}$ such that they start at~$0$.
Again, by heavily exploiting the Stretching Lemma~\ref{lem:stretching} and the Squeezing Lemma~\ref{lem:squeezing} as in the previous case, we obtain the desired bound on the height of this restructured packing (\Cref{lem:two_large_gaps_G_3_small}).

\section{The total width of \texorpdfstring{$\opt$}{OPT}-tall items is at least \texorpdfstring{$(1 - \frac{\eps}{5 + 4 \eps}) \D$}{(1-ε/(5+4ε))D}}\label{sec:case_a}

\begin{figure}
    \centering
    \resizebox{\textwidth}{!}{
    \begin{tikzpicture}
\pgfmathsetmacro{\w}{6}
\pgfmathsetmacro{\h}{3}
\pgfmathsetmacro{\os}{0.24}
\pgfmathsetmacro{\l}{0.35}
\pgfmathsetmacro{\r}{0.7}
\pgfmathsetmacro{\lamb}{0.04}

\begin{scope}[xshift = - 1.3*\w cm]
\draw (0,\h) -- (0,0) -- (\w,0) -- (\w,\h);
\draw (0*\w,-\os) node[below]{$0$}-- (0*\w,\h);
\draw (\w,-\os) node[below]{$\D$}-- (\w,\h);

\fill[colord] (0,0) rectangle (\w,\h);

\draw[fill=mediumItemColor] (0, .6*\h) rectangle (.18*\w, \h); 

\foreach \x in {.01,.03, 0.05, .08, .12, .15}{
    \draw (\x*\w, .6*\h) -- (\x*\w, \h);
}

\draw[fill = mediumItemColor] (.95*\w, 0) rectangle (\w, .85*\h);
\draw (.95*\w, .41*\h) -- (.97*\w, .41*\h);
\draw (.975*\w, .85*\h) -- (.975*\w, .41*\h);

\def \tallx {0}
\def \totalgap {0}

\foreach \processingtime/\demand/\gap [count = \i] in {
    .02/.92/0, .03/.56/0, .04/.7/0, .023/.768/0, .03/.64/0,
    .035/.67/.03, .045/.59/0, .04/.6/.02, .045/.54/0, .03/.6/0, 
    .023/.55/0, .015/.82/0, .02/.79/.01, .021/.72/0, .04/.67/.005, 
    .02/.65/0, .034/.73/0, .026/.79/0, .041/.78/0, .021/.66/0, 
    .015/.53/0, .017/.71/0, .028/.82/0, .037/.73/0, .032/.84/0, 
    .02/.82/0, .04/.7/0, .023/.76/0, .026/.69/0.005, .043/.7/.02, 
    .02/.51/0
}
{    
    \draw[fill = tallItemColor] (\tallx*\w, 0*\h) rectangle (\tallx*\w + \processingtime*\w, \demand*\h); 

    
    \pgfmathparse{\tallx + \processingtime + \gap}
    \global\let\tallx\pgfmathresult
    \pgfmathparse{\totalgap + \gap}
    \global\let\totalgap\pgfmathresult
}


\draw[fill = mediumItemColor] (.2*\w, .61*\h) rectangle (.4*\w, .87*\h);

\def \widey {0}
\def \wideyatend {1}

\foreach \processingtime/\demand in {.9/.02, .85/.04, .59/.02}
{   
    \draw[fill = coloraa] (\w - \processingtime*\w, \wideyatend*\h - \demand*2*\h) rectangle (\w, \wideyatend*\h);
    
    \pgfmathparse{\widey + \demand}
    \global\let\widey\pgfmathresult

    \pgfmathparse{\wideyatend - 2*\demand}
    \global\let\wideyatend\pgfmathresult
}

\draw[fill = mediumItemColor] (.178*\w, 0*\h) rectangle (.208*\w, .34*\h);
\draw[fill = mediumItemColor] (.178*\w, .34*\h) rectangle (.208*\w, .6*\h);
\draw[fill = mediumItemColor] (.293*\w, 0*\h) rectangle (.313*\w, .34*\h);
\draw[fill = mediumItemColor] (.293*\w, .34*\h) rectangle (.313*\w, .6*\h);

\draw[dashed] (-\os,0.75*\h) node[left]{$\frac{3\opt}{4}$} -- (\w+\os,0.75*\h) ;
\draw[dashed] (-\os,0.5*\h) node[left]{$\frac{\opt}{2}$} -- (\w+\os,0.5*\h) ;
\draw[dashed] (-\os,1*\h) node[left]{$\opt$} -- (\w+\os,1*\h) ;
\end{scope}


\draw (0,1.5*\h) -- (0,0) -- (\w,0) -- (\w,1.5*\h);
\draw (0*\w,-\os) node[below]{$0$}-- (0*\w,\h);
\draw (\w,-\os) node[below]{$\D$}-- (\w,\h);

\def \tallx {0}
\def \totalgap {0}

\foreach \processingtime/\demand/\gap in {.02/.92/0, .032/.84/0, .02/.82/0, .015/.82/0, .028/.82/0, .02/.79/.01, .026/.79/0, .041/.78/0, .023/.768/.01, .023/.76/0, .037/.73/0, .034/.73/0, .021/.72/0,  .017/.71/0, .04/.7/0, .04/.7/0,   .043/.7/.02, .026/.69/0.005, .035/.67/.02, .04/.67/.005, .021/.66/0, .02/.65/0, .03/.64/0, .04/.6/.02,  .03/.6/0, .045/.59/0, .03/.56/0, .023/.55/0, .045/.54/0,  .015/.53/0, .02/.51/0 }
{    
    \draw[fill = tallItemColor] (\tallx*\w, 0*\h) rectangle (\tallx*\w + \processingtime*\w, \demand*\h); 
    \pgfmathparse{\tallx + \processingtime}
    \global\let\tallx\pgfmathresult

    \pgfmathparse{\totalgap + \gap}
    \global\let\totalgap\pgfmathresult
}

\def \mediumx {0}

\foreach \processingtime/\demand/\gap in {.02/.89/0, .032/.84/0, .02/.82/0, .028/.82/0, .02/.79/.01, .026/.79/0, .041/.78/0, .023/.768/.01, .023/.76/0, .021/.72/0,  .017/.71/0, .024/.7/0,   .035/.67/.02, .04/.67/.005, .021/.66/0, .02/.65/0, .03/.64/0, .04/.6/.02,  .03/.6/0}
{    
    \draw[fill = mediumItemColor] (\mediumx*\w, 1.5*\h) rectangle (\mediumx*\w + \processingtime*\w, 1.5*\h-\demand/2.4*\h); 
    \pgfmathparse{\mediumx + \processingtime}
    \global\let\mediumx\pgfmathresult
}

\draw[fill = colorb] (.8*\w, .74*\h) rectangle (\w, \h); 
\node at (.9*\w, .86*\h) {$\bar i$};

\def \widey {0}
\def \wideyatend {1.5}

\foreach \processingtime/\demand in {.9/.02, .85/.04,  .54/.02} 
{
    \draw[fill = coloraa] (\w - \processingtime*\w, \h + \widey*\h) rectangle (.8*\w, \h + \demand*\h + \widey*\h); 
    \draw[fill = coloraa] (\w - \processingtime*\w, \h - \widey*\h) rectangle (.8*\w, \h - \demand*\h - \widey*\h);

    \draw[fill = coloraa] (.8*\w, \wideyatend*\h - \demand*2*\h) rectangle (\w, \wideyatend*\h);
    
    \pgfmathparse{\widey + \demand}
    \global\let\widey\pgfmathresult

    \pgfmathparse{\wideyatend - 2*\demand}
    \global\let\wideyatend\pgfmathresult
}

\draw[dashed] (.48*\w,-\os) node[below]{$\big(\frac12 - 2\eps'\big)\D$} -- (.48*\w,1.5*\h);

\draw[dashed] (-\os,1.5*\h) node[left]{$\frac{3}{2}\opt$} -- (\w+\os,1.5*\h) ;
\draw[dashed] (-\os,0.5*\h) node[left]{$\frac{1}{2}\opt$} -- (\w+\os,0.5*\h) ;
\draw[dashed] (-\os,1*\h) node[left]{$\opt$} -- (\w+\os,1*\h) ;
\draw[dashed] (-\os,1.25*\h) node[left]{$\frac{5}{4}\opt$} -- (\w+\os,1.25*\h) ;
\end{tikzpicture}
    }
    \caption{An example repacking generated by \cref{alg_a_pre_fit} when the tall items have a total width of at least $\big(1 - \frac{\eps}{5 + 4 \eps}\big) \D$. Here, the blue rectangles depict the same set of wide items.}
    \label{fig:tall_are_very_wide_1}
\end{figure}
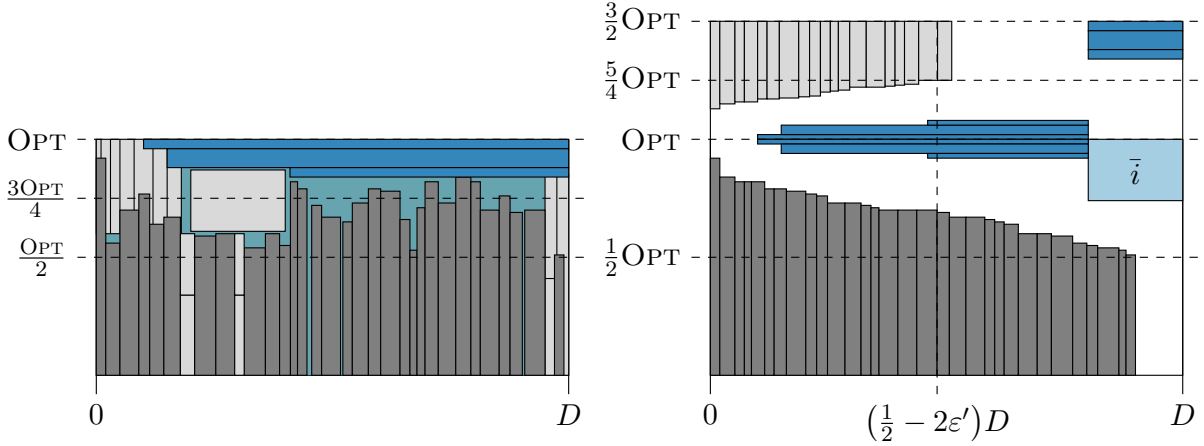

In this section, we consider the case that the total width of tall items is at least $(1 - \frac{\eps}{5 + 4 \eps}) \D$, i.e., $w\big(\items_{h > \frac{\opt}{2}}\big) \geq \big(1 - \frac{\eps}{5 + 4 \eps}\big) \D$, and prove the following lemma.
\begin{lemma}
\label{lem:caseA-main}
\label{lem:tall_items_are_wide}
    If the total width of tall items is at least $\big(1 - \frac{\eps}{5 + 4 \eps}\big) \D$, then $(\items,\D)$ admits a neat packing.
\end{lemma}

Note that we only need to prove that there is an $(\eps,\opt(\items))$-neat packing for $\items \setminus \items_{\mathrm{sq,\eps,\opt}}$.
By the squeezing-Lemma~\ref{lem:squeezing}, any $(\eps,\opt(\items))$-neat packing of $\items \setminus \items_{\mathrm{sq,\eps,\opt}}$ can be extended to an $(\eps,\opt(\items))$-neat packing of $\items$.
Hence, to simplify notation, we use $\items$ to denote $\items \setminus \items_{\mathrm{sq,\eps,\opt}}$. 

Throughout this section, we assume that we are given an optimal packing $\sopt$.  
For convenience, we set $\eps' \coloneqq \frac{\eps}{5 + 4 \eps}$ and work with $\eps'$ instead of $\frac{\eps}{5 + 4 \eps}$.
Note that the algorithm does not need to know the optimal packing. 
We only use the existence of the optimal packing $\sopt$ to prove the bound on the packing height.
We start by describing an algorithm that generates an $(\eps,\opt)$-neat packing.

The algorithm has two main steps. 
First, all items with a height of at least $\frac\opt4$ or a width of at least $\big(\frac{1}{2}+2\eps'\big)\D$ are preliminarily packed in a very structured way (\cref{alg_a_pre_fit}).  
\cref{alg_a} uses a technique similar to the Squeezing Algorithm to add the remaining items to this preliminary packing. 
To this end, first, all items with height at most $\frac\opt4$ and width at least $\big(\frac{1}{2}+2\eps'\big)\D$ are shifted as far to the left as possible without increasing the height of the packing above $\big(\frac{3}{2}+\eps\big)\opt$. 
Then, the remaining items are squeezed into the packing, each starting at the earliest point possible that does not increase the packing height above $\big(\frac{3}{2}+\eps\big)\opt$. 

\cref{alg_a_pre_fit} does the following: 
First, all tall items are packed next to each other starting at 0 and ordered non-increasingly by height.
Next, an item $\bar i$ is defined, which is the tallest non-tall item if its width is at most $\big(\frac 12 + 2\eps'\big) \D$.
If this tallest non-tall item is wider than $\big(\frac 12 + 2\eps'\big) \D$, then $\bar i$ is defined to be the second tallest non-tall item.
Then, all items in $\items_{h\in ( \frac{1}{4} \OPT, \frac{1}{2} \OPT]}$ except $\bar i$ are packed next to each other starting at 0 and ordered non-increasingly by height.
Finally, the item $\bar i$ and all wide items of width at least $\big(\frac 12 + 2\eps'\big) \D$ and height at most $\frac 14 \opt$ are packed such that they end at $\D$.

To analyze the packing height of the resulting packing, we rely on the following lemma about 
the relation between $H$-tall and wide items with regard to the packing height, proven in~\cite{DBLP:journals/algorithmica/DeppertJKRT23}.

\begin{algorithm}[t]
\DontPrintSemicolon
\caption{Pre-fit tall and wide items.}
\label{alg_a_pre_fit}
\textbf{Input:} $(\items\setminus \items_{\mathrm{sq},\eps,\opt}, \opt)$ \;
\smallskip 
$\start \leftarrow$ output of \Cref{alg_items_next_to_each_other} with $(\items_{h > \frac\opt2},0)$ \tcp*[r]{pack the tall items next to each other} 
$ \bar{i} \leftarrow \arg\max_{i \in \items_{h\in ( \frac{1}{4} \OPT, \frac{1}{2} \OPT]}} h(i)$ \tcp*[r]{consider tallest non-tall item}
\If(\tcp*[f]{check if tallest item is too wide}){$w(\bar i) > \big(\frac{1}{2}+ 2\eps'\big) \D$}{
    $ \bar{i} \leftarrow \arg\max_{i \in \items_{h\in ( \frac{1}{4} \OPT, \frac{1}{2} \OPT]} \setminus\{\bar i\}} h(i)$ \tcp*[r]{redefine $\bar i$ as second tallest item}
}
$\start \leftarrow $ output of \Cref{alg_items_next_to_each_other} with $(\items_{h\in ( \frac{1}{4} \OPT, \frac{1}{2} \OPT]} \setminus \{\bar{i}\},0)$ \tcp*[r]{pack $\items_{h\in ( \frac{1}{4} \OPT, \frac{1}{2} \OPT]}$}
\ForEach{$i \in \items_{h \leq \frac\opt4, \, w > (\frac12 + 2 \eps')\D} \cup \{ \bar i\}$}{$\start(i) \leftarrow \D - w(i)$}
\smallskip 
\textbf{return} $\start$ \; 
\end{algorithm}

\begin{lemma}[Corollary 12 in~\cite{DBLP:journals/algorithmica/DeppertJKRT23}]
    \label{lem:L-packing}
    Let $\Tilde{\items} \subseteq \items$ be a set of items such that 
    $w(\Tilde{\items}) \leq W$ and consider 
    $\items_W:= \items_{w>\max\{W-w(\Tilde{\items}), \frac W2\}}\setminus \Tilde{\items}$. 
    Furthermore, let $i_\bot \in \Tilde{\items}$ be the item with the smallest height. 
    Then it holds that $\min\big\{h(i_\bot) + \frac{h(\items_W)}2, 2 h(i_\bot)\big\} \leq \opt$.
\end{lemma}
We also use the notion of \emph{L-packings}.
The L-packing $\start_L(A,B)$ of two sets $A, B$ is defined as follows:
The items from $A$ are placed next to each other, sorted by non-increasing height such that the first item starts at $0$, while all items from $B$ end at $\D$.

We split the proof of \Cref{lem:caseA-main} into several claims targeted at the different steps in the algorithm and properties of the packing.  
We start by analyzing the packing $\start$ of the items $\items_{h>\frac{1}{4} \OPT} \cup \items_{w> (\frac{1}{2}+ 2 \eps') \D}$ given by \cref{alg_a_pre_fit}. An example repacking can be seen in \cref{fig:tall_are_very_wide_1}.

\begin{claim}
\label{claim:case-A-main-proof:1}
The packing $\start$ returned by \cref{alg_a_pre_fit} (of the items in $\items_{h>\frac{1}{4} \OPT} \cup \items_{w> (\frac{1}{2}+ 2 \eps') \D} $)
is $\D$-feasible and has a height of at most $\frac{3}{2} \OPT$.
\end{claim}

\begin{proof}
    Our first step is to prove that the items $\big(\items_{h \in (\frac\opt4, \frac\opt2]} \cup \items_{h \leq \frac\opt2, \, w> (\frac{1}{2} + 2 \eps') \D} \big) \setminus \{\bar i\}$ admit an L-packing of width $\D$ and height at most $\frac\opt2$ plus half of the height of the wide items.
    We then obtain a similar bound for the items $\big( \items_{h > \frac\opt2} \cup \items_{h \leq \frac\opt4, \, w> (\frac{1}{2} + 2 \eps') \D} \big) \setminus \{\bar i\}$. 
    By carefully combining these bounds, we obtain a bound of $\frac32\opt$ on the height of $\items_{h>\frac{1}{4} \OPT} \cup \items_{w> (\frac{1}{2}+ 2 \eps') \D} \setminus \{\bar i \}$. 
    We complete the proof by showing that adding $\bar i$ to the packing does not increase the height above $\frac32\opt$.

    We first prove that $h(\items_{w> (\frac{1}{2} + 2 \eps') \D}) \leq \frac{1}{2} \OPT$. 
    Note that in any feasible packing $\start'$, every item $i \in \items_{w> (\frac{1}{2}+ 2 \eps') \D}$ has to satisfy $\start'(i) < \big(\frac{1}{2}- 2 \eps'\big) \D$ and $\start'(i) + w(i) > \big(\frac{1}{2}+ 2 \eps'\big) \D$. 
    By assumption, the total width of tall items is at least $(1 - \eps') \D$. 
    Thus, in the optimal packing $\sopt$, there has to exist a tall item that is packed in parallel to all items in $\items_{w> (\frac{1}{2}+ 2 \eps') \D}$.
    Since the height of $\sopt$ is bounded by $\opt$, it has to hold that $h(\items_{w> (\frac{1}{2} + 2 \eps') \D}) \leq \frac{1}{2} \OPT$.
    
    As a consequence, there can be at most one item in $\items_{h \in (\frac\opt4, \frac\opt2]}$ that has a width greater than $\big(\frac{1}{2} + 2 \eps'\big) \D$ since such items belong to $\items_{w> (\frac{1}{2} + 2 \eps') \D}$ and two of them would increase the total height of the set above $\frac{\opt}{2}$.
    Further, observe that if there is an item in $\items_{h \in (\frac{\opt}{4},\frac{\opt}{2} ]}$ that has width at least $\big(\frac{1}{2} + 2 \eps'\big) \D$, then this item will be skipped when choosing the item $\bar{i}$ in \cref{alg_a_pre_fit}. 
    Hence, the item $\bar i$ has width at most $\big(\frac{1}{2} + 2 \eps'\big) \D$.

    In the next step, we show that we can restructure the packing of the items $(\items_{h \in (\frac\opt4, \frac\opt2]} \cup \items_{h \leq \frac\opt4, \, w> (\frac{1}{2} + 2 \eps') \D}) $ to an L-packing $\start_L$ and bound its height.
    Remember that there are no non-tall items of width less than $\frac{\eps}{1+\eps} \D$, the total width of tall items is at least $(1 - \eps') \D$ and $\eps' < \frac{\eps}{1+ \eps}$.
    Hence, using \Cref{alg_stretching}, we can stretch the packing of the non-tall items in the optimal solution~$\sopt$ by at most $\eps' \D$, such that the stretched packing $\start_S$ without the tall items has a height of at most $\frac{1}{2} \OPT$ and~$\items_S$ is empty, i.e., every non-tall item is packed by~$\start_S$.
    The items $\items_{h \in (\frac\opt4, \frac\opt2]}$ become the new set of tall items in this packing $\start_S$. 
    We place the items in $\items_{h \in (\frac\opt4, \frac\opt2]}$ and $\items_{h \leq \frac\opt4, \, w> ( \frac{1}{2} + 2 \eps') \D} $ as an L-packing $\start_{L,M} = \start_L(\items_{h \in (\frac\opt4, \frac\opt2]}, \items_{h \leq \frac\opt4, \, w> ( \frac{1}{2} + 2 \eps') \D} )$ with width $(1+\eps') \D$.
    Note that we can use \cref{lem:L-packing} to bound the height of $\start_{L,M}$:
    Given any point in time $\tau \leq w(\items_{h \in (\frac\opt4, \frac\opt2]})$, 
    we interpret the set $\Tilde{\items}$ as the set of items in $\items_{h \in (\frac\opt4, \frac\opt2]}$ that are packed at or left of $\tau$. 
    By \cref{lem:L-packing} and since the packing $\start_S$ has a height of at most $\opt' = \frac{\opt}{2}$ and hence $2h(i_\bot)> \opt'$ it holds that 
    \begin{equation}
    \begin{aligned}
            h(\itemsatt{\start_{L,M}}{\tau}) & \overset{\hphantom{\cref{lem:L-packing}}}{=}  h(\items^{\start_{L,M}}_{h \in (\frac\opt4, \frac\opt2]}(\tau)) + h(\items_{h \leq \frac\opt4, \, w> ( \frac{1}{2} + 2 \eps') \D}^{\start_{L,M}}(\tau)) \\
            & \overset{\hphantom{\cref{lem:L-packing}}}{=} h(\items^{\start_{L,M}}_{h \in (\frac\opt4, \frac\opt2]}(\tau)) + h(\items_{h \leq \frac\opt4, \, w> \max\{\D-\tau, ( \frac{1}{2} + 2 \eps') \D\}}) \\
            &\overset{\cref{lem:L-packing}}{\leq} \frac{\opt}{2} + \frac{1}{2}h(\items_{h \leq \frac\opt4, \, w> \max\{\D-\tau, ( \frac{1}{2} + 2 \eps') \D\}}).
    \end{aligned}\label{eq:first_L_packing}
    \end{equation}

    Next, we show that by removing $\bar{i}$ the width of the L-packing $\start_{L,M}$ can be reduced to be at most $\D$ without increasing the height of the packing.
    Consider $\bar{i}$ as defined by \cref{alg_a_pre_fit}.
    It is the tallest or second tallest item in $\items_{h \in (\frac\opt4, \frac\opt2]}$ and has a 
    width of at least $\frac{\eps}{1+\eps} \D \geq \eps' \D$.
    We remove $\bar{i}$ from the L-packing $\start_L$ and show that this allows us to decrease the packing width to be at most~$\D$.
    If the tallest item in $\items_{h \in (\frac\opt4, \frac\opt2]}$ has been removed, then it is clear that the resulting packing has a width of at most $\D$ since the wide items in $\items_{h \leq \frac\opt4, \, w> ( \frac{1}{2} + 2 \eps') \D}$ have a width of at most $\D$ and every item has a width of at least $\frac{\eps}{1+\eps} \D \geq \eps' \D$.
    If the second tallest item in $\items_{h \in (\frac\opt4, \frac\opt2]}$ is removed, the tallest item in $\items_{h \in (\frac\opt4, \frac\opt2]}$ has a width of at least $\big(\frac{1}{2} + 2 \eps'\big) \D$. 
    Observe that this tallest item must be packed in parallel to all wide items in $\items_{h \leq \frac\opt4, \, w> ( \frac{1}{2} + 2 \eps') \D}$ in the L-packing of width~$(1 + \eps')\D$.
    Hence, if we remove the second tallest item in $\items_{h \in (\frac\opt4, \frac\opt2]}$ and shift the tallest item in $\items_{h \in (\frac\opt4, \frac\opt2]}$ to the right by $\eps' \D$, the resulting packing has width at most $\D$ and height at most $\frac{1}{2} \OPT + \frac{1}{2} h(\items_{h \leq \frac\opt4, \, w> (\frac{1}{2} + 2 \eps') \D})$. 
    This is true as the removed second tallest item has a width of at least $\frac{\eps}{1 + \eps} \D > \eps' \D$ and the items in $\items_{h \leq \frac\opt4, \, w> ( \frac{1}{2} + 2 \eps') \D}$ have a width of at most~$\D$.
    We denote by $\start_{L,M}'$ the packing resulting from removing $\bar{i}$ and reducing the width of the packing $\start_{L,M}$ accordingly.
    
    Note that the above observations imply that the items in $\items_{h\in ( \frac{1}{4} \OPT, \frac{1}{2} \OPT]} \setminus \{\bar{i}\}$ can be packed such that they finish before $\D$ in $\start$. Therefore, all items considered (and packed) by \cref{alg_a_pre_fit} finish before $\D$ in $\start$.

    In the following, we prove that we can combine the packing $\start_{L,M}'$ with the tall items without increasing the height of the packing above $\frac{3}{2} \opt$.
    Consider the L-packing of the tall items and the items in $\items_{h \leq \frac\opt4, \, w> (\frac{1}{2} + 2 \eps') \D}$ of width $\D$. We denote this packing by $\start_{L,T} =$ $\start_L(\items_{h> \frac{\opt}{2}}, \items_{h \leq \frac\opt4, \, w> (\frac{1}{2} + 2 \eps') \D})$.
    By \Cref{lem:L-packing} it holds that
    \begin{equation}
    \begin{aligned}
        h(\itemsatt{\start_{L,T}}{\tau}) & = h(\items^{\start_{L,T}}_{h> \frac{\opt}{2}}(\tau)) + h(\items^{\start_{L,T}}_{h \leq \frac \opt4, \, w> (\frac{1}{2} + 2 \eps') \D}(\tau))\\
        & = h(\items^{\start_{L,T}}_{h> \frac{\opt}{2}}(\tau)) + h(\items_{h \leq \frac \opt4, \, w> \max\{\D-\tau, ( \frac{1}{2} + 2 \eps') \D\}})\\
        & \overset{\cref{lem:L-packing}}{\leq} \OPT + \frac{1}{2} h(\items_{h \leq \frac\opt4, \, w>\max\{\D-\tau, ( \frac{1}{2} + 2 \eps') \D\}})
    \end{aligned}\label{eq:second_L_packing}
    \end{equation}
    for any $\tau \in [0,\D]$.
    Now we combine both L-packings $\start_{L,T}$ and $\start_{L,M}'$ and claim that its height is bounded by $\frac 32 \OPT$. 
    Observe that the items in $\items_{h \leq \frac\opt4, \, w> (\frac{1}{2}+2\eps') \D} $ appear in both L-packings, and in both packings these items end at $\D$. 
    We cut each wide item $i \in \items_{h \leq \frac\opt4, \, w> (\frac{1}{2}+2\eps') \D}$ horizontally into two items $i_1$ and $i_2$ of the same height, i.e., $h(i_1) = h(i_2) = \frac{h(i)}{2}$.
    Let $W_1 \coloneqq \{ i_1 \mid i \in \items_{h \leq \frac\opt4, \, w> (\frac{1}{2}+2\eps') \D} \}$ and $W_2 \coloneqq \{ i_2 \mid i \in\items_{h \leq \frac\opt4, \, w> (\frac{1}{2}+2\eps') \D} \}$.

    By \cref{eq:second_L_packing}, an L-packing of the tall items and the items in $W_1$ has height at most $\OPT$, as the height of $W_1$ is half the height of the items in $\items_{h \leq \frac\opt4, \, w> (\frac{1}{2}+2\eps') \D}$.
    Similarly, by \cref{eq:first_L_packing}, an L-packing of the items in $W_2$ and $\items_{h \in (\frac\opt4, \frac\opt2]}$ (without item $\bar i \in \items_{h \in (\frac\opt4, \frac\opt2]}$ that was removed) has a height of at most $\frac\opt2$.
    Hence, both packings on top of each other have a height of at most $\frac{3}{2}\OPT$.
    As a consequence, the packing generated by \cref{alg_a_pre_fit} without $\bar{i}$, has a height of at most $\frac{3}{2}\OPT$.

    In the next step, we show that adding the item $\bar{i}$ to the packing does not increase its height above $\frac{3}{2}\opt$.
    Consider the item $\bar{i} \in \items_{h \in (\frac{1}{4} \OPT, \frac{1}{4} \OPT]}$ as defined by \cref{alg_a_pre_fit}. 
    We first argue that when placing the item $\bar{i}$ such that it ends at $\D$, the packing of this item and the sorted tall items has a height of at most $\OPT$: 
    Note that the tall items and the item $\bar{i}$ are packed together in $\sopt$ without exceeding the height of $\opt$.
    By sorting the tall items by non-increasing height and shifting the item $\bar{i}$, the height of $\opt$ cannot be exceeded.

    Next, we show that the item $\bar{i}$ also fits together with the items from $ \items_{h \in (\frac{1}{4} \OPT,\frac{1}{2} \OPT]} \cup \items_{w>(\frac{1}{2}+ 2 \eps') \D}$, i.e., the packing $\start$ returned by \cref{alg_a_pre_fit} of the items from $\items_{h > \frac\opt4} \cup \items_{w > (\frac{1}{2}+ 2 \eps') \D}$ has height at most $\frac 32 \OPT$.
    The analysis depends on the intersection of $\bar{i}$ with another item from the set $\items_{h \in (\frac{1}{4} \OPT,\frac{1}{2} \OPT]}$.
    If the total width of $\items_{h \in (\frac{1}{4} \OPT,\frac{1}{2} \OPT]}$ is at most one, the items from the set $\items_{w>(\frac{1}{2}+ 2 \eps') \D} \setminus \items_{h>\frac{1}{4} \OPT}$ are the only non-tall items overlapping with $\bar{i}$.
    Since $h(\items_{h \leq \frac\opt4, \, w>(\frac{1}{2}+ 2 \eps') \D} ) \leq \frac{1}{2} \OPT$, the packing is not larger than $\frac{3}{2} \OPT$ in this case.
    
    Otherwise, assume that $w(\items_{h \in (\frac{1}{4} \OPT,\frac{1}{2} \OPT]}) > \D$.
    We will prove that the smallest item $i_s$ (w.r.t.\ height) in~$\items_{h \in (\frac{1}{4} \OPT,\frac{1}{2} \OPT]}$ is the only item from this set that $\bar{i}$ overlaps with in $\start$ and that the total height of $i_s$ combined with the items in $\items_{h \leq \frac\opt4, \, w > (\frac{1}{2}+ 2 \eps') \D} $ is bounded by $\frac{\opt}{2}$.
    Since $w(\items_{h \in (\frac{1}{4} \OPT,\frac{1}{2} \OPT]}) > \D $, the item $\bar{i}$ overlaps with~$i_s$ in~$\start$.
    Our assumption on the total width of tall items and that $w(i) \geq \frac{\eps}{1+\eps}$ for $i \in \items \setminus \items_{h > \frac{\OPT}{2}}$ implies an upper bound of~$(1 + \eps') \D$ on the total width of the items $\items_{h \in (\frac{1}{4} \OPT,\frac{1}{2} \OPT]}$. 
    As a consequence, the total width where the items $\bar{i}$ and $i_s$ overlap is bounded by $\eps' \D$.
    Since $i_s$ has a width of at least $\frac{\eps}{1 + \eps} \D$, it is the only item from the set $\items_{h \in (\frac{1}{4} \OPT,\frac\opt2] }$ that overlaps with this item.
    
    Recall that the items in $\items_{h \leq \frac\opt4, \, w > (\frac{1}{2}+ 2 \eps') \D} $ completely overlap the segment $[(\frac{1}{2}- 2 \eps') \D,(\frac{1}{2}+ 2 \eps') \D]$ in any feasible packing due to their width. 
    Since the tall items have a total width of at least $(1-\eps')\D$, this segment contains a total width of at least $3\eps' \D$ of these items in $\sopt$.
    Let $w_T \in (3\eps' \D,4\eps' \D]$ be the total width of tall items overlapping the segment $\big[\big(\frac{1}{2}- 2 \eps'\big) \D,\big(\frac{1}{2}+ 2 \eps'\big) \D\big]$ and $w_M$ be the total width of items from $\items_{h \in (\frac{1}{4}\OPT,\frac\opt2]}$ overlapping the segment $\big[\big(\frac{1}{2}- 2 \eps'\big) \D,\big(\frac{1}{2}+ 2 \eps'\big) \D\big]$ in $\sopt$.
    It holds that $w(\items_{h \in (\frac{1}{4} \OPT,\frac{1}{2} \OPT]}) > \D$, at most 3 of them can be placed on top of each other, at most one per time can be placed on top of a tall item, and the total width of tall items is at least $(1-\eps')\D$. 
    Therefore, $w_M > \D - (1-4\eps')\D - 2(\eps'\D - (4\eps' \D -w_T)) = 4\eps'\D -2(w_T-3\eps'\D) \geq 2\eps'\D$.  
    Therefore, there exists an item $i' \in \items_{h \in (\frac{1}{4} \OPT,\frac{1}{2} \OPT]}$ that is packed in the packing $\sopt$ in parallel to some tall item and all items from the set $\items_{h \leq \frac\opt4, \, w > (\frac{1}{2}+ 2 \eps') \D} $.

    As a consequence, the items in $\items_{h \leq \frac\opt4, \, w > (\frac{1}{2}+ 2 \eps') \D}$ together have a height of at most $\big(\frac{1}{2} - h(i_s)\big) \OPT$ (as $i_s$ is the smallest item in $\items_{h \in (\frac{1}{4} \OPT,\frac{1}{2} \OPT]}$ w.r.t.\ height) and therefore fit above the item $\bar{i}$ and $i_s$.
    Hence, the packing of the items $\items_{h > \frac{1}{4} \OPT} \cup \items_{w>(\frac{1}{2}+ 2\eps') \D}$ has a height of at most $\frac{3}{2} \OPT$.
\end{proof}

\begin{algorithm}
\DontPrintSemicolon
\caption{Repacking when the total width of tall items is at least $(1 - \frac{\eps}{5 + 4 \eps}) \D$}
\label{alg_a}
\textbf{Input:} $(\items \setminus \items_{\mathrm{sq,\eps,\opt}}, \opt)$ \;
\smallskip 
$\start \leftarrow$ output of \Cref{alg_a_pre_fit} with $(\items, \D, \opt)$\; 
sort $\items_{ h \leq \frac \opt4, \, w > (\frac12 + 2 \eps')\D} $
by non-decreasing starting times \; 
\ForEach(\tcp*[f]{shift $i$ to the left without exceeding $\big(\frac32 + \eps\big) \opt$}){$i \in \items_{ h \leq \frac \opt4, \, w > (\frac12 + 2 \eps')\D}$}{
 $\start(i) 
 \leftarrow 
 \min\big\{t \in [0,\start(i)]\mid \forall t' \in [t, t + w(i)) : h(\itemsatt{\start}{t'} \setminus \{i\}) \leq \big(\frac32 + \eps\big)\opt - h(i)  \big \}$ 
}
$\tau \leftarrow \max\{\start(i)\mid i \in \items_{ h \leq \frac \opt4, \, w > (\frac12 + 2 \eps')\D
} \}$ \;
$\items_U \leftarrow \items_{w \leq (\frac12 + 2 \eps')\D , h \leq \frac\opt4}$ \tcp*[r]{items not yet packed} 
\While{$\items_U \neq \emptyset$}{
    \uIf{$\exists i \in \items_U$ with $h(i) \leq \big(\frac32 + \eps\big)\opt - h( \itemsatt{\start}{\tau}) $}{
        $\start(i) \leftarrow \tau$ \; 
    }
    \Else{
        $\tau \leftarrow \min\{ t > \tau : t = \start(i) + w(i) \text{ for } i \in \items\}$ \tcp*[r]{increase $\tau$ to next endpoint }
    }
}
\smallskip
\textbf{return} $\start$ \; 
\end{algorithm}

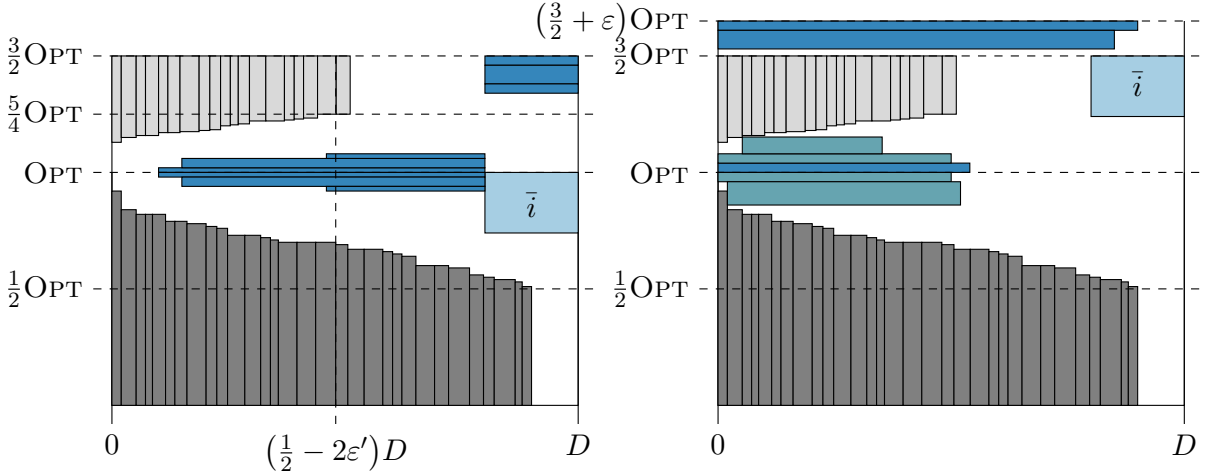
\begin{figure}
    \centering
    \resizebox{\textwidth}{!}{
    \begin{tikzpicture}
\pgfmathsetmacro{\w}{6}
\pgfmathsetmacro{\h}{3}
\pgfmathsetmacro{\os}{0.24}
\pgfmathsetmacro{\l}{0.35}
\pgfmathsetmacro{\r}{0.7}
\pgfmathsetmacro{\lamb}{0.04}

\draw (0,1.5*\h) -- (0,0) -- (\w,0) -- (\w,1.5*\h);
\draw (0*\w,-\os) node[below]{$0$}-- (0*\w,\h);
\draw (\w,-\os) node[below]{$\D$}-- (\w,\h);

\def \tallx {0}
\def \totalgap {0}

\foreach \processingtime/\demand/\gap in {.02/.92/0, .032/.84/0, .02/.82/0, .015/.82/0, .028/.82/0, .02/.79/.01, .026/.79/0, .041/.78/0, .023/.768/.01, .023/.76/0, .037/.73/0, .034/.73/0, .021/.72/0,  .017/.71/0, .04/.7/0, .04/.7/0,   .043/.7/.02, .026/.69/0.005, .035/.67/.02, .04/.67/.005, .021/.66/0, .02/.65/0, .03/.64/0, .04/.6/.02,  .03/.6/0, .045/.59/0, .03/.56/0, .023/.55/0, .045/.54/0,  .015/.53/0, .02/.51/0 }
{    
    \draw[fill = tallItemColor] (\tallx*\w, 0*\h) rectangle (\tallx*\w + \processingtime*\w, \demand*\h); 
    \pgfmathparse{\tallx + \processingtime}
    \global\let\tallx\pgfmathresult

    \pgfmathparse{\totalgap + \gap}
    \global\let\totalgap\pgfmathresult
}

\def \mediumx {0}

\foreach \processingtime/\demand/\gap in {.02/.89/0, .032/.84/0, .02/.82/0, .028/.82/0, .02/.79/.01, .026/.79/0, .041/.78/0, .023/.768/.01, .023/.76/0, .021/.72/0,  .017/.71/0, .024/.7/0,   .035/.67/.02, .04/.67/.005, .021/.66/0, .02/.65/0, .03/.64/0, .04/.6/.02,  .03/.6/0}
{    
    \draw[fill = mediumItemColor] (\mediumx*\w, 1.5*\h) rectangle (\mediumx*\w + \processingtime*\w, 1.5*\h-\demand/2.4*\h); 
    \pgfmathparse{\mediumx + \processingtime}
    \global\let\mediumx\pgfmathresult
}

\draw[fill = colorb] (.8*\w, .74*\h) rectangle (\w, \h); 
\node at (.9*\w, .86*\h) {$\bar i$};

\def \widey {0}
\def \wideyatend {1.5}

\foreach \processingtime/\demand in {.9/.02, .85/.04,  .54/.02} 
{
    \draw[fill = coloraa] (\w - \processingtime*\w, \h + \widey*\h) rectangle (.8*\w, \h + \demand*\h + \widey*\h); 
    \draw[fill = coloraa] (\w - \processingtime*\w, \h - \widey*\h) rectangle (.8*\w, \h - \demand*\h - \widey*\h);

    \draw[fill = coloraa] (.8*\w, \wideyatend*\h - \demand*2*\h) rectangle (\w, \wideyatend*\h);
    
    \pgfmathparse{\widey + \demand}
    \global\let\widey\pgfmathresult

    \pgfmathparse{\wideyatend - 2*\demand}
    \global\let\wideyatend\pgfmathresult
}

\draw[dashed] (.48*\w,-\os) node[below]{$\big(\frac12 - 2\eps'\big)\D$} -- (.48*\w,1.5*\h);

\draw[dashed] (-\os,1.5*\h) node[left]{$\frac{3}{2}\opt$} -- (\w+\os,1.5*\h) ;
\draw[dashed] (-\os,0.5*\h) node[left]{$\frac{1}{2}\opt$} -- (\w+\os,0.5*\h) ;
\draw[dashed] (-\os,1*\h) node[left]{$\opt$} -- (\w+\os,1*\h) ;
\draw[dashed] (-\os,1.25*\h) node[left]{$\frac{5}{4}\opt$} -- (\w+\os,1.25*\h) ;


\begin{scope}[xshift = 1.3*\w cm]
    \draw (0,1.65*\h) -- (0,0) -- (\w,0) -- (\w,1.65*\h);
    \draw (0*\w,-\os) node[below]{$0$}-- (0*\w,\h);
    \draw (\w,-\os) node[below]{$\D$}-- (\w,\h);

    \def \tallx {0}
    \def \totalgap {0}
    
    \foreach \processingtime/\demand/\gap in {.02/.92/0, .032/.84/0, .02/.82/0, .015/.82/0, .028/.82/0, .02/.79/.01, .026/.79/0, .041/.78/0, .023/.768/.01, .023/.76/0, .037/.73/0, .034/.73/0, .021/.72/0,  .017/.71/0, .04/.7/0, .04/.7/0,   .043/.7/.02, .026/.69/0.005, .035/.67/.02, .04/.67/.005, .021/.66/0, .02/.65/0, .03/.64/0, .04/.6/.02,  .03/.6/0, .045/.59/0, .03/.56/0, .023/.55/0, .045/.54/0,  .015/.53/0, .02/.51/0 }
    {    
        \draw[fill = tallItemColor] (\tallx*\w, 0*\h) rectangle (\tallx*\w + \processingtime*\w, \demand*\h); 
        \pgfmathparse{\tallx + \processingtime}
        \global\let\tallx\pgfmathresult
    
        \pgfmathparse{\totalgap + \gap}
        \global\let\totalgap\pgfmathresult
    }
    
    \def \mediumx {0}
    
    \foreach \processingtime/\demand/\gap [count=\i] in {.02/.89/0, .032/.84/0, .02/.82/0, .028/.82/0, .02/.79/.01, .026/.79/0, .041/.78/0, .023/.768/.01, .023/.76/0, .021/.72/0,  .017/.71/0, .024/.7/0,   .035/.67/.02, .04/.67/.005, .021/.66/0, .02/.65/0, .03/.64/0, .04/.6/.02,  .03/.6/0}
    {    
        \draw[fill = mediumItemColor] (\mediumx*\w, 1.5*\h) rectangle (\mediumx*\w + \processingtime*\w, 1.5*\h-\demand/2.4*\h);       
        \pgfmathparse{\mediumx + \processingtime}
        \global\let\mediumx\pgfmathresult
    }

    
    \draw[fill = colorb] (.8*\w, 1.5*\h) rectangle (\w, 1.24*\h); 
    \node at (.9*\w, 1.38*\h) {$\bar i$};

    \def \widey {1.61} 
    \draw[fill = coloraa] (0, \widey*\h) rectangle (.9*\w, \widey*\h + .04*\h); 
    \def \widey {1.53}   
    \draw[fill = coloraa] (0, \widey*\h) rectangle (.85*\w, \widey*\h + .08*\h); 
    \def \widey {1} 
    \draw[fill = coloraa] (0, \widey*\h) rectangle (.54*\w, \widey*\h + .04*\h);

    
    \foreach \processingtime/\demand/\smallx/\smally in {.5/.04/0/1.04, .5/.04/0/.96, .3/.072/.052/1.08, .5/.1/.02/.86}
    {
        \draw[fill = colord] (\smallx*\w, \smally*\h) rectangle (\smallx*\w + \processingtime*\w, \smally*\h + \demand*\h);
    }


    \draw[dashed] (-\os,1.65*\h) node[left]{$\big(\frac{3}{2}+\eps\big)\opt$} -- (\w+\os,1.65*\h) ;
    \draw[dashed] (-\os,1.5*\h) node[left]{$\frac{3}{2}\opt$} -- (\w+\os,1.5*\h) ;
    \draw[dashed] (-\os,0.5*\h) node[left]{$\frac{1}{2}\opt$} -- (\w+\os,0.5*\h) ;
    \draw[dashed] (-\os,1*\h) node[left]{$\opt$} -- (\w+\os,1*\h) ;
\end{scope}

\end{tikzpicture}
    }
    \caption{An example repacking generated by \cref{alg_a} when the tall items have a total width of at least $\big(1 - \frac{\eps}{5 + 4 \eps}\big) \D$.}
    \label{fig:tall_are_very_wide_2}
\end{figure}

Next, we analyze the packing returned by \cref{alg_a}.
\cref{alg_a} takes the output by \cref{alg_a_pre_fit} and first iteratively shifts all items of width at least $\big(\frac{1}{2}+2\eps'\big)\D$ and height at most $\frac 14 \OPT$ as far to the left as possible without exceeding a height of $\big(\frac 32 + \eps\big) \OPT$, where we consider these items by non-decreasing starting times.
Next, we iteratively pack all remaining items of height at most $\frac 14 \OPT$ and width at most $\big(\frac{1}{2}+2\eps'\big)\D$ by going from left to right and scheduling one of these items as soon as it fits in our current packing, i.e., adding this item to our current packing at this starting time if it does not increase the height of the packing above $\big(\frac 32 + \eps\big) \OPT$.
An example repacking can be seen in \cref{fig:tall_are_very_wide_2}.

Note that shifting the wide items to the left at the start of \cref{alg_a} does not increase the maximal height of the packing beyond $\big( \frac{3}{2}+\eps\big) \OPT$ by definition. For the remaining analysis, we will show in \cref{claim:case-A-main-proof:2} that the decision of \cref{alg_a} to pack one of the remaining items does not increase the packing height above $\big(\frac{3}{2}+\eps\big) \OPT$. Finally, in \cref{claim:case-A-main-proof:3}, we argue that all newly packed items finish before $\D$ as required by a feasible packing.

\begin{claim}
\label{claim:case-A-main-proof:2}
    When starting an item $i \in \items \setminus (\items_{h> \frac{1}{4} \OPT} \cup \items_{w> (\frac{1}{2}+ 2 \eps' ) \D})$ at $\tau$ in \cref{alg_a}, the packing height is never increased above $\big(\frac{3}{2}+\eps\big) \OPT$. Further, \cref{alg_a} packs all items.
\end{claim}

\begin{proof}
    Before proving the claim, we prove that the following invariant holds.
    Let $\tau$ be as defined in the algorithm at the start of each while-loop.
    If $\tau < \start(\bar{i})$, it holds that the packing height in the segment $[\tau, \start(\bar{i}))$ and the segment $[\start(\bar{i}),\D]$ is non-increasing.
    Otherwise, if $\tau \geq \start(\bar{i})$, the packing height in the segment $[\tau,\D]$ is non-increasing.

    To prove this, first note that $\tau \geq \max\{\start(i)\mid i \in \items_{h \leq \frac\opt4, \, w > (\frac12 + 2\eps')\D}\}$, as $\tau$ is set to this value before the while-loop and it never decreases. 
    Hence, for any $\tau$ considered by the algorithm, the height of the packing of the items in $\items_{h \leq \frac\opt4, \, w > (\frac12 + 2\eps')\D}$ is non-increasing. 
    Therefore, in the first loop, the statement is true since the items in $\items_{h> \frac{1}{4} \OPT}$ are packed sorted by non-increasing height and no other item, except $\bar{i}$, is starting after $\tau$. 
    Since the algorithm never starts an item after~$\tau$, the same reasoning also holds for every other iteration of the while-loop.

    We now have the tools to prove the claim.
    For contradiction, assume that there is an item~$i$ that is not packed by the algorithm. That is, starting~$i$ at the final value of $\tau$ would increase the height of the packing above $\big(\frac{3}{2}+\eps\big) \OPT$.
    Note that an increase above $\big(\frac{3}{2}+\eps\big) \OPT$ is only possible if the placed item overlaps with $\bar{i}$ because the height of the packing is non-increasing between $\tau$ and $\start(\bar{i})$ by the invariant and the height of the packing at $\tau$ was at most $\big(\frac{3}{2}+\eps\big) \OPT - h(i)$ before placing the item $i$. Furthermore, since the packing between $\bar{\tau} = \start(\bar{i})$ and $\D$ is non-increasing as well, it has to hold that the height of the packing at $\bar{\tau}$ is at least $\big(\frac{3}{2}+\eps\big) \OPT$. 
    
    At each point in time $t \in [0,\tau)$, the packing has a height of at least $\big(\frac{5}{4}+\eps\big) \OPT$: After we have placed all items in $\items_{h> \frac{1}{4} \OPT}$, the remaining items have a height of at most $\frac{1}{4} \OPT$.
    Thus, there cannot be a time point $t \in [0,\tau)$ such that the packing has a height less than $\big(\frac{5}{4}+\eps\big) \OPT$, as this contradicts \Cref{alg_a}.
    Observe that $\big(\frac{5}{4}+\eps\big) \OPT > \big(\frac{3}{2}+\eps\big) \OPT - h(\bar{i})$ as $\bar{i} \in \items_{h> \frac{1}{4} \OPT}$.

    Furthermore, at each point in time $t \in [\tau,\bar{\tau})$ the packing has a height of at least $\big(\frac{3}{2}+\eps\big) \OPT - h(\bar{i})$, as at $\bar{\tau}$ the packing has a height larger than $\big(\frac{3}{2} +\eps\big) \OPT$, and the packing height without the item $\bar{i}$ in $[\tau,\D)$ is monotonically decreasing.
    Finally, at each point in time $t \in [\bar{\tau},(1-\eps') \D]$, the packing has a height of at least $\frac{\opt}{2}  + h(\bar{i})$ since the tall items have a height of at least $\frac{\opt}{2}$. 
    Thus with $\bar \tau = \start(\bar{i}) = \D - w(\bar i)$, the total area of the packed items is at least 
    \begin{align*}
        \area(\items) 
        >& \big(\D -w(\bar{i})\big)\bigg(\bigg(\frac{3}{2}+\eps\bigg) \OPT -h(\bar{i})\bigg) +(w(\bar{i})-\eps' \D)\bigg(\frac\opt2 + h(\bar{i})\bigg)\\
        =& \bigg(\bigg(\frac{3}{2}+\eps\bigg) \OPT - h(\bar{i})\bigg) \D -w(\bar{i})\big((1+\eps) \OPT - 2h(\bar{i})\big) - \eps' \D \bigg(\frac{\opt}{2}  + h(\bar{i})\bigg). \\
        \shortintertext{Using $w(\bar{i}) < \big(\frac{1}{2} + 2 \eps'\big) \D$ and $\OPT -2h(\bar{i})> 0$, this implies}
        \area(\items) &>\bigg(\bigg(\frac{3}{2}+\eps\bigg) \OPT -h(\bar{i})\bigg) \D - \bigg(\frac{1}{2}+ 2 \eps'\bigg) \D \cdot \big((1+\eps) \OPT - 2h(\bar{i})\big) \\
        &- \eps' \D \bigg(\frac{\opt}{2} + h(\bar{i})\bigg)\\
        =& \bigg(\bigg(\frac{3}{2}+\eps\bigg) -\bigg(\frac{1}{2}+ 2 \eps'\bigg)(1+\eps) -\frac{\eps'}{2}\bigg)\D \cdot \OPT \\
        &+ \bigg(-1 +2\bigg(\frac{1}{2}+ 2 \eps'\bigg)-\eps'\bigg)\D \cdot h(\bar{i})\\
        =& \bigg(1+\frac{\eps}{2} -\bigg(\frac{5}{2}+2\eps\bigg)\eps'\bigg)\D \cdot \OPT
        + \eps'\D \cdot h(\bar{i})\\
        >& \bigg(1+\frac{\eps}{2} -\bigg(\frac{7}{4}+2\eps\bigg)\eps'\bigg)\D \cdot \OPT
    \end{align*}
    where we used $h(\bar i) > \opt/4$ in the last step. Since~\alertBound{$\eps' \leq \frac{2\eps}{7 + 8 \eps}$} we obtain $\area(\items)  > \D \cdot \OPT$,
    which is a contradiction since~$\area(\items) \leq \D \cdot \opt$. 
\end{proof}

Finally, we have to show that each of the remaining items $i \in \items_{h \leq  \frac{1}{4} \OPT, \, w \leq (\frac{1}{2}+ 2 \eps' ) \D}$ does not end later than $\D$ when packing it with the described algorithm.
In particular, this implies that the solution is $\D$-feasible.
    
\begin{claim}
\label{claim:case-A-main-proof:3}
    Each item $i \in \items_{h \leq  \frac{1}{4} \OPT, \, w \leq (\frac{1}{2}+ 2 \eps' ) \D}$ packed by \cref{alg_a} finishes before $\D$.
\end{claim}

\begin{proof}
    Consider some item $i \in \items_{h \leq  \frac{\opt}{4}, \, w \leq (\frac{1}{2}+ 2 \eps' ) \D}$. 
    Suppose for the sake of contradiction that there is an item $i$ that is started at $\start(i)$ ends after $\D$. Since $w(i) < \big( \frac12 + 2 \eps'\big) \D$ it holds that $\start(i) > \big(\frac{1}{2}- 2 \eps'\big) \D$.
    Let $\tau$ be the last start time of an item from $\items_{h \leq \frac\opt4, \, w>(\frac{1}{2}+ 2 \eps') \D}$ in~$\start$. 
    
    Let $\tau_1 \in [0, \tau]$ be the first point in time, such that at each point $t \in [ \tau_1,\start(i))$ it holds that $h(\itemsatt{\start}{t}) > \big(\frac{3}{2}+\eps\big) \OPT -h(i)$.
    Note that $\tau_1$ has to exist since otherwise the item $i$ would have been started earlier by the algorithm.
    However in the case that $\tau$ is the last start point of a wide item in $\items_{h \leq \frac{\opt}{4} , \, w> (\frac{1}{2}+ 2 \eps' ) \D} $ it can happen that $\tau_1 = \tau$.
    If $\tau_1 \not = 0$, there exists a $\tau' <\tau_1$ such that, at each point $t \in (\tau',\tau_1)$, the packing has a height of at most $h(\itemsatt{\start}{t}) \leq \big(\frac{3}{2}+\eps\big) \OPT - h(i)$.
    Hence, there has to be an item $i_1$  that starts at $\tau_1$.
    Note that this item has to be an item from $\items_{h \leq \frac{\opt}{4}, \, w > (\frac{1}{2}+ 2 \eps') \D}$ since otherwise the item $i$ would have been packed earlier and instead of $i_1$.
    Since $i_1$ was not started earlier and since $h(\itemsatt{\start}{t}) \leq \big(\frac{3}{2}+\eps\big) \OPT - h(i)$ for any $t \in (\tau',\tau_1)$, its height is at least $h(i_1) > h(i)$. 
    We define $h_1$ as the combined height of all wide items in $\items_{h \leq \frac\opt4, \, w > (\frac{1}{2}+ 2\eps') \D}$ starting at $\tau_1$.
    We recursively define $\tau_j$ as the first point in time such that at each point $t \in [ \tau_j,\tau_{j-1})$ it holds that $h(\itemsatt{\start}{t}) > \big(\frac{3}{2}+\eps\big) \OPT - h_{j-1}$.
    Analogously, at $\tau_j$ there has to start an item $i_j$ with a height $h(j) > h_{j-1}$. 
    We define $h_j$ as the combined height of all wide items starting at $\tau_j$.
    Note that there can be at most $n$ such points $\tau_j$.

    Let $k \in \mathbb{N}$ be the largest index such that $\tau_k \geq \max \big\{ \start(i) - \big(\frac{1}{2}+ 2 \eps'\big)\D, 0 \big\}$.
    Hence, we have $\tau_j \geq \start(i) - \big(\frac{1}{2}+ 2 \eps'\big)\D$ for all $1 \leq j \leq k$.
    Define $\tau_{k+1} = \max \big\{ \start(i) - \big(\frac{1}{2}+ 2 \eps'\big)\D, 0 \big\}$.
   
    Now, we can lower-bound the area of all items. 
    On a high level, we will carefully analyze the area covered by items intersecting certain segments of the packing, which is clearly a lower bound on the area of all items. We will gradually increase this lower bound to reach a contradiction to the fact $\area(\items) \leq \D \cdot \opt$.
    
    From time $0$ to time $\tau_{k+1}$ the packing has height at least $\big(\frac{5}{4} +\eps\big)\OPT$, due to the same reasoning as in the previous claim.
    From time $\tau_{j+1}$ to time $\tau_{j}$, the packing has height at least $\big(\frac{3}{2} +\eps\big) \OPT - h_j$, for $j \in \{ 1, ..., k \}$.
    In total, the area in the segment from $\tau_{k+1}$ to $\tau_1$ adds up to 
    \begin{align*}
        \area(\items)  >& \sum_{j = 1}^{k}(\tau_j-\tau_{j+1})\bigg(\bigg(\frac{3}{2} +\eps\bigg) \OPT -h_j\bigg)\\
        =& (\tau_1 - \tau_{k+1}) \bigg(\frac{3}{2} +\eps\bigg) \OPT + \sum_{j = 1}^{k} h_j (\tau_{j+1} - \tau_j).
    \end{align*}
    Also including the segment from $0$ to $\tau_{k+1}$ with an area of at least $\tau_{k+1}\big(\frac{5}{4} +\eps\big)\OPT$ in the lower bound, we obtain that the total area contained in the segment from $0$ to $\tau_1$ is at least 
    \begin{align*}
        \area(\items)  >  \tau_1\bigg(\frac{3}{2} +\eps\bigg) \OPT + \sum_{j = 1}^{k} h_j (\tau_{j+1} - \tau_j) - \frac{1}{4}\tau_{k+1}\opt.
    \end{align*}
    Similarly from $\tau_1$ on up to the starting time $\start(i)$ of item $i$, the packing has height at least $\big(\frac{3}{2} +\eps\big) \OPT -h(i)$, increasing the lower bound to
    \begin{align*}
        \area(\items)  >  \start(i)\bigg(\frac{3}{2} +\eps\bigg) \OPT + \sum_{j = 1}^{k} h_j (\tau_{j+1} - \tau_j) + h(i)(\tau_1-\start(i))- \frac{1}{4}\tau_{k+1}\opt .
    \end{align*}
    
    Next, we consider the area right of $\start(i)$. From $\start(i)$ up to time $(1 - \eps') \D$ the packing has height at least $\big(\frac\opt2 + h(i)\big)$ due to the tall items.
    Furthermore the items in $\items_{w > (\frac{1}{2}+ 2\eps') \D}$ after time $\start (i)$ add to the total area of items.
    These items have a length of at least $\big(\frac{1}{2}+ 2\eps'\big) \D$ and hence items starting at $\tau_j$ end after $\tau_j + \big(\frac{1}{2}+2\eps'\big) \D$.
    To the right of $\start(i)$, these item parts add up to an additional area of at least $\sum_{j = 1}^k (\tau_j + \big(\frac{1}{2}+2\eps'\big) \D - \start(i))h_j$.

    Hence, the total area of the items in $\items$ is lower-bounded by
    \begin{align*}
        \area(\items) 
         > & \start(i)\bigg(\frac{3}{2} +\eps\bigg) \OPT + \sum_{j = 1}^{k} (\tau_{j+1} - \tau_j)h_j  + (\tau_1-\start(i))h(i)- \frac{1}{4}\tau_{k+1} \opt\\
         & +  \big((1 - \eps') \D - \start(i)\big)\bigg(\frac1{2}\opt  + h(i)\bigg) + \sum_{j = 1}^{k} \bigg(\tau_j + \bigg(\frac{1}{2}+2\eps'\bigg) \D - \start(i)\bigg)h_j\\
         = & \start(i)((1 +\eps) \OPT -2h(i)) + \sum_{j = 1}^{k} \bigg(\tau_{j+1} + \bigg(\frac{1}{2}+2\eps'\bigg) \D - \start(i)\bigg)h_j\\& +(\tau_1 + (1 - \eps') \D)h(i)
          - \frac{1}{4}\tau_{k+1} \opt + \frac{1}{2}(1 - \eps') \D \cdot\opt
    \end{align*}

Recall that $\tau_{k+1} = \max \{ \start(i) - \big(\frac12 + 2 \eps'\big)\D, 0 \}$.
We now distinguish whether $\tau_{k+1} = 0$ or $\tau_{k+1} = \start(i) - \big(\frac12 + 2 \eps'\big)\D$ (which is then $> 0$).
First, we consider the case that $\tau_{k+1} = \start(i) - \big(\frac12 + 2 \eps'\big)\D$ and obtain
    \begin{align*}       
        \area(\items) > 
        & \start(i)((1 +\eps) \OPT -2h(i)) + \sum_{j = 1}^{k} \bigg(\tau_{j+1} + \bigg(\frac{1}{2}+2\eps'\bigg) \D - \start(i)\bigg)h_j\\ &+(\tau_1 + (1 - \eps') \D)h(i)
         - \frac{1}{4}(\start(i) - \big(\frac12 + 2 \eps'\big)\D) \opt + \frac{1}{2}(1 - \eps') \D \cdot\opt\\
        = &
        \start(i)\bigg(\bigg(\frac{3}{4} +\eps \bigg) \OPT - 2h(i)\bigg) + \sum_{j = 1}^{k} \bigg(\tau_{j+1} + \bigg(\frac{1}{2}+2\eps'\bigg) \D - \start(i)\bigg)h_j\\
        &+(\tau_1 + (1 - \eps') \D)h(i)
        + \frac{5}{8} \D \cdot \opt
    \end{align*}    
    Using that~$\start(i) > \big(\frac12 - 2 \eps'\big)\D$,  $h(i) \leq \frac{1}{4}\opt$, and~$\tau_j \geq \start(i) - \big(\frac12 + 2 \eps'\big)\D$ for all~$j \leq k+1$, this yields
    \begin{align*} 
        \area(\items) > & \bigg(\frac{1}{2} - 2 \eps'\bigg) \D \bigg(\bigg(\frac{3}{4} +\eps \bigg) \OPT - 2h(i)\bigg)  +(\tau_1 + (1 - \eps') \D)h(i) 
        + \frac{5}{8} \D \cdot \opt\\
        = & \bigg(\frac{1}{2} - 2 \eps'\bigg) \bigg(\frac{3}{4} +\eps \bigg) \D \cdot \OPT  +(\tau_1 + (1 - \eps')\D- (1-4\eps') \D)h(i) 
        + \frac{5}{8} \D \cdot \opt\\
        = & \bigg(1+\frac{\eps}{2}-\eps'\bigg(\frac{3}{2} +2\eps \bigg)\bigg)\D \cdot \opt  + (\tau_1 + 3\eps' \D)h(i) 
    \shortintertext{With~$h(i) \geq 0$ and~\alertBound{$\eps' \leq \frac{\eps}{3 + 4 \eps}$}, we obtain}
        \area(\items)
        > & \D \cdot \OPT, 
    \end{align*}
    a contradiction.

    In the other case, if $\tau_{k+1} = 0$, we obtain
    \begin{align*}       
        \area(\items) > & \start(i)((1 +\eps) \OPT -2h(i)) + \sum_{j = 1}^{k} \bigg(\tau_{j+1} + \bigg(\frac{1}{2}+2\eps'\bigg) \D - \start(i)\bigg)h_j\\ 
        &+(\tau_1 + (1 - \eps') \D)h(i)
        + \frac{1}{2}(1 - \eps') \D \cdot\opt\\
    \shortintertext{Using that~$\start(i) > \big(\frac12 - 2 \eps'\big)\D$,~$\tau_j \geq \start(i) - \big(\frac12 + 2 \eps'\big)\D$ for all~$j \leq k+1$ and $2 h(i) \leq \frac\opt2$, this yields
    }
        \area(\items) > & \bigg( \bigg(\frac{1}{2} - 2 \eps'\bigg) \D \bigg) ((1 +\eps) \OPT -2h(i)) +(\tau_1 + (1 - \eps') \D)h(i)\\
        &+ \frac{1}{2}(1 - \eps') \D \cdot\opt\\
        \geq & \bigg(1+\frac{\eps}{2} - \frac{\eps'}{2}\bigg(5+4\eps\bigg)\bigg)\D \cdot \OPT+(\tau_1 + 3 \eps'\D)h(i)\\
    \shortintertext{With~$h(i) \geq 0$ and~\alertBound{$\eps' \leq \frac{\eps}{5 + 4 \eps}$}, we obtain}
        \area(\items)
        > & \D \cdot \OPT, 
    \end{align*}
    a contradiction.   
\end{proof}

\begin{proof}[Proof of \Cref{lem:caseA-main}]
    Combining \Cref{claim:case-A-main-proof:1,claim:case-A-main-proof:2,claim:case-A-main-proof:3} proves the lemma. 
\end{proof}

\section{Fusing gaps}
\label{sec:case_b}

In this section, we restructure a given optimal packing $\sopt$ and transform it into a forgiving packing. Here, we require \alertBound{$\lambda\in \big(0,\frac{1}{60}\big]$}.
Our main lemmas of this section are the following.

\begin{restatable}{lemma}{lemsmallgapatborder}
    \label{lem:small-gap-at-border}
     $\sopt$ is an optimal packing of $(\items, \D)$ such that a tall item $T$ satisfies $\sopt(T) \leq \big(\frac 12 -  \lambda\big)\D$ 
    and the gaps to the left of $T$ have a combined width between $\lambda \D$ and $2 \lambda \D$. Then, $(\items, \D)$ admits a forgiving packing. 
\end{restatable}

\begin{restatable}{lemma}{lemsmallgapcenter}
    \label{lem:case_b_center}\label{lem:small-gap-center}
    Assume $\sopt$ is an optimal packing of $(\items, \D)$ such that there is a sequence of consecutive gaps $[\ell_1, r_1), [\ell_2, r_2), ..., [\ell_k, r_k)$ with the following properties: 
    \begin{enumerate}
        \item Their combined width satisfies $\sum_{i = 1}^k r_i - \ell_i \in [\lambda \D,2 \lambda \D]$ and
        \item $r_k - \ell_1 \leq \big(\frac 15 - 4 \lambda\big) \D $.
    \end{enumerate}
Then, $(\items, \D)$ admits a forgiving packing. 
\end{restatable}

We will highlight the general idea of the restructuring algorithm first before giving more details in the respective subsections.
Consider a segment of the packing in which a total width of at least $\lambda$ and at most $2\lambda$ does not contain tall items. We remark that it is possible that ``combining'' multiple gaps, i.e., adding their respective widths, guarantees this condition.
To restructure the packing, we remove all items completely contained in this segment, place the tall items as well as the additional item $\iadd$ next to each other in the segment, and place the other removed items (with height at most $\frac\opt2$) on top of the packing to the left and to the right of this segment using 
Steinberg's algorithm.

We consider two cases depending on the location of the segments that we will restructure. 
In the first subsection, the segment starts at $0$ or ends at $\D$. 
For reasons of symmetry, we only discuss the case in which the gap starts at $0$, i.e., the left border.
In the second subsection, the segment is located around the center of the packing.

\subsection{A segment of width at most \texorpdfstring{$\big(\frac12 - \lambda\big)\D$}{(1/2-λ)D} starting at \texorpdfstring{$0$}{0}}\label{subsec:small_space_at_border} 

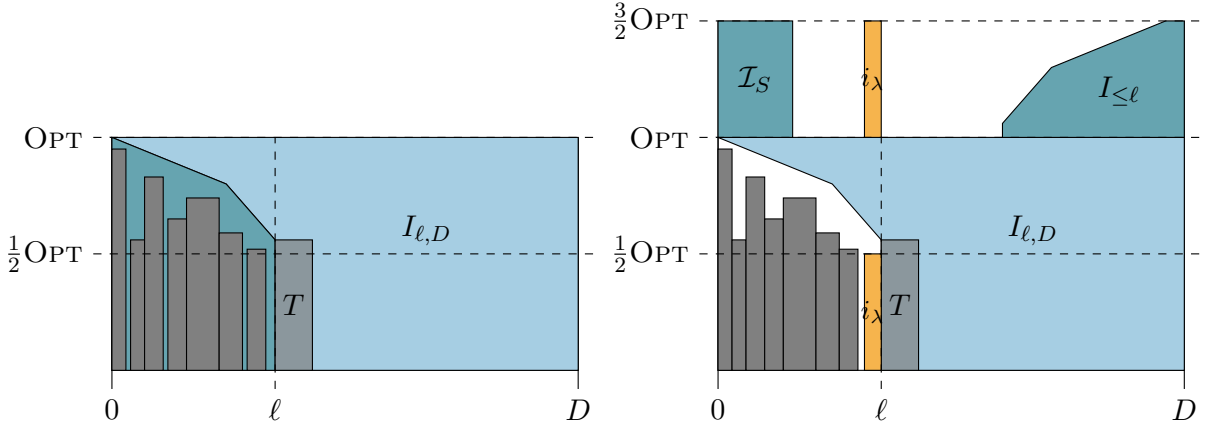
\begin{figure}
    \centering
    \resizebox{\textwidth}{!}{
    \begin{tikzpicture}
\pgfmathsetmacro{\w}{6}
\pgfmathsetmacro{\h}{3}
\pgfmathsetmacro{\os}{0.24}
\pgfmathsetmacro{\l}{0.35}
\pgfmathsetmacro{\r}{0.7}
\pgfmathsetmacro{\lamb}{0.04}

\draw (0,\h) -- (0,0) -- (\w,0) -- (\w,\h);
\draw (0*\w,-\os) node[below]{$0$}-- (0*\w,\h);
\draw (\w,-\os) node[below]{$\D$}-- (\w,\h);

\draw[fill = colorb] (\l*\w,0) -- (\w,0) --(\w,\h) --(0,\h) -- (0.7*\l*\w,0.8*\h)--(\l*\w,0.56*\h)--(\l*\w,0);
\node at (0.5*\l*\w+0.5*\w,0.6*\h){$I_{\ell,D}$};

\draw[fill = colord] (\l*\w,0) -- (0,0) --(0,\h)-- (0.7*\l*\w,0.8*\h) -- (\l*\w,0.56*\h) --(\l*\w,0);

\draw[fill = tallItemColor] (0,0) rectangle (0.03*\w,0.95*\h);
\draw[fill = tallItemColor] (0.04*\w,0) rectangle (0.07*\w,0.56*\h);
\draw[fill = tallItemColor] (0.07*\w,0) rectangle (0.11*\w,0.83*\h);
\draw[fill = tallItemColor] (0.12*\w,0) rectangle (0.16*\w,0.65*\h);
\draw[fill = tallItemColor] (0.16*\w,0) rectangle (0.23*\w,0.74*\h);
\draw[fill = tallItemColor] (0.23*\w,0) rectangle (0.28*\w,0.59*\h);
\draw[fill = tallItemColor] (0.29*\w,0) rectangle (0.33*\w,0.52*\h);
\draw[fill = tallItemColor!70!colorb] (\l*\w,0) rectangle node[midway]{$T$} (\l*\w+0.08*\w,0.56*\h);

\draw[dashed] (\l*\w,-\os) node[below]{$\ell$} -- (\l*\w,\h);

\draw[dashed] (-\os,0.5*\h) node[left]{$\frac{1}{2}\opt$} -- (\w+\os,0.5*\h) ;
\draw[dashed] (-\os,1*\h) node[left]{$\opt$} -- (\w+\os,1*\h) ;


\begin{scope}[xshift = 1.3*\w cm]

\draw (0,1.5*\h) -- (0,0) -- (\w,0) -- (\w,1.5*\h);
\draw (0*\w,-\os) node[below]{$0$}-- (0*\w,\h);
\draw (\w,-\os) node[below]{$\D$}-- (\w,\h);

\draw[fill = colorb] (\l*\w,0) -- (\w,0) --(\w,\h) --(0,\h) -- (0.7*\l*\w,0.8*\h)--(\l*\w,0.56*\h)--(\l*\w,0);
\node at (0.5*\l*\w+0.5*\w,0.6*\h){$I_{\ell,D}$};

\begin{scope}[xscale = -1, xshift=-\w cm +\lamb*\w cm,yshift=0.5*\h cm]
\draw[fill = colord] (\l*\w,0.5*\h) -- (-\lamb*\w,0.5*\h) --(-\lamb*\w,\h) -- (0,\h)-- (0.7*\l*\w,0.8*\h) -- (\l*\w,0.56*\h) --(\l*\w,0.5*\h);
\node at (0.1*\w,0.7*\h) {$I_{\leq \ell}$};
\end{scope}

\draw[fill = colord] (0,\h) rectangle node[midway]{$\items_S$}(4*\lamb*\w,1.5*\h);

\draw[fill = tallItemColor] (0,0) rectangle (0.03*\w,0.95*\h);
\draw[fill = tallItemColor] (0.03*\w,0) rectangle (0.06*\w,0.56*\h);
\draw[fill = tallItemColor] (0.06*\w,0) rectangle (0.10*\w,0.83*\h);
\draw[fill = tallItemColor] (0.10*\w,0) rectangle (0.14*\w,0.65*\h);
\draw[fill = tallItemColor] (0.14*\w,0) rectangle (0.21*\w,0.74*\h);
\draw[fill = tallItemColor] (0.21*\w,0) rectangle (0.26*\w,0.59*\h);
\draw[fill = tallItemColor] (0.26*\w,0) rectangle (0.3*\w,0.52*\h);
\draw[fill = tallItemColor!70!colorb] (\l*\w,0) rectangle node[midway]{$T$} (\l*\w+0.08*\w,0.56*\h);

\draw[fill = coloriadd] (\l*\w - 0.9*\lamb*\w,0) rectangle node[midway]{\small $\iadd$} (\l*\w,0.5*\h);
\draw[fill = coloriadd] (\l*\w - 0.9*\lamb*\w,\h) rectangle node[midway]{\small $\iadd$} (\l*\w,1.5*\h);

\draw[dashed] (\l*\w,-\os) node[below]{$\ell$} -- (\l*\w,\h);

\draw[dashed] (-\os,1.5*\h) node[left]{$\frac{3}{2}\opt$} -- (\w+\os,1.5*\h) ;
\draw[dashed] (-\os,0.5*\h) node[left]{$\frac{1}{2}\opt$} -- (\w+\os,0.5*\h) ;
\draw[dashed] (-\os,1*\h) node[left]{$\opt$} -- (\w+\os,1*\h) ;

\end{scope}

\end{tikzpicture}
    }
    \caption{An example repacking generated by \cref{alg_b_border} for a segment of width at most $\big(\frac12 - \lambda\big)\D$ starting at $0$ that has a width in $[\lambda \D,2\lambda \D]$ that is not covered by tall items.}
    \label{fig:many_tall_items_fig1}
\end{figure}

Given an optimal packing $\sopt$ of height~$\opt$, consider a tall item $T$ such that $\sopt(T) \leq \big(\frac12 - \lambda\big)\D$ and the total width in $[0,\sopt(T)]$ not covered by tall items is at least $\lambda \D$ and at most $2\lambda \D$.

Let $\ell = \sopt(T)$. We partition $\items$ into the following sets
\[I_{\leq \ell} = \iopt_{[0,\ell]} \setminus \items_{h > \frac\opt2}, \quad I_{\ell,\D} = \items_{\sopt(i)+w(i) > \ell}, \quad I_{\frac{\opt}{2},\leq\ell} = \iopt_{[0,\ell]} \cap \items_{h > \frac\opt2}\]
The algorithm starts by stretching the non-tall items $I_{\leq \ell}$ in the segment $[0,\sopt(T)]$ by at most $2 \lambda \D$ using the Left-Stretching Algorithm~\ref{alg_left_stretching}.
\Cref{alg_left_stretching} will remove the items $\items_S$ that~$\sopt$ packs entirely between tall items.
The stretched packing (for the remaining non-tall items) is then shifted to the right by $\D-\ell$ such that the last possible endpoint of an item in $I_{\leq \ell}$ is $\D$.
For the removed items $\items_S$, we use Steinberg's Algorithm to construct a packing $\start_S$ of height at most $\frac\opt2$ and width at most $8\lambda \D$.
This packing then is either started at $0$ or at $\ell$.
The tall items packed by~$\sopt$ in~$[0,\sopt(T))$ are shifted to the left to be packed consecutively, and the item $\iadd$ is packed starting at time~$\ell-\lambda\D$.
The items from the set $I_{\ell,\D}$ are not moved.
We summarize the algorithm in \cref{alg_b_border}, and an example-repacking can be found in \cref{fig:many_tall_items_fig1}.

\begin{algorithm}
\DontPrintSemicolon
\caption{Repacking when there is a suitable segment in $[0,\big(\frac12 - \lambda\big)\D]$}
\label{alg_b_border}
\textbf{Input:} $(\items, \sopt, T, \iadd)$ \tcp*[r]{using the above definitions}
\smallskip 
$\start \leftarrow \sopt$ \tcp*[r]{initialize the new packing}
$(\start', \items_S) \leftarrow$ output of \Cref{alg_left_stretching} with $\big(\sopt, \frac\opt2, \sopt(T), 0\big)$  \tcp*[r]{left-stretch $I_{\leq \ell}$}
\lForEach{$i \in I_{\leq \ell} \setminus \items_S$}{
    $\start(i) \leftarrow \start'(i) +\D-\ell$ 
}
$\start_S \leftarrow $ output of \steinberg$\big(\items_S,\frac\opt2\big)$ \;
\uIf{$\ell\geq 9\lambda\D$}{
    \lForEach(\tcp*[f]{place the items removed by the stretching}){$i \in \items_S$}{$\start(i) \leftarrow \start_S(i)$}
}
\Else{
    \lForEach(\tcp*[f]{place the items removed by the stretching}){$i \in \items_S$}{$\start(i) \leftarrow \start_S(i) + \ell$}
}
\smallskip 
\ForEach{$i \in I_{\frac{\opt}{2},\leq\ell}$}{ 
    $\start(i) \leftarrow w\big( \items_{h > \frac\opt2} \cap \iopt_{[0,\sopt(i)]} \big) $ \tcp*[r]{shift tall items to the left}
}
$\start(\iadd) \leftarrow \ell-\lambda\D$\;
return $\start$ \;
\end{algorithm}

\lemsmallgapatborder*

\begin{proof}
First observe that for any $t \in [0,\D]$ it holds that $h(\itemsatt{\start}{t} \cap (I_{\ell,\D} \cup I_{\frac{\opt}{2},\leq\ell})) \leq \opt$. This holds due to the fact that the height of $ \itemsatt{\start}{t} \cap I_{\ell,\D}$ is non-decreasing in the interval $[0,\ell]$. As the items in $I_{\frac{\opt}{2},\leq\ell}$ are shifted to the left and $h(\itemsatt{\sopt}{t} \cap (I_{\ell,\D} \cup I_{\frac{\opt}{2},\leq\ell})) \leq \opt$, the height of $\itemsatt{\start}{t} \cap (I_{\ell,\D} \cup I_{\frac{\opt}{2},\leq\ell})$ cannot be larger than $\opt$.

Next, consider the items from $I_{\leq \ell}$.
We start by observing that \Cref{lem:stretching} guarantees that~$\start'$ is indeed a packing for the items $I_{\leq \ell}\setminus\items_S$ in the segment $[-2\lambda\D, \ell]$ as the total width of gaps in the segment~$[0, \ell)$ is at most~$2 \lambda$.
As in $\sigma$ these items are shifted by $\D -\ell$ to the right compared to $\start'$, in $\sigma$ they are contained in the segment $[\D -\ell-2\lambda\D,\D] \subseteq \big[\big(\frac{1}{2}-\lambda\big)\D,\D\big]$ as $\ell \leq \big(\frac{1}{2}-\lambda\big)\D$.
Further, by \Cref{cor:steinberg_for_stretching}, $\start_S$ packs the items~$\items_S$ in the segment $[0, 8\lambda\D)$ with height~$\frac\opt2$ as their total area is bounded by $2\lambda\D\opt$ and the tallest item has a height of at most $\frac\opt2$. 

Depending on $\ell \geq 9\lambda \D$, the items~$\items_S$ are either not shifted or shifted by $\ell$ compared to $\start_S$.
Note that if $\ell \leq 9 \lambda \D$, it holds that $\ell +8\lambda\D \leq \D -\ell-2\lambda\D$ as \alertBound{$\lambda \leq \frac1{60} <  \frac{1}{28}$}.
Therefore, in any case, the items from $\items_S$ do not overlap with the items from $I_{\leq \ell}\setminus\items_S$.
Therefore at any time $t \in [0,\D]$ it holds that  $h(\itemsatt{\start}{t} \cap I_{\leq \ell}) \leq \frac{\opt}{2}$.
 
Finally, note that the last tall item from the set $I_{\frac{\opt}{2},\leq\ell}$ ends before $\ell-\lambda \D$ as their total width is at most $\ell-\lambda \D$. by the assumption on the combined width of the gaps in $[0,\ell)$. 
We separately analyze the height of~$\start$ in the segments 
$[0, \ell-\lambda\D)$, $[\ell-\lambda\D, \ell)$, and $[\ell, \D)$.

\noindent $\bullet \enspace \bm{[0, \ell-\lambda\D).} \enspace $ 
If $\ell < 9\lambda \D$ this segment only contains items from the set $I_{\ell,\D} \cup I_{\frac{\opt}{2},\leq\ell}$. As a consequence, by the above observation, the height of the packing in this segment is bounded by $\opt$.
On the other hand, if $\ell \geq 9\lambda \D$, the packing additionally contains the items from $\items_S$. As $\start_S$ has a height of at most $\frac{\opt}{2}$, the height of the packing in this segment is bounded by $\frac{3}{2}\opt$ in this case. 

\noindent $\bullet \enspace \bm{[\ell-\lambda\D,\ell).} \enspace $ 
After removing the tall items~$I_{\frac{\opt}{2},\leq\ell}$ (by shifting them to the left) and the items in $I_{\leq \ell}$, the only remaining items in the segment~$[\ell-\lambda\D,\ell)$ are the item $\iadd$ and the items belonging to~$\itemsoverlapt{\sopt}{\ell}$. 
\Cref{obs_height_G1G2} bounds the height of the latter set by~$\frac\opt2$.

\noindent $\bullet \enspace \bm{[\ell,\D).} \enspace $ 
This segment only contains the items from $I_{\leq \ell} \cup I_{\ell,\D}$. 
As $h(\itemsatt{\start}{t} \cap I_{\leq \ell}) \leq \frac{\opt}{2}$ and $h(\itemsatt{\start}{t} \cap I_{\ell,\D}) \leq \opt$, the height of the packing in this segment is bounded by $\frac{3}{2}\opt$.

\smallskip

    Observing that we have already argued that all items with $\start(i) \neq \sopt(i)$ are packed in $[0,\D)$ and using that $\sopt$ is an optimal and in particular $\D$-feasible packing, we conclude the proof of the lemma. 
\end{proof}

\subsection{A segment of width at most \texorpdfstring{$\big(\frac15 - 4 \lambda\big)\D$}{(1/5-4λ)D}}
\label{subsec:space_around_1/2}

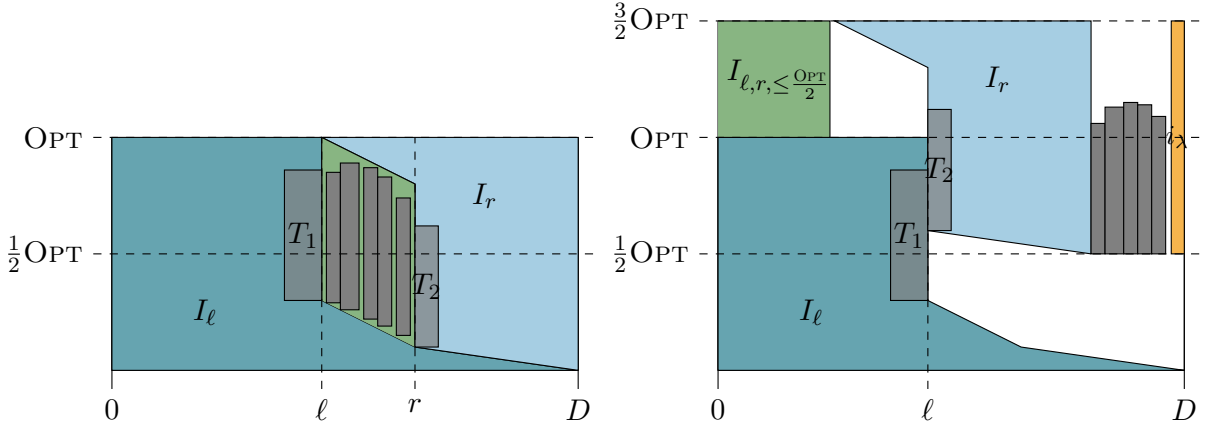
\begin{figure}
    \centering
    \resizebox{\textwidth}{!}{
    \begin{tikzpicture}
\pgfmathsetmacro{\w}{6}
\pgfmathsetmacro{\h}{3}
\pgfmathsetmacro{\os}{0.24}
\pgfmathsetmacro{\l}{0.45}
\pgfmathsetmacro{\r}{0.65}
\pgfmathsetmacro{\lamb}{0.04}

\draw (0,\h) -- (0,0) -- (\w,0) -- (\w,\h);
\draw (0*\w,-\os) node[below]{$0$}-- (0*\w,\h);
\draw (\w,-\os) node[below]{$\D$}-- (\w,\h);

\draw[fill = colord] (0,0)--(\w,0)--(\r*\w,0.1*\h) -- (\l*\w,0.3*\h) --(\l*\w,\h) --(0*\w,\h)--(0,0);
\node at (0.2*\w,0.25*\h) {$I_{\ell}$};
\draw[fill = colorb] (\w,0)--(\r*\w,0.1*\h)--(\r*\w,0.8*\h) -- (\l*\w,\h) --(\w,\h) --(\w,0);
\node at (0.8*\w,0.75*\h) {$I_{r}$};
\draw[fill = colore] (\r*\w,0.1*\h)--(\r*\w,0.8*\h)--(\l*\w,\h)--(\l*\w,0.3*\h);


\draw[fill = tallItemColor!70!colord] (\l*\w-0.08*\w,0+0.3*\h) rectangle node[midway]{$T_1$} (\l*\w,0.56*\h+0.3*\h);

\draw[fill = tallItemColor] (\l*\w+0.01*\w,+0.29*\h) rectangle (\l*\w+0.04*\w,0.56*\h+0.29*\h);
\draw[fill = tallItemColor] (\l*\w+0.04*\w,+0.26*\h) rectangle (\l*\w+0.08*\w,0.63*\h+0.26*\h);
\draw[fill = tallItemColor] (\l*\w+0.09*\w,+0.22*\h) rectangle (\l*\w+0.12*\w,0.65*\h+0.22*\h);
\draw[fill = tallItemColor] (\l*\w+0.12*\w,+0.19*\h) rectangle (\l*\w+0.15*\w,0.64*\h+0.19*\h);
\draw[fill = tallItemColor] (\l*\w+0.16*\w,+0.15*\h) rectangle (\l*\w+0.19*\w,0.59*\h+0.15*\h);

\draw[fill = tallItemColor!70!colorb] (\r*\w,+0.1*\h) rectangle node[midway]{$T_2$}(\r*\w+0.05*\w,0.52*\h+0.1*\h);

\draw[dashed] (\l*\w,-\os) node[below]{$\ell$} -- (\l*\w,\h);
\draw[dashed] (\r*\w,-\os) node[below]{$r$} -- (\r*\w,\h);

\draw[dashed] (-\os,0.5*\h) node[left]{$\frac{1}{2}\opt$} -- (\w+\os,0.5*\h) ;
\draw[dashed] (-\os,1*\h) node[left]{$\opt$} -- (\w+\os,1*\h) ;


\begin{scope}[xshift = 1.3*\w cm]

\draw (0,1.5*\h) -- (0,0) -- (\w,0) -- (\w,1.5*\h);
\draw (0*\w,-\os) node[below]{$0$}-- (0*\w,\h);
\draw (\w,-\os) node[below]{$\D$}-- (\w,\h);

\draw[fill = colord] (0,0)--(\w,0)--(\r*\w,0.1*\h) -- (\l*\w,0.3*\h) --(\l*\w,\h) --(0*\w,\h)--(0,0);
\draw[fill = tallItemColor!70!colord] (\l*\w-0.08*\w,0+0.3*\h) rectangle node[midway]{$T_1$} (\l*\w,0.56*\h+0.3*\h);
\node at (0.2*\w,0.25*\h) {$I_{\ell}$};

\begin{scope}[xshift=\l*\w cm -\r*\w cm, yshift=0.5*\h cm]
\draw[fill = colorb] (\w,0)--(\r*\w,0.1*\h)--(\r*\w,0.8*\h) -- (\l*\w,\h) --(\w,\h) --(\w,0);
\draw[fill = tallItemColor!70!colorb] (\r*\w,+0.1*\h) rectangle node[midway]{$T_2$}(\r*\w+0.05*\w,0.52*\h+0.1*\h);
\node at (0.8*\w,0.75*\h) {$I_{r}$};
\end{scope}

\draw[fill = colore] (0*\w,\h)--(\r*\w-\l*\w+\lamb*\w,\h)--(\r*\w-\l*\w+\lamb*\w,1.5*\h)--(0*\w,1.5*\h);
\node at (0.13*\w,1.25*\h) {$I_{\ell,r,\leq\frac{\opt}{2}}$};

\draw[fill = tallItemColor] (\w+\l*\w-\r*\w+0.0*\w,+0.5*\h) rectangle (\w+\l*\w-\r*\w+0.03*\w,0.56*\h+0.5*\h);
\draw[fill = tallItemColor] (\w+\l*\w-\r*\w+0.03*\w,+0.5*\h) rectangle (\w+\l*\w-\r*\w+0.07*\w,0.63*\h+0.5*\h);
\draw[fill = tallItemColor] (\w+\l*\w-\r*\w+0.07*\w,+0.5*\h) rectangle (\w+\l*\w-\r*\w+0.10*\w,0.65*\h+0.5*\h);
\draw[fill = tallItemColor] (\w+\l*\w-\r*\w+0.10*\w,+0.5*\h) rectangle (\w+\l*\w-\r*\w+0.13*\w,0.64*\h+0.5*\h);
\draw[fill = tallItemColor] (\w+\l*\w-\r*\w+0.13*\w,+0.5*\h) rectangle (\w+\l*\w-\r*\w+0.16*\w,0.59*\h+0.5*\h);

\draw[fill = coloriadd] (\w - 0.7*\lamb*\w,0.5*\h) rectangle node[midway]{\small $\iadd$} (\w,1.5*\h);

\draw[dashed] (\l*\w,-\os) node[below]{$\ell$} -- (\l*\w,\h);

\draw[dashed] (-\os,1.5*\h) node[left]{$\frac{3}{2}\opt$} -- (\w+\os,1.5*\h) ;
\draw[dashed] (-\os,0.5*\h) node[left]{$\frac{1}{2}\opt$} -- (\w+\os,0.5*\h) ;
\draw[dashed] (-\os,1*\h) node[left]{$\opt$} -- (\w+\os,1*\h) ;

\end{scope}

\end{tikzpicture}
    }
    \caption{An example repacking generated by \cref{alg_b_center} for a segment of width at most $\big(\frac15 - 4\lambda\big)\D$ such that the total width of gaps contained in the segment is in $[\lambda \D,2\lambda \D]$.}
    \label{fig:many_tall_items_fig2}
\end{figure}

Let $\sopt$ be an optimal packing of $(\items, \D)$ and suppose that there is a sequence of consecutive gaps $[\ell_1, r_1), [\ell_2, r_2), ..., [\ell_k, r_k)$ with the properties that (i) the combined width satisfies $\sum_{i = 1}^k r_i - \ell_i \in [\lambda \D,2 \lambda \D]$ and (ii) $r_k - \ell_1 \leq \frac 15 - 4 \lambda$. Our goal is to show that then $(\items, \D)$ admits a forgiving packing. 

The above assumptions imply that there are two tall items~$T_1$ and~$T_2$ such that, for~$\ell = \sopt(T_1) + w(T_1)$ and~$r = \sopt(T_2)$, it holds~$\eta := \frac{r - \ell}{\D} \leq \frac15 - 4 \lambda$,~$r \leq (1-\lambda)\D$, and~$\eta \D - w\big(\items_{h > \frac\opt2} \cap \iopt_{[\ell,r)}\big) \in [\lambda\D, 2 \lambda\D)$.
Note that \alertBound{$\lambda \leq \frac{1}{60} < \frac1{25}$} implies $\eta \geq \lambda$. 
We partition the items $\items$ into the following sets
\[
    I_{\ell}  := \items_{\sopt(i) \leq \ell}
    , \hspace{1ex}
    I_{r} := \items_{r < \sopt(i) +w(i)} \setminus \Gal
    , \hspace{1ex}
    I_{\ell,r,\leq\frac\opt2} := \items_{h\leq \frac\opt2} \cap \iopt_{[\ell, r]}
    , \hspace{1ex}
    I_{\ell,r,>\frac\opt2} := \items_{h>\frac\opt2} \cap \iopt_{[\ell, r]}.
\]

The high-level idea of \Cref{alg_b_center} is to remove all items between $\ell$ and $r$, shift everything that starts after $\ell$ and ends after $r$ to the left by $\eta\D$ to ``close'' the just created gap, and place the removed tall items and the extra item $\iadd$ into the
now constructed segment of height at most $\frac\opt2$ at the end of the packing. 
The non-tall items $I_{\ell,r,\leq\frac\opt2}$ between $\ell$ and $r$ are stretched with \cref{alg_stretching} such that their combined height does not exceed $\frac\opt2$ anywhere.
These items are packed at the beginning of the packing.

\begin{algorithm}[t]
\DontPrintSemicolon
\caption{Repacking when there is a space with width at most~$\big(\frac15 - 4 \lambda)\D$ around~$\frac\D2$}
\label{alg_b_center}
\textbf{Input:} $(\items, \sopt, \ell,r, \Gal, I_{r}, I_{\ell,r,<\frac\opt2}, I_{\ell,r,>\frac\opt2},\iadd)$ \tcp*[r]{using the definitions from above}
\smallskip
$\start \leftarrow \sopt$ \tcp*[r]{initialization (final starting times of $\Gal$)}
$(\start', \items_S) \leftarrow$ output of \Cref{alg_stretching} with $\big(\sopt, \frac12 \opt, \ell,r\big)$ \tcp*[r]{stretch $I_{\ell,r,<\frac\opt2}$}
\lForEach(\tcp*[f]{pack $I_{\ell,r,<\frac\opt2}$}){$i \in I_{\ell,r,<\frac\opt2} \setminus \items_S$}{
    $\start(i) \leftarrow \start'(i) -\ell $
}
$\start_S \leftarrow $ output of \steinberg$\big(\items_S,\frac\opt2\big)$ \;
\lForEach(\tcp*[f]{start $\items_S$ at $(\eta + 2 \lambda)\D$}){$i \in \items_S$}{$\start(i) \leftarrow \start_S(i) + (\eta + 2 \lambda)\D$}
\smallskip 
\lForEach(\tcp*[f]{shift $I_{r}$ to the left by $\eta \D$}){$i \in I_{r}$}{
    $\start(i) \leftarrow \sopt(i) - \eta \D$ 
}
\lForEach(\tcp*[f]{pack $I_{\ell,r,>\frac\opt2}$}){$i' \in I_{\ell,r,>\frac\opt2}$}{
    $\start(i') \leftarrow (1- \eta)\D + w\big(I_{\ell,r,>\frac\opt2} \cap \iopt_{\sopt(i) < \sopt(i')}\big)$  
}
$\start(\iadd)  \leftarrow (1- \eta)\D + w(I_{\ell,r,>\frac\opt2})$ \tcp*[r]{pack $\iadd$}
\smallskip
\textbf{return} $\start$ \;
\end{algorithm}

\lemsmallgapcenter*

\begin{proof}
    Without loss of generality we assume that $\D - r_k \leq \ell_1$ (otherwise we simply mirror the packing). Let $T_1$ be the tall item starting at $\ell_1$ and let $T_2$ be the tall item starting at $r_k$. Furthermore, define $\ell_1 = \ell$ and $r_k = r$.
    We show that \Cref{alg_b_center} with input~$(\items, \sopt, \ell,r, \Gal, I_{r}, I_{\ell,r,<\frac\opt2}, I_{\ell,r,>\frac\opt2},\iadd)$ creates a packing~$\start$ with total height bounded by~$\frac32 \opt$ that packs the extra item $\iadd$.

    By assumption~$\ell \geq \big(\frac12 - \frac\eta2)\D \geq \big(\frac25 + 2 \lambda)\D \geq \eta\D$. 
    Hence, for any item~$i \in I_r$, we have~$\start(i) = \sopt(i) - \eta\D \geq \ell - \eta \D \geq \big(\frac15 + 6 \lambda\big) \D$. 
    Further, the only items with $\sopt(i) = \start(i)$ are the items $\Gal$. 
    We separately analyze the height of the packing in $\big[0, \big(\frac15 + 6 \lambda\big)\D\big)$, $\big[\big(\frac15 + 6 \lambda\big)\D, \ell\big)$, and $[\ell, \D)$. 

    \noindent $\bullet \enspace \bm{\big[0, \big(\frac15 + 6 \lambda\big)\D\big).} \enspace $ 
    By assumption, the total width not covered by tall items in the segment~$[\ell, r)$ is at most~$2 \lambda \D$. 
    Hence, by \Cref{lem:stretching},~$\start'$ is a feasible packing for the items in~$I_{\ell,r,\leq\frac\opt2} \setminus \items_S$ with height~$\frac\opt2$ in the segment $[\ell, r + 2\lambda\D)$. 
    Further, \Cref{cor:steinberg_for_stretching} ensures that $\start_S$ packs~$\items_S$ with height~$\frac\opt2$ in the segment $[0, 8\lambda\D)$.      
    Hence, the items in~$I_{\ell,r,\leq\frac\opt2}$ satisfy~$\start(i) + w(i) \leq (\eta + 10 \lambda)\D \leq \big(\frac15 + 6 \lambda\big)\D$. 
    By the above observation on~$\start(i)$ for~$i \in I_r$, the only items (besides $I_{\ell,r,\leq\frac\opt2}$) that are packed in the current segment satisfy $\start(i) = \sopt(i)$ and such items have a height of $\opt$. 
    Therefore, the lemma statement holds in the current segment. 

    \noindent $\bullet \enspace \bm{\big[ \big(\frac15 + 6 \lambda\big)\D, \ell\big).} \enspace $ 
    By \Cref{obs_height_G1G2},~$h(I_r) \leq \frac12\opt$ for all $t < r$. 
    Hence, shifting~$I_r$ to the left by at most~$\eta\D = (r-\ell)\D$ increases the height in~$[\ell - \eta\D, \ell) \subseteq \big[\big(\frac15 + 6 \lambda)\D, \ell\big)$ by at most~$\frac12\opt$ when compared to~$\sopt$.
    
    \noindent $\bullet \enspace \bm{[\ell,\D).} \enspace $ 
    \Cref{obs_height_G1G2} guarantees~$h(\Gal) \leq \frac12 \opt$.     
    For~$t \in [\ell, (1-\eta)\D)$, we have~$h(\itemsatt{\start}{t}) \leq h(\Gal) + h(\itemsatt{\sopt}{t + \eta \D}) \leq  \frac32 \opt$. For~$t \in [(1-\eta)\D, (1-\lambda)\D)$, we have~$h(\itemsatt{\start}{t}) \leq h(\Gal) + \max_{i \in G_5} h(i) \leq \frac32 \opt$. 
    
    Further,~$w(I_{\ell,r,>\frac\opt2}) \leq (\eta - \lambda)\D$ by assumption, which implies that~$\start(i) + w(i) \leq (1 - \lambda)\D$ for~$i \in I_{\ell,r,>\frac\opt2}$. Hence without $\iadd$ the maximal height in the segment~$[(1-\lambda)\D, \D)$ is at most~$h(\Gal) \leq \frac12 \opt$.
    Therefore, adding the item $\iadd$ does not increase the height of the packing above $\frac{3}{2}\opt$ in the current segment. 

    \smallskip
    It remains to argue that $\start$ is a feasible packing, i.e., that all items are packed in $[0,\D)$. 
    For $\Gal \cup \itemsT{\sopt}_{[0,\ell]}$, the feasibility follows from the feasibility of $\sopt$ since $\sopt(i) = \start(i)$. 
    For $I_r$, we observe that $\start(i) = \sopt(i) - \eta \D \geq 0$ by the above observation on $\ell$. Since their starting times only decrease, they finish before $\D$ if $\sopt$ finishes them before~$\D$. 
    For $I_{\ell,r,\leq\frac\opt2}$, we have already argued that they are packed in $[0, (\eta + 10\lambda)\D) \subset [0,\D)$ by $\start$. 
    Similarly, $I_{\ell,r,>\frac\opt2}$ is packed in $[(1-\eta)\D, (1-\lambda)\D)$ as discussed above, which concludes the proof.    
\end{proof}

\section{A gap with width in \texorpdfstring{$\big[\lambda\D, (\frac12 - 3\lambda) \D \big]$}{[λD,(1/2-3λ)D]}}\label{subsec:between_lambda_and_1/2}

In this section, we prove the following result.

\begin{lemma}
    \label{lem:full-case-C}
    Assume $\sopt$ is an optimal packing of $(\items, \D)$ with a gap $[\ell,r)$ of width $r - \ell \in [\lambda \D, (\frac 12 - 3 \lambda) \D]$. 
    Then, $(\items, \D)$ admits a forgiving packing.
\end{lemma}
Throughout we consider the case that there is some optimal solution satisfying the condition of the above lemma.
These conditions imply that 
 there are two tall items $T_1$ and $T_2$ without any tall item between them such that $\frac{r - \ell}{\D} := \eta \in \big[\lambda, \frac12 - 3\lambda \big]$, where $\ell = \sopt(T_1) + w(T_1)$ and $r = \start(T_2)$. 
We assume w.l.o.g. that $\D - r \leq \ell$, i.e., the gap is -- if at all -- closer to $\D$ than to $0$. 
The proof of \Cref{lem:full-case-C} is divided into two cases: In the first we assume that $0 < \ell < r < \D$ (analyzed in \Cref{lem_case_C}) and in the second we assume that $\ell = 0$ or $r = \D$ (w.l.o.g.\ we assume $r = \D$; analyzed in \Cref{lem:medium-gap-at-border}).
\Cref{lem:full-case-C} immediately follows from \Cref{lem_case_C} and \Cref{lem:medium-gap-at-border}.

\paragraph*{Repacking a mountain.}

For some of the subcases, we use a new repacking routine. 
This new routine is based on a packing~$\sopt$ of height~$\opt$. It is used when there are items completely packed in the gap, i.e., in the segment $[\ell, r)$, that
have a total height of at least $\frac{\opt}{2}$ for a width of at least $\lambda \D$. 
Informally, this set of items is called a ``mountain".
We start by describing this repacking routine and then consider the general repacking if there is a gap as described above.

For some values $0\leq a<b \leq \D$ with $b - a \geq \lambda \D$ 
let $\itemsM \subseteq \items_{h \leq \frac\opt2} \cap \mountainininterval{\sopt}{a}{b}$ such that $h(\itemsM) \geq \frac\opt 2$. \Cref{alg_mountain_repack} is a procedure to repack $\itemsM$.  
The idea of the repacking method is the following:
We greedily take items from $\itemsM$ until the last item increases the total height of the chosen items above $\frac{\opt}{2}$. 
The first chosen items are then started at $0$ while the last chosen item is positioned elsewhere such that it neither intersects with the first set of items nor with the interval $[a,b]$. (Usually, it is shifted to the right).

\begin{algorithm}[t]
\DontPrintSemicolon
\caption{Repacking a mountain}
\label{alg_mountain_repack}
\textbf{Input:}  $(\items,\itemsM,\sopt,\tau_{\text{start}})$\;
\smallskip
$\start \leftarrow \sopt$\;
sort $\itemsM$ by non-decreasing starting points $\start(i)$ \;
\ForEach{$i \in \itemsM$}{
    $\start(i) \leftarrow 0$ \tcp*[r]{start item at $0$}
    \If(\tcp*[f]{test if exceeding height bound}){$h(\start) > \frac32 \opt$}{
        $\start(i) \leftarrow \tau_{\text{start}}$\;
        \textbf{break} \;
    }    
}
\textbf{return} $\start$ \;
\end{algorithm}

\begin{lemma}\label{lem:mountain_repack}
    Let $\bar \itemsM  \subseteq \itemsM := \items_{h \leq \frac\opt2} \cap \mountainininterval{\sopt}{a}{b}$ be the set of items repacked by \Cref{alg_mountain_repack} and let $\bar i \in \bar \itemsM$ be the last repacked item. If $\max_{i \in \itemsM \setminus \{\bar i\}} w(i) \leq a$ and $b \leq \start(\bar i) \leq \D - w(\bar i)$, then $h(\start) \leq \frac 32 \opt$ and $\max_{\tau \in [a,b)} h(\itemsT{\start}(\tau)) \leq \frac\opt2$. 
\end{lemma}

\begin{proof}
    We observe that by assumption, no repacked item in $\bar \itemsM$ overlaps the segment~$[a,b)$. 

    By construction, before repacking the last item, the height of~$\packing$ does not exceed~$\frac32\opt$ anywhere. This implies the claim for the segment $[0, a]$ as the last item~$\bar i$ is then actually packed to start at $\start(\bar i) \geq b > a$. Note that the last item~$\bar i$ starts before $\D - w(\bar i)$, implying that it can be feasibly packed. Since we only repack non-tall items, the maximal height in~$[b, \D)$ is at most $\frac32\opt$ in~$\packing$. 

    It remains to show the claim for the segment $[a,b)$. In~$\sopt$, the total height in $[a,b)$ was at most $\opt$. As placing all items in $\itemsM$ starting at $0$ would have increased the total height beyond $\frac32\opt$, this implies that the total height of repacked items is at least $\frac\opt2$. By assumption, all these items completely cover the segment $[a,b)$ in~$\sopt$, implying that after repacking the maximal height in this segment is at most $\frac\opt2$.
\end{proof}

\paragraph*{Repacking the optimal packing.}

In the following, we describe our repacking procedure \cref{alg_c}.
Suppose there is a segment $[\ell,r)$ that contains no tall item such that $\start(T_1)+w(T_1) = \ell$ and $\start(T_2) = r$ for two tall items $T_1$ and $T_2$. We assume further that $\frac{r - \ell}{\D} = \eta \in \big[\lambda, \frac12 - 3\lambda\big]$ and $\D - r \leq \ell$.

To describe the algorithm, we define three different sets of non-tall items 
\[
\Gal := \itemsoverlapt{\sopt}{\ell}
, \quad 
\Ga := \itemsoverlapt{\sopt}{r} \setminus \Gal 
, \text{ and } 
\Gb := \items_{h\leq \frac\opt2} \cap \iopt_{[\ell, r + \lambda \D)} .
\]
Note that for $r \geq  (1-\lambda)\D$ it holds that $\Ga = \emptyset$.

The algorithm tries to pack the extra item $\iadd$ in the segment $[\ell, r)$ or, if this is not possible, in the segment $[(1 - \lambda)\D, \D)$. 
The exact repacking procedure and, thus, the position of $\iadd$ depends on the items inside the gap as well as the ones overlapping the gap from the left or the right. 
We distinguish four different cases. 
For some cases, the algorithm uses \cref{alg_mountain_repack}. 

We define
\begin{align*}
\itemsM_1 &= \Gb \cap \mountainininterval{\sopt}{\big(\frac12 + \lambda\big)\D}{\big(\frac12 + 2\lambda\big)\D}   
\text{ and }\\
\itemsM_2 &= \Gb \cap  \mountainininterval{\sopt}{r - 2 \lambda \D}{r-\lambda\D} \, .
\end{align*}

If the items $\itemsM_1$ 
have a height of at least $\frac{\opt}{2}$, then the algorithm repacks these items with \cref{alg_mountain_repack}, creating an open space for $\iadd$ in the segment $\big[\big(\frac12 + \lambda\big)\D,\big(\frac12 + 2\lambda\big)\D\big)$.
Else, i.e., if their height is less than $\frac{\opt}{2}$, and if the items $\itemsM_2$
have a height of at least $\frac{\opt}{2}$, then the algorithm repacks these items with \cref{alg_mountain_repack}, creating an open space for $\iadd$ in the segment $[r-2 \lambda\D,r-\lambda\D)$.
Otherwise, the algorithm shifts the items $\itemsM_2$ by $\ell$ to the left, implying that no such item starts before $0$ and each such item ends at or before $(\eta+\lambda)\D$.
The items completely contained in the segment $[r-2 \lambda\D,r+\lambda\D]$ are removed and we use Steinberg's Algorithm to create a packing of height at most $\frac{\opt}{2}$ and width at most $12 \lambda \D$ for these iems. 
The only items remaining in the segment $[r- \lambda\D,r)$ are the items from $\Gal \cup \Ga$.
If for all $t \in [r - \lambda\D,r)$ their total height is at most $\frac{\opt}{2}$, the item $\iadd$ can be packed in $[r - \lambda\D,r)$. 
Otherwise, the algorithm shifts all the items that start right of $r$ to the left by $\lambda \D$.
In this case, the extra item $\iadd$ can be packed in $[(1-\lambda)\D, \D)$.

\begin{algorithm}[t]
\DontPrintSemicolon
\caption{Repacking when $\sopt$ has a gap with width in $\big[\lambda\D, \big(\frac12 - 3\lambda\big)\D\big]$}
\label{alg_c}
\textbf{Input:} $(\items,\sopt,\ell,r, \Gal, \Ga, \Gb,\itemsM_1,\itemsM_2,\iadd)$ \tcp*[r]{using above definitions} 
\smallskip
$\start \leftarrow \sopt$ \tcp*[r]{initialize the new packing}
\smallskip 
\nl\label{alg_c_midmountain} \uIf{$h(\itemsM_1) \geq \frac12 \opt$}{
$\start \leftarrow$ result of \cref{alg_mountain_repack} with 
$\big(\items, \itemsM_1, \sopt, \big(\frac12 + 2 \lambda\big)\D)$ \tcp*[r]{repack mountain $\itemsM_1$}
$\start(\iadd) = \big(\frac12 + \lambda\big)\D$
}
\nl\label{alg_c_rightmountain}\uElseIf{$h(\itemsM_2) \geq \frac12 \opt$}{
$\start \leftarrow$ result of \cref{alg_mountain_repack} with 
$\big(\items, \itemsM_2, \sopt, (\eta + \lambda)\D)$ \tcp*[r]{repack mountain $\itemsM_2$}
$\start(\iadd) = r-2 \lambda\D$\;
}
\Else{
    \lForEach(\tcp*[f]{shift items to the left by $\ell$ }){$i \in \itemsM_2
    $}{
        $\start(i) \leftarrow \sopt(i) - \ell$ 
    }
    $\start' \leftarrow$ \steinberg$\big(\Gb \cap \iopt_{[r-2\lambda \D, r+\lambda \D]}, \frac\opt2\big)$ \;
    \lForEach{$i \in \items_S$}{
        $\start(i) \leftarrow \start'(i) + (\eta + \lambda) \D$ 
    }
    $\start(\iadd) = r-\lambda\D$\;
    \nl\label{alg_c_largeoverlap}\If{$h\big( \itemsatt{\sopt}{t} \cap (\Gal \cup \Ga) \big) > \frac12\opt$ for some $t\in [r-2\lambda\D, r-\lambda\D)$ }{
        \ForEach{$i \in \items_{\sopt(i) \geq r}$}{
            $\start(i) \leftarrow \sopt(i) - \lambda \D$ \tcp*[r]{shift items to the left by $\lambda$ }
        }
        $\start(\iadd) = \big(1- \lambda\big)\D$
    }
}
\textbf{return} $\packing$ \; 
\end{algorithm}

\begin{figure}
    \centering
    \resizebox{\textwidth}{!}{
    \begin{tikzpicture}[every node/.style={font=\small}]
\pgfmathsetmacro{\w}{6}
\pgfmathsetmacro{\h}{3}
\pgfmathsetmacro{\os}{0.24}
\pgfmathsetmacro{\l}{0.4}
\pgfmathsetmacro{\r}{0.7}
\pgfmathsetmacro{\lamb}{0.04}

\draw [fill = white!95!black] (0,0) -- (\l*\w,0.3*\h) --(\l*\w,\h) --(0,\h) -- (0,0);
\draw [fill = white!95!black] (\w,0) -- (\r*\w,0.1*\h) --(\r*\w,0.62*\h) --(\r*\w+0.7*\lamb*\w,0.62*\h) -- (\r*\w+0.7*\lamb*\w,0.7*\h) -- (\r*\w+0.5*\lamb*\w,0.7*\h) -- (\r*\w+0.5*\lamb*\w,0.75*\h) -- (\r*\w+\lamb*\w,0.75*\h)-- (\r*\w+\lamb*\w,0.85*\h) -- (\w,\h)--(\w,0);
\draw [fill = colorb] (\l*\w,0.85*\h) -- (\l*\w,0.95*\h) -- (\r*\w,0.85*\h)--(\l*\w,0.85*\h);
\draw [fill = colorb] (\l*\w,0.62*\h) -- (\l*\w,0.75*\h) -- (\l*\w+0.1*\w,0.75*\h)--(\l*\w+0.1*\w,0.7*\h) -- (\l*\w+0.13*\w,0.7*\h)--(\l*\w+0.13*\w,0.62*\h) -- (\l*\w,0.62*\h);
\draw [fill = colorb] (\l*\w,0.48*\h) -- (\l*\w,0.3*\h) -- (\r*\w,0.1*\h) -- (\r*\w,0.62*\h) -- (\r*\w-0.1*\w,0.62*\h)--(\r*\w-0.1*\w,0.55*\h)--(\r*\w-0.01*\w,0.55*\h)--(\r*\w-0.01*\w,0.48*\h)--(\r*\w-0.06*\w,0.48*\h)--(\r*\w-0.06*\w,0.4*\h)--(\r*\w-0.04*\w,0.4*\h)--(\r*\w-0.04*\w,0.3*\h)--(\r*\w-0.02*\w,0.3*\h)--(\r*\w-0.02*\w,0.23*\h)--(0.5*\w+0.5*\lamb*\w,0.23*\h)--(0.5*\w+0.5*\lamb*\w,0.3*\h) --(\l*\w+0.08*\w,0.3*\h) --(\l*\w+0.08*\w,0.4*\h)--(\l*\w+0.03*\w,0.4*\h) --(\l*\w+0.03*\w,0.48*\h) --(\l*\w,0.48*\h);

\draw[fill = coloraa] (0,0) -- (\l*\w,0.3*\h) -- (\r*\w,0.1*\h) -- (\w,0)--(0,0);
\node at (0.8*\l*\w,0.1*\h) {$\Gal$};

\draw[fill = colorab] (\w,\h) -- (\l*\w,\h) -- (\l*\w,0.95*\h) -- (\r*\w,0.85*\h)-- (\r*\w+\lamb*\w,0.85*\h) -- (\w,\h);
\node at (\r*\w+0.08*\w,0.94*\h) {$\Ga$};

\draw[fill = colorm] (\l*\w,0.85*\h) rectangle (\r*\w+\lamb*\w,0.75*\h);
\draw[fill = colorm] (\l*\w+0.1*\w,0.75*\h) rectangle (\r*\w+0.5*\lamb*\w,0.7*\h);
\draw[fill = colorm] (\l*\w+0.13*\w,0.7*\h) rectangle (\r*\w+0.7*\lamb*\w,0.62*\h);
\draw[fill = colorm] (\l*\w,0.62*\h) rectangle (\r*\w-0.1*\w,0.55*\h);
\draw[fill = colorm] (\l*\w,0.55*\h) rectangle (\r*\w-0.01*\w,0.48*\h);
\draw[fill = colorm] (\l*\w+0.03*\w,0.48*\h) rectangle (\r*\w-0.06*\w,0.4*\h);
\draw[fill = colorm] (\l*\w+0.08*\w,0.4*\h) rectangle (\r*\w-0.04*\w,0.3*\h);
\draw[fill = colorm] (0.5*\w+0.5*\lamb*\w,0.3*\h) rectangle (\r*\w-0.02*\w,0.23*\h);
\node at (0.8*\l*\w+0.2*\r*\w,0.69*\h) {$\Gb$};

\draw[fill = gray] (\l*\w-0.01*\w,0.3*\h) rectangle (\l*\w,\h);

\draw[fill = gray] (\r*\w+0.02*\w,0.62*\h) rectangle (\r*\w,0.1*\h);

\draw (0,\h) -- (0,0) -- (\w,0) -- (\w,\h);
\draw (0*\w,-\os) node[below]{$0$}-- (0*\w,\h);
\draw (\w,-\os) node[below]{$\D$}-- (\w,\h);

\draw (\r*\w,-\os) node[below]{$r$}-- (\r*\w,\h);
\draw (\r*\w+\lamb*\w,0) -- (\r*\w+\lamb*\w,\h+\os) node[above]{$r+\lambda$};
\draw (\l*\w,-\os) node[below]{$\ell$}-- (\l*\w,\h);
\draw (0.5*\w+\lamb*\w,-\os) node[below]{$\big(\frac{1}{2} + \lambda\big)\D$}-- (0.5*\w+\lamb*\w,\h);
\draw (0.5*\w+2*\lamb*\w,0) -- (0.5*\w+2*\lamb*\w,\h+2*\os)node[above]{$\big(\frac{1}{2} + 2\lambda\big)\D$};

\draw[dashed] (-\os,0.5*\h) node[left]{$\frac{1}{2}\opt$} -- (\w+\os,0.5*\h) ;
\draw[dashed] (-\os,1*\h) node[left]{$\opt$} -- (\w+\os,1*\h) ;


\begin{scope}[xshift = 1.3*\w cm]

\draw [fill = white!95!black] (0,0) -- (\l*\w,0.3*\h) --(\l*\w,\h) --(0,\h) -- (0,0);
\draw [fill = white!95!black] (\w,0) -- (\r*\w,0.1*\h) --(\r*\w,0.62*\h) --(\r*\w+0.7*\lamb*\w,0.62*\h) -- (\r*\w+0.7*\lamb*\w,0.7*\h) -- (\r*\w+0.5*\lamb*\w,0.7*\h) -- (\r*\w+0.5*\lamb*\w,0.75*\h) -- (\r*\w+\lamb*\w,0.75*\h)-- (\r*\w+\lamb*\w,0.85*\h) -- (\w,\h)--(\w,0);
\draw [fill = colorb] (\l*\w,0.85*\h) -- (\l*\w,0.95*\h) -- (\r*\w,0.85*\h)--(\l*\w,0.85*\h);
\draw [fill = colorb] (\l*\w,0.62*\h) -- (\l*\w,0.75*\h) -- (\l*\w+0.1*\w,0.75*\h)--(\l*\w+0.1*\w,0.7*\h) -- (\l*\w+0.13*\w,0.7*\h)--(\l*\w+0.13*\w,0.62*\h) -- (\l*\w,0.62*\h);
\draw [fill = colorb] (\l*\w,0.48*\h) -- (\l*\w,0.3*\h) -- (\r*\w,0.1*\h) -- (\r*\w,0.62*\h) -- (\r*\w-0.1*\w,0.62*\h)--(\r*\w-0.1*\w,0.55*\h)--(\r*\w-0.01*\w,0.55*\h)--(\r*\w-0.01*\w,0.48*\h)--(\r*\w-0.06*\w,0.48*\h)--(\r*\w-0.06*\w,0.4*\h)--(\r*\w-0.04*\w,0.4*\h)--(\r*\w-0.04*\w,0.3*\h)--(\r*\w-0.02*\w,0.3*\h)--(\r*\w-0.02*\w,0.23*\h)--(0.5*\w+0.5*\lamb*\w,0.23*\h)--(0.5*\w+0.5*\lamb*\w,0.3*\h) --(\l*\w+0.08*\w,0.3*\h) --(\l*\w+0.08*\w,0.4*\h)--(\l*\w+0.03*\w,0.4*\h) --(\l*\w+0.03*\w,0.48*\h) --(\l*\w,0.48*\h);

\draw[fill = coloraa] (0,0) -- (\l*\w,0.3*\h) -- (\r*\w,0.1*\h) -- (\w,0)--(0,0);
\node at (0.8*\l*\w,0.1*\h) {$\Gal$};

\draw[fill = colorab] (\w,\h) -- (\l*\w,\h) -- (\l*\w,0.95*\h) -- (\r*\w,0.85*\h)-- (\r*\w+\lamb*\w,0.85*\h) -- (\w,\h);
\node at (\r*\w+0.08*\w,0.94*\h) {$\Ga$};

\draw[fill = colorm] (0, 1*\h) rectangle (\r*\w+\lamb*\w-\l*\w,1.1*\h);
\draw[fill = colorm] (0,1.1*\h) rectangle (\r*\w-0.1*\w-\l*\w,1.17*\h);
\draw[fill = colorm] (0,1.17*\h) rectangle (\r*\w-0.01*\w-\l*\w,1.24*\h);
\draw[fill = colorm] (0,1.24*\h) rectangle (\r*\w-0.06*\w-\l*\w-0.03*\w,1.32*\h);
\draw[fill = colorm] (0,1.32*\h) rectangle (\r*\w-0.04*\w-\l*\w-0.08*\w,1.42*\h);
\draw[fill = colorm] (0,1.42*\h) rectangle (\r*\w+0.5*\lamb*\w-\l*\w-0.1*\w,1.47*\h);
\draw[fill = colorm] (0.5*\w+2*\lamb*\w,1*\h) rectangle (\r*\w-0.02*\w-0.5*\w-0.5*\lamb*\w+0.5*\w+2*\lamb*\w,1.07*\h);

\draw[fill = colorm] (\l*\w+0.13*\w,0.7*\h) rectangle (\r*\w+0.7*\lamb*\w,0.62*\h);
\node at (0.8*\l*\w+0.2*\r*\w,0.69*\h) {$\Gb$};

\draw[fill = gray] (\l*\w-0.01*\w,0.3*\h) rectangle (\l*\w,\h);

\draw[fill = gray] (\r*\w+0.02*\w,0.62*\h) rectangle (\r*\w,0.1*\h);

\draw[fill = coloriadd] (0.5*\w+\lamb*\w,1*\h) rectangle (0.5*\w+2*\lamb*\w,1.5*\h); 
\draw[fill = coloriadd] (0.5*\w+\lamb*\w,0.7*\h) rectangle coordinate[midway](iadd) (0.5*\w+2*\lamb*\w,0.85*\h);
\draw[fill = coloriadd] (0.5*\w+\lamb*\w,0.62*\h) rectangle (0.5*\w+2*\lamb*\w,0.23*\h); 
\draw (iadd) -- (\l*\w/2 + 0.5*\w/2 + \lamb*\w/2, 5/4*\h) node[fill=white, inner sep = 1pt] {\small $i_{\lambda}$};

\draw (0,1.5*\h) -- (0,0) -- (\w,0) -- (\w,1.5*\h);
\draw (0*\w,-\os) node[below]{$0$}-- (0*\w,\h);
\draw (\w,-\os) node[below]{$\D$}-- (\w,\h);

\draw (\r*\w,-\os) node[below]{$r$}-- (\r*\w,1.5*\h);
\draw (\r*\w+\lamb*\w,0) -- (\r*\w+\lamb*\w,1.5*\h+\os) node[above]{$r+\lambda\D$};
\draw (\l*\w,-\os) node[below]{$\ell$}-- (\l*\w,1.5*\h);
\draw (0.5*\w+\lamb*\w,-\os) node[below]{$\big(\frac12+\lambda\big)\D$}-- (0.5*\w+\lamb*\w,1.5*\h);
\draw (0.5*\w+2*\lamb*\w,0) -- (0.5*\w+2*\lamb*\w,1.5*\h+2*\os)node[above]{$\big(\frac12+2\lambda\big)\D$};

\draw[dashed] (-\os,1.5*\h) node[left]{$\frac{3}{2}\opt$} -- (\w+\os,1.5*\h) ;
\draw[dashed] (-\os,0.5*\h) node[left]{$\frac{1}{2}\opt$} -- (\w+\os,0.5*\h) ;
\draw[dashed] (-\os,1*\h) node[left]{$\opt$} -- (\w+\os,1*\h) ;

\end{scope}
\end{tikzpicture}
    }
    \caption{A possible repacking generated by \Cref{alg_c} if $\sopt$ satisfies Line \ref{alg_c_midmountain}.}
    \label{fig:enter-label1}
\end{figure}
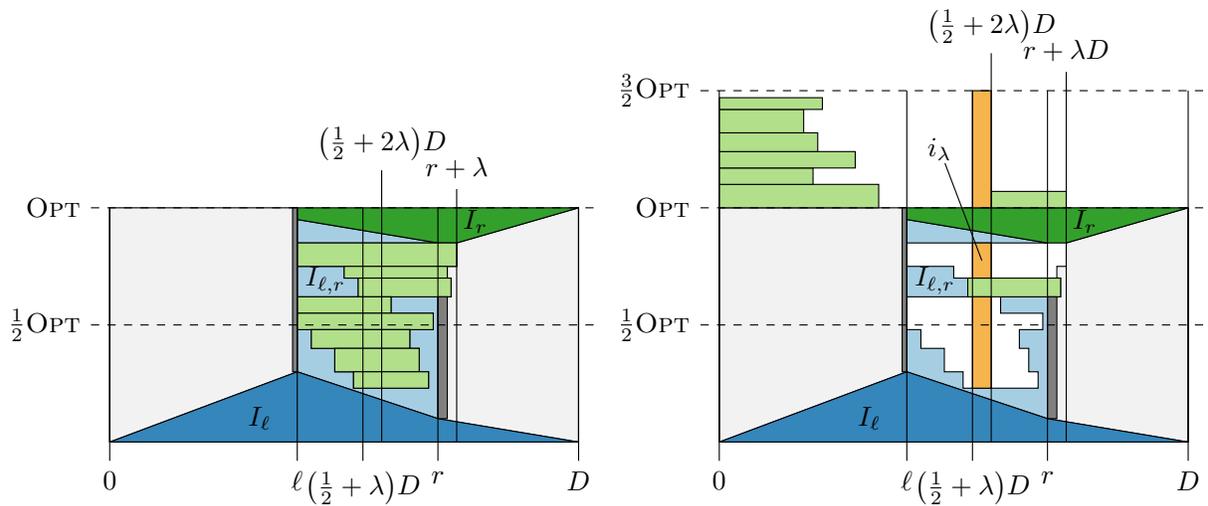

\begin{lemma}
    \label{lem_case_C}
    \label{lem:medium-wide-gap}
    Assume $\sopt$ is an optimal packing of $(\items, \D)$ with a gap $[\ell,r)$ of width $r - \ell \in \big[\lambda \D, \big(\frac 12 - 3 \lambda\big) \D\big]$ satisfying $\D - r \leq \ell$ and $r < \D$. 
    Then, $(\items, \D)$ admits a forgiving packing.
\end{lemma}

We require \alertBound{$\lambda \leq \frac1{50}$}. 
The assumptions in the lemma above imply the existence of the following two tall items: There are two tall items~$T_1$ and~$T_2$ with $\ell = \sopt(T_1) + w(T_1)$ and $r =  \sopt(T_2)$ with $\D - r \leq \ell$, and there are no tall items packed in $[\ell,r)$. 
Furthermore, let $\eta := \frac{r - \ell}{\D} \in \big[\lambda, \frac12 - 3 \lambda\big]$. 
We show that \cref{alg_c} returns a forgiving packing~$\start$, i.e., a packing that packs an extra item $\iadd$ of height~$\opt$ and width~$\lambda \D$ and has height at most $\frac32\opt$.
We will prove the lemma by separately proving the statement for each different case encountered by \Cref{alg_c} and arguing in the end that this indeed proves the lemma.

\begin{claim}\label{claim:c_midmountain}
    If $\sopt$ satisfies Line \ref{alg_c_midmountain}, \Cref{alg_c} returns a feasible packing $\packing$ with total height bounded by $\frac32\opt$.
\end{claim}

\begin{proof}
    We observe that $\max \{ w(i) : i \in \Gb \} \leq (\eta + \lambda)\D \leq \big(\frac12 - 2\lambda\big)\D$. Hence, no repacked item in $\itemsM_1 \subseteq \Gb$ overlaps the segment $\big[\big(\frac12 + \lambda\big)\D, \big(\frac12 + 2 \lambda\big)\D\big)$ in~$\packing$. Further, the last item that \Cref{alg_mountain_repack} repacks starts at~$\big(\frac12 + 2 \lambda\big)\D$ and, thus, finishes not later than~$\D$. 
    Since all items in $\items \setminus \itemsM_1$ are packed in~$\start$ as in~$\sopt$, the observation on the placement of~$\itemsM_1$ implies that all items are packed in~$[0,\D]$. 
    By \Cref{lem:mountain_repack}, we know that the height of the packing $\start$ without $\iadd$ is bounded by $\frac{3}{2}\opt$ and is bounded by $\frac{\opt}{2}$ in the segment  $\big[\big(\frac{1}{2}+\lambda)\D,\big(\frac{1}{2}+2\lambda\big)\D\big)$. 
    Hence, packing $\iadd$ in $\big[\big(\frac{1}{2}+\lambda\big)\D,\big(\frac{1}{2}+2\lambda\big)\D\big)$ does not increase the height of the packing beyond $\frac{3}{2}\opt$. 
\end{proof}

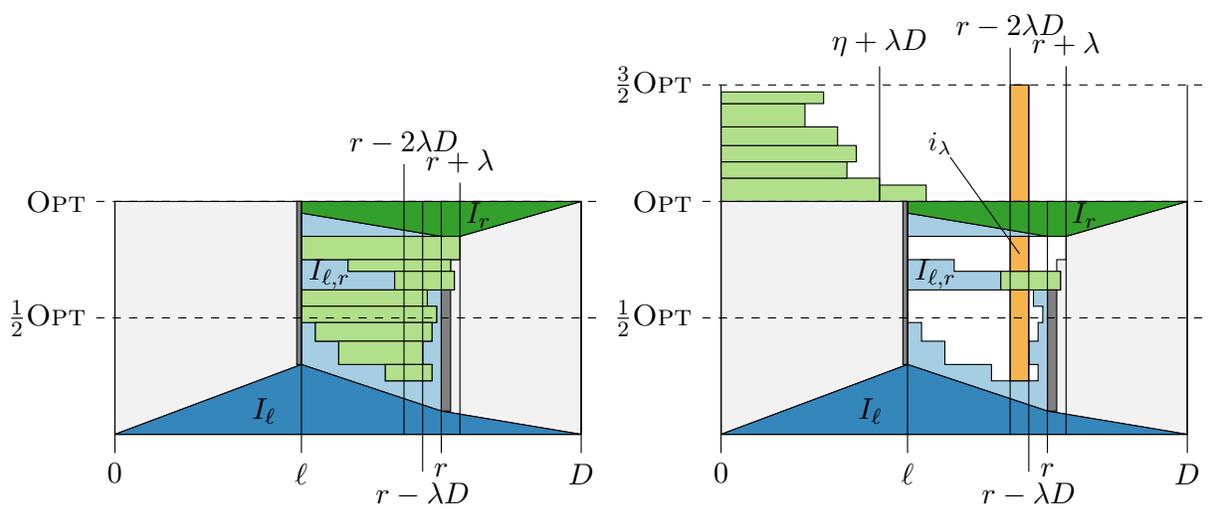
\begin{figure}
    \centering
    \resizebox{\textwidth}{!}{
    \begin{tikzpicture}
\pgfmathsetmacro{\w}{6}
\pgfmathsetmacro{\h}{3}
\pgfmathsetmacro{\os}{0.24}
\pgfmathsetmacro{\l}{0.4}
\pgfmathsetmacro{\r}{0.7}
\pgfmathsetmacro{\lamb}{0.04}

\draw [fill = white!95!black] (0,0) -- (\l*\w,0.3*\h) --(\l*\w,\h) --(0,\h) -- (0,0);
\draw [fill = white!95!black] (\w,0) -- (\r*\w,0.1*\h) --(\r*\w,0.62*\h) --(\r*\w+0.7*\lamb*\w,0.62*\h) -- (\r*\w+0.7*\lamb*\w,0.7*\h) -- (\r*\w+0.5*\lamb*\w,0.7*\h) -- (\r*\w+0.5*\lamb*\w,0.75*\h) -- (\r*\w+\lamb*\w,0.75*\h)-- (\r*\w+\lamb*\w,0.85*\h) -- (\w,\h)--(\w,0);
\draw [fill = colorb] (\l*\w,0.85*\h) -- (\l*\w,0.95*\h) -- (\r*\w,0.85*\h)--(\l*\w,0.85*\h);
\draw [fill = colorb] (\l*\w,0.62*\h) -- (\l*\w,0.75*\h) -- (\l*\w+0.1*\w,0.75*\h)--(\l*\w+0.1*\w,0.7*\h) -- (\l*\w+0.2*\w,0.7*\h)--(\l*\w+0.2*\w,0.62*\h) -- (\l*\w,0.62*\h);
\draw [fill = colorb] (\l*\w,0.48*\h) -- (\l*\w,0.3*\h) -- (\r*\w,0.1*\h) -- (\r*\w,0.62*\h) -- (\r*\w-0.03*\w,0.62*\h)--(\r*\w-0.03*\w,0.55*\h)--(\r*\w-0.01*\w,0.55*\h)--(\r*\w-0.01*\w,0.48*\h)--(\r*\w-0.02*\w,0.48*\h)--(\r*\w-0.02*\w,0.4*\h)--(\r*\w-0.04*\w,0.4*\h)--(\r*\w-0.04*\w,0.3*\h)--(\r*\w-0.02*\w,0.3*\h)--(\r*\w-0.02*\w,0.23*\h)--(\r*\w-3*\lamb*\w,0.23*\h)--(\r*\w-3*\lamb*\w,0.3*\h) --(\l*\w+0.08*\w,0.3*\h) --(\l*\w+0.08*\w,0.4*\h)--(\l*\w+0.03*\w,0.4*\h) --(\l*\w+0.03*\w,0.48*\h) --(\l*\w,0.48*\h);

\draw[fill = coloraa] (0,0) -- (\l*\w,0.3*\h) -- (\r*\w,0.1*\h) -- (\w,0)--(0,0);
\node at (0.8*\l*\w,0.1*\h) {$\Gal$};

\draw[fill = colorab] (\w,\h) -- (\l*\w,\h) -- (\l*\w,0.95*\h) -- (\r*\w,0.85*\h)-- (\r*\w+\lamb*\w,0.85*\h) -- (\w,\h);
\node at (\r*\w+0.08*\w,0.94*\h) {$\Ga$};

\draw[fill = colorm] (\l*\w,0.85*\h) rectangle (\r*\w+\lamb*\w,0.75*\h);
\draw[fill = colorm] (\l*\w+0.1*\w,0.75*\h) rectangle (\r*\w+0.5*\lamb*\w,0.7*\h);
\draw[fill = colorm] (\l*\w+0.2*\w,0.7*\h) rectangle (\r*\w+0.7*\lamb*\w,0.62*\h);
\draw[fill = colorm] (\l*\w,0.62*\h) rectangle (\r*\w-0.03*\w,0.55*\h);
\draw[fill = colorm] (\l*\w,0.55*\h) rectangle (\r*\w-0.01*\w,0.48*\h);
\draw[fill = colorm] (\l*\w+0.03*\w,0.48*\h) rectangle (\r*\w-0.02*\w,0.4*\h);
\draw[fill = colorm] (\l*\w+0.08*\w,0.4*\h) rectangle (\r*\w-0.04*\w,0.3*\h);
\draw[fill = colorm] (\r*\w-3*\lamb*\w,0.3*\h) rectangle (\r*\w-0.02*\w,0.23*\h);
\node at (0.8*\l*\w+0.2*\r*\w,0.69*\h) {$\Gb$};

\draw[fill = gray] (\l*\w-0.01*\w,0.3*\h) rectangle (\l*\w,\h);

\draw[fill = gray] (\r*\w+0.02*\w,0.62*\h) rectangle (\r*\w,0.1*\h);

\draw (0,\h) -- (0,0) -- (\w,0) -- (\w,\h);
\draw (0*\w,-\os) node[below]{$0$}-- (0*\w,\h);
\draw (\w,-\os) node[below]{$\D$}-- (\w,\h);

\draw (\r*\w,-\os) node[below]{$r$}-- (\r*\w,\h);
\draw (\r*\w+\lamb*\w,0) -- (\r*\w+\lamb*\w,\h+\os) node[above]{$r+\lambda$};
\draw (\l*\w,-\os) node[below]{$\ell$}-- (\l*\w,\h);
\draw (\r*\w-\lamb*\w,-2*\os) node[below]{$r-\lambda\D$}-- (\r*\w-\lamb*\w,\h);
\draw (\r*\w-2*\lamb*\w,0) -- (\r*\w-2*\lamb*\w,\h+2*\os)node[above]{$r-2\lambda\D$};

\draw[dashed] (-\os,0.5*\h) node[left]{$\frac{1}{2}\opt$} -- (\w+\os,0.5*\h) ;
\draw[dashed] (-\os,1*\h) node[left]{$\opt$} -- (\w+\os,1*\h) ;


\begin{scope}[xshift = 1.3*\w cm]

\draw [fill = white!95!black] (0,0) -- (\l*\w,0.3*\h) --(\l*\w,\h) --(0,\h) -- (0,0);
\draw [fill = white!95!black] (\w,0) -- (\r*\w,0.1*\h) --(\r*\w,0.62*\h) --(\r*\w+0.7*\lamb*\w,0.62*\h) -- (\r*\w+0.7*\lamb*\w,0.7*\h) -- (\r*\w+0.5*\lamb*\w,0.7*\h) -- (\r*\w+0.5*\lamb*\w,0.75*\h) -- (\r*\w+\lamb*\w,0.75*\h)-- (\r*\w+\lamb*\w,0.85*\h) -- (\w,\h)--(\w,0);
\draw [fill = colorb] (\l*\w,0.85*\h) -- (\l*\w,0.95*\h) -- (\r*\w,0.85*\h)--(\l*\w,0.85*\h);
\draw [fill = colorb] (\l*\w,0.62*\h) -- (\l*\w,0.75*\h) -- (\l*\w+0.1*\w,0.75*\h)--(\l*\w+0.1*\w,0.7*\h) -- (\l*\w+0.2*\w,0.7*\h)--(\l*\w+0.2*\w,0.62*\h) -- (\l*\w,0.62*\h);
\draw [fill = colorb] (\l*\w,0.48*\h) -- (\l*\w,0.3*\h) -- (\r*\w,0.1*\h) -- (\r*\w,0.62*\h) -- (\r*\w-0.03*\w,0.62*\h)--(\r*\w-0.03*\w,0.55*\h)--(\r*\w-0.01*\w,0.55*\h)--(\r*\w-0.01*\w,0.48*\h)--(\r*\w-0.02*\w,0.48*\h)--(\r*\w-0.02*\w,0.4*\h)--(\r*\w-0.04*\w,0.4*\h)--(\r*\w-0.04*\w,0.3*\h)--(\r*\w-0.02*\w,0.3*\h)--(\r*\w-0.02*\w,0.23*\h)--(\r*\w-3*\lamb*\w,0.23*\h)--(\r*\w-3*\lamb*\w,0.3*\h) --(\l*\w+0.08*\w,0.3*\h) --(\l*\w+0.08*\w,0.4*\h)--(\l*\w+0.03*\w,0.4*\h) --(\l*\w+0.03*\w,0.48*\h) --(\l*\w,0.48*\h);

\draw[fill = coloraa] (0,0) -- (\l*\w,0.3*\h) -- (\r*\w,0.1*\h) -- (\w,0)--(0,0);
\node at (0.8*\l*\w,0.1*\h) {$\Gal$};

\draw[fill = colorab] (\w,\h) -- (\l*\w,\h) -- (\l*\w,0.95*\h) -- (\r*\w,0.85*\h)-- (\r*\w+\lamb*\w,0.85*\h) -- (\w,\h);
\node at (\r*\w+0.08*\w,0.94*\h) {$\Ga$};

\draw[fill = colorm] (0, 1*\h) rectangle (\r*\w+\lamb*\w-\l*\w,1.1*\h);
\draw[fill = colorm] (0,1.1*\h) rectangle (\r*\w-0.03*\w-\l*\w,1.17*\h);
\draw[fill = colorm] (0,1.17*\h) rectangle (\r*\w-0.01*\w-\l*\w,1.24*\h);
\draw[fill = colorm] (0,1.24*\h) rectangle (\r*\w-0.02*\w-\l*\w-0.03*\w,1.32*\h);
\draw[fill = colorm] (0,1.32*\h) rectangle (\r*\w-0.04*\w-\l*\w-0.08*\w,1.42*\h);
\draw[fill = colorm] (0,1.42*\h) rectangle (\r*\w+0.5*\lamb*\w-\l*\w-0.1*\w,1.47*\h);
\draw[fill = colorm] (\r*\w+\lamb*\w-\l*\w,1*\h) rectangle (\r*\w-0.02*\w-\r*\w+3*\lamb*\w+\r*\w+\lamb*\w-\l*\w,1.07*\h);

\draw[fill = colorm] (\l*\w+0.2*\w,0.7*\h) rectangle (\r*\w+0.7*\lamb*\w,0.62*\h);
\node at (0.8*\l*\w+0.2*\r*\w,0.69*\h) {$\Gb$};

\draw[fill = gray] (\l*\w-0.01*\w,0.3*\h) rectangle (\l*\w,\h);

\draw[fill = gray] (\r*\w+0.02*\w,0.62*\h) rectangle (\r*\w,0.1*\h);

\draw[fill = coloriadd] (\r*\w-2*\lamb*\w,1*\h) rectangle (\r*\w-\lamb*\w,1.5*\h); 
\draw[fill = coloriadd] (\r*\w-\lamb*\w,0.7*\h) rectangle coordinate[midway] (iadd) (\r*\w-2*\lamb*\w,0.85*\h);
\draw (iadd) -- (\l*\w/2 + 0.5*\w/2 + \lamb*\w/2, 5/4*\h) node[fill=white, inner sep = 1pt] {\small $i_{\lambda}$};
\draw[fill = coloriadd] (\r*\w-\lamb*\w,0.62*\h) rectangle (\r*\w-2*\lamb*\w,0.23*\h); 

\draw (0,1.5*\h) -- (0,0) -- (\w,0) -- (\w,1.5*\h);
\draw (0*\w,-\os) node[below]{$0$}-- (0*\w,\h);
\draw (\w,-\os) node[below]{$\D$}-- (\w,\h);

\draw (\r*\w,-\os) node[below]{$r$}-- (\r*\w,\h);
\draw (\r*\w+\lamb*\w,0) -- (\r*\w+\lamb*\w,1.5*\h+\os) node[above]{$r+\lambda$};
\draw (\l*\w,-\os) node[below]{$\ell$}-- (\l*\w,\h);
\draw (\r*\w-\lamb*\w,-2*\os) node[below]{$r-\lambda\D$}-- (\r*\w-\lamb*\w,1.5*\h);
\draw (\r*\w-2*\lamb*\w,0) -- (\r*\w-2*\lamb*\w,1.5*\h+2*\os)node[above]{$r-2\lambda\D$};
\draw (\r*\w-\l*\w+\lamb*\w,\h) -- (\r*\w-\l*\w+\lamb*\w,1.5*\h+\os)node[above]{$\eta+\lambda \D$};

\draw[dashed] (-\os,1.5*\h) node[left]{$\frac{3}{2}\opt$} -- (\w+\os,1.5*\h) ;
\draw[dashed] (-\os,0.5*\h) node[left]{$\frac{1}{2}\opt$} -- (\w+\os,0.5*\h) ;
\draw[dashed] (-\os,1*\h) node[left]{$\opt$} -- (\w+\os,1*\h) ;

\end{scope}

\end{tikzpicture}
    }
    \caption{A possible repacking generated by \Cref{alg_c} if $\sopt$ does not satisfy Line \ref{alg_c_midmountain} but Line \ref{alg_c_rightmountain}.}
    \label{fig:enter-label2}
\end{figure}

\begin{claim}\label{claim:c_rightmountain}
    If $\sopt$ does not satisfy Line \ref{alg_c_midmountain}, but satisfies Line \ref{alg_c_rightmountain}, \Cref{alg_c} returns a feasible packing $\packing$ with total height bounded by $\frac32 \opt$.
\end{claim}

\begin{proof}
    We observe that $\max_{i \in \itemsM_2} w(i) \leq (\eta + \lambda)\D$, implying that the last item~$i$ that \Cref{alg_mountain_repack} repacks does not overlap with the previously repacked items from~$\itemsM_2$ since they finish not later than $\max_{i \in \itemsM_2} w(i) $. 
    Further, its starting time~$\sopt(i)$ is greater than~$\big(\frac12 + \lambda\big)\D$ as otherwise $h(\itemsM_1) \geq \frac12\opt$, contradicting the assumption. By definition, $w(i) \leq r + \lambda\D - \sopt(i)$. Hence, after repacking, the last item satisfies 
    $$
        \start(i) + w(i) 
            \leq (\eta + \lambda)\D + \big( r + \lambda\D - \sopt(i) \big)
            \leq \bigg(\frac12 - 2 \lambda\bigg)\D + \bigg( r + \lambda\D - \bigg(\frac12 + \lambda\bigg)\D \bigg) 
            \leq r - 2 \lambda \D
    $$ 
    and therefore finishes before $r - 2 \lambda \D$. 
    Since $\start$ packs all items in~$\items \setminus \itemsM_2$ as~$\sopt$, the above calculation implies that all items are packed in~$[0,\D]$. 
    By \Cref{lem:mountain_repack} we know that the height of the packing $\start$ without $\iadd$ is bounded by $\frac{3}{2}\opt$ and is bounded by $\frac{\opt}{2}$ in the segment $[r-2\lambda\D,r-\lambda\D)$. 
    Hence, packing $\iadd$ in $[r-2\lambda\D,r-\lambda\D)$ does not increase the height of the packing beyond $\frac{3}{2}\opt$. 
\end{proof}

\begin{claim}\label{claim:c_no_mid-right_mountain}
    If~$\sopt$ does not satisfy lines \ref{alg_c_midmountain} and \ref{alg_c_rightmountain}, before line \ref{alg_c_largeoverlap}, the maximal height of the packing~$\packing$ in the segments~$[0, (\eta + 13\lambda)\D)$ and~$[(\eta + 13 \lambda)\D, \D)$ is at most~$\frac32\opt$ and~$\opt$, respectively. Packing~$\start$ is $\D$-feasible, and~$(\eta + 13\lambda)\D \leq r - \lambda \D$ if~\alertBound{$\lambda \leq \frac1{50}$}. 
\end{claim}

\begin{proof} We start by observing that due to the lemma assumption, only items in $\itemsM_2$ and in $\Gb \cap \itemsT{\sopt}_{[r- 2\lambda\D, r + \lambda\D]}$ are packed differently by $\start$ than by~$\sopt$.
We analyze the height of~$\start$ separately for the three segments $[0, (\eta + \lambda)\D)$, $[(\eta + \lambda)\D, (\eta + 13 \lambda)\D)$, and $[(\eta + 13\lambda)\D, \D)$ while also arguing about the $\D$-feasibility of~$\start$.  

    \noindent $\bullet \enspace \bm{[0, (\eta + \lambda)\D).} \enspace $
    By assumption, the items $\itemsM_2 = \Gb \cap \mountainininterval{\sopt}{r - 2\lambda\D}{r - \lambda\D}$ satisfy~$\sopt(i) + w(i) \leq r + \lambda \D$ and~$\sopt(i) \geq \ell$.  
    Hence, by shifting these items to the left by~$\ell$, they satisfy~$0 \leq \start(i) \leq r + \lambda \D - \ell = (\eta + \lambda)\D$, which implies that they are feasibly packed. Further, the total height of these items is at most~$\frac12\opt$ since~$\sopt$ does not satisfy the if-condition of Line~\ref{alg_c_rightmountain}. 
    Therefore, the statement on the maximal height is satisfied in the segment $[0, (\eta + \lambda)\D)$. 

    \noindent $\bullet \enspace \bm{[(\eta + \lambda)\D, (\eta + 13 \lambda)\D).} \enspace $
    For the items $\Gb \cap \iopt_{[r - 2 \lambda \D, r + \lambda) \D} $, we observe that the maximal height of such an item is at most $\frac12\opt$ by definition of $\Gb$ and the maximal width is at most $3\lambda\D$. Further, their total area is at most $3\lambda\D\opt$. Hence, by \Cref{lem:steinberg},~$\start'$ will have a width of~$12 \lambda \D$ and height $\frac12\opt$. This implies that these items are feasibly packed in~$\start$ and the statement on the height of~$\start$ is also satisfied in the segment $[(\eta + \lambda)\D, (\eta + 13 \lambda)\D)$. 

    \noindent $\bullet \enspace \bm{[(\eta + 13\lambda)\D, \D).} \enspace$ The height of the packing~$\packing$ did only decrease compared to~$\sopt$. 

    For the last part of the claim, we observe that~$\D - r \leq \ell = r - \eta\D$ by our initial assumption on the gap. Hence, $r \geq \frac{(1 + \eta)\D}{2}$. Thus,~$\eta + 13\lambda \leq \frac{1 + \eta}{2} - \lambda$ implies~$(\eta + 13\lambda)\D \leq r - \lambda \D$. After simple algebraic reformulations, it is easy to see that the choice~\alertBound{$\lambda \leq \frac1{50}$} implies the former inequality. 
\end{proof}

\begin{figure}
    \centering
    \resizebox{\textwidth}{!}{
    \begin{tikzpicture}
\pgfmathsetmacro{\w}{6}
\pgfmathsetmacro{\h}{3}
\pgfmathsetmacro{\os}{0.24}
\pgfmathsetmacro{\l}{0.45}
\pgfmathsetmacro{\r}{0.8}
\pgfmathsetmacro{\lamb}{0.029}

\draw [fill = white!95!black] (0,0) -- (\l*\w,0.3*\h) --(\l*\w,\h) --(0,\h) -- (0,0);

\draw [fill = white!95!black] (\r*\w,0.25*\h) -- (\r*\w + \lamb*\w, 0.215*\h) -- (\r*\w + \lamb*\w, 0.77*\h) -- (\r*\w + 0.015*\w, 0.77*\h) -- (\r*\w + 0.015*\w, 0.25*\h) -- (\r*\w,0.25*\h); 

\draw [fill = white!95!black] (\r*\w + \lamb*\w, 0.215*\h) -- (\w, 0) -- (\w, \h) -- (\r*\w + \lamb*\w, 0.77*\h) -- (\r*\w + \lamb*\w, 0.215*\h);


\draw [fill = colorb] (\l*\w, 0.95*\h) -- (\r*\w - \lamb*\w, 0.785*\h) -- (\r*\w - \lamb*\w, 0.77*\h) -- (\l*\w, 0.77*\h); 
\draw [fill = colorb] (\r*\w - \lamb*\w, 0.785*\h) -- (\r*\w, 0.77*\h) -- (\r*\w - \lamb*\w, 0.77*\h); 
\draw [fill = colorb] (\l*\w, 0.59*\h) -- (\l*\w +0.2*\w, 0.59*\h) -- (\l*\w +0.2*\w, 0.67*\h) -- (\l*\w + 0.1*\w, 0.67*\h) -- (\l*\w + 0.1*\w, 0.72*\h) -- (\l*\w, 0.72*\h) -- (\l*\w, 0.59*\h);
\draw [fill = colorb] (\l*\w, 0.45*\h) -- (\r*\w - \lamb*\w, 0.4*\h) -- (\r*\w - \lamb*\w, 0.45*\h) -- (\l*\w, 0.45*\h); 
\draw [fill = colorb] (\r*\w,0.4*\h) -- (\r*\w - \lamb*\w, 0.405*\h) -- (\r*\w - \lamb*\w, 0.45*\h) --  (\r*\w - 0.01*\w, 0.45*\h) -- (\r*\w - 0.01*\w, 0.52*\h) -- (\r*\w - \lamb*\w, 0.52*\h) -- (\r*\w - \lamb*\w,0.59*\h) -- (\r*\w,0.59*\h)  -- (\r*\w,0.4*\h);

\draw[fill = coloraa] (0,0) -- (\l*\w,0.45*\h) -- (\r*\w,0.4*\h) -- (\r*\w,0.25*\h) -- (\w,0) -- (0,0);
\node at (0.8*\l*\w,0.1*\h) {$\Gal$};

\draw[fill = colorab] (\w,\h) -- (\l*\w,\h) -- (\l*\w,0.95*\h) -- (\r*\w,0.77*\h)-- (\r*\w+\lamb*\w,0.77*\h) -- (\w,\h);
\node at (\r*\w+0.08*\w,0.94*\h) {$\Ga$};

\begin{scope}[yshift = -.09 cm]
    \draw[fill = colorm] (\l*\w,0.8*\h) rectangle (\r*\w,0.75*\h);
    \draw[fill = colorm] (\l*\w+0.1*\w,0.75*\h) rectangle (\r*\w,0.7*\h);
    \draw[fill = colorm] (\l*\w+0.2*\w,0.7*\h) rectangle (\r*\w,0.62*\h);
    \draw[fill = colorm] (\l*\w,0.62*\h) rectangle (\r*\w-0.03*\w,0.55*\h);
    \draw[fill = colorm] (\l*\w,0.55*\h) rectangle (\r*\w-0.01*\w,0.48*\h);
\end{scope}
\node at (0.8*\l*\w+0.2*\r*\w,0.69*\h) {$G_3$};

\draw[fill = tallItemColor] (\l*\w-0.015*\w,0.45*\h) rectangle (\l*\w,\h);

\draw[fill = tallItemColor] (\r*\w + 0.015*\w, 0.77*\h) rectangle  (\r*\w,0.25*\h);

\draw (0,\h) -- (0,0) -- (\w,0) -- (\w,\h);
\draw (0*\w,-\os) node[below]{$0$}-- (0*\w,\h);
\draw (\w,-\os) node[below]{$\D$}-- (\w,\h);

\draw (\r*\w,-\os) node[below]{$r$}-- (\r*\w,\h);
\draw (\r*\w+\lamb*\w,0) -- (\r*\w+\lamb*\w,\h+\os) node[above]{$r+\lambda$};
\draw (\l*\w,-\os) node[below]{$\ell$}-- (\l*\w,\h);
\draw (\r*\w-\lamb*\w,-2*\os) node[below]{$r-\lambda\D$}-- (\r*\w-\lamb*\w,\h);
\draw (\r*\w-2*\lamb*\w,0) -- (\r*\w-2*\lamb*\w,\h+2*\os)node[above]{$r-2\lambda\D$};

\draw[dashed] (-\os,0.5*\h) node[left]{$\frac{1}{2}\opt$} -- (\w+\os,0.5*\h) ;
\draw[dashed] (-\os,1*\h) node[left]{$\opt$} -- (\w+\os,1*\h) ;


\begin{scope}[xshift = 1.3*\w cm]

\draw [fill = white!95!black] (0,0) -- (\l*\w,0.3*\h) --(\l*\w,\h) --(0,\h) -- (0,0);

\draw [fill = white!95!black, xshift = -\lamb*\w cm, yshift = .5*\h cm] (\r*\w,0.25*\h) -- (\r*\w + \lamb*\w, 0.215*\h) -- (\r*\w + \lamb*\w, 0.77*\h) -- (\r*\w + 0.015*\w, 0.77*\h) -- (\r*\w + 0.015*\w, 0.25*\h) -- (\r*\w,0.25*\h); 

\draw [fill = white!95!black, yshift = .5*\h cm, xshift = - \lamb*\w cm] (\r*\w + \lamb*\w, 0.215*\h) -- (\w, 0) -- (\w, \h) -- (\r*\w + \lamb*\w, 0.77*\h) -- (\r*\w + \lamb*\w, 0.215*\h);

\draw [fill = colorb] (\l*\w, 0.95*\h) -- (\r*\w - \lamb*\w, 0.785*\h) -- (\r*\w - \lamb*\w, 0.77*\h) -- (\l*\w, 0.77*\h); 
\draw [fill = colorb] (\l*\w, 0.59*\h) -- (\l*\w +0.2*\w, 0.59*\h) -- (\l*\w +0.2*\w, 0.67*\h) -- (\l*\w + 0.1*\w, 0.67*\h) -- (\l*\w + 0.1*\w, 0.72*\h) -- (\l*\w, 0.72*\h) -- (\l*\w, 0.59*\h);
\draw [fill = colorb] (\l*\w, 0.45*\h) -- (\r*\w - \lamb*\w, 0.4*\h) -- (\r*\w - \lamb*\w, 0.45*\h) -- (\l*\w, 0.45*\h); 

\draw[fill = coloraa] (0,0) -- (\l*\w,0.45*\h) -- (\r*\w,0.4*\h) -- (\r*\w,0.25*\h) -- (\w,0)--(0,0);
\node at (0.8*\l*\w,0.1*\h) {$\Gal$};


\draw[fill = colorab] (\r*\w - \lamb*\w, 0.785*\h) -- (\r*\w - \lamb*\w, \h) -- (\l*\w,\h) -- (\l*\w,0.95*\h) -- (\r*\w - \lamb*\w, 0.785*\h); 

\draw[fill = colorab, yshift = .5*\h cm] (\r*\w - \lamb*\w, 0.785*\h) -- (\r*\w, 0.77*\h) -- (\r*\w, \h) -- (\r*\w - \lamb*\w, \h);

\draw[fill = colorab, yshift = - .5*\h cm] (\w,\h) -- (\r*\w, \h) -- (\r*\w, 0.77*\h) -- (\r*\w + \lamb*\w, 0.77*\h) -- (\w,\h) -- (\r*\w, \h); 

\begin{scope}[yshift = 0.52*\h cm]
    \draw[fill = colorm] (0,0.8*\h) rectangle (\r*\w-\l*\w,0.75*\h);
    \draw[fill = colorm] (0,0.75*\h) rectangle (\r*\w-\l*\w-0.1*\w,0.7*\h);
    \draw[fill = colorm] (0,0.7*\h) rectangle (\r*\w-\l*\w-0.2*\w,0.62*\h);
    \draw[fill = colorm] (0,0.62*\h) rectangle (\r*\w-0.03*\w-\l*\w,0.55*\h);
    \draw[fill = colorm] (0,0.55*\h) rectangle (\r*\w-0.01*\w-\l*\w,0.48*\h);
\end{scope}
\node at (0.8*\l*\w+0.2*\r*\w,0.69*\h) {$G_3$};

\draw[fill = tallItemColor] (\l*\w-0.015*\w,0.45*\h) rectangle (\l*\w,\h);

\draw[fill = tallItemColor, xshift = -\lamb*\w cm, yshift = .5*\h cm] (\r*\w+0.015*\w,0.77*\h) rectangle (\r*\w,0.25*\h);


\draw[fill = colorm] (\r*\w-\l*\w+\lamb*\w,\h) rectangle node[midway]{Steinberg} (\r*\w-\l*\w+13*\lamb*\w,1.5*\h);

\draw[fill = coloriadd] (\w - \lamb*\w, .5*\h) rectangle (\w, 1.5*\h) coordinate[midway, yshift = .05*\h cm] (iadd);
\draw (iadd) -- (\r*\w/2 + \w/2, 1.25*\h) node[fill =  white!95!black, inner sep = 1pt] {$\iadd$};

\draw (0,1.5*\h) -- (0,0) -- (\w,0) -- (\w,1.5*\h);
\draw (0*\w,-\os) node[below]{$0$}-- (0*\w,\h);
\draw (\w,-\os) node[below]{$\D$}-- (\w,\h);

\draw (\r*\w,-\os) node[below]{$r$}-- (\r*\w,\h);
\draw (\r*\w+\lamb*\w,0) -- (\r*\w+\lamb*\w,1.5*\h+\os) node[above]{$r+\lambda$};
\draw (\l*\w,-\os) node[below]{$\ell$}-- (\l*\w,\h);
\draw (\r*\w-\lamb*\w,-2*\os) node[below]{$r-\lambda\D$}-- (\r*\w-\lamb*\w,1.5*\h);
\draw (\r*\w-2*\lamb*\w,0) -- (\r*\w-2*\lamb*\w,1.5*\h+2*\os)node[above]{$r-2\lambda\D$};
\draw (\r*\w-\l*\w+\lamb*\w,\h) -- (\r*\w-\l*\w+\lamb*\w,1.5*\h+\os)node[above]{$\eta+\lambda \D$};

\draw[dashed] (-\os,1.5*\h) node[left]{$\frac{3}{2}\opt$} -- (\w+\os,1.5*\h) ;
\draw[dashed] (-\os,0.5*\h) node[left]{$\frac{1}{2}\opt$} -- (\w+\os,0.5*\h) ;
\draw[dashed] (-\os,1*\h) node[left]{$\opt$} -- (\w+\os,1*\h) ;

\end{scope}

\end{tikzpicture}
    }
    \caption{A possible repacking generated by \Cref{alg_c} if $\sopt$ does neither satisfy Line~\ref{alg_c_midmountain} nor Line~\ref{alg_c_rightmountain}, but Line~\ref{alg_c_largeoverlap}.}
    \label{fig:}
\end{figure}
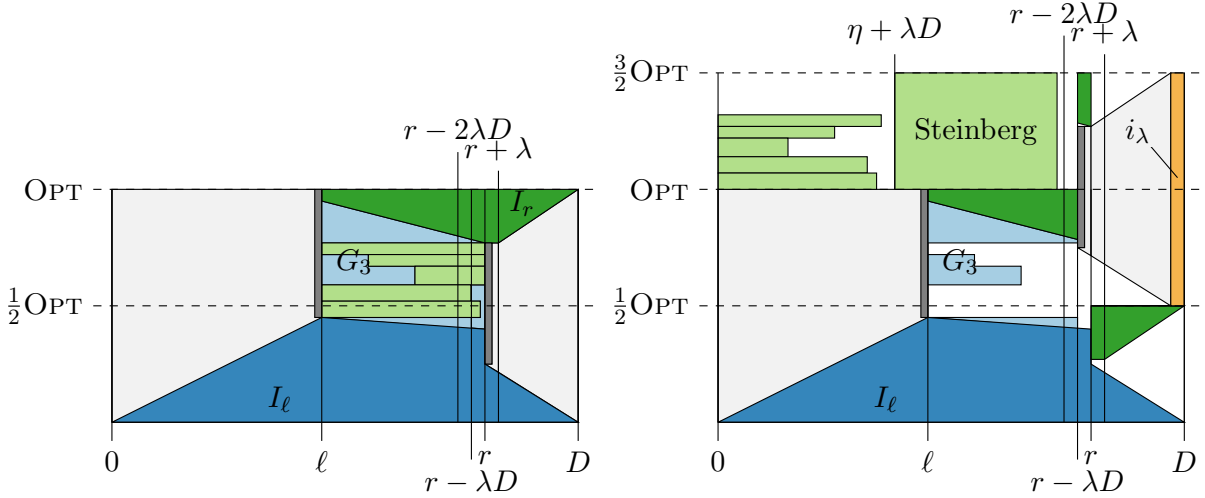

\begin{claim}\label{claim:c_largeoverlap}
    Suppose \alertBound{$\lambda \leq \frac1{50}$.} 
    If $\sopt$ does not satisfy lines \ref{alg_c_midmountain} and \ref{alg_c_rightmountain} but line \ref{alg_c_largeoverlap}, \Cref{alg_c} returns a $\D$-feasible packing $\packing$ with total height at most $\frac32 \opt$.
\end{claim}

\begin{proof} 
    We start by arguing about the feasibility of~$\start$ before separately analyzing its height in the segments $[0, r - \lambda\D)$ and $[r - \lambda\D, \D)$. 
    Before line \ref{alg_c_largeoverlap}, the previous claim guarantees the $\D$-feasibility of~$\start$. 
    After line \ref{alg_c_largeoverlap}, only items with $\sopt(i) \geq r \geq \ell + \lambda\D \geq \lambda \D$ are packed differently; they are shifted by~$\lambda \D$ to the left and hence satisfy~$\start(i) \geq 0$. 
    This proves that all items are packed in $[0,\D]$ as desired.

    \noindent $\bullet \enspace \bm{[0, r - \lambda\D).}\enspace$ By \Cref{claim:c_no_mid-right_mountain}, the maximal height of~$\packing$ in the segments~$[0, (\eta+13\lambda)\D)$ and~$[(\eta + 13\lambda)\D,\D)$ is at most~$\frac32\opt$ and~$\opt$, respectively. We observe that the only segment where~$\start$ is changed after line \ref{alg_c_largeoverlap} is in the segment~$[r - \lambda\D,\D)$. Using \Cref{claim:c_no_mid-right_mountain} for bounding $(\eta + 13\lambda)\D \leq r - \lambda \D$ guarantees the bound on the maximal height in the segment~$[0, r - \lambda\D)$.    

    \noindent $\bullet \enspace \bm{[r - \lambda\D, \D) .}\enspace$ 
    We note that the condition $h\big( \itemsatt{\sopt}{t}\cap (\Gal \cup \Ga) \big) > \frac12\opt$ for some $t\in [r-2\lambda\D, r-\lambda\D)$ implies that~$r \leq (1 - \lambda)\D$. 
    Indeed, $r > (1 - \lambda)\D$ implies $\Ga = \emptyset$ and $\Gal \cup \Ga = \Gal$. By definition, $\Gal = \itemsoverlapt{\sopt}{\ell}$ and $\ell = \sopt(T_1) + p(T_1)$ for a tall item $T_1$, implying $h(\Gal) \leq \frac12$ by \Cref{obs_height_G1G2}.

    \noindent $\circ \enspace \bm{[r, (1-\lambda)\D).}\enspace $ We observe that the items in~$\packing$ have maximal height at most $\opt$ and the non-shifted items have height at most $\frac12\opt$ by \Cref{obs_height_G1G2} because they belong to $\itemsoverlapt{\sopt}{r}$ by definition. 
    Hence, the maximal height in~$\packing$ in the segment $[r, (1-\lambda)\D)$ is at most~$\frac32\opt$. 

    \noindent $\circ \enspace \bm{[r - \lambda\D, r).} \enspace$ We apply a similar but stronger argument. The height of $\Ga$ is only increasing in $[r - \lambda\D, r)$ and remains constant in $[r, r+ \lambda\D)$. 
    All items with $\start(i) \in [r - \lambda, r)$ have~$\sopt(i) \in [r, r + \lambda\D)$. Hence, the maximal height of these items and the items in $\Ga$ is at most $\opt$ as $\sopt$ feasibly packed them. Therefore, in~$\packing$, the maximal height in the segment~$[r - \lambda\D, r)$ is at most $\opt + h(\Gal) \leq \frac32 \opt$ (using again \Cref{obs_height_G1G2} to bound the height of~$\Gal$). 

    \noindent $\circ \enspace \bm{[(1 - \lambda)\D, \D) .}\enspace$  Observe that, at the end of \Cref{alg_c}, the only items with~$[\start(i), \start(i) + w(i)) \cap [(1-\lambda)\D, \D) \neq \emptyset$ are items belonging to~$\Gal \cup \Ga \cup \{\iadd\}$. Using \Cref{obs_height_G1G2}, we bound the height of $\Gal \cup \Ga$ by~$\frac12\opt$. By definition, $h(\iadd) \leq \opt$, which concludes the proof of the claim.
\end{proof}

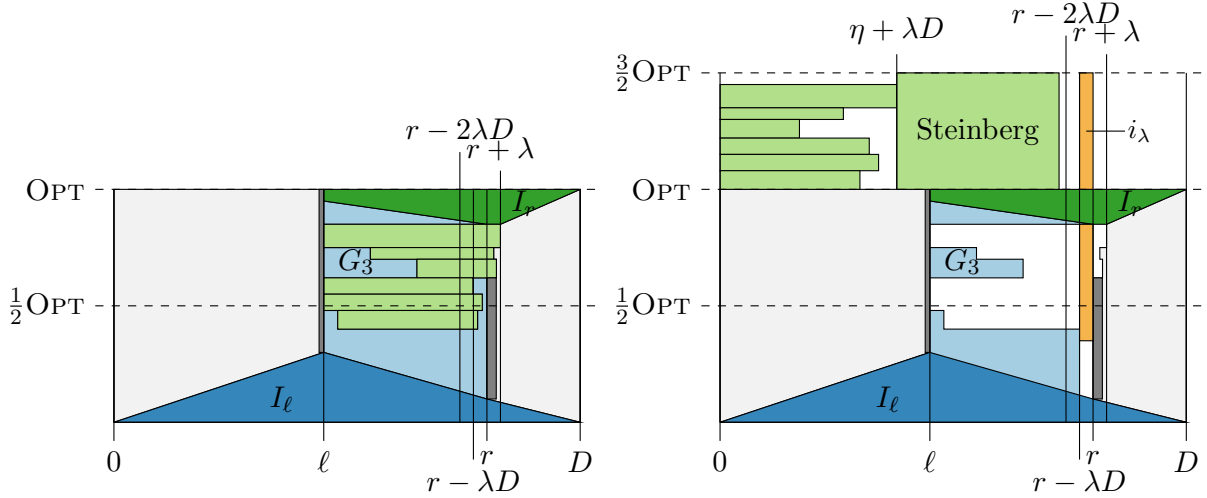
\begin{figure}
    \centering
    \resizebox{\textwidth}{!}{
    \begin{tikzpicture}
\pgfmathsetmacro{\w}{6}
\pgfmathsetmacro{\h}{3}
\pgfmathsetmacro{\os}{0.24}
\pgfmathsetmacro{\l}{0.45}
\pgfmathsetmacro{\r}{0.8}
\pgfmathsetmacro{\lamb}{0.029}

\draw [fill = white!95!black] (0,0) -- (\l*\w,0.3*\h) --(\l*\w,\h) --(0,\h) -- (0,0);
\draw [fill = white!95!black] (\w,0) -- (\r*\w,0.1*\h) --(\r*\w,0.62*\h) --(\r*\w+0.7*\lamb*\w,0.62*\h) -- (\r*\w+0.7*\lamb*\w,0.7*\h) -- (\r*\w+0.5*\lamb*\w,0.7*\h) -- (\r*\w+0.5*\lamb*\w,0.75*\h) -- (\r*\w+\lamb*\w,0.75*\h)-- (\r*\w+\lamb*\w,0.85*\h) -- (\w,\h)--(\w,0);
\draw [fill = colorb] (\l*\w,0.85*\h) -- (\l*\w,0.95*\h) -- (\r*\w,0.85*\h)--(\l*\w,0.85*\h);
\draw [fill = colorb] (\l*\w,0.62*\h) -- (\l*\w,0.75*\h) -- (\l*\w+0.1*\w,0.75*\h)--(\l*\w+0.1*\w,0.7*\h) -- (\l*\w+0.2*\w,0.7*\h)--(\l*\w+0.2*\w,0.62*\h) -- (\l*\w,0.62*\h);
\draw [fill = colorb] (\l*\w,0.48*\h) -- (\l*\w,0.3*\h) -- (\r*\w,0.1*\h) -- (\r*\w,0.62*\h) -- (\r*\w-0.03*\w,0.62*\h)--(\r*\w-0.03*\w,0.55*\h)--(\r*\w-0.01*\w,0.55*\h)--(\r*\w-0.01*\w,0.48*\h)--(\r*\w-0.02*\w,0.48*\h)--(\r*\w-0.02*\w,0.4*\h)--(\r*\w-0.04*\w,0.4*\h)--(\l*\w+0.08*\w,0.4*\h)--(\l*\w+0.03*\w,0.4*\h) --(\l*\w+0.03*\w,0.48*\h) --(\l*\w,0.48*\h);

\draw[fill = coloraa] (0,0) -- (\l*\w,0.3*\h) -- (\r*\w,0.1*\h) -- (\w,0)--(0,0);
\node at (0.8*\l*\w,0.1*\h) {$\Gal$};

\draw[fill = colorab] (\w,\h) -- (\l*\w,\h) -- (\l*\w,0.95*\h) -- (\r*\w,0.85*\h)-- (\r*\w+\lamb*\w,0.85*\h) -- (\w,\h);
\node at (\r*\w+0.08*\w,0.94*\h) {$\Ga$};

\draw[fill = colorm] (\l*\w,0.85*\h) rectangle (\r*\w+\lamb*\w,0.75*\h);
\draw[fill = colorm] (\l*\w+0.1*\w,0.75*\h) rectangle (\r*\w+0.5*\lamb*\w,0.7*\h);
\draw[fill = colorm] (\l*\w+0.2*\w,0.7*\h) rectangle (\r*\w+0.7*\lamb*\w,0.62*\h);
\draw[fill = colorm] (\l*\w,0.62*\h) rectangle (\r*\w-0.03*\w,0.55*\h);
\draw[fill = colorm] (\l*\w,0.55*\h) rectangle (\r*\w-0.01*\w,0.48*\h);
\draw[fill = colorm] (\l*\w+0.03*\w,0.48*\h) rectangle (\r*\w-0.02*\w,0.4*\h);
\node at (0.8*\l*\w+0.2*\r*\w,0.69*\h) {$G_3$};

\draw[fill = gray] (\l*\w-0.01*\w,0.3*\h) rectangle (\l*\w,\h);

\draw[fill = gray] (\r*\w+0.02*\w,0.62*\h) rectangle  (\r*\w,0.1*\h);

\draw (0,\h) -- (0,0) -- (\w,0) -- (\w,\h);
\draw (0*\w,-\os) node[below]{$0$}-- (0*\w,\h);
\draw (\w,-\os) node[below]{$\D$}-- (\w,\h);

\draw (\r*\w,-\os) node[below]{$r$}-- (\r*\w,\h);
\draw (\r*\w+\lamb*\w,0) -- (\r*\w+\lamb*\w,\h+\os) node[above]{$r+\lambda$};
\draw (\l*\w,-\os) node[below]{$\ell$}-- (\l*\w,\h);
\draw (\r*\w-\lamb*\w,-2*\os) node[below]{$r-\lambda\D$}-- (\r*\w-\lamb*\w,\h);
\draw (\r*\w-2*\lamb*\w,0) -- (\r*\w-2*\lamb*\w,\h+2*\os)node[above]{$r-2\lambda\D$};

\draw[dashed] (-\os,0.5*\h) node[left]{$\frac{1}{2}\opt$} -- (\w+\os,0.5*\h) ;
\draw[dashed] (-\os,1*\h) node[left]{$\opt$} -- (\w+\os,1*\h) ;


\begin{scope}[xshift = 1.3*\w cm]

\draw [fill = white!95!black] (0,0) -- (\l*\w,0.3*\h) --(\l*\w,\h) --(0,\h) -- (0,0);
\draw [fill = white!95!black] (\w,0) -- (\r*\w,0.1*\h) --(\r*\w,0.62*\h) --(\r*\w+0.7*\lamb*\w,0.62*\h) -- (\r*\w+0.7*\lamb*\w,0.7*\h) -- (\r*\w+0.5*\lamb*\w,0.7*\h) -- (\r*\w+0.5*\lamb*\w,0.75*\h) -- (\r*\w+\lamb*\w,0.75*\h)-- (\r*\w+\lamb*\w,0.85*\h) -- (\w,\h)--(\w,0);
\draw [fill = colorb] (\l*\w,0.85*\h) -- (\l*\w,0.95*\h) -- (\r*\w,0.85*\h)--(\l*\w,0.85*\h);
\draw [fill = colorb] (\l*\w,0.62*\h) -- (\l*\w,0.75*\h) -- (\l*\w+0.1*\w,0.75*\h)--(\l*\w+0.1*\w,0.7*\h) -- (\l*\w+0.2*\w,0.7*\h)--(\l*\w+0.2*\w,0.62*\h) -- (\l*\w,0.62*\h);
\draw [fill = colorb] (\l*\w,0.48*\h) -- (\l*\w,0*\h) -- (\r*\w-\lamb*\w,0) -- (\r*\w-\lamb*\w,0.4*\h) --(\l*\w+0.08*\w,0.4*\h)--(\l*\w+0.03*\w,0.4*\h) --(\l*\w+0.03*\w,0.48*\h) --(\l*\w,0.48*\h);

\draw[fill = coloraa] (0,0) -- (\l*\w,0.3*\h) -- (\r*\w,0.1*\h) -- (\w,0)--(0,0);
\node at (0.8*\l*\w,0.1*\h) {$\Gal$};

\draw[fill = colorab] (\w,\h) -- (\l*\w,\h) -- (\l*\w,0.95*\h) -- (\r*\w,0.85*\h)-- (\r*\w+\lamb*\w,0.85*\h) -- (\w,\h);
\node at (\r*\w+0.08*\w,0.94*\h) {$\Ga$};

\begin{scope}[yshift = 0.6*\h cm]
\draw[fill = colorm] (0,0.85*\h) rectangle (\r*\w+\lamb*\w-\l*\w,0.75*\h);
\draw[fill = colorm] (0,0.75*\h) rectangle (\r*\w+0.5*\lamb*\w-\l*\w-0.1*\w,0.7*\h);
\draw[fill = colorm] (0,0.7*\h) rectangle (\r*\w+0.7*\lamb*\w-\l*\w-0.2*\w,0.62*\h);
\draw[fill = colorm] (0,0.62*\h) rectangle (\r*\w-0.03*\w-\l*\w,0.55*\h);
\draw[fill = colorm] (0,0.55*\h) rectangle (\r*\w-0.01*\w-\l*\w,0.48*\h);
\draw[fill = colorm] (0,0.48*\h) rectangle (\r*\w-0.02*\w-\l*\w-0.03*\w,0.4*\h);
\end{scope}
\node at (0.8*\l*\w+0.2*\r*\w,0.69*\h) {$G_3$};

\draw[fill = gray] (\l*\w-0.01*\w,0.3*\h) rectangle (\l*\w,\h);

\draw[fill = gray] (\r*\w+0.02*\w,0.62*\h) rectangle (\r*\w,0.1*\h);

\draw[fill = coloriadd] (\r*\w-\lamb*\w,1*\h) rectangle coordinate[midway] (iadd) (\r*\w,1.5*\h); 
\draw (iadd) -- (\r*\w/2 + \w/2, 5/4*\h)  node[fill=white, inner sep = 1pt] {\small $i_{\lambda}$};
\draw[fill = coloriadd] (\r*\w-\lamb*\w,0.85*\h) rectangle (\r*\w,0.35*\h); 

\draw[fill = colorm] (\r*\w-\l*\w+\lamb*\w,\h) rectangle node[midway]{Steinberg} (\r*\w-\l*\w+13*\lamb*\w,1.5*\h);

\draw (0,1.5*\h) -- (0,0) -- (\w,0) -- (\w,1.5*\h);
\draw (0*\w,-\os) node[below]{$0$}-- (0*\w,\h);
\draw (\w,-\os) node[below]{$\D$}-- (\w,\h);

\draw (\r*\w,-\os) node[below]{$r$}-- (\r*\w,\h);
\draw (\r*\w+\lamb*\w,0) -- (\r*\w+\lamb*\w,1.5*\h+\os) node[above]{$r+\lambda$};
\draw (\l*\w,-\os) node[below]{$\ell$}-- (\l*\w,\h);
\draw (\r*\w-\lamb*\w,-2*\os) node[below]{$r-\lambda\D$}-- (\r*\w-\lamb*\w,1.5*\h);
\draw (\r*\w-2*\lamb*\w,0) -- (\r*\w-2*\lamb*\w,1.5*\h+2*\os)node[above]{$r-2\lambda\D$};
\draw (\r*\w-\l*\w+\lamb*\w,\h) -- (\r*\w-\l*\w+\lamb*\w,1.5*\h+\os)node[above]{$\eta+\lambda \D$};

\draw[dashed] (-\os,1.5*\h) node[left]{$\frac{3}{2}\opt$} -- (\w+\os,1.5*\h) ;
\draw[dashed] (-\os,0.5*\h) node[left]{$\frac{1}{2}\opt$} -- (\w+\os,0.5*\h) ;
\draw[dashed] (-\os,1*\h) node[left]{$\opt$} -- (\w+\os,1*\h) ;

\end{scope}

\end{tikzpicture}
    }
    \caption{A possible repacking generated by \Cref{alg_c} if $\sopt$ does neither satisfy Line~\ref{alg_c_midmountain}, Line~\ref{alg_c_rightmountain} nor Line~\ref{alg_c_largeoverlap}.}
    \label{fig:enter-label3}
\end{figure}

\begin{claim}\label{claim:c_rest}
    Assume \alertBound{$\lambda \leq \frac{1}{50}$}.
    If $\sopt$ does neither satisfy lines \ref{alg_c_midmountain},  \ref{alg_c_rightmountain} nor \ref{alg_c_largeoverlap}, \Cref{alg_c} returns a $\D$-feasible packing $\packing$ with total height bounded by $\big(\frac32 + \eps\big) \opt$, and we can place an extra item of height $\opt$ and width $\lambda \D$ in the segment $[r - \lambda \D, r)$. 
\end{claim}

\begin{proof}
    By \Cref{claim:c_no_mid-right_mountain}, $\start$ is $\D$-feasible, and the maximal height of~$\packing$ in the segments~$[0, (\eta+\lambda)\D)$ and~$[(\eta + 13\lambda)\D,\D)$ is at most~$\frac32\opt$ and~$\opt$, respectively. 
    
    For the segment $[r - \lambda\D, r)$, we observe again that~$(\eta + 13\lambda)\D \leq r - \lambda\D$ by \Cref{claim:c_no_mid-right_mountain}.
    Further, in~$\packing$, we removed any item in~$\Gb$ that had a non-empty intersection with the segment~$[r - \lambda \D, r)$ in~$\sopt$. Therefore, the only items with a non-empty intersection with the segment $[r - \lambda\D, r)$ in the packing~$\packing$ belong to $\Gal \cup \Ga \cup \{\iadd\}$. By our assumption, the maximal height of $\Gal \cup \Ga$ in the segment is at most $\frac12 \opt$ and $h(\iadd) \leq \opt$, implying the last statement of the lemma. 
\end{proof}

\begin{proof}[Proof of \Cref{lem_case_C}]
    \Cref{claim:c_midmountain,claim:c_rightmountain,claim:c_largeoverlap,claim:c_rest} cover the case where the optimal packing~$\sopt$ admits a gap with width~$\eta \in \big[\lambda, \big(\frac12 - 3 \lambda\big)\D\big]$, which implies that \Cref{alg_c} correctly constructs the packing~$\packing$ as required by the lemma.  
\end{proof}

If~$\max_{i \in \items_{h > \frac{\OPT}{2}}} \{\sopt(i) + w(i) \} \in \big[\big(\frac12 + 3\lambda\big)\D, (1-\lambda)\D\big]$, we let~$T_1$ be the tall item attaining this maximum and set~$r = \D$. 
Then, $\Ga = \emptyset$, and the following lemma can be proven along the lines of the proof of \Cref{lem:medium-wide-gap}, showing that \Cref{alg_c} with $r = \D$ returns a forgiving packing.

\begin{lemma}
    \label{lem:medium-gap-at-border}
    Assume $\sopt$ is an optimal packing of $(\items, \D)$ with a gap $[\ell, \D)$ of width $\D - \ell \in [\lambda \D, (\frac 12 - 3 \lambda) \D]$. 
    Then, $(\items, \D)$ admits a forgiving packing.
\end{lemma}

\section{Exactly one gap of width at least \texorpdfstring{$(\frac12 - 3\lambda)\D$}{(1/2-3λ)D}}
\label{sec:one-wide-gap}

In this section, we discuss how to reorder an optimal packing $\sopt$ with height at most~$\opt$ in the case that there is a gap with width at least $\big(\frac12 - 3\lambda\big)\D$, and tall items almost completely cover the residual packing.
Formally, we prove the following lemma.

\begin{lemma}
    \label{lem:one-wide-gap}
    Assume $\sopt$ is an optimal packing of $(\items, \D)$ with a gap $[\ell,r)$ such that
    \begin{enumerate}
        \item the gap width satisfies $r - \ell \geq (\frac 12 - 3 \lambda) \D$, 
        \item the gaps to the left of $\ell$ have a combined width of at most $\lambda \D$, and
        \item the gaps to the right of $r$ have a combined width of at most $\lambda \D$.
    \end{enumerate}
Then $(\items, \D)$ admits a neat packing.
\end{lemma}

Throughout this section we assume that the assumptions on $\sopt$ of the above lemma hold.
In order to prove the lemma, 

we consider three cases of reordering the packing depending on the exact position of the gap. 
The proof of \Cref{lem:one-wide-gap} then directly follows from the Lemmas~\ref{lem:repacking_l-small_r-at-least-1/2}, \ref{lem:repacking_large-gap_ell-large}, and~\ref{lem:l-close-to-border_r-smaller-than-half}.

We will sort the tall items by height and place them left-aligned, tallest first, in all three cases (as required in a neat packing).
This allows us to ignore all squeezable items, i.e., non-tall items with a width of at most $\frac{\eps }{(1+\eps)}\D$, since we can pack items with height at most~$\frac{\opt}{2}$ and width at most~$\frac{\eps}{1+\eps}\D$ using the Squeezing Algorithm~\ref{alg:iterated-squeezing}. 
As observed in \cref{cor:stretching_without_narrow_items}, this implies that~$\items_S$ in the output of the stretching algorithms \cref{alg_right_stretching,alg_left_stretching} is empty.

Similar to previous sections, we assume that
there exists a segment $[\ell,r)$ in $\sopt$ that contains no tall item,  
there exists a tall item $T_1$ with $\start(T_1)+w(T_1) = \ell$ or $\ell = 0$, there exists a tall item $T_2$ with $\start(T_2) = r$ or $r = \D$, $r - \ell \in \big(\big(\frac12 - 3\lambda\big)\D,\D\big]$, and $\ell \leq \D - r$. 
Let $d_\ell := \frac{\ell - w\big(\items^{\sopt}_{[0,\ell]} \cap \items_{h > \frac\opt2}\big)}{\D}$ and $d_r := \frac{(\D - r) - w\big(\items^{\sopt}_{[r,\D]} \cap \items_{h > \frac\opt2}\big)}{\D}$ be the relative width of the segments~$[0,\ell)$ and~$[r,\D]$, respectively, not covered by tall items.

In all three cases, we assume that $\alertBound{d_\ell\leq \frac{\eps}{1+\eps}}$ and $\alertBound{d_r\leq \frac{\eps}{1+\eps}}$. 
However, depending on the case, we might require stronger bounds.
This bound implies that, in $\sopt$, each non-tall item with a width larger than $\frac{\eps}{1+\eps}\D$ that is not completely contained in the gap $[\ell,r)$ is processed in parallel to some tall item. 
Additionally, we assume that $\ell \leq \D-r$ in all three cases.

We will partition the packing into several sets of items depending on how they are positioned relative to the gap. 
Roughly speaking, we have the set that intersects the right border of the gap, the set that intersects the left border of the gap, the set that is wholly contained in the gap, the set of items left of the gap, and the set of items right of the gap.
However, the exact definition will slightly differ for each case. 
In general, we will make sure that sets that are treated similarly by our algorithms receive the same name across the sub-sections.

In each of the three cases, the analysis follows the same idea: We partition~$[0,\D]$ into suitably chosen sub-segments (depending on the case assumptions, the exact definitions of the item sets, and the algorithm) and show for each of the sub-segments that the height of the new packing is at most $\frac32\opt$. 
As in the previous sections, we provide figures highlighting the repacking and giving an overview of the analysis. 
We separately prove that all items are packed in~$[0,\D]$, i.e., no item starts before~$0$ and no item finishes after~$\D$. 

The structure of this section is as follows.
First, in the next paragraph, we show a useful lemma that we use for shifting non-tall items over tall items in the subsequent subsections.
We then consider three different cases depending on the boundaries $\ell$ and $r$ of the gap: 
First, we consider the case that $\ell \leq \frac{\eps}{1+\eps} \D$ and $r \geq \frac 1 2 \D$.
Second, we consider the case that $\ell \geq \frac{\eps}{1+\eps} \D$.
Finally, we consider the case that $\ell \leq \frac{\eps}{1+\eps} \D$ and $r \leq \frac 1 2 \D$.
We remark that we have arranged the three cases by order of increasing complexity of their analyses.
We emphasize that throughout this section, we assume that the gap has a width of at least $\big(\frac{1}{2} - 3 \lambda\big) \D$.

\paragraph*{Shifting non-tall items over tall items.}
In the following, we design an intermediate repacking procedure that is needed by the repacking algorithms described in this section and the next. It outputs a packing of height~$H$ where the $H$-tall items are packed in order of non-increasing height, and a subset of the non-$H$-tall items is shifted to the right (compared to the original packing).

Let~$\start$ be a packing of height~$H$, where the $H$-tall items~$\items_{h > \frac H2}$ can be partitioned into
\[
    \itemsT{\start}_{[0,\ell]} \cap  \items_{h > \frac H2}, \quad  \text{ and } \quad \itemsT{\start}_{[r,\D]} \cap  \items_{h > \frac H2}
\]
for $0 \leq \ell < r \leq \D$. 
Let $d_{\ell} = \frac{\ell - w(\itemsT{\start}_{[0,\ell]} \cap  \items_{h > \frac H2})}{\D}$ denote the relative total width of segments in $[0, \ell)$ not covered by items with height at least~$\frac H2$ and $d_r = \frac{(\D-r) - w(\itemsT{\start}_{[r,\D]} \cap  \items_{h > \frac H2})}{\D}$ denote the relative total width of segments in $[r, \D)$ not covered by items with height at least~$\frac H2$.  
Let 
\[
\Gb = {\items_{\sopt(i) + w(i) \in[\ell, r + d_r\D]}} \setminus \items_{h> \frac{\opt}{2}}
\]
denote the items that are either completely packed in the segment $[\ell,r + d_r \D)$ and have a height of at most $\frac H2$ or start before $\ell$ and finish before $r+ d_r\D$.

\begin{algorithm}
    \DontPrintSemicolon
    \caption{Shifting non-tall items over tall items}
    \label{alg:non-tall-over-tall}
    \textbf{Input:}  $(\start,\items_{h > \frac H2},I_{\mathrm{shift}},\ell,r,d_r)$ \tcp*[r]{packing~$\start$ with height~$H$} 
    \smallskip    
    $\start' \leftarrow $ output of \Cref{alg_items_next_to_each_other} with $(\items_{h > \frac H2},0)$ \tcp*[r]{sort $\items_{h > \frac H2}$ by height}  
    \lForEach(\tcp*[f]{shift $I_{shift}$ to the right in $\start'$}){$i \in I_{\mathrm{shift}}$}{$\start'(i) \leftarrow \start(i) + (\D - r) - d_r\D$} 
    \textbf{return} $\start'$ \;     
\end{algorithm}

An example of how \cref{alg:non-tall-over-tall} shifts the items $\items_{h > \frac H2} \cup I_{\ell,r} \cup \itemsatt{\start}{\ell}$ can be found in \cref{fig:shifting_non-tall_items_over_tall_items}.

\begin{figure}
\resizebox{\textwidth}{!}{
\begin{tikzpicture}
\pgfmathsetmacro{\w}{6}
\pgfmathsetmacro{\h}{3}
\pgfmathsetmacro{\os}{0.24}
\pgfmathsetmacro{\l}{0.28}
\pgfmathsetmacro{\r}{0.78}
\pgfmathsetmacro{\dFive}{0.03}
\pgfmathsetmacro{\dFour}{0.02}
\pgfmathsetmacro{\lOffset}{\r*\w-\dFour*\w-\dFive*\w}

\begin{scope}[xscale = -1, xshift = -\w cm]
\begin{scope}[yscale=-1,yshift= -\h cm]
\draw[fill = white!85!black] (\w,0) -- (\r*\w+\dFour*\w,0.2*\h) -- (\r*\w,0.4*\h)--(\r*\w,1*\h) -- (\w,\h)--(\w,0);
\draw[fill = gray] (\r*\w+0.02*\w,0.42*\h) rectangle (\r*\w,\h);
\draw[fill = gray] (\w,0.4*\h) rectangle (\r*\w+0.04*\w,\h);

\draw[fill = white!95!black] (0,0) -- (\l*\w,0.1*\h)-- (\r*\w,0.1*\h) -- (\l*\w,0.3*\h)--(\l*\w,\h) -- (0,\h)--(0,0);

\draw[fill = gray] (\l*\w-0.02*\w,0.4*\h) rectangle (\l*\w,\h);
\draw[fill = gray] (\l*\w-0.05*\w,0.45*\h) rectangle (\l*\w-0.02*\w,\h);
\draw[fill = gray] (0.05*\w,0.4*\h) rectangle (\l*\w-0.06*\w,\h);
\draw[fill = gray] (0.0*\w,0.35*\h) rectangle (0.01*\w,\h);
\draw[fill = gray] (0.02*\w,0.35*\h) rectangle (0.04*\w,\h);

\draw[fill = colorab] (\w,0) --(0*\w,0.0*\h) -- (\l*\w,0.1*\h)--(\r*\w,0.1*\h)--(\w,0);

\draw[fill = colorb](\w,0*\h)--(\r*\w,0.42*\h) --(\r*\w,\h)--(\l*\w,\h)-- (\l*\w,0.4*\h) -- (\l*\w-\dFive*\w,0.37*\h) -- (\l*\w,0.3*\h)  -- (\r*\w,0.1*\h) --(\w,0*\h);
\node at (0.5*\r*\w+0.5*\l*\w,0.7*\h) {$\Gb$};

\end{scope}

\draw (0,\h) -- (0,0) -- (\w,0) -- (\w,\h);
\draw (0*\w,-\os) node[below]{$\D$}-- (0*\w,\h);
\draw (\w,-\os) node[below]{$0$}-- (\w,\h);

\draw[thick,dashed] (\l*\w,-\os) node[below,black]{$r$}-- (\l*\w,\h);
\draw[thick,dashed] (\l*\w-\dFive*\w,0) -- (\l*\w-\dFive*\w,\h+\os) node[above,black]{$r+d_r\D$};
\draw[thick,dashed] (\r*\w,-\os) node[below,black]{$\ell$}-- (\r*\w,\h);

\draw[thick,dashed] (-\os,0.5*\h) -- (\w+\os,0.5*\h) node[left]{$\frac{1}{2}\opt$};
\draw[thick,dashed] (-\os,1*\h) -- (\w+\os,1*\h) node[left]{$\opt$};
\end{scope}

\begin{scope}[xscale = -1, xshift = -\w cm]
\begin{scope}[xshift = -1.3*\w cm, yshift=-0*\h cm]

\begin{scope}[xshift=-\l*\w cm+\dFive*\w cm,yscale=-1,yshift= -\h cm]
\draw[fill = colorb](\w,0*\h)--(\r*\w,0.42*\h) --(\r*\w,\h)--(\l*\w,\h)-- (\l*\w,0.4*\h) -- (\l*\w-\dFive*\w,0.37*\h) -- (\l*\w,0.3*\h)  -- (\r*\w,0.1*\h) --(\w,0*\h);
\node at (0.5*\r*\w+0.5*\l*\w,0.7*\h) {$\Gb$};
\draw[fill = colorab] (\w,0) --(0*\w,0.0*\h) -- (\l*\w,0.1*\h)--(\r*\w,0.1*\h)--(\w,0);
\end{scope}

\draw[dashed] (-\os,0.5*\h)  -- (\w+\os,0.5*\h)node[left]{$\frac{1}{2}\opt$};
\draw[dashed] (-\os,1*\h) -- (\w+\os,1*\h)  node[left]{$\opt$};

\draw (0,1*\h) -- (0,0) -- (\w,0) -- (\w,1*\h);


\draw[fill = gray] (0.05*\w - \l*\w+\r*\w,\h-0.45*\h) rectangle (0.08*\w - \l*\w+\r*\w,0*\h);
\draw[fill = gray] (0.08*\w - \l*\w+\r*\w,\h-0.42*\h) rectangle (0.10*\w - \l*\w+\r*\w,0*\h);
\draw[fill = gray] (0.10*\w - \l*\w+\r*\w,\h-0.4*\h) rectangle (1.08*\w - \l*\w,0*\h);
\draw[fill = gray] (1.08*\w - \l*\w,\h-0.4*\h) rectangle (1.10*\w - \l*\w,0*\h);
\draw[fill = gray] (1.10*\w - \l*\w,\h-0.4*\h) rectangle (0.97*\w,0*\h);
\draw[fill = gray] (0.97*\w,\h-0.35*\h) rectangle (0.98*\w,0*\h);
\draw[fill = gray] (0.98*\w,\h-0.35*\h) rectangle (1.00*\w,0*\h);

\draw[thick,dashed] (\dFive*\w,0) -- (\dFive*\w,1*\h+\os) node[above]{$\D-d_r \D$};
\draw[thick,dashed] (\w-\l*\w+\dFive*\w,-\os) node[below]{$\D-r+d_r\D$}-- (\w-\l*\w+\dFive*\w,1*\h);

\draw (0*\w,-\os) node[below]{$\D$}-- (0*\w,\h);
\draw (\w,-\os) node[below]{$0$}-- (\w,\h);
\end{scope}
\end{scope}
\end{tikzpicture}
}
\caption{An example for a repacking done by \cref{alg:non-tall-over-tall} for the set $I_{\mathrm{shift}} = I_{\ell,r} \cup \items^{\start}(\ell)$ where 
the blue items represent $\items_{\start(i) < \ell, \, \start(i) + w(i) > r + d_r\D}$.
}
\label{fig:shifting_non-tall_items_over_tall_items}
\end{figure}

\begin{lemma}
    \label{lem:G2-shift-over-tall-fit} 
    Let~$\start$ be a packing of height~$H$ where the items $\items_{h > \frac H2}$ can be partitioned into $\itemsT{\start}_{[0,\ell]} \cap  \items_{h > \frac H2}$ and  $\itemsT{\start}_{[r,\D]} \cap  \items_{h > \frac H2}$ for some $0 \leq \ell < r \leq \D$.    
    \Cref{alg:non-tall-over-tall} called with $(\start,\items_{h > \frac H2},I_{\ell,r},\ell,r,d_r)$ and called with $(\start,\items_{h > \frac H2},I_{\ell,r} \cup \itemsatt{\start}{\ell},\ell,r,d_r)$ creates two packings $\start_1$ and $\start_2$ of height at most $H$ for the items $\items_{h > \frac H2} \cup I_{\ell,r}$ and $\items_{h > \frac H2} \cup I_{\ell,r} \cup \itemsatt{\start}{\ell}$, respectively. 
    Further, in $\start_1$, all items finish at or before $\D$. 
\end{lemma}

\begin{proof}
    We observe that the $H$-tall items are packed in the segment~$[0, (1 - d_\ell - d_r) \D + \ell - r)$ as their total width is given by $(1 - d_\ell - d_r) \D + \ell - r$. Hence, in both packings, $\start_1$ and $\start_2$, the~$H$-tall items are packed in~$[0,\D)$. 
    
    We start by analyzing the structure of~$\start_1$. The items~$\Gb$ finish before~$r + d_r\D$ in~$\start$ and are shifted by $(D-r) - d_r\D$ to the right. Hence, 
    \[
        \start_1(i) + w(i) = \start(i) + w(i) + (D-r) - d_r\D \leq r + d_r\D + (D -r) - d_r\D \leq \D \, .
    \] 
    Since $\start_1(i) \geq \start(i) \geq 0$, this implies that $\Gb$ is packed in~$[0,\D)$ in $\start_1$ as required. (Observe that the lemma does not claim such a statement on $\start_2$.)

    For bounding the height of~$\start_1$ and $\start_2$, we observe that the earliest start time in~$\start$ of an item in $\Gb$ and in $\Gb \cup \itemsatt{\start}{\ell}$ is~$0$. 
    Hence, in the segment~$[0, (1-d_r)\D - r )$ no item from~$\Gb$ is packed in~$\start_1$ and no item from $\Gb \cup \itemsatt{\start}{\ell}$ is packed in~$\start_2$. 
    Thus, the height bound is trivially satisfied. 
    Similarly, in $[(1 - d_\ell - d_r) \D + \ell - r, \D)$ no $H$-tall items are packed as observed above, which also implies that the heights of~$\start_1$ and~$\start_2$ are bounded by~$H$.

    It remains to bound the height of~$\start_1$ and $\start_2$ in the segment~$[(1 - d_r) \D - r, (1 - d_\ell - d_r) \D + \ell - r)$. 
    We observe that in both packings, only items from~$\itemsoverlapt{\start}{\ell}$ overlap this segment. 
    Further,~$\start_1$ and~$\start_2$ coincide in the packing of the $H$-tall items and the items from {$\items_{\start(i) < \ell, \, \start(i) + w(i) \leq r + d_r \D}$}, 
    while~$\start_2$ additionally packs the items $\items_{\start(i) < \ell, \, \start(i) + w(i) > r + d_r \D}$. 
    Hence, showing the height bound for~$\start_2$ directly implies the same bound for~$\start_1$. 

    To this end, we consider an auxiliary packing~$\start'$ that packs the items~$\itemsT{\start}_{[0,\ell]} \cap  \items_{h > \frac H2}$ ordered by non-increasing height, starting at~$0$ and packs~$A := \itemsoverlapt{\start}{\ell}$ as~$\start$. 
    We will show that the height of~$\start'$ in the segment $[0, \ell)$ is at most~$H$. 
    We note that $h(A^{\start}(t))$ is non-decreasing in~$t$ for $t \in [0, \ell)$. 
    Consider $\itemsT{\start}_{[0,\ell]} \cap  \items_{h > \frac H2}$ and suppose that~$\start'(i) \leq \start(i)$. If~$t \in[\start'(i), \start'(i) + w(i))$, then 
    \[  
        h(A^{\start'}(t)) = h(A^{\start}(t)) \leq h\big( A^{\start} (t + \start(i) - \start'(i)) \big) \leq H - h(i)
    \]
    since $t + \start(i) - \start'(i) \in[\start(i),\start(i) + w(i))$ and the height of~$A$ is non-decreasing. 
    For $i \in \itemsT{\start}_{[0,\ell]} \cap  \items_{h > \frac H2}$ with $\start'(i) > \start(i)$, we know that there is a set of items $\items_i \subseteq \itemsT{\start}_{[0,\ell]} \cap  \items_{h > \frac H2}$ of total width at least $\start'(i) - \start(i)$ with $h(i') \geq h(i)$ and $\start(i) < \start(i')$ for all~$i' \in \items_i$. 
    Due to their total width, the last such item~$i_l$ finishes at or after $\start'(i) + w(i)$ in~$\start$. 
    Thus, 
    \[
        H \geq h \big( \{i_l\} \cup A^{\start}(\start(i_l) + w(i_l)) \big)\geq h(i) + h(A^{\start}(\start'(i) + w(i)) \, ,
    \]
    where we used that the height of~$A$ is non-decreasing. Using this fact again and rearranging this bound shows that 
    \[
        h(A^{\start'}(t)) \leq H - h(i)
    \]
    for all $t \in [\start'(i), \start'(i) + w(i))$, which in turn implies that $h(\itemsatt{\start'}{t}) \leq H$ for all $t \in [0, \ell - d_\ell\D) $. 

    Recall that the width of~$G_r$ is $(\D-r) - d_r \D$. 
    Since~$\start_2$ packs the items in $\items^{\start'}_{h>\frac\opt2}$ in order of non-increasing height, this implies that 
    \[
        h(\itemsatt{\start_2}{t}\cap \items_{h>\frac\opt2}) \leq h \big(\itemsatt{\start'}{t - (\D - r - d_r\D)}\cap\items_{h>\frac\opt2}\big)   
    \]
    for $t \in [(1 - d_r) \D - r, (1 - d_\ell - d_r) \D + \ell - r)$. 
    Recall that the items in $\itemsatt{\start}{\ell}$ are shifted by $(\D-r) - d_r\D$ to the right in $\start_2$ when compared to $\start'$. Hence, using the above bounds, we obtain that 
    \[
        h(\itemsatt{\start_2}{t}) \leq h(\itemsatt{\start'}{t - (\D - r - d_r\D)}) \leq H \, ,
    \] 
    which concludes the proof of the lemma. 
\end{proof}

\subsection{A gap with  \texorpdfstring{$\ell \leq \frac{\eps}{1+\eps}\D$}{ℓ ≤ ε/(1+ε)D} and \texorpdfstring{$r\geq \frac{\D}{2}$}{r ≥ D/2}}

In this section, we consider the case where there is a gap of width at least~$\big(\frac 12 -3\lambda\big)\D$ and $\ell$ is close to 0, i.e., $\ell \leq \frac{\eps}{1+\eps}\D$. 
Furthermore, we assume that $r \geq \frac{\D}{2}$.
We design \cref{alg_l_close_to_border_r_larger_than_half} and prove that it returns a feasible packing of height at most~$\frac32\opt$.

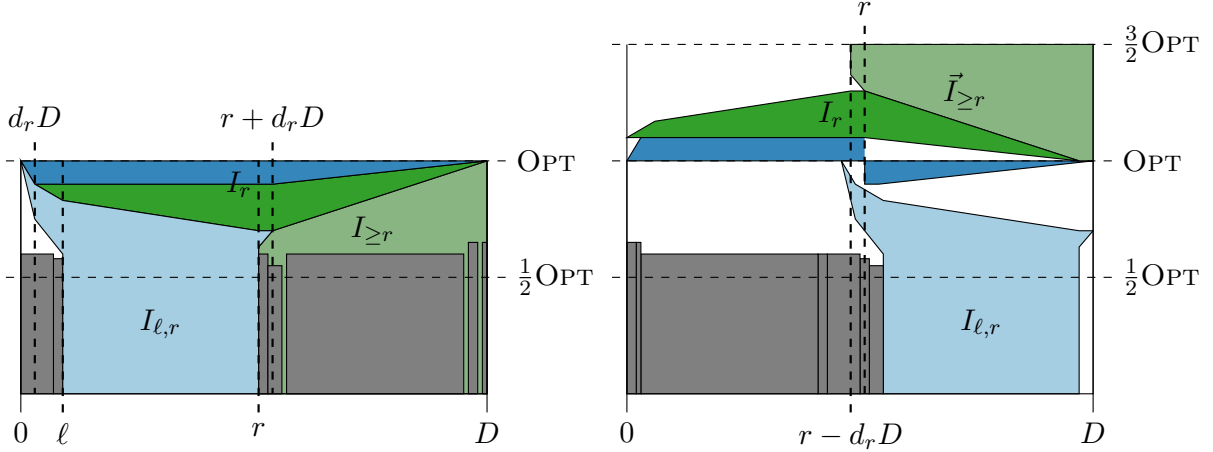
\begin{figure}
\resizebox{\textwidth}{!}{
\begin{tikzpicture}
\pgfmathsetmacro{\w}{6}
\pgfmathsetmacro{\h}{3}
\pgfmathsetmacro{\os}{0.24}
\pgfmathsetmacro{\l}{0.49}
\pgfmathsetmacro{\r}{0.91}
\pgfmathsetmacro{\dFive}{0.03}
\pgfmathsetmacro{\g1}{0.5 +\l -0.015}
\pgfmathsetmacro{\hBlue}{0.1}
\pgfmathsetmacro{\hYellow}{0.07}
\pgfmathsetmacro{\hGrayR}{0.07}
\pgfmathsetmacro{\hGrayL}{0.2}
\pgfmathsetmacro{\hGTwo}{0.4-\hGrayR-\hYellow-\hBlue}

\begin{scope}[xscale=-1,xshift = -\w cm]

\begin{scope}[yscale=-1,yshift= -\h cm]
\draw[fill = gray] (\r*\w+0.02*\w,0.42*\h) rectangle (\r*\w,\h);
\draw[fill = gray] (\w,0.4*\h) rectangle (\r*\w+0.02*\w,\h);

\draw[fill = colore] (0,0)-- (\l*\w-\dFive*\w,\hGrayL*\h+\hBlue*\h) --(\l*\w,\hGrayL*\h+\hBlue*\h+\hYellow*\h)--(\l*\w,\h) -- (0,\h)--(0,0);
\node at (0.5*\l*\w,0.3*\h) {$\Ge$};

\draw[fill = gray] (\l*\w-0.02*\w,0.4*\h) rectangle (\l*\w,\h);
\draw[fill = gray] (\l*\w-0.05*\w,0.45*\h) rectangle (\l*\w-0.02*\w,\h);
\draw[fill = gray] (0.05*\w,0.4*\h) rectangle (\l*\w-0.06*\w,\h);
\draw[fill = gray] (0.0*\w,0.35*\h) rectangle (0.01*\w,\h);
\draw[fill = gray] (0.02*\w,0.35*\h) rectangle (0.04*\w,\h);

\draw[fill = colorab] (0,0) -- (\l*\w-\dFive*\w,\hGrayL*\h+\hBlue*\h) -- (\l*\w,\hGrayL*\h+\hBlue*\h) -- (\r*\w,\hGrayR*\h+\hBlue*\h)--(\w-\dFive*\w,\hBlue*\h) -- (\l*\w-\dFive*\w,\hBlue*\h)--(0,0);
\draw[fill = coloraa] (0,0) -- (\l*\w-\dFive*\w,\hBlue*\h) -- (\l*\w,\hBlue*\h) -- (\r*\w,\hBlue*\h)--(\w-\dFive*\w,\hBlue*\h)--(\w,0)--(\w-\dFive*\w,0)--(\r*\w,0) -- (\l*\w,0) -- (\l*\w-\dFive*\w,0)--(0,0);
\node at (0.9*\l*\w + 0.1*\r*\w,0.1*\h) {$\Ga$};
\draw[fill = colorb] (\l*\w-\dFive*\w,\hGrayL*\h+\hBlue*\h) -- (\l*\w,\hGrayL*\h+\hBlue*\h+\hYellow*\h) -- (\l*\w,\h)-- (\r*\w,\h)-- (\r*\w,\hGrayR*\h+\hBlue*\h+\hYellow*\h+\hGTwo*\h) --(\w-\dFive*\w,\hBlue*\h+\hYellow*\h+0.5*\hGTwo*\h)--(\w,0*\h)--(\w-\dFive*\w,\hBlue*\h)--(\r*\w,\hGrayR*\h+\hBlue*\h)--(\l*\w,\hGrayL*\h+\hBlue*\h)--(\l*\w-\dFive*\w,\hGrayL*\h+\hBlue*\h);
\node at (0.5*\l*\w+0.5*\r*\w,0.7*\h) {$\Gb$};
\end{scope}
\end{scope}



\begin{scope}[xscale=-1,xshift = -\w cm]
\draw (0,\h) -- (0,0) -- (\w,0) -- (\w,\h);
\draw (0,-\os) node[below]{$\D$}-- (0,\h);
\draw (\w,-\os) node[below]{$0$}-- (\w,\h);
\draw[thick,dashed] (\l*\w,-\os) node[below]{$r$}-- (\l*\w,\h);
\draw[thick,dashed] (\r*\w,-\os) node[below]{$\ell$}-- (\r*\w,\h);
\draw[thick,dashed] (\l*\w-\dFive*\w,\h+\os) node[above]{$r+d_r\D$}-- (\l*\w-\dFive*\w,0);
\draw[thick,dashed] (\w-\dFive*\w,\h+\os) node[above]{$d_r\D$}-- (\w-\dFive*\w,0);

\draw[dashed] (-\os,0.5*\h) node[right]{$\frac{1}{2}\opt$} -- (\w+\os,0.5*\h) ;
\draw[dashed] (-\os,1*\h) node[right]{$\opt$} -- (\w+\os,1*\h) ;
\end{scope}

\begin{scope}[xscale=-1,xshift = -\w cm]
\begin{scope}[xshift=-1.3*\w cm]

\begin{scope}[xshift=-\l*\w cm+\dFive*\w cm,yscale=-1,yshift= -\h cm]
\draw[fill = colorb] (\l*\w-\dFive*\w,\hGrayL*\h+\hBlue*\h) -- (\l*\w,\hGrayL*\h+\hBlue*\h+\hYellow*\h) -- (\l*\w,\h)-- (\r*\w,\h)-- (\r*\w,\hGrayR*\h+\hBlue*\h+\hYellow*\h+\hGTwo*\h) --(\w-\dFive*\w,\hBlue*\h+\hYellow*\h+0.5*\hGTwo*\h)--(\w,0*\h)--(\w-\dFive*\w,\hBlue*\h)--(\r*\w,\hGrayR*\h+\hBlue*\h)--(\l*\w,\hGrayL*\h+\hBlue*\h)--(\l*\w-\dFive*\w,\hGrayL*\h+\hBlue*\h);
\node at (0.5*\l*\w+0.5*\r*\w,0.7*\h) {$\Gb$};
\end{scope}

\begin{scope}[yscale=-1,yshift= -\h cm]
\draw[fill = coloraa] (0,0) -- (\l*\w-\dFive*\w,\hBlue*\h) -- (\l*\w,\hBlue*\h) -- (\l*\w,0) -- (\l*\w-\dFive*\w,0)--(0,0);
\end{scope}

\begin{scope}[yscale=1, yshift= 1*\h cm, xshift= \dFive*\w cm]
\draw[fill = colore] (-0.03*\w,0) --(0,0)-- (\l*\w-\dFive*\w,\hGrayL*\h+\hBlue*\h) --(\l*\w,\hGrayL*\h+\hBlue*\h+\hYellow*\h)--(\l*\w,0.5*\h) -- (-0.03*\w,0.5*\h)--(-0.03*\w,0);
\node at (0.5*\l*\w,0.3*\h) {$\Vec{I}_{\geq r}$};
\draw[fill = colorab] (0,0) -- (\l*\w-\dFive*\w,\hGrayL*\h+\hBlue*\h) -- (\l*\w,\hGrayL*\h+\hBlue*\h) -- (\r*\w,\hGrayR*\h+\hBlue*\h)--(\w-\dFive*\w,\hBlue*\h) -- (\l*\w-\dFive*\w,\hBlue*\h)--(0,0);
\node at (0.9*\l*\w + 0.1*\r*\w,0.2*\h) {$\Ga$};
\end{scope}

\begin{scope}[yscale=1, yshift= 1*\h cm]
\draw[fill = coloraa] (\l*\w,\hBlue*\h) -- (\r*\w,\hBlue*\h)--(\w-\dFive*\w,\hBlue*\h)--(\w,0)--(\w-\dFive*\w,0)--(\r*\w,0) -- (\l*\w,0);
\end{scope}

\begin{scope}[yscale=-1, yshift= -1*\h cm]
\draw[fill = gray] (0.03*\w - \l*\w+\r*\w,0.45*\h) rectangle (0.06*\w - \l*\w+\r*\w,\h);
\draw[fill = gray] (0.06*\w - \l*\w+\r*\w,0.42*\h) rectangle (0.08*\w - \l*\w+\r*\w,\h);
\draw[fill = gray] (0.08*\w - \l*\w+\r*\w,0.4*\h) rectangle (1.06*\w - \l*\w,\h);
\draw[fill = gray] (1.06*\w - \l*\w,0.4*\h) rectangle (1.08*\w - \l*\w,\h);
\draw[fill = gray] (1.08*\w - \l*\w,0.4*\h) rectangle (0.97*\w,\h);
\draw[fill = gray] (0.97*\w,0.35*\h) rectangle (0.98*\w,\h);
\draw[fill = gray] (0.98*\w,0.35*\h) rectangle (1.00*\w,\h);
\end{scope}

\draw[dashed] (-\os,0.5*\h) node[right]{$\frac{1}{2}\opt$} -- (\w+\os,0.5*\h) ;
\draw[dashed] (-\os,1*\h) node[right]{$\opt$}-- (\w+\os,1*\h) ;
\draw[dashed] (-\os,1.5*\h) node[right]{$\frac{3}{2}\opt$} -- (\w+\os,1.5*\h) ;

\draw (0,1.5*\h) -- (0,0) -- (\w,0) -- (\w,1.5*\h);

\draw[thick,dashed] (\l*\w+\dFive*\w,-\os) node[below]{$r-d_r\D$}-- (\l*\w+\dFive*\w,1.5*\h);
\draw[thick,dashed] (\l*\w,1.5*\h+\os) node[above]{$r$}-- (\l*\w,0*\h);
\draw (0,-\os) node[below]{$\D$}-- (0,\h);
\draw (\w,-\os) node[below]{$0$}-- (\w,\h);
\end{scope}
\end{scope}

\end{tikzpicture}
}
\caption{An example for a repacking done by \cref{alg_l_close_to_border_r_larger_than_half} for $\ell \leq \frac{\eps}{1+\eps}\D$ and $r\geq \frac{\D}{2}$.}
\label{fig:l_close_to_border_r_larger_than_half}
\end{figure}

Given an optimal packing $\sopt$, we partition the set $\items_{h\leq \frac{\opt}{2}} $ as follows:
\[
\Ga := {\items_{ \sopt(i) < r, \, \sopt(i) + w(i) > r + d_r\D}}
, \hspace{1ex} 
\Gb := {\items_{\sopt(i) + w(i) \in[\ell, r + d_r\D]}} \setminus \items_{h> \frac{\opt}{2}}
, \hspace{1ex}
\Ge := \items^{\sopt}_{[r,\D]}\setminus \items_{h > \frac{\opt}{2}} 
\]

We observe that the two sets~$\Gb$ and~$\Ge$ are disjoint since any non-tall item starting at or after~$r$ does not finish before~$r + \frac\eps{1+\eps}\D$ due to our assumption on the minimal width of such an item and~$d_r < \frac\eps{1+\eps}$ by assumption. 
We will later use the notation of $\Gd:= \items^{\sopt}_{[0,\ell]}\setminus \items_{h > \frac{\opt}{2}}$. However, in this particular case, $\Gd$ is empty since $\ell \leq \frac{\eps}{1+\eps}\D$ and we assume that all non-tall items have a width of at least $\frac{\eps}{1+\eps}\D$.

\begin{algorithm}[ht]
\DontPrintSemicolon
\caption{Repacking for $\ell \leq \frac{\eps}{(1+\eps)}$ and $r \geq \frac{\D}{2}$}
\label{alg_l_close_to_border_r_larger_than_half}
\textbf{Input:}  $(\sopt,\items  \setminus \items_{\mathrm{sq,\eps,\opt}},\Ga,\Gb,\Ge, \ell,r, d_r)$\;
\smallskip
$\start \leftarrow$ output of \cref{alg:non-tall-over-tall} with $\big(\sopt,\items_{h > \frac \opt2},\Gb,\ell,r,d_r\big)$\tcp*[t]{shift $\Gb$, sort $\items_{h>\frac{\opt}{2}}$}
$\start_{r} \leftarrow$ output of \Cref{alg_stretching} with $\big(\sopt,  \frac{\opt}{2}, \tau_{\min} = r,\D\big)$  \tcp*[r]{right-stretch $\Ge$}

\lForEach(\tcp*[f]{shift $\Ge$ by $-d_r\D$}){$i \in \Ge$}{
$\start(i) \leftarrow \start_{r}(i) -d_r \D$
}
\lForEach(\tcp*[f]{start $\Ga$ at $0$}){$i \in \Ga$}{
    $\start(i) \leftarrow 0$ 
}
\smallskip 
\textbf{return} $\start$ \;
\end{algorithm}

\Cref{alg_l_close_to_border_r_larger_than_half} does the following: Given a packing $\sopt$ of height $\opt$ and width $\D$, the algorithm first uses the packing $\start_1$ output by \Cref{alg:non-tall-over-tall} that sorts the tall items by non-increasing height, starting at $0$ and shifts the items in $\Gb$ to the right by $(1-d_r)\D-r$.
Then, the algorithm right-stretches the items in $\Ge$ by at most $d_r\D$ and shifts them to the left by $d_r\D$ compared to $\sopt$ such that the first item from $\Ge$ starts not earlier than $r-d_r\D$ and the last item ends no later than $\D$.
Finally, the remaining items, which are the items in $\Ga$, are packed to start at $0$.
\Cref{fig:l_close_to_border_r_larger_than_half} pictures a repacking done by \Cref{alg_l_close_to_border_r_larger_than_half}. 
Here, the items in $\Ga$ are light blue and dark blue (depending on their starting time in $\sopt$), and the items in $\Gb$ are pink.
Note that the items from $\Ga$ are not packed as described by the algorithm. 
Instead, they are pictured closer to the argument used to show that the packing does not exceed the height of $\frac{3}{2}\opt$, as the dark blue part of $\Ga$ is shifted to the left by at least $d_r\D$ by the algorithm.
We now prove the following lemma.

\begin{lemma}
\label{lem:repacking_l-small_r-at-least-1/2}
Assume $\sopt$ is an optimal packing of $(\items, \D)$ such that there is a gap $[\ell,r)$ with $0 \leq \ell \leq \frac{\eps}{1+ \eps} \D$ and $r \geq \frac{\D}{2}$ and the gaps to the right of $r$ have a combined width of at most $\frac{\eps}{1 + \eps}$. 
Then $(\items, \D)$ admits a neat packing.
\end{lemma}

\begin{proof}
Let $T_1, T_2, \ell, r$ be according to the above definitions and satisfying the conditions of the lemma.
We show that \cref{alg_l_close_to_border_r_larger_than_half} returns a neat packing~$\start$ of the items $\items \setminus \items_{h \leq \frac\opt2, w \leq \frac\eps{1+\eps}\D}$.
First note that there is no non-tall item ending before $\ell$ since $\ell \leq \frac{\eps}{1+\eps}\D$ and all non-tall items have a width at least $\frac{\eps}{1+\eps}\D$.
For the same reason, no non-tall item is completely contained in the segment $[r,r+d_r\D)$.
Furthermore, note that for any $t \in [0,\D]$ it holds by \cref{lem:G2-shift-over-tall-fit}, that 
\begin{equation}
\label{eq:repacking_l-small_r-at-least-1/2_1}
h(\itemsatt{\start}{t} \cap (\Gb \cup \items_{h>\frac{\opt}{2}})) \leq \opt.
\end{equation}

We first prove the height bound of $\frac 32 \opt$ individually for the three segments $[0,r-d_r\D)$, $[r-d_r\D, r)$, and $[r, \D)$ and then prove that all items are completely packed in $[0, D)$.

\bigskip 

\noindent $\bullet \enspace  \bm{[0,r-d_r\D).} \enspace $ Let~$t \in [0,r-d_r\D)$. 
First note that $\itemsatt{\start}{t}$ does not contain any item from $\Ge$ as their starting point in $\sopt$ is at least $r$, the right-stretching does not decrease their starting times, and the algorithm shifts them to the left by $d_r \D$. 
Hence, the only items packed in this segment are from the set $\Ga \cup \Gb \cup \items_{h>\frac{\opt}{2}}$. 
By \cref{obs_height_G1G2}, $h(\Ga) \leq \frac{\opt}{2}$. Combining with \eqref{eq:repacking_l-small_r-at-least-1/2_1}, we bound the height of the packing at $t$ by $\frac{3}{2}\opt$.

\bigskip 

\noindent $\bullet \enspace  \bm{[r-d_r\D ,r).} \enspace $ Since \cref{alg_l_close_to_border_r_larger_than_half} shifts~$\Ge$ by~$d_r\D$ to the left when compared to $\start_{r}$, we have that 
$\diffitemsatt{\Ge}{\start}{t} = \diffitemsatt{\Ge}{\start_r}{t + d_r\D}$ for $t \in [r-d_r\D, \D)$. 
Recall that the height of $\start_{r}$ in $[r, (1+d_r) \D)$ restricted to the non-tall items is at most $\frac\opt2$ by \cref{lem:right_stretching}. 
(Note that $\start_{r}$ is also defined for~$\Ga$ and satisfies $\start_{r}(i) = \sopt(i)$ for $i \in \Ga$ because $\sopt(i) < r = \tau_{\min}$.) 

By definition, any item in $\Ga$ intersects~$r$ and~$r + d_r\D$ in~$\sopt$, and such items do not start later in~$\start$ than in~$\sopt$. 
Hence, for $t \in [r, r + d_r\D)$, $ \diffitemsatt{\Ga}{\start_r}{t} = \Ga$, and, in particular, $h(\diffitemsatt{\Ge}{\start_r}{t}) \leq  \frac\opt2 - h(\Ga)$. 
Hence, for $t \in [r-d_r\D, r)$, we have 
\[
    h\big( \diffitemsatt{(\Ga \cup \Ge)}{\start}{t} \big)\leq h(\Ga) +h(\diffitemsatt{\Ge}{\start_{r}}{t+d_r\D}) \leq \frac{\opt}{2} \, . 
\] 
Combining again with \eqref{eq:repacking_l-small_r-at-least-1/2_1}, the packing at $t$ has a height of at most $\frac{3}{2}\opt$ in total.

\bigskip 

\noindent $\bullet \enspace  \bm{[r,\D).} \enspace $ 
We partition the set $\Ga$ into two parts 
$\Gaa := \{i \in \Ga \mid \sopt(i) \leq d_r\D\}$ and 
$\Gab := \{i \in \Ga \mid \sopt(i) > d_r\D\}$. 
As indicated in \Cref{fig:l_close_to_border_r_larger_than_half}, in this analysis, we will treat the items in $\Gaa$ and $\Gab$ differently. 

First, we observe that $h(\itemsatt{\start_r}{t}  \cap (\Ge\cup \Ga)) \leq \frac{\opt}{2}$ for all $t \in [0,\D]$ by \cref{lem:right_stretching}. 
By definition, $\sopt(i) > d_r\D$ for $i \in \Gab$. 
Therefore, the algorithm shifts each item $i \in \Gab$ to the left by at least $d_r\D$ when defining $\start(i) = 0$. 
Thus, 
\[h(\itemsatt{\start}{t} \cap (\Ge\cup \Gab)) \leq h(\itemsatt{\start_{r}}{t + d_r\D} \cap (\Ge\cup \Ga)) \leq \frac{\opt}{2},\]
for all $t \in [0,\D]$. 

Next, we prove that the height of $\Gaa  \cup \Gb \cup \items_{h > \frac{\opt}{2}}$ is bounded by $\opt$ in the segment $[r,\D)$. 
To this end, fix $t \in [r,\D) \subseteq [d_r\D + (\D - r - d_r\D),r+d_r\D + (\D - r - d_r\D))$ as $(\D-r)\leq r$ since $r \geq \frac{\D}{2}$ by assumption. 
We upper bound $h((\Gaa \cup \Gb \cup \items_{h>\frac{\opt}{2}})^{\start}(t))$ as follows. 
For analysis purposes, let $\hat\start$ be the output of \cref{alg:non-tall-over-tall} when called with $(\sopt, \items_{h > \frac\opt2}, \Gb \cup \itemsatt{\sopt}{\ell}, \ell, r, d_r)$. (Note that $\start$ as used in \cref{alg_l_close_to_border_r_larger_than_half} only shifts $\Gb$.) 
By \cref{lem:G2-shift-over-tall-fit}, $\start$ and $\hat\start$ coincide for $\Gb \cup \items_{h>\frac{\opt}{2}}$.
\begin{align*}
h(\diffitemsatt{(\Gaa \cup \Gb \cup \items_{h>\frac{\opt}{2}})}{\start}{t}) & \leq 
    h(\Gaa)+h(\diffitemsatt{(\Gb \cup \items_{h>\frac{\opt}{2}})}{\start}{t}) \\
    & = h(\Gaa)+h(\diffitemsatt{(\Gb \cup \items_{h>\frac{\opt}{2}})}{\hat{\start}}{t}) \\ 
     & \leq h(\diffitemsatt{(\Ga \cup \Gb \cup \items_{h>\frac{\opt}{2}})}{\hat{\start}}{t}) \\ 
     & \leq \opt. 
\end{align*}
The second inequality holds because the items in $\Gaa$ overlap the segment $[d_r\D,r+d_r\D]$ in $\sopt$  and the last inequality follows from \Cref{lem:G2-shift-over-tall-fit}. With this, we conclude the analysis of the height of our packing~$\start$ in~$[0,\D)$.

It remains to prove that all items are completely packed in~$[0,\D)$. To this end, we observe that in~$\start$, the items~$\Ga$ start at~$0$, implying that they finish at or before $\D$. \cref{lem:G2-shift-over-tall-fit} guarantees that~$\items_{h> \frac\opt2} \cup \Gb$ are also completely contained in~$[0,\D)$. For the items~$\Ge$, we observe that by \cref{lem:right_stretching} those items start at or after~$r \geq \frac\D2 > d_r\D$ and finish at or before $(1+d_r)\D$ in the intermediate packing~$\start_{r}$. As \cref{alg_l_close_to_border_r_larger_than_half} shifts them by $d_r\D$ to the left, the items in~$\Ge$ are packed in~$[r - d_r\D, \D)$ in~$\start$, which concludes the proof.
\end{proof}

\subsection{A gap of width at least \texorpdfstring{$(\frac 12 -3\lambda)\D$}{(1/2-3λ)D} and \texorpdfstring{$\frac{\eps}{1+\eps} \D  \leq \ell$}{ε/(1+ε)D ≤ ℓ}}

In this section, we consider the case where there is a gap of width at least~$\big(\frac 12 -3\lambda\big)\D$ and $\ell$ is not too close to 0. We design \cref{alg:gap_almost_1/2_not_at_border} and prove that it returns a feasible packing of height at most~$\frac32\opt$.
We partition the set of non-tall items as follows: 
\[
    \Ga = {\items_{ \sopt(i) < r, \, \sopt(i) + w(i) > r + d_r\D}}, \quad \quad \Gb := {\items_{\sopt(i) + w(i) \in[\ell, r + d_r\D]}} \setminus \items_{h> \frac{\opt}{2}},
\] 
\[ \Gd = {\items_{\sopt(i) + w(i) \leq \ell}}\setminus \items_{h> \frac{\opt}{2}}, \quad \quad  \Ge = {\items_{\sopt(i) \geq r}}\setminus \items_{h> \frac{\opt}{2}}.\]
Note that these sets are disjoint since all items have a width larger than $\frac{\eps }{1+\eps}\D \geq d_r\D$ and therefore no non-tall item is completely contained in the segment $[r,r+d_r\D)$.

The high-level idea of \Cref{alg:gap_almost_1/2_not_at_border} is to pack the tall items sorted by decreasing height starting at~$0$, shift the non-tall items in~$\Gb$ completely to the right, stretch the sets $\Gd \cup \Ge$, and carefully position $\Ga$ in order to not increase the height of the resulting packing beyond $\frac32\opt$. To this end, we further partition $\Ga$ according to the starting and end points of the items into the following  three subsets
\begin{align*}
\Gaa & = {\items_{\sopt(i) < \ell + (d_\ell + d_r)\D, \, r+ d_r\D < \sopt(i) + w(i) \leq (1-d_\ell)\D}} \, , \\
\Gab & = {\items_{\sopt(i) < \ell + (d_\ell + d_r)\D, \, \sopt(i) + w(i) > (1-d_\ell)\D}} \, \text{ and} \\
\Gac & = {\items_{ \ell + (d_\ell + d_r)\D \leq \sopt(i) < r, \, \sopt(i) + w(i) > r + d_r\D}}    
\end{align*}

\begin{figure}
    \centering
    \resizebox{\textwidth}{!}{
    \begin{tikzpicture}
\pgfmathsetmacro{\w}{6}
\pgfmathsetmacro{\h}{3}
\pgfmathsetmacro{\os}{0.24}
\pgfmathsetmacro{\l}{0.28}
\pgfmathsetmacro{\r}{0.78}
\pgfmathsetmacro{\dFive}{0.03}
\pgfmathsetmacro{\dFour}{0.02}
\pgfmathsetmacro{\lOffset}{\r*\w-\dFour*\w-\dFive*\w}

\begin{scope}[xscale = -1, xshift = -\w cm]
\begin{scope}[yscale=-1,yshift= -\h cm]
\draw[fill = colord] (\w,0) -- (\r*\w+\dFour*\w,0.2*\h) -- (\r*\w,0.4*\h)--(\r*\w,1*\h) -- (\w,\h)--(\w,0);
\node at (0.5*\w+0.5*\r*\w,0.3*\h) {$\Gd$};
\draw[fill = gray] (\r*\w+0.02*\w,0.42*\h) rectangle (\r*\w,\h);
\draw[fill = gray] (\w,0.4*\h) rectangle (\r*\w+0.04*\w,\h);

\draw[fill = colore] (0,0) -- (\dFour*\w,0.1*\h) -- (\l*\w,0.4*\h)--(\l*\w,\h) -- (0,\h)--(0,0);
\node at (0.4*\l*\w,0.3*\h) {$\Ge$};

\draw[fill = gray] (\l*\w-0.02*\w,0.4*\h) rectangle (\l*\w,\h);
\draw[fill = gray] (\l*\w-0.05*\w,0.45*\h) rectangle (\l*\w-0.02*\w,\h);
\draw[fill = gray] (0.05*\w,0.4*\h) rectangle (\l*\w-0.06*\w,\h);
\draw[fill = gray] (0.0*\w,0.35*\h) rectangle (0.01*\w,\h);
\draw[fill = gray] (0.02*\w,0.35*\h) rectangle (0.04*\w,\h);

\draw[fill = colorac] (0,0) -- (\dFour*\w,0.1*\h)-- (\l*\w-\dFive*\w,0.37*\h)-- (\l*\w,0.37*\h) -- (\lOffset,0.3*\h)--(\l*\w,0.3*\h)--(\dFour*\w,0.07*\h)--(0,0);
\draw[fill = coloraa] (0,0) -- (\dFour*\w,0.07*\h)-- (\lOffset,0.07*\h)--(\r*\w,0.05*\h)--(\w,0);
\draw[fill = colorab] (\dFour*\w,0.07*\h) -- (\l*\w-\dFive*\w,0.3*\h)-- (\lOffset,0.3*\h)--(\r*\w,0.1*\h)--(\w,0) -- (\r*\w,0.05*\h) -- (\lOffset,0.07*\h)--(\dFour*\w,0.07*\h);
\node at (0.83*\l*\w + 0.17*\r*\w,0.2*\h) {$\Gaa$};
\draw (0.45*\w,0.04*\h) -- (0.5*\w,-4*\os) node[above]{$\Gab$};
\draw (0.45*\w,0.32*\h) -- (0.37*\w,-4*\os) node[above]{$\Gac$};



\draw[fill = colorb] (\l*\w,0.4*\h) -- (\l*\w-\dFive*\w,0.37*\h) -- (\l*\w,0.37*\h) -- (\lOffset,0.3*\h) -- (\r*\w,0.1*\h) --(\w,0*\h)--(\r*\w,0.4*\h) -- (\r*\w,\h) -- (\l*\w,\h) -- (\l*\w,0.4*\h);
\node at (0.5*\l*\w+0.5*\r*\w,0.6*\h) {$\Gb$};

\end{scope}

\draw (0,\h) -- (0,0) -- (\w,0) -- (\w,\h);
\draw (0*\w,-\os) node[below]{$\D$}-- (0*\w,\h);
\draw (\w,-\os) node[below]{$0$}-- (\w,\h);

\draw[thick,dashed] (\l*\w,-\os) node[below,black]{$r$}-- (\l*\w,\h);
\draw[thick,dashed] (\l*\w-\dFive*\w,0) -- (\l*\w-\dFive*\w,\h+\os) node[above,black]{$r+d_r\D$};
\draw[thick,dashed] (\r*\w,-\os) node[below,black]{$\ell$}-- (\r*\w,\h);
\draw[thick,dashed] (\dFour*\w,\h+2.5*\os) node[above,black]{$\D-d_\ell \D$}-- (\dFour*\w,0);
\draw[thick,dashed] (\lOffset,\h+\os) node[above,black]{$\ell+ (d_\ell+d_r) \D$}-- (\lOffset,0);

\draw[thick,dashed] (-\os,0.5*\h) -- (\w+\os,0.5*\h) node[left]{$\frac{1}{2}\opt$};
\draw[thick,dashed] (-\os,1*\h) -- (\w+\os,1*\h) node[left]{$\opt$};
\end{scope}

\begin{scope}[xscale = -1, xshift = -\w cm]
\begin{scope}[xshift = -1.3*\w cm, yshift=-0*\h cm]

\begin{scope}[yscale=-1,yshift= -\h cm]
\draw[fill = coloraa] (0,0) -- (\dFour*\w,0.07*\h)-- (\r*\w-\dFour*\w-\l*\w,0.07*\h)--(\r*\w-\dFour*\w-\l*\w,0.0*\h) -- (0,0);
\end{scope}

\begin{scope}[yscale=-1,yshift= -\h cm,xshift=-\dFour*\w cm]
\draw[fill = colorab] (\dFour*\w,0.07*\h) -- (\l*\w,0.3*\h)-- (\r*\w-\dFour*\w-\l*\w,0.3*\h)-- (\r*\w-\dFour*\w-\l*\w,0.07*\h)--(\dFour*\w,0.07*\h);
\node at (0.75*\l*\w + 0.25*\r*\w,0.18*\h) {$\Gaa$};
\end{scope}

\draw[fill = coloraa] (\r*\w-2*\dFour*\w-\dFive*\w,\h-0.07*\h)-- (\lOffset,\h-0.07*\h)--(\lOffset,\h-0.07*\h)-- (\r*\w,\h-0.05*\h)--(\w,\h)--(\r*\w-\dFour*\w-\dFive*\w,\h)--(\r*\w-\dFour*\w-\dFive*\w,\h-0.07*\h);
\draw[white, fill=white] (\r*\w-\dFour*\w-3*\dFive*\w,\h) rectangle (\r*\w-\dFour*\w,0.8*\h);

\begin{scope}[xshift=-\l*\w cm+\dFive*\w cm,yscale=-1,yshift= -\h cm]

\draw[fill = colorb] (\l*\w,0.4*\h) -- (\l*\w-\dFive*\w,0.37*\h) -- (\l*\w,0.37*\h) -- (\lOffset,0.3*\h) -- (\r*\w,0.1*\h) --(\w,0*\h)--(\r*\w,0.4*\h) -- (\r*\w,\h) -- (\l*\w,\h) -- (\l*\w,0.4*\h);
\node at (0.4*\l*\w+0.5*\r*\w,0.6*\h) {$\Gb$};
\end{scope}

\begin{scope}[yscale=1, yshift= 1.5*\h cm]
\draw[fill = coloraa] (\r*\w-\dFour*\w-\l*\w,0)--(\r*\w-\dFour*\w-\l*\w,-0.07*\h)-- (\r*\w-\dFour*\w-\dFive*\w,-0.07*\h)--(\r*\w,-0.05*\h) --(\r*\w,0*\h) --(\r*\w-\dFour*\w-\l*\w,0);
\end{scope}
\begin{scope}[yscale=1, yshift= 1*\h cm]
\begin{scope}[xshift=-\dFour*\w cm]
\draw[fill = colorab] (\r*\w-\dFour*\w-\l*\w,0.3*\h)-- (\lOffset,0.3*\h)--(\r*\w,0.1*\h)--(\w,0) -- (\r*\w,0.05*\h) -- (\lOffset,0.07*\h)--(\r*\w-\dFour*\w-\l*\w,0.07*\h)--(\r*\w-\dFour*\w-\l*\w,0.3*\h);
\node at (0.25*\l*\w + 0.75*\r*\w,0.18*\h) {$\Gaa$};
\draw[fill = colord] (\w,0) -- (\r*\w+\dFour*\w,0.2*\h) -- (\r*\w,0.4*\h)--(\r*\w,0.5*\h) -- (\w+\dFour*\w,0.5*\h)--(\w+\dFour*\w,0);
\node at (0.5*\w+0.57*\r*\w,0.3*\h) {\reflectbox{$\vec{\vphantom{I}}$}\hspace{-0.9ex}$I_{\leq\ell}$};
\end{scope}
\end{scope}

\begin{scope}[yscale=1, yshift= 1*\h cm, xshift= \dFive*\w cm]
\draw[fill = colore] (-0.03*\w,0) --(0,0) -- (\dFour*\w,0.1*\h) -- (\l*\w,0.4*\h)--(\l*\w,0.5*\h) -- (-0.03*\w,0.5*\h)--(-0.03*\w,0);
\node at (0.5*\l*\w,0.4*\h) {$\vec{I}_{\geq r}$};
\draw[fill = colorac] (0,0) -- (\dFour*\w,0.1*\h)-- (\l*\w-\dFive*\w,0.37*\h)-- (\l*\w,0.37*\h) -- (\lOffset,0.3*\h)--(\l*\w,0.3*\h)--(\dFour*\w,0.07*\h)--(0,0);
\end{scope}

\draw[dashed] (-\os,0.5*\h) node[right]{$\frac{1}{2}\opt$} -- (\w+\os,0.5*\h) ;
\draw[dashed] (-\os,1*\h) node[right]{$\opt$} -- (\w+\os,1*\h) ;
\draw[dashed] (-\os,1.5*\h) node[right]{$\frac{3}{2}\opt$} -- (\w+\os,1.5*\h) ;

\draw (0,1.5*\h) -- (0,0) -- (\w,0) -- (\w,1.5*\h);


\draw[fill = gray] (0.05*\w - \l*\w+\r*\w,\h-0.45*\h) rectangle (0.08*\w - \l*\w+\r*\w,0*\h);
\draw[fill = gray] (0.08*\w - \l*\w+\r*\w,\h-0.42*\h) rectangle (0.10*\w - \l*\w+\r*\w,0*\h);
\draw[fill = gray] (0.10*\w - \l*\w+\r*\w,\h-0.4*\h) rectangle (1.08*\w - \l*\w,0*\h);
\draw[fill = gray] (1.08*\w - \l*\w,\h-0.4*\h) rectangle (1.10*\w - \l*\w,0*\h);
\draw[fill = gray] (1.10*\w - \l*\w,\h-0.4*\h) rectangle (0.97*\w,0*\h);
\draw[fill = gray] (0.97*\w,\h-0.35*\h) rectangle (0.98*\w,0*\h);
\draw[fill = gray] (0.98*\w,\h-0.35*\h) rectangle (1.00*\w,0*\h);

\draw[pattern=north west lines] (\dFour*\w+\dFive*\w,\h) rectangle (\w+\l*\w-\r*\w-\dFour*\w,1.07*\h);

\draw[thick,dashed] (\l*\w+\dFive*\w,0) -- (\l*\w+\dFive*\w,1.5*\h+\os) node[above]{$r - d_r \D$};
\draw[thick,dashed] (\dFour*\w,0) -- (\dFour*\w,1.5*\h+\os) node[above]{$\D-d_\ell \D$};
\draw[thick,dashed] (\r*\w,-\os) node[below]{$\ell$}-- (\r*\w,1.5*\h);
\draw[thick,dashed] (\r*\w-\dFour*\w,0) -- (\r*\w-\dFour*\w,1.5*\h+\os) node[above]{$\ell+d_\ell \D$};
\draw[thick,dashed] (\r*\w-\dFour*\w-\l*\w,1.5*\h) -- (\r*\w-\dFour*\w-\l*\w,-\os) node[below]{$\ell+\D-r+d_\ell \D$};
\draw[thick,dashed] (\r*\w+0.1*\w,-\os) node[below]{$\tau$}-- (\r*\w+0.1*\w,1.5*\h);

\draw (0*\w,-\os) node[below]{$\D$}-- (0*\w,\h);
\draw (\w,-\os) node[below]{$0$}-- (\w,\h);
\end{scope}
\end{scope}
\end{tikzpicture}
    }
    \caption{An example for a repacking done by \cref{alg:gap_almost_1/2_not_at_border}. \reflectbox{$\vec{\vphantom{I}}$}\hspace{-0.9ex}$I_{\leq\ell}$ and $\vec{I}_{\geq r}$ represent the stretched packings of $\Gd$ and $\Ge$ respectively.
    Note that in this example, it holds that $\tau' = 0$. The hatched area represents the part of the packing where the second part of $\Gd$ would be packed if $\tau' > 0$.} 
    \label{fig:case_d_part_2}
    \label{fig:gap_almost_1/2_not_at_border}
\end{figure}
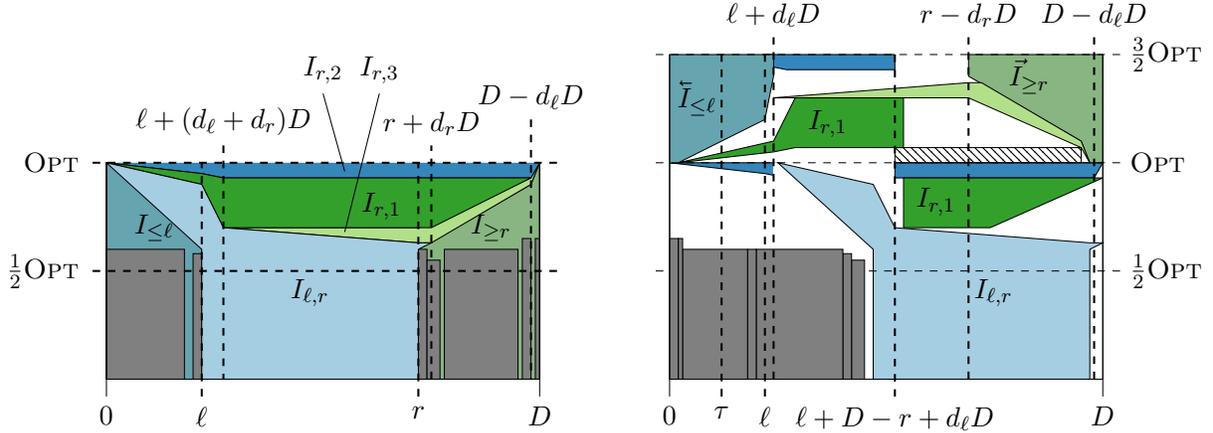

For an overview of the above-defined sets and the packing constructed by \Cref{alg:gap_almost_1/2_not_at_border}, refer to \Cref{fig:gap_almost_1/2_not_at_border}. 
Therein, the green, red and light gray set of items crossing $r$ are $\Gaa$, $\Gab$. and $G_{1,c}$, respectively, and together they form $\Ga$.

\begin{algorithm}[t]
    \DontPrintSemicolon
    \caption{Repacking for a gap with width at least $\big(\frac 12 - 3 \lambda\big) \D$, and $\ell \geq \frac\eps{1+\eps}$}
    \label{alg_D}
    \label{alg:gap_almost_1/2_not_at_border}
    \textbf{Input:}  $(\sopt,\items \setminus \items_{h \leq \frac\opt2, w \leq \frac\eps{1+\eps}\D},\Gaa, \Gab, \Gac, \Gb, \Gd, \Ge, \ell,r,d_\ell,d_r)$ \;
    \smallskip 
$\start \leftarrow$ output of \cref{alg:non-tall-over-tall} with $\big(\sopt,\items_{h > \frac \opt2},\Gb,\ell,r,d_r\big)$\tcp*[t]{shift $\Gb$, sort $\items_{h>\frac{\opt}{2}}$}

    \lForEach(\tcp*[f]{place $\Gaa$}){$i \in \Gaa$}{
        $\start(i) \leftarrow \sopt(i)+d_\ell\D$ 
    }
    \lForEach(\tcp*[f]{place $\Gab$}){$i \in \Gab 
    $}{
        $\start(i) \leftarrow \sopt(i)$ 
    }
    \lForEach(\tcp*[f]{place $\Gac$}){$i \in \Gac$}{
        $\start(i) \leftarrow \sopt(i)-d_r\D$ 
    }
    $\start_{r} \leftarrow$ output of \Cref{alg_stretching} with $\big(\sopt, \frac\opt2,r,\D\big)$  \tcp*[r]{right-stretch $\Ge$}
    \lForEach(\tcp*[f]{shift $\Ge$}){$i \in \Ge$}{
        $\start(i) \leftarrow \start_{r}(i)-d_r\D$ 
    }
    $\tau \leftarrow \min\{\tau \in [0,\D ] \mid \forall t > \tau$:  $h(\itemsatt{\start}{t}\cap(\items_{h> \frac{\opt}{2}} \cup \Gab)\leq \opt\big\}$ \;
    $\tau'  \leftarrow \max \big\{\tau' \in \big[0,\min\{ \tau, \ell + d_\ell \D \}\big] \mid \itemsatt{\sopt}{\tau'}\cap\items_{h\geq \opt- h(\Gab^{\sopt}(\tau))}\not = \emptyset\big \}$ \tcp*[r]{$\tau' \leq \ell + d_\ell \D$}
    $\start_{\ell} \leftarrow$ output of \Cref{alg_left_stretching} with $\big(\sopt, \frac\opt2,\ell,0\big)$  \tcp*[r]{left-stretch $\Gd$ left of $\ell$}
    $\start_{\ell}' \leftarrow$ output of \Cref{alg_left_stretching} with $\big(\sopt, \opt -h(\Gab^{\sopt}(\tau)),\tau',0\big)$  \tcp*[r]{left-stretch $\Gd$ left of $\tau'$}
    \lForEach(\tcp*[f]{shift $\Gd \cap \iopt_{\sopt(i) \leq \tau'}$}){$i \in \Gd \cap \iopt_{\sopt(i) \leq \tau'}$}{
        $\start(i) \leftarrow \start_{\ell}'(i)  + \ell+(1+2d_\ell) \D-r $ 
    } 
    \lForEach(\tcp*[f]{shift $\Gd \cap \items_{\sopt(i) > \tau'}$}){$ i \in \Gd \cap \items_{\sopt(i) > \tau'}$}{
        $\start(i) \leftarrow \start_{\ell}(i) -\tau' +d_\ell \D $ 
    }    
    \textbf{return} $\start$ \;
\end{algorithm}

The remainder of this section is to analyze \cref{alg:gap_almost_1/2_not_at_border} and prove the following lemma.

\begin{lemma}
\label{lem:alg_large_gap}
\label{lem:repacking_large-gap_ell-large}
    Assume $\sopt$ is an optimal packing of $(\items, \D)$ such that there is a gap $[\ell,r)$ such that
    \begin{enumerate}
        \item $\frac{\eps}{1 + \eps} \D \leq \ell \leq \D - r$,
        \item the gap width satisfies $r - \ell \geq (\frac 12 - 3 \lambda) \D$, 
        \item the gaps to the left of $\ell$ have a combined width of at most $\lambda \D$, and
        \item the gaps to the right of $r$ have a combined width of at most $\lambda \D$.
    \end{enumerate}
Then $(\items, \D)$ admits a neat packing.
\end{lemma}

\begin{proof}
    Let $T_1, T_2, \ell, r$ be according to the above definitions and satisfying the conditions of the lemma.
    We show that \cref{alg_D} returns a neat packing~$\start$ of the items $\items \setminus \items_{h \leq \frac\opt2, w \leq \frac\eps{1+\eps}\D}$.
    The proof consists of two parts: First, we show that the maximal height of the packing~$\start$ \emph{without}~$\Gd$ is bounded by~$\frac32\opt$, that is, $h(\itemsatt{\start}{t} \setminus \Gd) \leq \frac{3}{2}\opt$ for each $t \in [0,\D]$. Second, using that the algorithm packs~$\Gd$ in two ways depending on the starting time of an item with respect to~$\tau'$, we show that adding~$\Gd$ to the packing does not increase the maximal height. 

    We bound the height $h(\itemsatt{\start}{t} \setminus \Gd)$ by $\frac32\opt$ separately for $t \in [0,r-d_r \D]$ and~$t \in [r-d_r\D,\D]$.  

    \bigskip 

    \noindent $\bullet \enspace \bm{[0,r-d_r \D).} \enspace$ We observe that by using \Cref{alg:non-tall-over-tall} with $\big(\sopt,\items_{h > \frac \opt2},\Gb,\ell,r,d_r\big)$, \cref{lem:G2-shift-over-tall-fit} implies that the maximal height of the tall items together with~$\Gb$ is bounded by~$\opt$, i.e.,
    \begin{equation}
        h\big( \itemsatt{\start}{t} \cap (\Gb \cup \items_{h> \frac{\opt}{2}})\big) \leq \opt  
    \end{equation}
    for any $t \in [0,\D]$.
    As the earliest starting time of an item in~$\Ge$ in the new packing~$\start$ is at least~$r - d_r\D$ and the height of~$\Ga$ is at most~$\frac\opt2$ by \cref{obs_height_G1G2}, this implies that, indeed, $h(\itemsatt{\start}{t}\setminus \Gd) \leq \frac{3}{2}\opt$ for each $t \in [0,r-d_r \D)$. 

    \bigskip

    \noindent $\bullet \enspace \bm{[r-d_r\D,\D).} \enspace $ 
    By the Stretching Lemma~\ref{lem:stretching}, we have that 
    $h(\itemsatt{\start_r}{t}\cap (\Ga \cup \Ge)) \leq \frac{\opt}{2}$ for all~$t \in [0,\D)$, and, as the sets~$\Gac$ and~$\Ge$ are both shifted to the left by $d_r\D$, this implies the same bound in $\start$, i.e., for all $t \in [0,\D)$ 
    \[
        h(\itemsatt{\start}{t}\cap (\Gac \cup \Ge)) \leq \frac{\opt}{2}\, . 
    \] 
    By definition, the items in~$\Gaa\cup \Gab$ start before $\ell +(d_\ell+d_r)\D$ and finish after $r + d_r\D$ in $\sopt$.  Hence, for all $t \in [\ell + (d_\ell+d_r)\D ,r+ d_r\D]$, 
    \[
        \itemsatt{\sopt}{t} \cap (\Gaa \cup \Gab)= \Gaa \cup \Gab \, . 
    \]
    \cref{alg:non-tall-over-tall} shifts the items in $\Gb$ by $(1 - d_r)\D-r$ to the right. Thus, for each $t \in [\ell + (d_\ell+d_r)\D + (1- d_r)\D - r, \D)$,
    \begin{equation}
    \label{eq:alg_large_gap_1}
        h\big(\itemsatt{\start}{t}\cap(\Gb \cup \Gaa \cup \Gab)\big) \leq  h\big(\itemsatt{\sopt}{t-(1-d_r)\D+r} \cap \Gb\big) + h(\Gaa \cup \Gab) \leq \opt.
    \end{equation}
     
    Combining this with the bound on the height of $\Ge \cup \Gac$ would guarantee the desired bound on $h(\items^\start (t)\setminus \Gd)$ for $t \in [r - d_r\D, \D)$ if we could show that $\ell + (d_\ell+d_r)\D + (1- d_r)\D - r \leq r - d_r\D$ which is equivalent to proving that $\ell + (1+d_\ell+d_r)\D \leq 2r$.
    Indeed, using the assumptions of the lemma, we know $r-\ell \geq (\frac{1}{2}-3\lambda)\D$, \alertBound{$d_r \leq 2\lambda$}, and $2\ell \geq \ell +d_\ell\D$. Rearranging and combining those inequalities gives
    \[ 
        2 r \geq \D + 6\lambda \D +2\ell \geq \ell + (1 + d_\ell + d_r)\D.
    \]   
    Hence, 
    \begin{equation}
    \label{eq:alg_large_gap_2}
        \ell +(1 + d_\ell)\D -r\leq r-d_r \D . 
    \end{equation} 
    Therefore, the height of~$\start$ without the items in~$\Gd$ is indeed bounded: $h(\itemsatt{\start}{t}\setminus \Gd) \leq \frac{3}{2}\opt$ for each $t \in [0,\D)$.

\bigskip

    As described before, the second part of the proof is to show that ``adding" $\Gd$ does not increase the height beyond $\frac32 \opt$. 
    To this end, recall that $\tau$ is the first time point where the height of the items in $\items_{h > \frac\opt2}$ together with $\Gab$ does not exceed \opt\, anymore, that is 
    \begin{equation*} 
        \tau = \min\{\tau \in [0,\D ] \mid \forall t > \tau:  h(\itemsatt{\start}{t}\cap(\items_{h> \frac{\opt}{2}} \cup \Gab))\leq \opt\big\} 
    \end{equation*}
    As $\Gab$ completely overlaps the segment $[\ell + (d_\ell + d_r)\D, (1-d_\ell)\D)$ in the optimal packing $\sopt$, items of height larger than $\opt - h(\Gab)$ can only be packed before or after that segment, i.e., in $[0,\ell + (d_\ell + d_r)\D) \cup [(1-d_\ell)\D, \D)$. Further, by \cref{obs_height_G1G2}, $h(\Gab) \leq h(\Ga) \leq \frac\opt2$ as $\Ga$ is packed in parallel to the tall item~$T_1$. By assumption, no item with height larger than $\frac\opt2 \leq \opt - h(\Gab)$ is packed in the segment~$[\ell, r)$. Hence, the total width of items of height larger than $\opt - h(\Gab)$ is bounded by $\ell + d_\ell\D$. Thus, $\tau \leq \ell + d_\ell \D$.    
    We observe that~$\tau$ is the first time point for which we can safely add the stretched $\Gd$ of height at most $\frac\opt2$ without exceeding $\frac32\opt$ in total. 
    
    The time point $\tau'$ is the latest point in time (before~$\tau$) where in $\sopt$ there is an item of height at least $\opt - h(\Gab^{\sopt}(\tau))$ in $\opt$. 
    Note that since $\tau \leq \ell+d_\ell\D$ we can redefine $\tau'$ from the algorithm as
    \begin{equation*}
        \tau' = \max \big\{\tau' \in \big[0,\tau\big] \mid \items^{\sopt}_{h\geq \opt- h(\Gab^{\sopt}(\tau))}(\tau') \not = \emptyset\big \} \, .
    \end{equation*}
    Note that, by definition of $\ell$ and the above discussed upper bound of $\ell + d_\ell\D$ on~$\tau$, we know that $\tau' \leq \ell$ since no item from $\items_{h > \frac\opt2} \supseteq \items_{h > \opt - h( \Gab^{\sopt}(\tau))}$ is packed in~$[\ell,r)$.
    
    The algorithm uses $\tau'$ to partition~$\Gd$ into the two sets $\Gd \cap \items_{\sopt(i) \leq \tau'}$ and $\Gd \cap \items_{\sopt(i) > \tau'}$. 
    We first argue that this partition is reflected in the packing $\start$, that is, the items from these two sets do not overlap in $\start$. This allows us to split the segment $[0,\D)$ again into two parts when arguing about the maximal height of $\start$, including $\Gd$.

    The Left-Stretching \Cref{lem:left_stretching} guarantees that the maximal packing height of the former set is at most $h(\Gab^{\sopt}(\tau))$ in~$\start_{\ell}'$ and that the maximal packing height of the latter set is at most  $\frac\opt2$ in~$\start_{\ell}$.
    Further, by definition of $d_\ell$, we additionally have that no item starts earlier than $- d_\ell \D$ in $\start_{\ell}$.  
    
    We can prove the same bound for $\start_{\ell}'$ (with more technical details): The items $\Gab^{\sopt}(\tau)$ completely overlap the segment $[\tau,(1-d_\ell)\D)$ in $\sopt$ and $\start$. Hence, $\sopt$ can pack the items with height greater than $\opt - h(\Gab^{\sopt}(\tau))$ only in the segments $[0,\tau)$ and $[(1-d_\ell)\D, \D)$.
    By definition of $\tau$ and $\start$, the tall item ending at~$\tau$ (in $\start$) has a height greater than $\opt - h(\Gab^{\sopt}(\tau))$. Hence, the total width of items with a height greater than $\opt - h(\Gab^{\sopt}(\tau))$ is at least $\tau$. 
    Thus, such items of total width at most $d_\ell\D$ can be packed in $\sopt$ in $[(1-d_\ell)\D,\D)$, implying that such items of total width at least $\tau - d_\ell\D$ have to be packed in~$[0,\tau)$. Equivalently, a total width of at most $d_\ell\D$ in $[0,\tau)$ is not covered by such items. This has two implications, the first one is 
    \begin{equation}
        \label{eq:alg_large_gap_distance-tau-and-tau'}
        \tau - \tau' \leq d_\ell \D \, , 
    \end{equation} 
    and the second is the desired bound on the earliest starting time in $\start_{\ell}'$: \Cref{alg_left_stretching} left-stretches items only up to time $\tau'$, and thus, the Left-Stretching \Cref{lem:left_stretching} guarantees $\start_{\ell}'(i) \geq - d_\ell\D$ for all $i \in \Gd$ as desired.

    We observe that the stretched items in $\Gd \cap \items_{\sopt(i) > \tau'}$ are shifted by $-\tau' +d_\ell\D$, which implies that they finish before time $\ell - \tau' + d_\ell\D \leq \ell + d_\ell\D$ in $\start$. Conversely, the stretched items in $\Gd \cap \items_{\sopt(i) \leq \tau'}$ are shifted by $ \ell + (1+2d_\ell)\D - r = \ell + (\D - r) + 2 d_\ell\D \geq \ell + 2d_\ell \D$ and satisfy $\start_{\ell}'(i) \geq - d_\ell \D$ as established above. Thus, they start after time~$\ell + d_\ell \D$ in $\start$. Hence, in~$\start$, the two parts of $\Gd$ do not overlap, and, in order to bound the height of $\start$, it suffices to consider the segments~$[0, \ell - \tau' + d_\ell \D)$ and $[\ell + (1+d_\ell)\D - r, \D)$ separately. 

    \bigskip
     
    \noindent $\bullet \enspace \bm{[\ell + (1+d_\ell)\D - r, \D).} \enspace$ 
    We observe that $\start'_\ell(i) + w(i) \leq \sopt(i) + w(i) \leq \ell$ for all $i \in \Gd$.   
    By shifting each item in $\Gd \cap \items^{\sopt}_{\sopt(i)\leq \tau'}$ to the right by $\ell+(1+2d_\ell) \D-r$, the first shifted item starts at or after $-d_\ell\D + (\ell+(1+2d_\ell)\D-r) = \ell+(1+d_\ell)\D-r$ and the last item ends at or before $\ell + (\ell+(1+2 d_\ell) \D - r)$. We can upper bound this time point by $(1 - 2d_r) \D$ as follows: 
    By the assumptions of the lemma, $\ell \leq \D-r$ and $r-\ell \geq \big(\frac{1}{2}-3\lambda\big)D$, and hence $\alertBound{\ell \leq \big(\frac{1}{4}+\frac{3}{2}\lambda\big)D}$. 
    Additionally using \alertBound{$\lambda \leq \frac1{42}$}, \alertBound{$d_r\leq 2\lambda$}, and \alertBound{$d_\ell \leq \lambda$}, we have $ \alertBound{\frac{9}{2}\lambda  +2d_\ell + 2d_r \leq \frac{1}{4}}$. Thus, indeed
    \[
        \ell + (\ell+(1+2d_\ell) \D - r) \leq \bigg(\frac{1}{4}+\frac{3}{2}\lambda  +1 + 2d_\ell- \bigg(\frac{1}{2}-3\lambda\bigg)\bigg)\D = \bigg( \frac{3}{4} + \frac{9}{2}\lambda + 2d_\ell\bigg) \D \leq (1-2d_r)\D,
    \]
   proving that $\Gd \cap \items_{\sopt(i)\leq \tau'}$ is packed in the segment $[\ell+(1+d_\ell)\D-r, (1 - 2d_r) \D)$.

    Recall that the maximal height of $\start_{\ell}'$ restricted to $\Gd \cap \itemsT{\sopt}_{\sopt(i) \leq \tau'}$ is bounded by $h(\Gab^{\sopt}(\tau))$. 
    Thus, in order to establish the desired bound on $\start$ in the segment $[\ell + (1+2d_\ell) \D - r, \D)$, it is sufficient to show  $h(\itemsatt{\start}{t} \setminus \Gd)\leq \frac{3}{2}\opt -  h(\Gab^{\sopt}(\tau))$ for all $t \in [\ell + (1+d_\ell) D - r, (1-2d_r)\D)$. 

    First, we prove this for the segment $t \in [r-d_r\D,(1-2d_r)\D)$ before arguing on how to extend this to the whole segment.
    Recall that all items from $\Gab$ start before $r$ and end after $(1-d_r)\D$ in $\start_{r}$ and hence $\Gab^{\start_{r}}(t) = \Gab$ for all $t \in [r,(1-d_r)\D)$.
    Therefore, due to the Stretching Lemma~\ref{lem:right_stretching} applied to $\Ge$ and $\Ga$ we have for any $t \in [r,(1-d_r)\D)$ that 
    \[
        \frac{\opt}{2} \geq h((\Ga \cup \Ge)^{\start_{r}}(t)) \geq h((\Gac \cup \Ge)^{\start_{r}}(t)) + h(\Gab) =  h((\Gac \cup \Ge)^{\start}(t-d_r\D)) + h(\Gab).
    \] 
    Hence, $h((\Gac \cup \Ge)^{\start}(t)) \leq \frac{\opt}{2}-h(\Gab)$ for any $t \in [r-d_r\D,(1-2d_r)\D)$.
    Furthermore, in the segment $[0,r-d_r\D)$, the packing $\start$ does not pack any item from $\Ge$, and $\arg \max_{t}(h(\Gac^{\start}(t))) = r-d_r\D$.
    Therefore, the bound $h((\Gac \cup \Ge)^{\start}(t)) \leq \frac{\opt}{2}-h(\Gab)$ actually holds for any $t \in [\ell+(1+d_\ell)\D-r,(1-2d_r)\D)$, which is a superset of $[r-d_r\D,(1-2d_r)\D)$ by \cref{eq:alg_large_gap_2}.
    
    Further, \cref{eq:alg_large_gap_1} guarantees for all $t \in [\ell+(1+d_\ell)D-r,\D)$ that
    \[
        h\big((\Gb \cup \Gaa \cup \Gab)^{\start}(t)\big)\leq \opt.
    \]
    Combining these two bounds, we obtain for all $t \in [\ell+D-r+d_\ell, D-2d_r)$
    \[
        h(\items^{\start}(t) \setminus \Gd)\leq \frac{3}{2}\opt -  h(\Gab)\leq \frac{3}{2}\opt -  h(\Gab^{\sopt}(\tau)).
    \] 
    Recall that for $t \in [\ell+(1+d_\ell)\D-r, (1-2d_r) \D)$ we can upper bound the height of $\Gd \cap \items_{\sopt(i) \leq \tau'}$ by $h(\Gab^{\sopt}(\tau))$ as argued above, which bounds the height of $\itemsT{\start}(t)$ for $t \in [\ell + (1+d_\ell)\D - r, \D)$ as desired.

    \bigskip
    \noindent $\bullet \enspace  \bm{[0, \ell - \tau'  + d_\ell\D).} \enspace $ 
    The items $\Gd \cap \items_{\sopt(i) > \tau'}$ are first stretched to the left by \cref{alg_left_stretching}, generating $\start_{\ell}$, and then shifted by $d_\ell\D - \tau'$. 
    First, we show that these items are packed in $[\tau'-d_\ell\D , \ell)$ in $\start_{\ell}$. 
    By assumption, the total width in $[0,\ell)$ not overlapped by tall items is at most $d_\ell \D $. 
    Thus, the smallest starting time of any item in $\Gd \cap \items_{\sopt(i) > \tau'}$ is at least 
    \begin{equation}\label{eq:alg_large_gap_bound_on_sigma_4_prime}
        \min_{i \in \Gd \cap \items_{\sopt(i) > \tau'} } \start_{\ell}(i) \geq \tau'-d_\ell \D 
    \end{equation} by the Left-Stretching \cref{lem:left_stretching}.
    Furthermore, for all $t \in [-d_\ell \D,\ell)$, the same lemma guarantees 
    \begin{align*}
        \frac{\opt}{2} & \geq h\big( \big( \Gd \cup \Gaa \cup \Gab \cup \Gb\big)^{\start_{\ell}} (t) \big). 
    \shortintertext{By definition, the items in $\Ga \cup \Gb$ \emph{end} after time $\ell$, which implies that \cref{alg_left_stretching} does not change the packing of the items in $\Gaa \cup \Gab \cup \Gb$ relative to $\sopt$. Thus, for these items, we can replace $\start_{\ell}$ by $\sopt$ without changing the right hand side of the above inequality and obtain}
        \frac{\opt}{2} & \geq h\big( \Gd^{\start_{\ell}} (t) \cup \big(\Gaa \cup \Gab \cup \Gb\big)^{\sopt} (t) \big).
    \shortintertext{Finally using that the items in $\big(\Gaa \cup \Gab \cup \Gb\big)$ do not start earlier in $\start$ than in $\sopt$ and are non-decreasing in $[-d_\ell \D,\ell)$, this gives}
        \frac{\opt}{2} & \geq h\big( \Gd^{\start_{\ell}} (t) \cup \big(\Gaa \cup \Gab \cup \Gb\big)^{\start} (t) \big)
    \end{align*}
    For the remaining proof, we distinguish the two cases $\tau' \geq d_\ell\D$ and $\tau' < d_\ell \D$.
    
    If $\tau' \geq d_\ell \D$, the items from $\Gd \cap \items_{\sopt(i) > \tau'}$ are packed further to the left in $\start$ than in $\start_{\ell}$. Hence, the above inequality implies
    \[
        \frac{\opt}{2} \geq h\big( \big( \Gd \cup \Gaa \cup \Gab \cup \Gb\big)^{\start} (t) \big) \ ,
    \]    
    as again the height of $\Gaa \cup \Gab \cup \Gb$ is non-decreasing in $[0, \ell)$.
    Since the segment $[0,\ell - \tau' + d_\ell\D)$ only contains items from $\items_{h>\frac{\opt}{2}}\cup \Gd \cup \Gaa \cup \Gab \cup \Gb$, we obtain the desired upper bound on the height, i.e., $h(\items^{\start}(t)) \leq \frac{3}{2}\opt$ for each $t \in [0,\ell - \tau' + d_\ell\D)$.
    
    In the other case, i.e., $\tau' < d_\ell\D$, the starting time $\start(i)$ of an item $i \in \Gd \cap \items^{\sopt}_{\sopt(i) > \tau'}$ increases by at most $d_\ell\D$ when compared to $\start_{\ell}$. 
    Further, the starting time of the items $\Gaa \cup \Gb$ in $\start$ increases by at least $d_\ell\D$ when compared to $\sopt$ (and thus $\start_{\ell}$ as discussed above). 
    Thus, for each $t \in [0,\ell - \tau' + d_\ell\D)$
    \[
        h((\Gd \cup \Gaa \cup \Gb)^{\start}(t)))\leq \frac{\opt}{2} \, ,
    \]
    where we used again that $\Ga \cup \Gb$ is non decreasing in $[0,\ell)$ in $\sopt$ by definition. 
    By definition of $\tau$, we know that $h(\Gab^{\start}(t) \cup \items_{h> \frac{\opt}{2}}^{\start}) \leq \opt$ for $t \in [\tau,\ell+d_\ell\D)$. 
    Combining this with the above bound on the height of $\Gd\cup \Gaa\cup \Gb$, we obtain the desired bound on the total height of $\start$ in the segment $[\tau, \ell - \tau' + d_\ell\D)$. 

    For the remaining segment $[0, \tau)$, we can bound $\tau \leq  \tau' + d_\ell\D < \alertBound{2 d_\ell \D \leq \frac \eps{1+\eps}\D}$ by using \eqref{eq:alg_large_gap_distance-tau-and-tau'} for the first inequality, the current-case assumption for the second inequality, and our assumption on $d_\ell$ for the third inequality. 
    As the minimal width of a non-tall item is at least $\frac \eps{1+\eps}\D > 2 d_\ell \D $, this implies for $i \in \Gd \cap \items^{\sopt}_{\sopt(i) > \tau'}$ that
    \[
        \start_{\ell}(i) + w(i) \geq (\tau' - d_\ell\D) + w(i) \geq \tau - 2d_\ell\D + w(i) > \tau, 
    \]
    where we used \eqref{eq:alg_large_gap_bound_on_sigma_4_prime} to bound $\start_{\ell}(i)$.
    
    Using this observation, we can show that $\Gd^{\start}(t) \subseteq \Gd^{\start_{\ell}}(t)$ for all $t \in [0,\tau)$ as follows: We fix an item $i \in \Gd^{\start}(t)$ for some $t \in [0,\tau)$. Then,  
    \[
        \start_{\ell}(i) \leq \start(i) < t \text{ and } \start_{\ell}(i) + w(i) > \tau, 
    \]
    which implies that $i \in \Gd^{\start_{\ell}}(t)$ as desired. Thus, for $t \in [0,\tau)$,
    \[
        h((\Gd \cup \Ga \cup \Gb)^{\start}(t)) + h(\items_{h >\frac\opt2}^{\start} (t))  \leq  h(\Gd^{\start_{\ell}}(t)\cup (\Ga \cup \Gb)^{\start}(t)) + h(\items_{h >\frac\opt2}^{\start} (t)) \leq \frac\opt2 + \opt \, .
    \]

    \bigskip

    It remains to argue that~$\start$ packs all items in~$[0,\D)$. 
    For arguing about~$\Ga$, we distinguish again~$\Gaa$, $\Gab$, and $\Gac$.  
    By definition,  $\Gaa = {\items_{\sopt(i) < \ell + (d_\ell + d_r)\D, \, r+ d_r\D < \sopt(i) + w(i) \leq (1-d_\ell)\D}}$, i.e., these items finish before~$(1-d_\ell)\D$ in~$\sopt$. \cref{alg:gap_almost_1/2_not_at_border} shifts them by $d_\ell\D$ to the right, implying that they finish before~$\D$ in $\start$. 
    The items in~$\Gab = {\items_{\sopt(i) < \ell + (d_\ell + d_r)\D, \, \sopt(i) + w(i) > (1-d_\ell)\D}}$ are packed in~$\start$ as in~$\sopt$. 
    Hence, their feasible packing follows from the feasibility of~$\sopt$. 
    By definition, the items $\Gac = {\items_{ \ell + (d_\ell + d_r)\D \leq \sopt(i) < r, \, \sopt(i) + w(i) > r + d_r\D}}$ do not start before~$\ell + (d_\ell + d_r)\D \geq d_r\D$ in~$\sopt$ and \cref{alg:gap_almost_1/2_not_at_border} shifts them by~$d_r\D$ to the left, implying that they start at or after~$0$ in~$\start$. 

    The feasibility of the packing of~$\Gb$ follows from \cref{lem:G2-shift-over-tall-fit}. 
    When ``adding" $\Gd$ to the packing, we have already argued that $\Gd \cap \items^{\sopt}_{\sopt(i)\leq \tau'}$ is packed in the segment $[\ell+(1+d_\ell)\D-r, (1 - 2d_r) \D)$ in~$\start$ and that $\Gd \cap \items^{\sopt}_{\sopt(i) > \tau'}$ is packed in $[\tau'-d_\ell\D , \ell)$ in $\start_{\ell}$. 
    The latter set is shifted by~$-\tau' + d_\ell\D$ in~$\start$ when compared to~$\start_{\ell}$. 
    Hence, also~$\Gd$ is packed in~$[0,\D)$ in~$\start$. 
    For~$\Ge$, we observe that these items are packed in~$[r, (1+d_r)\D)$ in $\start_{r}$  by \cref{lem:right_stretching}. 
    Since \cref{alg:gap_almost_1/2_not_at_border} shifts them by $d_r\D$ to the left and $r \geq \ell \geq \frac\eps{1+\eps} \D \geq d_r\D$, they are packed in~$[0,\D)$ in~$\start$.     
    This concludes the proof of the lemma.     
\end{proof}

\subsection{A gap of width at least~\texorpdfstring{$\big(\frac 12 - 3\lambda\big)\D$}{(1/2 - 3λ)D} with \texorpdfstring{$\ell \leq \frac\eps{1+\eps}\D$}{ℓ ≤ ε/(1+ε)D} and \texorpdfstring{$r\leq \frac{\D}{2}$}{r ≤ D/2}}

In this section, we consider the case where there is a gap of width at least~$\big(\frac 12 - 3\lambda\big)\D$ and $r \leq \frac{\D}{2}$. Therefore, $\ell \leq 3\lambda \D$. 
We design \cref{alg_l_close_to_border_r_smaller_than_half} and prove that it returns a feasible packing of height at most~$\frac32\opt$.

Given an optimal packing $\sopt$, we partition the set $\items_{h\leq \frac{\opt}{2}} $ as follows:
\[
\Gal := {\items_{\sopt(i) < \ell - d_\ell\D , \sopt(i) + w(i) > \ell}}
\, , \quad 
\Gb := {\items_{\sopt(i) \in [\ell - d_\ell\D ,r] } } \setminus \items_{h> \frac{\opt}{2}} \, 
, \quad 
\Ge := \items^{\sopt}_{[r,\D]}\setminus \items_{h > \frac{\opt}{2}} \, .
\]
Note that the set we previously defined as $\Gd := \items^{\sopt}_{[0,\ell]}\setminus \items_{h > \frac{\opt}{2}}$ is empty since $\ell \leq \frac{\eps}{1+\eps}\D$ and we assume that all non-tall items have a width of at least $\frac{\eps}{1+\eps}\D$. 
Further, note that we kept the same definitions of the item sets as in the previous section. However, in this section, the set of items $\Ge$ will play a similar role as the set $\Gd$ from the previous section.
Therefore, in this section, we need to handle~$\Ge$ more carefully. 
To do so, we denote the total width that is not covered by tall items in~$\sopt$ in the segment $\big[ r + \big(\frac\eps{1+\eps}-d_r\big)\D, \D\big)$ by $d_r'\D$.

\Cref{alg_l_close_to_border_r_smaller_than_half} works as follows. 
We first mirror the packing $\sopt$ and then apply \cref{alg:non-tall-over-tall} to $\Gb$ and $\items_{h>\frac{\opt}{2}}$, i.e., we sort the items in $\items_{h>\frac{\opt}{2}}$ in non-increasing order, starting at $0$ and shift $\Gb$.
We then right-stretch $\Ge$ in the non-mirrored packing of $\sopt$ and (partially) shift these items by $-d_r'\D$ afterward.
Finally, we place the items in $\Gal$ as they were placed in $\sopt$, i.e., in the non-mirrored packing.

\begin{figure}
\resizebox{\textwidth}{!}{

\begin{tikzpicture}
\pgfmathsetmacro{\w}{6}
\pgfmathsetmacro{\h}{3}
\pgfmathsetmacro{\os}{0.24}
\pgfmathsetmacro{\l}{0.05}
\pgfmathsetmacro{\r}{0.46}
\pgfmathsetmacro{\dFive}{0.03}
\pgfmathsetmacro{\dFivePrime}{0.02}
\pgfmathsetmacro{\dFour}{0.03}
\pgfmathsetmacro{\lamb}{0.04}


\draw[fill = colore] (\w,0) -- (\r*\w,0)--(\r*\w,\h) --(\w-\r*\w,0.75*\h)-- (\w,\h)--(\w,0);
\node at (0.5*\w+0.5*\r*\w,0.75*\h) {$\Ge$};

\draw[fill = coloraa] (0,\h) -- (\dFour*\w,0.6*\h) -- (\l*\w,0.52*\h) -- (\r*\w,0.8*\h)--(\w-\r*\w +\l*\w-\dFour*\w-\dFive*\w,0.84*\h)--(\w,\h) -- (0,\h);
\node at (0.5*\r*\w,0.85*\h) {$\Gal$};



\draw[fill = colorb] (\w,\h) --(\w-\r*\w +\l*\w-\dFour*\w-\dFive*\w,0.75*\h) -- (\r*\w+0.03*\w,0.7*\h) -- (\r*\w, 0.52*\h) -- (\r*\w,0) -- (\l*\w,0) -- (\l*\w,0.55*\h) --  (\r*\w,0.8*\h) --(\w-\r*\w,0.84*\h)-- (\w,\h);
\node at (0.5*\l*\w+0.5*\r*\w,0.3*\h) {$\Gb$};

\draw[fill = gray] (\l*\w-0.01*\w,0.55*\h) rectangle (\l*\w,0);
\draw[fill = gray] (0.0*\w,0.65*\h) rectangle (0.01*\w,0);

\draw[fill = gray] (\r*\w+0.02*\w,0.52*\h) rectangle (\r*\w,0);
\draw[fill = gray] (\r*\w+0.2*\w,0.7*\h) rectangle (\r*\w+0.03*\w,0);
\draw[fill = gray] (\r*\w+0.4*\w,0.6*\h) rectangle (\r*\w+0.22*\w,0);
\draw[fill = gray] (\w,0.75*\h) rectangle (\r*\w+0.4*\w,0);

\draw (0,\h) -- (0,0) -- (\w,0) -- (\w,\h);
\draw (0*\w,-\os) node[below]{$0$}-- (0*\w,\h);
\draw (\w,-\os) node[below]{$\D$}-- (\w,\h);

\draw[thick,dashed] (\r*\w,-\os) node[below]{$r$}-- (\r*\w,\h);
\draw[thick,dashed] (\l*\w,-\os) node[below]{$\ell$}-- (\l*\w,\h);
\draw[thick,dashed] (0.5*\w,-\os) node[below]{$\frac{D}{2}$}-- (0.5*\w,\h);

\draw[dashed] (-\os,0.5*\h) node[left]{$\frac{1}{2}\opt$} -- (\w+\os,0.5*\h) ;
\draw[dashed] (-\os,1*\h) node[left]{$\opt$} -- (\w+\os,1*\h) ;


\begin{scope}[xscale = -1, xshift = -2.3*\w cm]
\begin{scope}[xshift = -\l*\w cm+\dFive*\w cm,yshift = 0.4*\h cm]
\draw[fill = colorb] (\w,\h) --(\w-\r*\w +\l*\w-\dFour*\w-\dFive*\w,0.75*\h) -- (\r*\w+0.03*\w,0.7*\h) -- (\r*\w, 0.52*\h) -- (\l*\w,0.55*\h) -- (\r*\w,0.8*\h) --(\w-\r*\w,0.84*\h)-- (\w,\h);
\node at (\r*\w-0.1*\w,0.7*\h) {$G_2$};
\draw[white,fill=white] (\w,\h) rectangle (\w-\r*\w+\dFivePrime*\w ,0.5*\h);
\draw[white,fill=white] (0,\h) rectangle (\r*\w -\dFivePrime*\w ,0.5*\h);
\end{scope}

\begin{scope}[xshift = -\l*\w cm+\dFour*\w cm]
\draw[fill = colorb] (\w,\h) --(\w-\r*\w +\l*\w-\dFour*\w-\dFive*\w,0.75*\h) -- (\r*\w+0.03*\w,0.7*\h) -- (\r*\w, 0.52*\h) -- (\r*\w,0) -- (\l*\w,0) -- (\l*\w,0.55*\h) --  (\r*\w,0.8*\h) --(\w-\r*\w,0.84*\h)-- (\w,\h);
\draw[white,fill=white] (\r*\w +\dFive*\w,\h) rectangle (\w-\r*\w+\dFivePrime*\w,0.5*\h);
\node at (0.5*\l*\w+0.5*\r*\w,0.3*\h) {$\Gb$};
\end{scope}

\begin{scope}[xscale= -1, xshift=-\w cm-0.5*\dFour*\w cm, yshift = 0.5*\h cm]
\draw[fill = colore] (\r*\w,0.5*\h)--(\w-\r*\w +\l*\w-0.5*\dFour*\w-\dFive*\w,0.5*\h) --(\w-\r*\w +\l*\w-0.5*\dFour*\w-\dFive*\w,0.75*\h) -- (\w,\h) --(\w+0.5*\dFour*\w,\h)--(\w+0.5*\dFour*\w,0.5*\h);
\node at (0.5*\w+0.5*\r*\w,0.75*\h) {$\Ge$};
\end{scope}
\begin{scope}[xscale= -1, yscale= -1,xshift=-\w cm-0.5*\dFour*\w cm, yshift = -1.5*\h cm]
\draw[fill = colore] (\r*\w,0.5*\h) -- (\r*\w, 0.52*\h)-- (\r*\w+0.03*\w,0.7*\h) --(\w-\r*\w +\l*\w-0.5*\dFour*\w-\dFive*\w,0.75*\h) --(\w-\r*\w +\l*\w-0.5*\dFour*\w-\dFive*\w,0.5*\h) -- (\r*\w,0.5*\h);
\end{scope}

\begin{scope}[xscale= -1, xshift=-\w cm, yshift = 0.5*\h cm]
\draw[fill = coloraa] (0,\h) -- (\dFour*\w,0.6*\h) -- (\l*\w,0.52*\h) -- (\r*\w,0.8*\h)--(\w-\r*\w +\l*\w-\dFour*\w-\dFive*\w,0.84*\h)--(\w-\r*\w +\l*\w-\dFour*\w-\dFive*\w,\h) -- (0,\h);
\node at (0.5*\r*\w,0.85*\h) {$\Gal$};
\end{scope}

\begin{scope}[xscale= -1, xshift=-\w cm]
\draw[fill = coloraa] (\w,\h)--(\w-\r*\w +\l*\w-\dFour*\w-\dFive*\w,0.84*\h)--(\w-\r*\w +\l*\w-\dFour*\w-\dFive*\w,\h) -- (\w,\h);
\end{scope}

\draw[fill = gray] (\r*\w+0.01*\w,0.52*\h) rectangle (\r*\w+0.03*\w,0);
\draw[fill = gray] (\r*\w+0.03*\w,0.55*\h) rectangle (\r*\w+0.04*\w,0);
\draw[fill = gray] (\r*\w+0.04*\w,0.6*\h) rectangle (\r*\w+0.22*\w,0);
\draw[fill = gray] (\r*\w+0.22*\w,0.65*\h) rectangle (\r*\w+0.23*\w,0);
\draw[fill = gray] (\r*\w+0.23*\w,0.7*\h) rectangle (\r*\w+0.40*\w,0);
\draw[fill = gray] (\r*\w+0.40*\w,0.75*\h) rectangle (\w,0);

\draw[dashed] (-\os,0.5*\h) -- (\w+\os,0.5*\h) node[left]{$\frac{1}{2}\opt$};
\draw[dashed] (-\os,1*\h)  -- (\w+\os,1*\h) node[left]{$\opt$};
\draw[dashed] (-\os,1.5*\h)  -- (\w+\os,1.5*\h) node[left]{$\frac{3}{2}\opt$};

\draw (0,1.5*\h) -- (0,0) -- (\w,0) -- (\w,1.5*\h);

\draw[thick,dashed] (\r*\w-\l*\w+\dFour*\w+\dFive*\w,-3*\os) node[below]{$(1-d_\ell-d_r)\D-r+\ell$} -- (\r*\w-\l*\w+\dFour*\w+\dFive*\w,1.5*\h);
\draw[thick,dashed] (\w-\r*\w,-\os) node[black,below]{$r$}  -- (\w-\r*\w,1.5*\h);
\draw[thick,dashed] (\w-\r*\w+0.5*\dFour*\w,0)-- (\w-\r*\w+0.5*\dFour*\w,1.5*\h+\os)  node[black,above]{$r-d_r'\D$} ;
\draw[thick,dashed] (\w,-\os) node[below]{$0$}-- (\w,\h);
\draw[thick,dashed] (0*\w,-\os) node[below]{$\D$}-- (0*\w,\h);
\end{scope}
\end{tikzpicture}
}
\caption{An example for a repacking done by \cref{alg_l_close_to_border_r_smaller_than_half}.}
\end{figure}
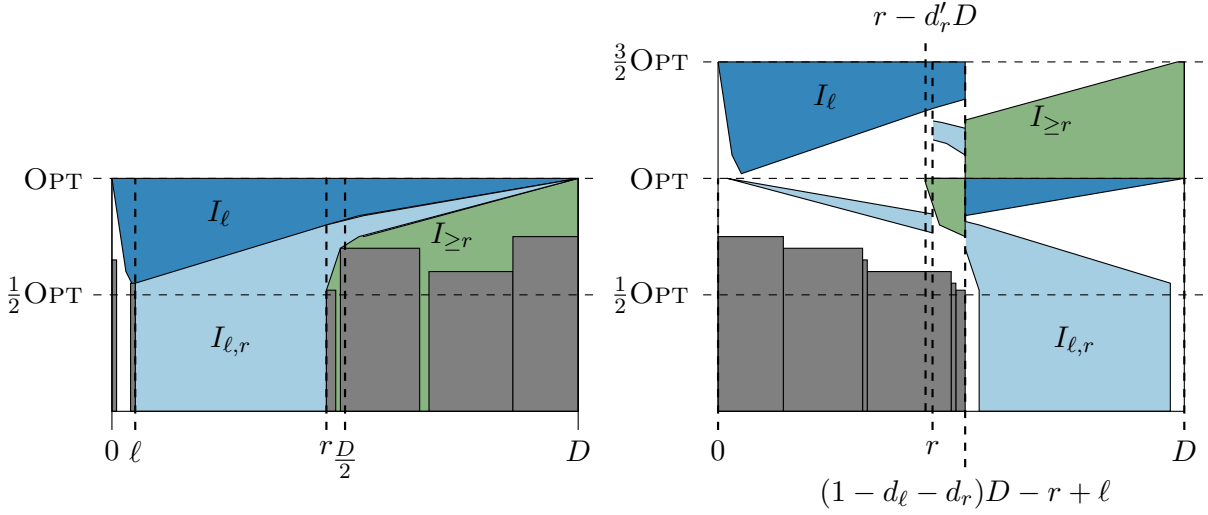

\begin{algorithm}[ht]
\DontPrintSemicolon
\caption{Repacking for $\ell \leq 3 \lambda$ and $r\in \big[(\frac 12 - 3\lambda)\D,\frac{\D}{2}\big]$}
\label{alg_l_close_to_border_r_smaller_than_half}
\textbf{Input:}  $(\sopt, \items \setminus \items_{\mathrm{sq,\eps,\opt}}, \Gal, \Gb, \Ge, \ell, r, d_\ell,d_r, d_r')$\;
\smallskip
$\bar{\start} \leftarrow$ output of \Cref{alg_mirroring} with $\sopt$ \tcp*[r]{mirror $\sopt$}
$\start  \leftarrow$ output of \cref{alg:non-tall-over-tall} with $\big(\bar{\start},\items_{h > \frac \opt2},\Gb,\D-r,\D-\ell,d_\ell\big)$\tcp*[t]{shift $\Gb$, sort $\items_{h>\frac{\opt}{2}}$}
$\start_{r} \leftarrow$ output of \Cref{alg_stretching} with $\big(\sopt, \frac\opt2, r+(\frac{\eps }{1+\eps} - d_r)\D,\D \big)$  \tcp*[r]{right-stretch $\Ge$}
\lForEach(\tcp*[f]{shift $\Ge$ by $-d_r'\D$}){$i \in \Ge$}{
    $\start(i) \leftarrow  \start_{r}(i) - d_r' \D$ 
}
\lForEach(\tcp*[f]{pack $\Gal$ as in $\sopt$}){$i \in \Gal$}{
    $\start(i) \leftarrow  \sopt(i)$ 
}
\smallskip 
\textbf{return} $\start$ \;
\end{algorithm}

\begin{lemma}
\label{lem:l-close-to-border_r-smaller-than-half}
Assume $\sopt$ is an optimal packing of $(\items, \D)$ such that there is a gap $[\ell,r)$ with $r \in [(\frac 12 -3 \lambda) \D, \frac{\D}{2}]$ and the gaps to the right of $r$ have a combined width of at most $2 \lambda$. 
Then $(\items, \D)$ admits a neat packing.
\end{lemma}

\begin{proof} 
Let $T_1, T_2, \ell, r$ be according to the above definitions and satisfying the conditions of the lemma.
We show that \cref{alg_l_close_to_border_r_smaller_than_half} returns a neat packing~$\start$ of the items $\items \setminus \items_{h \leq \frac\opt2, w \leq \frac\eps{1+\eps}\D}$.
We start by making some crucial observations about where the item sets $\Gal$, $\Gb$, $\Ge$, and $\items_{ h > \frac\opt2}$ are packed and what their packing looks like.  
\begin{enumerate}

    \item\label{enum:tall-items} $\items_{h > \frac\opt2}$ is packed in $\big[0, (1-d_\ell-d_r)\D-r+\ell \big)$ in~$\start$ as their total width is at most $(\ell - d_\ell \D) + (D-r -d_r\D)$. 

    \item The items $\Gal$ are packed in~$[0,\D)$ in~$\start$ since~$\start$ packs these items exactly as~$\sopt$. 

    \item\label{enum:packing-G3} By \Cref{lem:G2-shift-over-tall-fit}, the items $\Gb$ are packed in $[0,\D)$ and $h((\Gb \cup \items_{h>\frac{\opt}{2}})^{\start}(t)) \leq \opt$ for any $t \in [0,\D)$.

    \item\label{enum:packing-G2}
    For any $t \in [0,\D-r+\ell-d_\ell]$, it holds that $\Gb^\start(t) \subseteq \itemsoverlapt{\sopt}{r}$ and $\Gb^{\sopt}(t)$ is non-increasing in $[r,D]$.

    \item\label{enum:packing-G5} The items $\Ge$ are packed in $[r - d_r'\D, \D)$ in~$\start$ as
    \cref{alg_l_close_to_border_r_smaller_than_half} shifts them by~$d_r'\D$ to the left compared to~$\start_{r}$ and in~$\start_{r}$, these items are packed in $[r, (1+d_r')\D)$ by \cref{lem:right_stretching}.

    \item\label{enum:G5-in-sigma-vs-sopt}
    Observe that \cref{alg_l_close_to_border_r_smaller_than_half} does not stretch the first part of $\Ge$, but shifts it by $d_r'\D$ to the left.
    Hence, for $t \in \big[r- d_r' \D, r + \big(\frac\eps{1+\eps} - d_r - d_r' \big) \D \big)$, we have  $\Ge^{\start}(t) = \Ge^{\sopt}(t + d_r' \D)$.
    
    \item\label{enum:height-G5} 
    In $\sopt$, there is no item from $\Ge$ ending before $r +\frac{\eps \D}{1+\eps}$ since they have a width of at least $\frac{\eps \D}{1+\eps}$. 
    Therefore, the height of $\Ge^{\sopt}(t)$  for~$t \in [r,r+\frac{\eps}{1+\eps}\D)$ is non-decreasing for increasing~$t$.
    We now show that $\Ge^{\sopt}(t)$ is bounded by $\frac{\opt}{2}$ for $t \in \big[r,r+ \big(\frac{\eps}{1+\eps}-d_r\big) \D\big)$.
    Assume the contrary, i.e., $\Ge^{\sopt}(t) > \frac{\opt}{2}$ for some $t \in \big[r,r+ \big(\frac{\eps}{1+\eps}-d_r\big) \D\big)$. Let $t'$ be the earliest such time point.
    Then, as the height of $\Ge^{\sopt}(t)$  for~$t \in [r,r+\frac{\eps}{1+\eps}\D)$ is non-decreasing for increasing $t$, we have that $\Ge^{\sopt}(t) > \frac{\opt}{2}$ for all $t \in \big[t', r+\frac{\eps}{1+\eps}\D))$.
    Since $t' < r+ \big(\frac{\eps}{1+\eps}-d_r) \D$, the width of the interval $\big[t', r+\frac{\eps}{1+\eps}\D))$ is strictly larger than $d_r \D$.
    Recall that the total width in $[r,\D]$ not covered by tall items is bounded by $d_r \D$.  
    Hence, there must be some non-tall item $T$ such that $\opt(T) \leq t'' < \opt(T) + w(T)$ for some $t'' \in \big[t', r+\frac{\eps}{1+\eps}\D))$.
    But then $h(\items^{\sopt}(t'')) > \opt$, a contradiction.

    Combining this observation with \cref{lem:right_stretching} for $t \in \big[r+\frac{\eps \D}{1+\eps} \D, \D\big)$ bounds the height of $\Ge^\start(t)$ by~$\frac\opt2$ for all~$t \in [0,\D)$. 
\end{enumerate}
Combining the first five observations, we have proven that, indeed,~$\start$ packs all items in $[0,\D)$ as required.

It remains to bound the height of $\start$. To this end, we separately analyze the height of $\start$ in the following five segments: $[0, r- d_r'\D)$, $[r-d_r'\D,r)$, $[r,(1 - d_\ell - d_r) \D-r+\ell)$, $[(1-d_\ell-d_r)\D-r+\ell,(1-d_\ell)\D-r+\ell)$, and $[(1-d_\ell)\D-r+\ell,\D)$. 
We note that due to our assumptions on~$r$ and~$\ell$, we only know that 
$r < (1-d_\ell)\D-r+\ell$. 
In particular, the second and the fourth segment, i.e., $[r-d_r'\D,r)$ and $[(1-d_\ell-d_r)\D-r+\ell,(1-d_\ell)\D-r+\ell)$ are not necessarily disjoint; if $r > (1-d_\ell-d_r)\D-r+\ell$, then $[r, (1-d_\ell-d_r)\D-r+\ell)$ is empty.
Hence, during the analysis of the height of~$\start$ in $[r-d_r'\D,r)$ we further distinguish $r \leq (1-d_\ell-d_r)\D-r+\ell)$ and $r > (1-d_\ell-d_r)\D-r+\ell)$. In any case, we also provide an analysis for $[r, (1-d_\ell-d_r)\D-r+\ell)$ under the implicit assumption that the segment is non-empty.

\smallskip 

\noindent $\bullet \enspace \bm{[0, r- d_r'\D).} \enspace$ By \ref{enum:packing-G5}, no $\Ge$-item is packed in the current segment. Hence, for $t \in [0,r-d_r'\D)$, $\itemsatt{\start}{t} \subset \Gal \cup \Gb \cup \items_{h>\frac{\opt}{2}}$. Thus, the height of the packing is bounded by $\frac{3}{2}\opt$ due to \ref{enum:packing-G3} and \cref{obs_height_G1G2} for $\Gal$, i.e., $h(\Gal) \leq \frac{\opt}{2}$. 
\bigskip

\noindent $\bullet \enspace \bm{[r-d_r' \D ,r).} \enspace$ 
If $r - d_r' \D \geq (1-d_\ell-d_r)\D - r + \ell$, the current segment is contained in $\big[ (1-d_\ell-d_r)\D -r+\ell, \D\big)$, and hence, we refer the reader to the analyses for the segments $\big[ (1-d_\ell-d_r)\D -r+\ell, (1-d_\ell)\D -r+\ell \big)$ and $\big[(1-d_\ell)\D -r+\ell, \D\big)$.
Therefore, we assume 
\begin{equation}
    \label{eq:assumption_on_r-d5'd}
    r - d_r' \D < (1-d_\ell-d_r)\D - r + \ell
\end{equation}
for the height analysis of this segment $[r-d_r' \D ,r)$. 

Since $h(\Gal) \leq \frac{\opt}{2}$ by \cref{obs_height_G1G2}, it remains to prove that  $h((\Gb \cup \Ge \cup \items_{h>\frac{\opt}{2}})^\start(t)) \leq \opt$ for each $t \in [r-d_r' \D ,r)$. 
By \ref{enum:packing-G3}, we have $h((\Gb \cup \items_{h>\frac{\opt}{2}})^\start(t) ) \leq \opt$. However, in the current segment, we need to show a stronger bound also involving $\Ge^\start(t)$, and thus, we need to argue more carefully. Instead of directly arguing about~$\start$, we construct an auxiliary packing~$\start'$ closely related to~$\opt$ for which we can bound the total height of the items in $\Gb \cup \items_{h>\frac{\opt}{2}}(t')$ by~$\opt - h(\Ge^{\start}(t))$ for suitably chosen $t$ and~$t'$.

We start by making some useful observations on the height (and origin in~$\sopt$) of~$\Ge$ and~$\Gb$ in the current segment.
Using \ref{enum:height-G5} and \ref{enum:G5-in-sigma-vs-sopt}, we can upper bound the height of~$\Ge$ by 
\begin{equation}
\label{eq:hG5_in_r-d5D_r}
    h(\Ge^{\start}(t)) = h(\Ge^{\sopt}(t + d_r' \D) \leq h(\Ge^{\sopt}(r + d_r' \D))  =  h(\Ge^{\start}(r))  
\end{equation}
for $t \in [r-d_r' \D, r)$.

\cref{alg_l_close_to_border_r_smaller_than_half} first mirrors the items in $\Gb$ and then shifts them to the right by $\ell - d_\ell \D$. 
Hence, for $t \in [r- d_r' \D, r)$, we have 
\begin{equation}
    \label{eq:G2-in-sigma-vs-sopt}
    \Gb^{\start}({t}) = \Gb^{\sopt}({(D-t) + (\ell - d_\ell \D)}) \, . 
\end{equation}
By \eqref{eq:G2-in-sigma-vs-sopt}, we have for the two borders of the current segment 
\[
    \Gb^{\start}({r-d_r'\D}) = \Gb^{\sopt}({ (1 - d_\ell +d_r') \D - r +\ell})
\]
and 
\[
    \Gb^{\start}({r}) = \Gb^{\sopt}({(1 - d_\ell) \D - r +\ell}) \, . 
\]
That is, the items in~$\Gb$ that~$\start$ packs in $[r - d_r'\D)$ originate from the segment $[(1 - d_\ell) \D - r +\ell, (1 - d_\ell +d_r') \D - r +\ell)$ in~$\sopt$.

Our goal is to construct an auxiliary packing~$\start'$ for the (by \eqref{eq:assumption_on_r-d5'd} and $d_r'\leq d_r$) slightly larger segment $[r + d_r'\D, (1 - d_\ell +d_r') \D - r +\ell)$.
To this end, we consider the tall items packed in $[r + d_r'\D, (1 - d_\ell +d_r') \D - r +\ell)$ in~$\sopt$. 
If tall items intersect the borders of the segment, we create new items by cutting the respective tall item vertically at the respective segment border. 
We denote the set of the tall items (including the two potentially new auxiliary items) contained in $[r + d_r'\D, (1 - d_\ell +d_r') \D - r +\ell)$ by $\items_T$.
Remember that by \ref{enum:packing-G2}, $h(\Gb^{\sopt}({t}))$ is non-increasing in~$t$ in the segment $[r, \D)$. 
Therefore, we sort~$\items_T$ by non-decreasing order of height and pack them next to each other at the end of the segment. 
That is, a tallest item from $\items_T$ ends at $(1-d_\ell+d_r')\D +\ell-r$, and there might be a segment that does not contain tall items at the beginning of the segment $[r+d_r'\D, (1-d_\ell+d_r')\D +\ell-r)$.

We calculate the width of this potentially uncovered segment next.
To this end, we observe that by the assumptions of the lemma on~$r$ and~$d_r'$, i.e., $r \geq (\frac 12 - 3\lambda)\D$ and \alertBound{$d_r' + d_r + 9 \lambda \leq 2 d_r + 9 \lambda \leq \frac\eps{1+\eps}$}, it holds that 
\begin{equation}
    \label{eq:l-close-to-border_r-smaller-than-half_1}
    (1 + d_r' - d_\ell) \D - r + \ell \leq r + \Big( \frac{\eps}{1+\eps} - d_r\Big) \D 
    \,.
\end{equation}	 
Hence, $\big[r, r+ \big(\frac{\eps }{1+\eps} -d_r\big) \D\big) \supseteq [r+d_r'\D, (1-d_\ell+d_r')\D +\ell-r)$. 
Recall that the total width that is not covered by tall items in $\sopt$ (and thus in $\start'$) in the larger segment is bounded by $(d_r- d_r') \D$ by definition.
Therefore, in $\start'$, the segment $\big[r+d_r\D, (1-d_\ell+d_r')\D +\ell-r\big)$ is completely covered by tall items from $\items_T$. 
(Note the different left border compared to the original segment we defined~$\start'$ in.)

We will analyze the height of the items in $\items_T$ in order to show 
\begin{equation}
    \label{eq:bound-extra-packing-first-case}
    h(\items_T^{\start'}(t) \cup \Gb^{\sopt}(t)) + h(\Ge^{\sopt}(r + d_r' \D)) \leq \opt 
\end{equation}
for all $t \in [r + d_r\D, (1 - d_\ell +d_r') \D - r +\ell)$. 
Before proving \eqref{eq:bound-extra-packing-first-case}, we give a brief overview why this inequality is helpful: 
By \eqref{eq:hG5_in_r-d5D_r}, $h(\Ge^{\sopt}(r + d_r'\D))$ is an upper bound on $h(\Ge^{\start}(t))$ for $t \in [r-d_r', r)$. We will show that $h(\items_T^{\start'}((D-t) + (\ell - d_\ell\D))$ is an upper bound on $h( \items^\start_{h > \frac\opt2} (t))$. 
Recalling that $\Gb^{\start}(t) = \Gb^{\sopt}((\D - t) + (\ell - d_\ell\D))$ by \eqref{eq:G2-in-sigma-vs-sopt} allows us to bound $h\big( (\Gb \cup \Ge \cup \items_{h > \frac\opt2})^{\start}(t)\big)$ by $\opt$ for $t \in [r-d_r'\D, r )$ as desired.

In order to show \eqref{eq:bound-extra-packing-first-case}, we need the following observation on $h(\Ge^{\sopt} (t))$ for $t \in [r + d_r'\D, (1 - d_\ell +d_r') \D - r +\ell)$. 
By \eqref{eq:l-close-to-border_r-smaller-than-half_1}, the height of~$\Ge$ in~$\sopt$ is non-decreasing in $[r + d_r'\D, (1-d_\ell + d_r')\D - r + \ell)$ by~\ref{enum:height-G5}, and more precisely, we can lower bound this height by 
\begin{equation}
\label{eq:hG5_in_r+d5'D-some-term}
    h(\Ge^{\sopt}(t)) \geq h(\Ge^{\sopt}(r + d_r'\D)) 
\end{equation}
for~$t \in [r + d_r'\D, (1-d_\ell + d_r')\D - r + \ell)$.  

We next fix an item $i \in \items_T$ with $r + d_r' \D \leq \sopt(i) < r + d_r\D$. 
As $h(\Gb^{\sopt}(t))$ is non-increasing in the segment $[r + d_r'\D, r + d_r\D)$ and due to \eqref{eq:hG5_in_r+d5'D-some-term}, we have that 
\[
    h(i) \leq \opt - h(\Ge^{\sopt}(r + d_r' \D)) -  h(\Gb^{\sopt}(r + d_r \D)) \, . 
\]
Hence, \eqref{eq:bound-extra-packing-first-case} holds for all~$t \in [\start'(i), \start'(i) + w(i))$, where we used that $h(\Gb^{\sopt}(r + d_r \D)) \geq h(\Gb^{\sopt}(t)) $ for $t \in [r+d_r\D, \D )$. 

Next, we fix an item $i \in \items_T$ with $r + d_r \D \leq \sopt(i)$. It holds 
\[
    h(i) \leq \opt - h\big( (\Gb \cup \Ge)^{\sopt} (\sopt(i)) \big) \leq \opt - h(\Gb^{\sopt}(\sopt(i))) -  h(\Ge^{\sopt}(r + d_r' \D)),
\]
where we used again that the height of $\Ge$ is non-decreasing and $\sopt$ has height at most $\opt$. 

If $\start'(i) \geq \sopt(i)$, then \eqref{eq:bound-extra-packing-first-case} holds for $t \in [\start'(i), \start'(i) + w(i))$ because in $\sopt$ the height of $\Gb$ is non-increasing in~$[r,\D)$ by \ref{enum:packing-G2}. 

If $\start'(i) < \sopt(i)$, we need to argue more carefully: By definition of~$\start'$, there are tall items~$\items_i$ with a total width $w(\items_i) \geq \sopt(i) - \start'(i)$ such that $h(i') \geq h(i)$ and $\sopt(i') < \sopt(i)$ holds for all~$i' \in \items_i$. 
Due to their width, the first such item~$i_f \in \items_i$ starts at or before $\start'(i)$ in~$\sopt$. 
Thus, 
\[
  \opt \geq h\big( \{i_f\} \cup (\Gb\cup \Ge)^{\sopt}(\sopt(i_f)) \big) 
    \geq 
    h(i) + h(\Gb^{\sopt}(\start'(i)) + h(\Ge^{\sopt}(r + d_r' \D)) \, ,
\]
where we used again that the height of~$\Gb$ is non-increasing in $[r,\D]$. 
Relying on this fact again and rearranging proves \eqref{eq:bound-extra-packing-first-case} for $t \in [\start'(i), \start'(i) + w(i))$. This completes the proof of \eqref{eq:bound-extra-packing-first-case}. 

We are now ready to bound the height of~$\start$ in the segment $[r-d_r'\D, r)$. 
To this end, we observe that $(1-d_\ell - d_r)\D - r + \ell < r$ is possible, that is, tall items are only packed in a sub-segment of $[r-d_r'\D, r)$. To make the exposition easier to follow, we distinguish the two cases $(1-d_\ell - d_r)\D - r + \ell \geq r$ and $(1-d_\ell - d_r)\D - r + \ell < r$ in the following, although the argumentation is similar.

\smallskip 
\noindent $\bm{\circ \enspace r \leq (1-d_\ell-d_r)\D-r+ \ell}$, i.e., the right border of the current segment is earlier than the endpoint of the last tall item.
To bound the height of~$\start$, recall that $h(\Ge^{\start}(t)) \leq h(\Ge^{\sopt}(r + d_r' \D))$ for $t \in [r-d_r'\D, r)$ by \eqref{eq:hG5_in_r-d5D_r}. 
For bounding the height of $\Gb\cup \items_{h > \frac\opt2}$, we recall that $\Gb^{\start}(t) = \Gb^{\sopt}\big( (\D-t) + (\ell - d_\ell\D) \big)$ by \eqref{eq:G2-in-sigma-vs-sopt}, and hence, $\sopt$ packs the relevant part of the~$\Gb$ items in the segment $\big[(1-d_\ell)\D +\ell-r , (1-d_\ell+d_r')\D +\ell-r\big)$ (where \eqref{eq:bound-extra-packing-first-case} holds due to our case assumption $r \leq (1 - d_\ell - d_r)\D - r + \ell$). 

Recall that~$\start$ packs in $\big[r - d_r' \D , (1-d_\ell-d_r)\D-r+\ell\big)$ the smallest tall items sorted by \emph{non-increasing} height of total width 
\[
    (1-d_\ell-d_r)\D-r+\ell - (r + d_r' \D) =  (1-d_\ell-d_r+d_r')\D-2r+\ell
\]
while in $\start'$ the segment $\big[r+d_r\D, (1-d_\ell+d_r')\D +\ell-r\big)$ is completely covered by tall items sorted by \emph{non-decreasing} height; these tall items have a total width of 
\[
    (1-d_\ell+d_r')\D +\ell-r - (r + d_r\D) =  (1-d_\ell-d_r+d_r')\D-2r+\ell \, . 
\]
We observe that the bijection $f(t) = (D-t) + (\ell - d_\ell\D)$ maps the former segment onto the latter segment and, importantly, ``reverses" time. 
Hence, for $t \in \big[r - d_r' \D , (1-d_\ell-d_r)\D-r+\ell\big) $, we have 
$h(\items_{h > \frac\opt2}^{\start}(t)) \leq h\big(\items_{h > \frac\opt2}^{\start'}\big((D-t) + (\ell - d_\ell\D)\big)\big)$. With the case assumption $r \leq (1 - d_\ell - d_r)\D - r + \ell$ and \eqref{eq:bound-extra-packing-first-case}, we obtain 
\begin{align*}
    h\big( (\Gb \cup \Ge \cup \items_{h > \frac\opt2})^{\start} (t) \big)  \leq   &  h  \big(\Gb^{\sopt}( (D-t) + (\ell - d_\ell\D)) \big) \\
    & + h\big( \items_{h > \frac\opt2}^{\start'}( (D-t) + (\ell - d_\ell\D) )  \big)  +  h(\Ge^{\sopt}(r + d_r' \D)) \\
     \leq  & \opt
\end{align*}
for $t \in [r-d_r'\D,r)$ as desired.

\smallskip
\noindent $\bm{\circ \enspace r > (1-d_\ell-d_r)\D-r+\ell}$, i.e., only parts of the current segment are overlapped by tall items. 
For $t \in \big[(1-d_\ell-d_r)\D-r+\ell, r\big)$, we refer the reader to the analyses of the two subsequent segments covering $\big[(1-d_\ell-d_r)\D-r+\ell,\D\big)$.
It remains to consider the sub-segment $\big[r-d_r'\D,(1-d_\ell-d_r)\D-r+\ell\big)$. 
To this end, recall that $h(\Ge^{\start}(t)) \leq h(\Ge^{\sopt}(r + d_r' \D))$ for $t \in \big[r-d_r'\D, (1-d_\ell-d_r)\D-r+\ell\big)$ by \eqref{eq:hG5_in_r-d5D_r}. 
For bounding the height of $\Gb \cup \items_{h > \frac\opt2}$, we recall that $\Gb^{\start}(t) = \Gb^{\sopt}\big( (\D-t) + (\ell - d_\ell\D) \big)$ by \eqref{eq:G2-in-sigma-vs-sopt}, and hence, $\sopt$ packs the relevant~$\Gb$ items in the segment $\big[r+d_r\D, (1-d_\ell+d_r')\D +\ell-r\big)$ (where \eqref{eq:bound-extra-packing-first-case} holds). 
Since~$\start$ packs the smallest tall items in $\big[r-d_r'\D, (1-d_\ell-d_r)\D-r+\ell\big)$ sorted by \emph{non-increasing} height and in $\start'$ the segment $\big[r+d_r\D, (1-d_\ell+d_r')\D +\ell-r\big)$ is completely covered by tall items sorted by \emph{non-decreasing} height, we have that 
$h(\items_{h > \frac\opt2}^{\start}(t)) \leq h\big(\items_{h > \frac\opt2}^{\start'}\big((D-t) + (\ell - d_\ell\D)\big)\big)$. 
Combining everything with \eqref{eq:bound-extra-packing-first-case}, we obtain 
\begin{align*}
    h\big( (\Gb \cup \Ge \cup \items_{h > \frac\opt2})^{\start} (t) \big) \leq &   h  \big(\Gb^{\sopt}( (D-t) + (\ell - d_\ell\D)) \big)\\ 
    & + h\big( \items_{h > \frac\opt2}^{\start'}( (D-t) + (\ell - d_\ell\D) )  \big)  +  h(\Ge^{\sopt}(r + d_r' \D)) \\
 \leq & \opt
\end{align*}
for $t \in \big[r-d_r'\D,(1-d_\ell-d_r)\D-r+\ell\big)$ as desired.

\bigskip

\noindent $\bullet \enspace \bm{[r,(1 - d_\ell - d_r) \D-r+\ell).} \enspace$ By \ref{enum:packing-G3}, $\items^{\start}(t) \subseteq \Gal \cup \Gb \cup \Ge \cup \items_{h> \frac\opt2}$ for $t \in [r,(1 - d_\ell - d_r) \D-r+\ell)$. For each $t \geq r$, $\Gal^{\start}(t) = \Gal^{\sopt}(t) \subseteq \itemsoverlapt{\sopt}{r}$, 
and $\itemsatt{\sopt}{t} \cap \Gb \subseteq \itemsoverlapt{\sopt}{r}$ for any $t \geq r$ by \ref{enum:packing-G2}. 
Hence, with \cref{obs_height_G1G2}, we can bound the height of $\Gal \cup \Gb$ for $t \in [r,(1 - d_\ell - d_r) \D-r+\ell)$ by 
\begin{equation}
    \label{eq:third-interval_bound-on-g1-g2}
    h(\Gal^{\start}(t) \cup \Gb^{\start}(t)) \leq h(\itemsoverlapt{\sopt}{r})\leq \frac{\opt}{2}.
\end{equation}

It remains to show that the height of the items in $\items_{h > \frac\opt2}(t) \cup \Ge^{\start}(t)$ does not exceed $\opt$ for all $t \in [r,(1 - d_\ell - d_r) \D-r+\ell)$. 
To this end, we consider the items from $\Ge$ intersecting the slightly larger segment $[r-d_r'\D,\D-r+\ell)$ in $\start$. 
Due to our assumptions $\ell \leq 3 \lambda \D$,  $r \geq(\frac 12 - 3 \lambda)\D$, and \alertBound{$2 d_r' + 9 \lambda \leq 2 d_r + 9 \lambda \leq \frac\eps{1+\eps}$}, this segment has a width of 
\[\D-r+\ell - r+ d_r'\D \leq \ell+(d_r'+6\lambda)\D \leq (9\lambda +d_r')\D \leq \Big(\frac{\eps}{1+\eps} -d_r'\Big)\D.\]
Using again that the width of every non-tall item is at least $\frac\eps{1+\eps} \D$, we know that, in $\start$, no item in $\Ge$ ends before $r-d_r'\D +\frac{\eps D}{1+\eps}$. 
Hence, the items from $\Ge$ intersecting the slightly larger segment $[r-d_r'\D,\D-r+\ell)$ in $\start$ have not been stretched by \Cref{alg_stretching}. Therefore, it holds that \(\Ge^{\start}(t) = \Ge^{\sopt}(t+d_r'D)\) for each $t \in [r-d_r'\D,\D-r+\ell)$ and $h(\Ge^{\start}(t))$ is non-decreasing in~$t$ in the segment $[r-d_r'\D,\D-r+\ell)$.

In order to analyze the height of $\Ge \cup \items_{h>\frac\opt2}$ in $[r-d_r'\D,\D-r+\ell)$ in $\start$, we create an auxiliary packing~$\start'$ based on~$\sopt$ that packs only the items in $\Ge \cup \items_{h>\frac\opt2}$. 
In~$\start'$, the tall items are sorted by height and $\Ge$ is packed as in $\sopt$. 
We will show that its height is at most~$\opt$. The starting times of both item sets decrease in $\start$ compared to $\start'$ by at least~$d_r' \D$  and, more importantly, the tall items are (potentially) started even earlier than the items from $\Ge$. 

For constructing~$\start'$, we consider the items $\items_{h>\frac\opt2}$ that are packed in the segment $[r, (1+d_r')\D-r+\ell)$ in the optimal packing $\sopt$; potentially cutting an item vertically at $(1+d_r') \D-r+\ell$ if it overlaps that time point. We define~$\start'$ by sorting the tall items non-increasingly by height and packing~$\Ge$ as in $\sopt$. We now argue that 
the combined height of the items in $\Ge$ and the tall items does not exceed $\opt$ in $\start'$ at any point $t \in [r, (1+d_r') \D-r+\ell)$, i.e.,  $h(\Ge^{\start'}(t)) + h(I_{h>\frac{\opt}{2}}^{\start'}(t)) \leq \opt$. 

For the sake of contradiction, suppose that there is some $\tau \in [r,(1+d_r')\D-r+\ell)$ with $h(\Ge^{\start'}(\tau)) + h(I_{w>\frac{\opt}{2}}^{\start'}(\tau)) > \opt$. Recall that the height of $\Ge$ is non-decreasing in $[r,(1+d_r')\D-r+\ell)$ in both, $\sopt$ and $\start'$. Hence, this implies that the tall item $T$ intersecting $\tau$ must have been started earlier in $\sopt$ than in $\start'$ as otherwise, $\sopt$ would have a height greater than $\opt$ at some point in $[\tau, (1+d_r')\D-r+\ell)$, which contradicts the definition of $\sopt$. 
Since $\start'$ sorts the tall items by non-increasing height, this upper bound on the starting time of $T$ implies that there is an item taller than $T$ whose packing in $\sopt$ intersects $[\tau, (1+d_r')\D-r+\ell)$, which implies again that the height of $\sopt$ exceeds $\opt$ at some point later than~$\tau$; a contradiction.

Next, we want to compare~$\start'$ and $\start$. 
Recall that $[r,(1+d_r')\D-r+\ell) \subseteq \big[r,r+\big(\frac{\eps}{1+\eps}-d_r\big)\D\big)$ by \eqref{eq:l-close-to-border_r-smaller-than-half_1} and the total width not covered by tall items in the larger segment is at most $(d_r-d_r')\D$ by assumption. 
Hence, in the auxiliary packing $\start'$, the last tall item ends at or after $((1+d_r')\D-r+\ell)-(d_r-d_r')\D$. By \ref{enum:tall-items}, the last tall item ends at $(1-d_r-d_\ell)\D +\ell - r$ in~$\start$.
Compared to $\start'$, in $\start$, the tall items from the segment  $[r,(1+d_r')\D-r+\ell)$ are shifted to the left by at least 
\[
    \big((1+2d_r'-d_r) \D - r + \ell\big) - \big((1-d_r-d_\ell)\D +\ell -r\big) =  (2d_r' +d_\ell)\D \geq d_r' \D.
\] 
Conversely, compared to $\start'$, in $\start$, the items from $\Ge$ are shifted to the left by exactly $d_r'\D$, implying 
\[
h(\Ge^{\start}(t)) + h(I_{h>\frac{\opt}{2}}^{\start}(t)) \leq h(\Ge^{\start'}(t + d_r'\D)) + h(I_{h>\frac{\opt}{2}}^{\start'}(t + d_r'\D)) \leq \opt
\]
for each $t \in [r,(1+d_r')\D-r+\ell)$.
Combining with the bound on $h(\Gal^\start(t) \cup \Gb^\start(t)) \leq \frac\opt2$ in \eqref{eq:third-interval_bound-on-g1-g2}, this implies the desired bound on $h(\items^\start(t))$ for each $t \in [r,(1+d_r')\D-r+\ell)$.  

\bigskip
\noindent $\bullet \enspace \bm{[(1-d_\ell-d_r)\D-r+\ell,(1- d_\ell)\D-r+\ell).} \enspace$ No tall item and no item from $G_3$ intersects this segment.
Hence, 
\begin{align*}
    h(\items^{\start}(t)) & = h(\Gal^{\start}(t))+h(\Gb^{\start}(t))+h(\Ge^{\start}(t))
    \leq \frac{\opt}{2}+\frac{\opt}{2}+\frac{\opt}{2} = \frac{3}{2}\opt,
\end{align*}
where we bound $h(\Gal^{\start}(t))$ and $h(\Gb^{\start}(t))$ with \cref{obs_height_G1G2} and $h(\Ge^\start(t))$ with the Stretching Lemma~\ref{lem:stretching}.  

\bigskip

\noindent $\bullet \enspace \bm{ [(1-d_\ell)\D-r+\ell, \D).} \enspace$ By \ref{enum:tall-items}, no tall item overlaps this segment. Thus, $\items^{\start}(t) \subseteq \Gal\cup \Gb \cup \Ge$ for $t \in [(1-d_\ell-d_r)\D - r + \ell,(1-d_\ell) \D)$. By \cref{lem:stretching}, 
$h(\Ge(t')) \leq \frac{\opt}{2}$ for any $t' \in [0,\D)$. 

Therefore, bounding the height of $\Gal \cup \Gb$ by $\opt$ suffices to show the lemma statement in the considered segment. 
\cref{alg:non-tall-over-tall} shifts the mirrored image of $\Gb$ by $\ell - d_\ell\D$ by \cref{lem:G2-shift-over-tall-fit}. Thus,  
\begin{align*}
    (\Gb)^{\sopt}(t)  &= (\Gb)^{\start}(D-t +\ell -d_\ell\D) \,.  
    \shortintertext{As $t$ appears on the right-hand side with reversed sign, this implies in particular}
    (\Gb)^{\start}(t)  & = (\Gb)^{\sopt}(D-t +\ell -d_\ell\D) \, . 
\end{align*}
If we could argue that $h(\Gal^\start(t)) \leq h(\Gal^{\sopt}((1-d_\ell)\D-t+\ell))$ for $t \in [(1-d_\ell)\D-r+\ell,\D)$, the upper bound~$\opt$ on the height of $\sopt$ would imply the desired bound on the height of $\Gal \cup \Gb$. 

Recall that $\start$ packs $\Gal$ as in $\sopt$ and $\Gal$ is non-increasing for $t' \geq \ell - d_\ell \D $ (and, hence, in the current segment). 
By definition, $d_\ell \D \leq \ell$, and by assumption,~$r \leq \frac\D2$. Hence, $t \geq (1-d_\ell) \D - r + \ell \geq \D - r \geq r$, and thus, 
$t \geq (1-d_\ell) \D - t + \ell \geq \ell - d_\ell \D $ for $t \in [(1-d_\ell)\D-r+\ell, \D)$. This implies that indeed $h(\Gal^\start(t)) \leq h(\Gal^{\sopt}((1-d_\ell)\D-t+\ell))$ holds in the current segment. Combining everything, we obtain 
\begin{align*}
    h(\items^\start(t) & = \big((\Gal \cup \Gb \cup \Ge)^{\start}(t)\big) \\
    & = h((\Gb)^{\sopt}((1-d_\ell)\D-t+\ell)) + h(\Gal^{\sopt}(t)) + h(\Ge^\start(t)) \\ 
    &\leq h((\Gb)^{\sopt}((1-d_\ell)\D-t+\ell)) + h(\Gal^{\sopt}((1-d_\ell)\D-t+\ell)) + \frac\opt2 \\
    &\leq \opt + \frac\opt2 \, .
\end{align*}

\bigskip 
This bounds the height of~$\start$ in $[0,\D)$ by $\frac32 \opt$ as required. Combining with our observation at the beginning of the proof that indeed all items are packed in the segment~$[0,\D)$ concludes the proof. 
\end{proof}

\section{Exactly two gaps of width at least \texorpdfstring{$\big(\frac{1}{2}-3\lambda\big)\D$}{(1/2-3λ)D}}
\label{sec:two-large-gaps}
In this section, we assume that there is an optimal solution $\sopt$ that has more than one large gap, each with a width greater than $\big(\frac{1}{2}-3\lambda\big)\D$.
We then show that then $(\items, \D)$ admits a neat packing, formalized in the following lemma.

\begin{restatable}{lemma}{lemtwolargegaps}
\label{lem:two_large_gaps_G_3_small}
If $\sopt$ is an optimal packing of $(\items, \D)$ with two gaps of width at least $\big(\frac 12 - 3 \lambda \D\big)$ each, then $(\items, \D)$ admits a neat packing.
\end{restatable}

If \alertBound{$\lambda < \frac1{18}$}, there can be at most two gaps with a width greater than $\big(\frac{1}{2}-3\lambda\big)\D$.
Throughout this section we assume that the first gap spans from $\leftl$ to $\leftr$, while the second gap spans from $\rightl$ to $\rightr$, i.e., we assume that $\leftl < \leftr < \rightl < \rightr$, no tall item intersects the segment $[\leftl,\leftr)$ or the segment $[\rightl,\rightr)$, and there are tall items ending at $\leftl$ and $\rightl$ as well as tall items starting at $\leftr$ and $\rightr$.
Furthermore, we assume that $\leftr-\leftl > (\frac{1}{2}-3\lambda)\D$ and $\rightr-\rightl > (\frac{1}{2}-3\lambda)\D$. 
Note that the lower bound on the width of one gap implies an upper bound of $\big(\frac{1}{2}+3\lambda\big)\D$ on the width of the other gap. 
That is, $\leftr-\leftl, \rightr - \rightl \in \big( \big(\frac{1}{2}-3\lambda\big)\D , \big(\frac{1}{2}+3\lambda\big)\D \big)$.

Finally, we assume that $\leftr + \rightl \geq D$. 
This can be achieved by mirroring the packing if it is not the case in the given packing.
We define three parameters: $d_1 \D = \leftl$, $d_2 \D = \rightl - \leftr$, and $d_3 \D = \D - \rightr$.
Note that by the assumptions on $\leftl,\leftr,\rightl,\rightr$, it holds that $d_1+d_2+d_3 \leq 6\lambda$.

In the following, we assume that the packing does not contain squeezable items, i.e., non-tall item with a width smaller than $\frac{\eps \D}{(1+\eps)}$.
Since we will sort the tall items in both cases, we can add all items with a width smaller than $\frac{\eps \D}{(1+\eps)}$ in a later step.
By this assumption, we know that no non-tall item is wholly contained in one of the segments $[0,\leftl)$, $[\leftr,\rightl)$ or $[\rightr, \D)$.
We use this fact in the following way: In \Cref{sec:one-wide-gap}, we needed to stretch the non-tall items that are packed completely between the tall items. 
However, in this section, we do not need to stretch similarly defined sets ($G_L$ and $G_{3,R}$, defined below). 
This implies that we ``lose'' less width and are able to repack the packing.

Similar to previous repackings,
given an optimal packing $\sopt$, we partition the set $\items_{h\leq \frac{\opt}{2}}$ as follows:
\[
G_{1,R} := \itemsoverlapt{\sopt}{\srightr}
,\quad 
G_{2,R} := \items_{\sopt(i) < \srightl, \, \srightl < \sopt(i) + w(i) \leq \srightr} 
, \quad 
G_{3,R} := \items^{\sopt}_{[\srightl,\srightr]} \setminus \items_{h> \frac{\opt}{2}}
, 
\]
\[
G_L := \items_{\sopt(i) \leq \sleftr, \, \sopt(i) + w(i) \leq \srightl} \setminus \items_{h> \frac{\opt}{2}}
\]

\begin{figure}[t]
    \centering
    \resizebox{\textwidth}{!}{
    \input{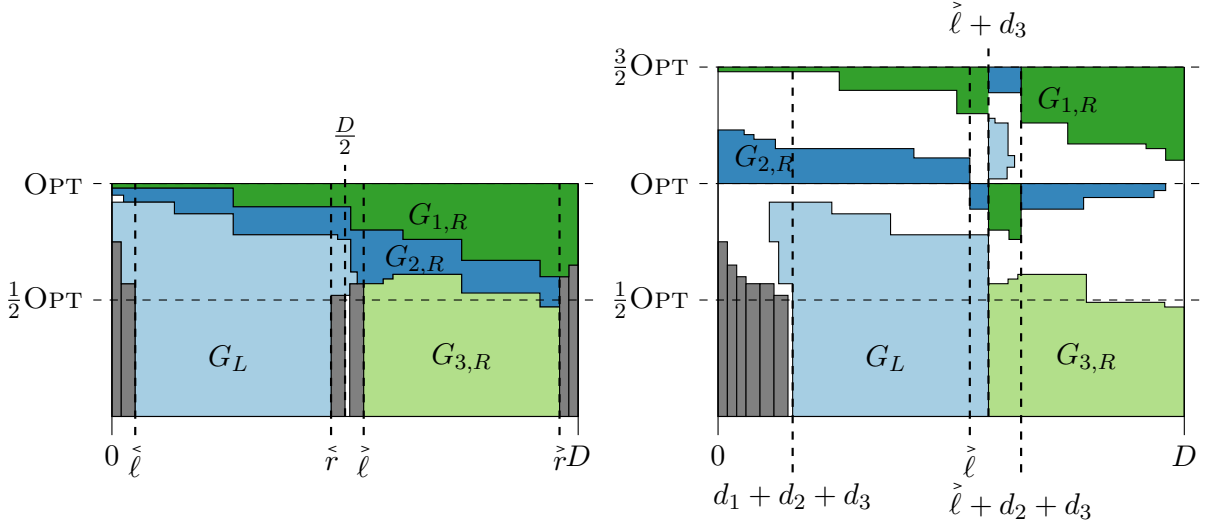}
    }
    \caption{An example for the repacking when there are two wide gaps}
    \label{fig:Two_Gaps}
\end{figure}

The algorithm works as follows. 
First, we sort the tall items by height and packing them non-increasingly by height starting at $0$.
Next, we let $G_{1,R}$ start such that each item ends at $\D$ and then pack the items in $G_{2,R}$ such that they start at $0$.
Finally, we shift $G_{3,R}$ to the right by $d_3 \D$ compared to $\sopt$ and then shift $G_L$ to the right by $(d_2 + d_3) \D$ compared to $\sopt$.
This is formalized in \Cref{alg_two_large_gaps_G_3_small}, and an example can be seen in \Cref{fig:Two_Gaps}.

\begin{algorithm}[ht]
\DontPrintSemicolon
\caption{Repacking for two wide gaps with width at least $\left(\frac{1}{2}-3\lambda\right)\D$}
\label{alg_two_large_gaps_G_3_small}
\textbf{Input:}  $(\sopt, \items, G_{1,R}, G_{2,R}, G_{3,R}, G_L, \leftl, \leftr, \rightl, \rightr)$\;
\smallskip
$\start \leftarrow$ output of \cref{alg_items_next_to_each_other} with $(\items_{h > \frac{\opt}{2}},0)$ \tcp*[t]{sort tall items}
\lForEach(\tcp*[f]{finish $G_{1,R}$ at $\D$}){$i \in G_{1,R}$}{
    $\start(i) \leftarrow  \D - w(i)$ 
}
\lForEach(\tcp*[f]{start $G_{2,R}$ at $0$}){$i \in G_{2,R}$}{
    $\start(i) \leftarrow  0$ 
}
\lForEach(\tcp*[f]{shift $G_{3,R}$ to the right}){$i \in G_{3,R}$}{
    $\start(i) \leftarrow  \sopt(i) + d_3\D$ 
}
\lForEach(\tcp*[f]{shift $G_{L}$ to the right}){$i \in G_L$}{
    $\start(i) \leftarrow  \sopt(i) + d_2\D + d_3\D$ 
}
\smallskip 
\textbf{return} $\start$ \;
\end{algorithm}

\lemtwolargegaps*

Let $[\leftl, \leftr)$ and $[\rightl, \rightr)$ be the gaps as specified in the lemma.
For the remainder of the section let $T_2$ and~$T_3$ be the tall items such that $\leftrm = \sopt(T_2)$ and $\rightlm = \sopt(T_3) + w(T_3)$. 
We then have that $\leftlm = 0$ or $\leftlm : = \sopt(T_1) + w(T_1)$ for a tall item $T_1$ and $\rightrm := \D$ or $\rightrm := \sopt(T_4)$ for a tall item $T_4$.
We note that the items $T_2$ and $T_3$ from the lemma can potentially coincide. This does neither influence the algorithm nor the proof. 

\begin{proof}
We show that \cref{alg_two_large_gaps_G_3_small} returns a neat packing~$\start$ of the items $\items \setminus \items_{h \leq \frac\opt2, w \leq \frac\eps{1+\eps}\D}$. We prove the lemma by considering the following segments separately: $[0,(d_1+d_2+d_3)\D)$, $[(d_1+d_2+d_3)\D,\rightl)$, $[\rightl,\rightl +d_3\D)$, $[\rightl +d_3\D),\rightl +(d_2+ d_3)\D)$, and $[\rightl +(d_2+ d_3)\D, \D]$.
Before going into the details, we make some useful observations about the different item sets. 
\begin{enumerate}
    \item\label{enum:G1R-G2r-at-ell} For $t \leq \rightl$, we know that any item from $G_{1,R}^{\start}(t)$ intersects $\rightl$ in $\sopt$ since their starting time does not decrease. By definition, $G_{2,R} \subseteq \itemsoverlapt{\sopt}{\srightl}$ and combining with \cref{obs_height_G1G2}, we get $ h( (G_{1,R} \cup G_{2,R})^{\start}(t)) \leq \frac{\opt}{2}$ for $t \leq \rightl$. 
\end{enumerate}

\noindent $\bullet \enspace \bm{[0,(d_1+d_2+d_3)\D). }$ \enspace
This segment does not contain items from $G_{3,R}$ but contains all the tall items as well as items from $G_{1,R}$, $G_{2,R}$, and $G_{L}$. 
The items from the set $G_L$ have been shifted to the right by $(d_2+d_3)\D$. 
Thus, only those items of $G_L$ intersecting $\leftl$ in $\sopt$ overlap the current segment in $\start$.
Therefore, these items, together with the tall items, add up to a height of at most $\opt$ in $\start$ as the tall items are sorted non-increasingly by height and start at $0$. 
With \ref{enum:G1R-G2r-at-ell}, this implies that the height is at most $\frac32\opt$. 

\bigskip
\noindent $\bullet \enspace \bm{[(d_1+d_2+d_3)\D,} \bm{\ell}\hspace{-.7ex}\raisebox{0.8ex}{\textrptr} \bm{ ).} $ \enspace 
This segment does not contain any tall item since they have a total width of at most $(d_1+d_2+d_3)\D$, and no item from the set $G_{3,R}$ since they start at or after $\rightl +d_3\D$. 
With \ref{enum:G1R-G2r-at-ell}, we obtain for each $t \in [(d_1+d_2+d_3)\D,\rightl]$ 
\begin{align*}
    h(\itemsatt{\start}{t}) &= h((G_{L} \cup G_{1,R} \cup G_{2,R})^{\start}(t))\\
    &= h(G_{L}^{\sopt}(t -d_2\D-d_3\D)) + h((G_{1,R} \cup G_{2,R})^{\start}(t)) \\
    & \leq \opt + \frac12 \opt = \frac{3}{2}\opt \ .
\end{align*}
This bounds the height of the current segment.
\bigskip

\noindent $\bullet \enspace \bm{[}\bm{\ell}\hspace{-.7ex}\raisebox{0.8ex}{\textrptr}\bm{,}\bm{\ell}\hspace{-.7ex}\raisebox{0.8ex}{\textrptr} \bm{+d_3\D)}.$\enspace
Since the items from $G_{3,R}$ start at or after $\rightl +d_3\D$, for each $t \in [\rightl,\rightl +d_3\D]$, 
\[ h(\itemsatt{\start}{t}) = h((G_{L} \cup G_{1,R} \cup G_{2,R})^{\start}(t)).\]

We know $h(G_{1,R}^{\start}(t)) \leq \frac{\opt}{2}$ by \cref{obs_height_G1G2}, and hence showing $h((G_{L} \cup G_{2,R})^{\start}(t)) \leq \opt$ suffices to prove the desired height bound. 
Note that $G_{L}^{\start}(t) = G_{L}^{\sopt}(t-d_2\D-d_3\D)$.
We will prove that $G_{2,R}^{\start}(t) \subseteq \items^{\sopt}(t-d_2\D-d_3\D)$ for $t \in [\rightl,\rightl +d_3\D)$.
To this end, note that those items have a width of at least $\rightl$ since they start at $0$ and end after $\rightl$ in $\start$. 
Thus, they have to start before $\rightr - \rightl$ in $\sopt$ as they end before $\rightr$ by definition of $G_{2,R}$.
Using $D \leq \leftr + \rightl$, we calculate 
\begin{equation}
\label{eq:bound_on_checkr-checkl}
    \rightr - \rightl = D-d_3\D-   \rightl \leq \leftr -d_3\D = \rightl-d_2\D-d_3\D  \leq t' -d_2\D-d_3\D \, 
\end{equation}
for $t' \geq \rightl$. 
Hence, $G_{2,R}^{\start}(t) \subseteq \items^{\sopt}(t-d_2\D-d_3\D)$ for $t \in [\rightl, \rightl + d_3 \D)$ as required. 
Thus, $(G_{2,R} \cup G_L)^{\start}(t) \subseteq \items^{\sopt}(t-d_2\D-d_3\D)$, which bounds their height by $\opt$.  
Therefore, $h(\itemsatt{\start}{t}) = h((G_{L} \cup G_{1,R} \cup G_{2,R})^{\start}(t)) \leq \frac32 \opt$. 

\bigskip

\noindent $\bullet \enspace \bm{[}\bm{\ell}\hspace{-.7ex}\raisebox{0.8ex}{\textrptr} \bm{+d_3\D},\bm{\ell}\hspace{-.7ex}\raisebox{0.8ex}{\textrptr} \bm{+(d_2 + d_3)\D).}$ \enspace 
As in the previous segments, there are no tall items packed in this segment. Hence, $
    \itemsatt{\start}{t} = (G_{L} \cup G_{1,R} \cup G_{2,R} \cup G_{3,R})^{\start}(t)$ for $t \in [\rightl +d_3\D,\rightl +(d_2 + d_3)\D)$.

First, we prove that $(G_{L} \cup G_{2,R})^{\start}(t) \subseteq \itemsoverlapt{\sopt}{\sleftr}$ (which implies a bound of $\frac\opt2$ on their height by \cref{obs_height_G1G2}). 
To this end, we observe that the items in $G_{2,R}^{\start}(\rightl)$ have a width of at least $\rightl$. 
Since by definition, they finish before $\rightr$ in $\sopt$, they start before $\rightr - \rightl$ in $\sopt$.  
Using $\rightr - \rightl \leq \leftr - d_3 \D$ by \eqref{eq:bound_on_checkr-checkl}, we get $G_{2,R}^{\start}(\rightl) \subseteq G_{2,R}^{\sopt}(\leftr -d_3 \D)$. 
Therefore,
\[
    G_{2,R}^{\start}(t) \subseteq G_{2,R}^{\start}(\rightl) \subseteq G_{2,R}^{\sopt}(\leftr -d_3\D) \subseteq \itemsoverlapt{\sopt}{\sleftr} 
\]
for $t \in [\rightl +d_3\D,\rightl +(d_2 + d_3)\D)$. 
For $G_L$, it holds that $G_{L}^{\start}(t) = G_{L}^{\sopt}(t-(d_2+d_3)\D)$. 
Since $t-(d_2+d_3)\D \geq \rightl - d_2\D = \leftr$ for $t$ in the current segment and no item in $G_{L}$ starts after $\leftr$ in $\sopt$, it holds that 
$G_{L}^{\start}(t) = G_{L}^{\sopt}(t-(d_2+d_3)\D) \subseteq  \itemsoverlapt{\sopt}{\sleftr}$.
With \cref{obs_height_G1G2}, we get
\begin{equation}
    \label{eq:h-GL-G2R}
    h(G_{L} \cup G_{2,R})^{\start}(t)) \leq h(\itemsoverlapt{\sopt}{\sleftr}) \leq \frac{\opt}{2} \, . 
\end{equation}

Next, we consider $(G_{1,R} \cup G_{3,R})^{\start}(t)$ and show that $h((G_{1,R} \cup G_{3,R})^{\start}(t)) \leq \opt$.
Note that \alertBound{$d_2+d_3 \leq \frac{\eps}{(1+\eps)}$}  as $d_1+d_2+d_3\leq 6 \lambda$ and \alertBound{$\lambda \leq \frac{\eps}{6(1+\eps)}$} by assumption. 
As the width of the current segment is $d_2 \D$ and each item in $G_{3,R}$ has a width larger than $(d_2+d_3)\D$, this implies that neither in $\sopt$ nor in $\start$ an item from $G_{3,R}$ ends in $[\rightl,\rightl +(d_2 + d_3)\D)$. 
Therefore, 
\[
    G_{3,R}^{\start}(t) \subseteq G_{3,R}^{\start}(\rightl +(d_2 + d_3)\D) = G_{3,R}^{\sopt}(\rightl +d_2 \D) \subseteq G_{3,R}^{\sopt}(\rightl +(d_2+d_3)\D),
\]
for $t \in [\rightl +d_3\D,\rightl +(d_2 + d_3)\D)$.
In~$\start$, the items in~$G_{1,R}$ end at $\D$, and, in particular, if they overlap $t < \rightl +(d_2 + d_3)\D$, they also overlap $\rightl + (d_2 + d_3)\D$. Compared to~$\sopt$, their starting time does not decrease. 
Hence, 
\[
    G_{1,R}^{\start}(t) \subseteq G_{1,R}^{\start}(\rightl +(d_2 + d_3)\D) \subseteq G_{1,R}^{\sopt}(\rightl +(d_2+ d_3)\D )
\]
for $t < \rightl +(d_2 + d_3)\D$.
Therefore,
\[ 
    (G_{3,R} \cup G_{1,R} )^{\start}(t) \subseteq (G_{3,R} \cup G_{1,R} )^{\sopt}(\rightl +(d_2+d_3)\D).
\]
for $t \in [\rightl +d_3\D,\rightl +(d_2 + d_3)\D)$. Thus, $h( (G_{3,R} \cup G_{1,R} )^{\start}(t) ) \leq \opt$, and combining with \eqref{eq:h-GL-G2R}, this bounds the height of $\start$ in the current segment. 

\bigskip

\noindent $\bullet \enspace \bm{[}\bm{\ell}\hspace{-.7ex}\raisebox{0.8ex}{\textrptr} \bm{+(d_2+ d_3)\D, \D].}$ \enspace 
All items from $G_L$ end at or before $\rightl$ in $\sopt$ and, thus, at or before $\rightl +(d_2+d_3)\D$ in $\start$. 
Therefore, for $t \in [\rightl +(d_2 + d_3)\D, \D]$, 
\[ 
    h(\itemsatt{\start}{t}) = h((G_{1,R} \cup G_{2,R} \cup G_{3,R})^{\start}(t)).
\]

Since the items in $G_{2,R}$ start at~$0$ in~$\start$, we have $G_{2,R}^{\start}(t')\subseteq G_{2,R}^{\start}(t'-d_3\D)$ for any $t' \geq d_3 \D$. Further, as those items have been shifted to the left in $\start$ compared to~$\sopt$, we know that $G_{2,R}^{\start}(t') \subseteq G_{2,R}^{\sopt}(t')$ for all $t' \in [\check \ell, \D)$. Combining, we get  
$G_{2,R}^{\start}(t)\subseteq G_{2,R}^{\start}(t-d_3\D) \subseteq G_{2,R}^{\sopt}(t-d_3\D)$.
Furthermore, it holds that
$ G_{3,R}^{\start}(t) = G_{3,R}^{\sopt}(t-d_3\D)$. Therefore,
\[
    h((G_{2,R} \cup G_{3,R})^{\start}(t)) \leq h((G_{2,R} \cup G_{3,R})^{\sopt}(t-d_3\D)) \leq \opt \, .
\]
Since $h(G_{1,R}) \leq \frac{\opt}{2}$ by \cref{obs_height_G1G2},
it holds that 
\[ 
    h(\itemsatt{\start}{t}) = h((G_{1,R} \cup G_{2,R} \cup G_{3,R})^{\start}(t)) \leq \frac{3}{2}\opt \, .
\]

\bigskip

It remains to argue that indeed all items are packed in~$[0,\D]$. Since~$\sopt$ is feasible, the maximal width of any item is bounded by~$\D$. 
By construction, all tall items are packed in the segment~$[0, d_1 + d_2 + d_3]$. 
In order to argue about the packing of the non-tall items, we use again the partition $G_{1,R} \cup G_{2,R} \cup G_{3,R} \cup G_L$ and argue separately. 
The items~$G_{1,R}$ are packed by~$\start$ such that they finish at~$\D$, implying that they start at or after~$0$.
Similarly, the items~$G_{2,R}$ start at~$0$ in~$\start$, implying that they finish at or before~$\D$.  
The items~$G_{3,R}$ finish before~$\rightr$ in~$\sopt$ and are shifted to the right by~$d_3 \D = D - \rightr$ in~$\start$, compared to~$\opt$.
Hence, they finish at or before~$\D$ and start after $0$ in~$\start$. 
The items in~$G_L$ finish before $\rightl \leq (\frac 1 2 +3 \lambda) \D$ in~$\sopt$ by definition. In~$\start$, they are shifted to the right by $d_2\D + d_3\D \leq 6 \lambda$.
Hence, they finish before~$\rightl + (d_2 + d_3)\D < \D$ and start after $0$ in~$\start$. 
This concludes the proof of the lemma. 
\end{proof}

\section{Proof of structural theorem}
\label{sec:proof-thm1}

In this section, we argue how to derive a packing~$\start$ with maximal height~$\big(\frac32+\eps\big)\opt$ and a lot of structure from an optimal packing~$\sopt$ of height~$\opt$. That is, we prove \cref{thm:restructuring}.

\thmstructuralresult*

\newcounter{assumptions}

\begin{proof}
    In order to prove the theorem, we show that (up to mirroring) all optimal solutions of every possible instance satisfy at least one of the cases considered in the previous sections. 
    Let $\sopt$ be an optimal packing of $\items$. If there are no tall items in $\items$, then $\sopt$ is already neat. So, we assume 
    \begin{equation}
    \tag{A\theassumptions}\stepcounter{assumptions}
    \label{eq:at-least-one-tall}
    \begin{gathered}
        \items_{h > \frac\opt2} \neq \emptyset \, , \\
        \text{i.e., there is at least one tall item.}  
    \end{gathered}    
    \end{equation}

    If the total width of tall items is at least $\big(1-\frac{\eps}{5 + 4\eps}\big)\D$ then there is a neat packing by \cref{lem:tall_items_are_wide}. Hence, we may assume the total width of all gaps is at least $\frac{\eps}{5 + 4\eps}\D$.  If $\lambda = \Theta(\eps)$ is chosen small enough, then this implies
    \begin{equation}
    \tag{A\theassumptions}\stepcounter{assumptions}
    \label{eq:width-tall-upper-bounded}    
    \begin{gathered}
        \D - w \big( \items_{h > \frac\opt2} \big) > 3\lambda \D\, , \\
        \text{i.e., the total width of all gaps is at least $3\lambda \D$.}
    \end{gathered}     
    \end{equation}
    
    If there exists a gap $[\ell,r)$ such that $r- \ell \in [ \lambda \D, (\frac 12 - 3\lambda) \D]$, then \Cref{lem:full-case-C} guarantees the existence of a forgiving packing.
    Therefore, we may assume that 
    \begin{equation}
    \tag{A\theassumptions}\stepcounter{assumptions}
    \label{eq:distance-two-tall}
    \begin{gathered}
        r - \ell < \lambda \D \quad \text{ or } \quad r - \ell > (\tfrac 12 -3\lambda) \D \, , \\
        \text{i.e., every gap has width less than $\lambda\D$ or more than $\textstyle{\big(\frac 12 -3\lambda\big) \D}$.}
    \end{gathered}
    \end{equation}
    We call a gap with $r - \ell < \lambda \D$ and with $r- \ell > \big(\frac 12 -3\lambda\big) \D$ \emph{narrow} and \emph{wide}, respectively. 

    Moreover, call a gap $(\ell,r)$ \emph{early} if its endpoint is left of $\big(\frac12-3\lambda\big)\D$, i.e., $r \leq \big(\frac12-3\lambda\big)\D$ and \emph{late} if its starting point is right of $\big(\frac12+3\lambda\big)\D$, i.e., $\ell \geq \big(\frac12+3\lambda\big)\D$. 
    By \eqref{eq:distance-two-tall}, all early and all late gaps must be narrow. 
    If the early gaps have a combined width of at least $\lambda\D$, then some prefix of the early gaps has a combined width between $\lambda\D$ and $2\lambda\D$. 
    Analogously, if the late gaps have a total width of at least $\lambda\D$, then some suffix of the late gaps has a total width between $\lambda\D$ and $2\lambda\D$.     
    In both cases, \cref{lem:small-gap-at-border} guarantees a forgiving packing. 
    We may therefore assume 
    \begin{equation}        
        \tag{A\theassumptions}\stepcounter{assumptions}
        \label{eq:early-and-late-gaps}
        \begin{gathered}
            \tau_1 - w\big( \items_{h > \frac\opt2} \cap \itemsT{\sopt}_{[0,\tau_1]} \big) < \lambda \D \, \\ 
            \text{i.e., the early gaps have a total width of less than $\lambda\D$, and} \\
            (\D - \tau_2) - w\big( \items_{h > \frac\opt2} \cap \itemsT{\sopt}_{[\tau_2, \D]} \big) <  \lambda \D \, , \\
            \text{i.e., the late gaps have a total width of less than $\lambda\D$,}
        \end{gathered}
    \end{equation}
    where $\tau_1$ is the latest starting point of a tall item that starts before $\big(\frac12-3\lambda\big)\D$ (or $0$ if no such item exists) and $\tau_2$ is the earliest endpoint of a tall item $T_2$ that ends after $\big(\frac12+3\lambda\big)\D$ (or $\D$ if no such item exists). 

    Using \eqref{eq:width-tall-upper-bounded}, we conclude that the \emph{intermediate} gaps, i.e., gaps that are neither early nor late, satisfy 
    \begin{equation}  
    \tag{O\theassumptions}\stepcounter{assumptions}
    \label{eq:intermediate-gaps}
    \begin{gathered}
        (\tau_2 - \tau_1) - w\big( \items_{h > \frac\opt2} \cap \itemsT{\sopt}_{[\tau_1,\tau_2]} \big) > \lambda \D \, , \\
        \text{i.e., the intermediate gaps have a combined width exceeding $\lambda\D$.}
    \end{gathered}
    \end{equation}

    By definition of narrow and intermediate gaps, any narrow intermediate gap has to start after $\big(\frac12-4\lambda\big)\D$ and finish before $\big(\frac12+4\lambda\big)\D$. 
    If we can find a sequence $(\ell_1, r_1),\ldots, (\ell_k, r_k)$ of narrow intermediate gaps with combined width between $\lambda\D$ and $2 \lambda\D$, \cref{lem:small-gap-center} guarantees a forgiving packing if $\lambda \in \Theta(\eps)$ small enough. 
    Hence, we may assume that  
    \begin{equation}  
    \tag{O\theassumptions}\stepcounter{assumptions}
    \label{eq:narrow-intermediate-gaps}
    \begin{gathered}
        \text{the narrow intermediate gaps have a combined width less than $\lambda\D$.}
    \end{gathered}
    \end{equation}
    Therefore, we may assume     
    \begin{equation}  
    \tag{A\theassumptions}\stepcounter{assumptions}
    \label{eq:gap_at_least_one_wide_gap}
        \begin{gathered}
              \text{there exists at least one wide gap.}
        \end{gathered}
    \end{equation}

    Three or more wide gaps would have a combined width exceeding $\D$ assuming $\lambda$ is small enough, so there are either one or two wide gaps. Suppose there are two wide gaps $(\leftlm, \leftrm)$ and $(\rightlm, \rightrm)$ with $\leftrm - \leftlm > \big(\frac12-3\lambda\big)\D$, $\rightrm - \rightlm > \big(\frac12-3\lambda\big)\D$, and $\leftrm < \rightlm$. Since all gaps completely contained in $[\leftrm, \rightlm]$ are intermediate and narrow by \eqref{eq:distance-two-tall} and hence, by \eqref{eq:narrow-intermediate-gaps}, they have a combined width less than $\lambda\D$. 
    Similarly, all gaps in $[0, \leftlm]$ (in $[\rightrm,\D]$) are early (late) and narrow and hence, their combined width is less than $\lambda\D$ by \eqref{eq:early-and-late-gaps}. 
    Thus, \cref{lem:two_large_gaps_G_3_small} guarantees a neat packing.
    Therefore, we may assume 
    \begin{equation}  
    \tag{A\theassumptions}\stepcounter{assumptions}
    \label{eq:exactly_one_wide_gap}
        \begin{gathered}
              \text{there exists exactly one wide gap $[\ell,r)$.}
        \end{gathered}
    \end{equation}
    
    We now have the assumptions that (i) the gap width satisfies $r - \ell \geq (\frac 12 - 3 \lambda) \D$, 
    (ii) the gaps to the left of $\ell$ have a combined width of at most $\lambda \D$, and (iii)
     the gaps to the right of $r$ have a combined width of at most $\lambda \D$.
     Note that these are exactly the assumption for \Cref{lem:one-wide-gap}. 
     Hence, \Cref{lem:one-wide-gap} guarantees the existence of a neat packing.
\end{proof}

\section{\texorpdfstring{$(3/2+\eps)$}{(3/2+ε)}-approximation algorithm}
\label{sec:algorithm}

This section is dedicated to showing the existence of a $\big(\frac32+\eps\big)$-approximation algorithm, that is, proving \Cref{theo:approxalg}. 
Based on our structural result, we design two algorithms depending on whether $(\items,\D)$ admits an $\lambda$-forgiving or an $(\eps,\opt)$-neat packing.

If $(\items,\D)$ admits a forgiving packing, we use the following result by~\cite{DBLP:journals/algorithmica/DeppertJKRT23}, which we restate here.
\begin{lemma}[Theorem 19 in \cite{DBLP:journals/algorithmica/DeppertJKRT23}]
\label{lem:APTAS_known}
    Let $\bar{\eps} > 0$ and $\bar{\items}$ be a set of items. 
    There is a polynomial-time algorithm $\mathcal{A}(\bar\items, \bar\D, \bar\eps)$ that returns two packings $(\start,\bar{\start})$, such that each item is packed either in $\start$ or in $\bar{\start}$, $\start$ is $\bar\D$-feasible and $\bar\start$ is $\bar\eps\bar\D$-feasible, 
    and height $h(\bar{\start}), h(\start) \leq (1+c \bar{\eps})\opt(\bar{\items})$, where $c > 0$ is some global constant. 
\end{lemma}

Our first algorithm uses the algorithm $\mathcal A$ guaranteed by \cref{lem:APTAS_known}. 
To this end, it sets $h_{\max} := \max\{h(i) \mid i \in \items\} \leq \opt$ and creates a new item $i_{\lambda}$ with height $h_{\max} $ 
and width~$\lambda\D$.
Further, let $\bar\eps = \min\{ \frac{\lambda}{12+12c},\eps'\}$ where $\eps' \leq \frac{\eps}{2(3c+1)}$.
Now, we call $\mathcal A (\items \cup \{\iadd\}, \D, \bar\eps)$ to obtain the packings $\start$ and $\bar \start$. 
The items currently packed by $\bar\start$ are packed using Steinberg's algorithm before this packing is placed where the additional item $\iadd$ is packed by $\start$. 
We summarize the algorithm in \cref{alg_add_item_i_lambda}.
 
\begin{algorithm}[ht]
\DontPrintSemicolon
\caption{Introducing $\iadd$ to the packing}
\label{alg_add_item_i_lambda}
    \textbf{Input:}  $(\items, \eps',\lambda)$\;
    \smallskip
    $H \leftarrow \max\{\max\{h(i) \mid i \in \items\},\frac{1}{\D}\sum_{i \in \items}h(i)w(i)\}$\;
    create $\iadd$ with $h(\iadd) \leftarrow H$ and $w(\iadd) \leftarrow \lambda \D$ \;
    $(\start,\bar{\start}) \leftarrow$ output of algorithm $\mathcal A\big(\items \cup \{\iadd\}, \D,  \min\big\{ \frac{\lambda}{12+12c},\eps'\big\}\big)$  from \cref{lem:APTAS_known} \;
    $\start_S  \leftarrow$ Steinberg$( \{i \in \items: \bar\start(i) \text{ defined}\} ,H)$\;
    \lForEach{$i \in \{i \in \items: \bar\start(i) \text{ defined}\}$}{$\start(i) = \start(\iadd) +\start_S(i)$}
    \textbf{return} $\start$ \;
\end{algorithm}

\begin{lemma}
\label{lem:algorithm_if_extra_item}
    Let $\eps>0 $ and $\eps' \leq \frac{\eps}{2(3c+1)}$, where $c$ is the constant from \cref{lem:APTAS_known}.
    If $(\items,\D)$ admits a $\lambda$-forgiving packing, then \cref{alg_add_item_i_lambda} is a polynomial-time algorithm that finds a packing with height at most $\big(\frac{3}{2}+\eps\big)\OPT$ for any $\eps \in (0,1)$.   
\end{lemma}

\begin{proof}
If $(\items,\D)$ admits a $\lambda$-forgiving packing, then $\opt(\items \cup \{i_{\lambda}\})\leq \left(\frac{3}{2} +\eps'\right)\opt(\items)$.
As guaranteed by \cref{lem:APTAS_known}, the packings $\start$ and $\bar\start$ have height at most
\[  
    (1+c\bar{\eps})\opt(\items\cup \{i_{\lambda}\}) \leq (1+c\bar{\eps})\left(\frac{3}{2}+\eps'\right)\opt(\items) \leq \left(\frac{3}{2} +\eps\right)\opt
\]
using $\eps' + c\bar{\eps}\left(\frac{3}{2}+\eps'\right) \leq \eps' + c\eps'\left(\frac{3}{2}+\eps'\right) \leq \eps$ by our assumption $\eps' \leq \frac{\eps}{2(3c+1)}$.
Further, the items packed by $\bar\start$ have width at most $\bar{\eps}D \leq \frac{\lambda}{2} \D$ and height at most $h_{\max} \leq H$ and their total area is bounded by $\bar{\eps} \D \cdot \frac{3}{2} (1+c\bar{\eps})\opt \leq \frac{3}{2} \cdot \frac{\lambda}{12+12c}\left(1+c\right)\D\opt  = \frac{\lambda}{8}\D\opt $. 
As $H\geq \opt/2$, \cref{lem:steinberg} implies that $\start_S$ has width at most $\lambda\D$ and height at most $H$. 
Therefore, we can indeed set $\start(i) = \start(\iadd) + \start_S(i)$ for these items without increasing the width or the height of~$\start$.

It remains to bound the running time of \cref{alg_add_item_i_lambda}: \cref{lem:APTAS_known} ensures that $(\start,\bar\start)$ can be found in polynomial time while \cref{lem:steinberg} bounds the running time of Steinberg's algorithm. 
\end{proof}

We can formulate a similar lemma for the case that $(\items,\D)$ admits an $(\eps,\opt)$-neat packing. 
The corresponding algorithm combines ideas from \cite{Galvez0AK21} and \cite{DBLP:journals/algorithmica/DeppertJKRT23}. 
Through a careful analysis, we significantly improve the running time of the algorithm proposed in \cite{Galvez0AK21}.
For the sake of completeness, we prove this lemma in \cref{sec:proof-of-alg-if-tall-sorted}.

\begin{restatable}{lemma}{lemalgiftallsorted}\label{lem:algorithm_if_tall_sorted}
    Let $\eps > 0$ and  \alertBound{$\eps' \leq \frac{\eps}{15}$}.
    If $(\items,\D)$ admits an $(\eps',H)$-neat packing,
    then there is an algorithm $A(\items,\eps',H)$ that can find a $(\eps,H)$-neat packing $\start$ with $h(\start) \leq \big(\frac{3}{2}+\eps\big)H$ in time $\bigO\Big(n \cdot \min\{\log(n),\frac{1}{\eps}\log\frac1{\eps}\}\Big) + \frac{1}{\eps^{\mathcal{O}(1/\eps^5)}}$.   
\end{restatable}

By \cref{thm:restructuring}, all instances admit either a forgiving or a neat packing. 
Hence, we are now ready to state our $\left(\frac{3}{2}+\eps\right)$-approximation algorithm. 

To this end, we choose \alertBound{$\eps' = \frac12\min\{\frac{\eps}{2(3c+1)},\frac{\eps}{15}\}$}, where $c$ is the constant from \cref{lem:APTAS_known}, and \alertBound{$\lambda \leq \min\{\frac{\eps'}{3(5+4\eps')}, \frac{1}{80}\}$} and set $H_{LB} \leftarrow \max\big\{\frac{\area(\items)}\D, \max\{h(i) \mid i \in \items\}\big\}$. 
We start by using \cref{alg_add_item_i_lambda} to obtain a packing $\start_\lambda$.

This packing implies an upper bound $H_{UB}= h(\start_\lambda)$ on the optimal packing height.
If $H_{UB} \leq \big(\frac32 + \eps\big) H_{LB}$, the algorithm returns the packing.
Otherwise, we use binary search to find an approximation of $\opt(\items)$: 
In binary search fashion, we check for $H \in [H_{LB}, H_{UB}]$ if there is a packing of height at most $\big(\frac{3}{2}+\frac{\eps}{2}\big)H$ using the algorithm guaranteed by \cref{lem:algorithm_if_tall_sorted}.
The algorithm stops when the search interval has a size of at most $\frac{\eps}{4} H_{LB}$.
We summarize this procedure in \Cref{alg_three_half}.

\begin{algorithm}[ht]
\DontPrintSemicolon
\caption{$\left(\frac{3}{2}+\eps\right)$ Approximation}
\label{alg_three_half}
\textbf{Input:}  $(\items, \eps)$\;
\smallskip
$\eps' \leftarrow \frac{1}{2}\min\{\frac{\eps}{2(3c+1)},\frac{\eps}{15}\}$\;
$\lambda \leftarrow \min\{\frac{\eps'}{3(5+4\eps')},\frac{\eps'}{13(1+\eps')}, \frac{1}{80}\}$\;
$H_{LB} \leftarrow \max\big\{\frac{\area(\items)}\D, \max\{h(i) \mid i \in \items\}\big\}$\;
$\start_{f} \leftarrow$ output of \cref{alg_add_item_i_lambda} with input $(\items,\eps',\lambda)$\;
$H_{UB} = h(\start_{f} )$\;
$l \leftarrow H_{LB}$\; 
$u\leftarrow H_{UB}$\;
\While{$u-l > \frac{\eps}{4} H_{LB}$}{
$\start_{n} \leftarrow $ output of Algorithm from \cref{lem:algorithm_if_tall_sorted} with $(\items,\eps',\frac{u+l}{2})$\;
\uIf{$\start_n$ is defined}{
    $u \leftarrow \frac{u+l}{2}$\;
}
\Else{$l \leftarrow \frac{u+l}{2}$}
}
\textbf{return} $\arg\min\{h(\start) : \start \in \{\start_f, \start_n\}\}$ \;
\end{algorithm}

\thmthreehalfapproxalg*

\begin{proof}
    Observe that all intermediate algorithms run in polynomial time as guaranteed by \cref{lem:algorithm_if_extra_item} and \cref{lem:algorithm_if_tall_sorted}.
    Further, by applying Steinberg's \cref{lem:steinberg} to $\items$ and simply starting $\iadd$ at $0$, we observe that $\opt(\items \cup \{\iadd\}) \leq 3 H_{LB}$. 
    Hence, $H_{UB} = h(\start_\lambda) \leq (1 + c\eps') 3 H_{LB}$. 
    It is well known that binary search runs in time logarithmic in the relative length of the interval. 
    Therefore, \Cref{alg_three_half} finishes in polynomial time as required. 
    
    Next, we show that the packing generated by the algorithm has a height of at most $\big(\frac{3}{2}+\eps\big)\opt$. We start by observing that $H_{LB} \leq \opt$ since $\area(\items) \leq \D \cdot \opt$ and $\max\{h(i) | i \in \items\} \leq \opt$.    
 
    At the end of the binary search process, we find a value $H$ such that the algorithm from \cref{lem:algorithm_if_tall_sorted} returns an $\big(\frac\eps2, H\big)$-neat packing and cannot find an $\big(\frac\eps2, H-\frac{\eps}{4} H_{LB}\big)$-neat packing. 

    If $h(\start_f) \leq \big(\frac32+\eps\big)\opt$, the packing returned by the algorithm satisfies
    \[
        h(\start) \leq h(\start_f) \leq \bigg(\frac32+\eps\bigg)\opt \, ,
    \]
    as required. 

    It remains to consider the case that $h(\start_f) > \big(\frac32+\eps\big)\opt$. 
    This implies that, by \cref{lem:algorithm_if_extra_item}, $(\items,\D)$ does not admit a $\lambda$-forgiving packing. 
    Hence, by \cref{thm:restructuring}, $(\items,\D)$ admits an $(\eps',\opt)$-neat packing. 
    Thus, the algorithm from \cref{lem:algorithm_if_tall_sorted} finds an $\big(\frac\eps2,\opt\big)$-neat packing. (Observe the extra factor $\frac12$ in the definition of $\eps'$.) 
    Since this algorithm cannot find an $\big(\frac\eps2, H-\frac{\eps}{4} H_{LB}\big)$-neat packing, this implies $\big(\frac32 + \frac\eps2\big) \big(H-\frac{\eps}{4} H_{LB}\big) < \big(\frac32 + \frac\eps2\big) \opt$. 
    Hence, $H-\frac{\eps}{4} H_{LB} \leq \opt$ as $\eps' \leq \frac\eps2$. 
    Thus, 
    \begin{align*}
        h(\start) \leq h(\start_n) &= \bigg(\frac{3}{2}+\frac{\eps}{2}\bigg)H  \leq \bigg(\frac{3}{2}+\frac{\eps}{2}\bigg)\bigg(\opt + \frac{\eps}{4} H_{LB}\bigg) \\
        &\leq \bigg(\frac{3}{2}+\frac{\eps}{2}\bigg)\bigg(1 + \frac{\eps}{4}\bigg)\opt \leq \bigg(\frac{3}{2} +\eps\bigg)\opt,
    \end{align*}
    which concludes the proof. 
\end{proof}

\appendix
\section{Some helpful packing procedures}\label{app:packing_algorithms}

The following algorithm simply sorts a set of items by non-increasing height and packs them next to each other, that is for an item~$i$ that is the direct successor of an item~$i'$ in the ordered set, we have~$\start(i) = \start(i') + w(i')$. 

\begin{algorithm}
\DontPrintSemicolon
\caption{Packing items next to each other}
\label{alg_items_next_to_each_other}
\textbf{Input:} $(\items, S)$ \tcp*[r]{items and intended starting date}
\smallskip
sort~$\items$ by non-increasing height \; 
\ForEach{$i \in \items$}{
    $\start(i) \leftarrow S$ \; 
    $S \leftarrow S + w(i)$ \; 
}
\textbf{return} $\start$ \;
\end{algorithm}

The following algorithm is useful for "mirroring" a packing~$\start$. Formally, an item~$i$ with start time~$\start(i)$ and end time~$\start(i) + w(i)$ starts in the mirrored packing at~$W - (\start(i) + w(i)$ and ends at~$W - \start(i)$. 

\begin{algorithm}
    \DontPrintSemicolon
    \caption{Mirroring a packing}
    \label{alg_mirroring}
    \textbf{Input: } $(\start)$ \tcp*[r]{packing $\start$ of width $W$} 
    \ForEach{$i \in \items$}{
        $\start'(i) \leftarrow W - (\start(i) + w(i))$ \; 
    }
    \textbf{return} $\start'$ \; 
\end{algorithm}

The next algorithm does the same as \Cref{alg_stretching}, except it extends the original packing~$\start$ to the left. 
\begin{algorithm}
    \DontPrintSemicolon
    \caption{Left-stretching Algorithm}
    \label{alg_left_stretching}
    \textbf{Input:} $(\start, H', \tau_{\max}, \tau_{\min})$\;
    \smallskip 
    $\mathcal{T} \leftarrow \items_{h > H'}$ \tcp*[r]{tall items}
$G  \leftarrow$ the set of maximal segments $[\ell,r) \subseteq [\tau_{\min},\tau_{\max}]$ where $\sigma$ packs no item from $\mathcal{T}$\;
$\items_S \leftarrow \{j \in \items \mid (\sigma(j),\sigma(j)+w(j)) \text{ is subsegment of an element in $G$}\}$ \tcp*[r]{removed items}
$\items' \leftarrow \items_{h \leq H', [\start(i), \start(i) + w(i)) \cap [\tau_{\min},\tau_{\max}] \neq \emptyset}$ \tcp*[r]{items overlapping $[\tau_{\min}, \tau_{\max}]$}
\For{$i \in\items' \setminus \items_S$}{
  $\sigma'(i) \leftarrow \sigma(i)$\;
}
\For(\tcp*[f]{iterate by decreasing size of $\ell$}){$(\ell,r) \in G$}{
  \For{$i \in \items'_{\sigma(i) +w(i) \leq r}$}{
    $\sigma'(i) \leftarrow \sigma'(i) - (r-\ell)$\tcp*[r]{shift items that started after $\ell$ by the gap width $(r-\ell)$}
  }
}
\textbf{return} $(\sigma',\items_S)$ \;
\end{algorithm}
The proof of the following lemma is analogue to \Cref{lem:stretching} and we omit it. 

\begin{lemma}[Left-Stretching Lemma]
\label{lem:left_stretching}
    Let $\sigma$ a packing of a set $\items$ of items and $H$ a number with $\frac{h(\sigma)}2 \leq H \leq h(\sigma)$. 
     Let $\mathcal{T} = \items_{h > H}$, and let $[\tau_{\min},\tau_{\max}]$ be a segment of $\sigma$ where an item from $\mathcal{T}$ ends at $\tau_{\min}$ and an item from $\mathcal{T}$ starts at $\tau_{\max}$. 
     Let $\items' = \items^{\start}_{h \leq H, [\start(i), \start(i) + w(i)) \cap [\tau_{\min}, \tau_{\max}] \neq \emptyset} $ and let $d$ be the total width in $[\tau_{\min},\tau_{\max}]$ where $\sigma$ packs no item from $\mathcal{T}$.

    Then, there exists a set $\items_S \subseteq \items'$ with $\area(\items_S) \leq d\cdot h(\sigma)$ such that there exists a packing $\sigma'$ of $\items' \setminus \items_S$ with $h(\sigma') \leq h(\sigma) - H$ and $\sigma(i) - d \leq \sigma'(i) \leq \sigma(i)$ for each $i \in \items' \setminus \items_S$. 
\end{lemma}

\section{Proof of Lemma \ref{lem:algorithm_if_tall_sorted}}\label{sec:proof-of-alg-if-tall-sorted}

In this section, we describe an algorithm for the case that the instance admits an $(\eps,\opt)$-neat packing. 
Instead of directly using the algorithm from \cref{lem:APTAS_known}, we present an algorithm with an improved running time.
We carefully revisit some of the techniques used for the algorithm in \cref{lem:APTAS_known} and adapt them to fit our needs.
Note that in \cite{Galvez0AK21} as well as in \cite{DBLP:journals/algorithmica/DeppertJKRT23} a set of medium items is introduced. It is required that the total area of these medium items is small, i.e., smaller than $\mathcal{O}(\eps)$. 
In the worst case, the definition of this set leads to a running time that is larger than $\big(\frac1\eps\big)^{c^{\mathcal{O}(\frac{1}{\eps})}}$ for some constant $c \geq 2$.
We combine a reduced number of starting points for horizontal items due to techniques in \cite{DBLP:journals/algorithmica/DeppertJKRT23} with the fact that narrow items can be squeezed in in a final step (as done in \cite{Galvez0AK21}). 
This allows us to find an algorithm without introducing a set of medium items.

\lemalgiftallsorted*

Define a lower bound $H_{LB} = \max\big\{\frac{\area(\items)}\D, h_{\max}\big\}$, where $h_{\max} := \max\{h(i) \mid i \in \items\}$. 
Note that $H_{LB} \leq \opt \leq 2H_{LB}$ by Steinberg's Lemma~\ref{lem:steinberg}.

Given an estimate $H$ for $\opt$ with $2H_{LB} \geq H \geq H_{LB}$, 
we partition the set of items into the following sets:
\[
    \cS = \left\{i \in \items \mid h(i) \leq \frac{H}{2}, w(i) \leq \delta \D \right\}, 
    \quad \cT := \left\{i \in \items \mid h(i) > \frac{H}{2}\right\},
\]
\[
    \cH := \left\{i \in \items \mid h(i) \leq \mu  H_{LB}\right\} \setminus \mathcal{S}, 
    \quad \cL := \items \setminus (\mathcal{T}\cup \mathcal{H}  \cup \mathcal{S}),
\]
called squeezable, tall, horizontal, and large, respectively, where \alertBound{$\delta = \frac{\eps}{1+\eps}$} and \alertBound{$\mu = \frac{\eps'^3}{\lceil\log(1/\delta)\rceil}$}.
By losing additional constants, we assume that $\frac1\delta, \log\frac1\delta, \frac1{\eps'}, \frac1\mu\in \mathbb N$.

\newcommand{\rpara}[1]{\ensuremath{#1^{R}}}
\newcommand{\fstart}{\varphi}
\newcommand{\fStart}{F^\fstart}
\newcommand{\fStartprime}{F^{\fstart'}}

We first observe that $|\cL| \leq \frac{1}{\delta \mu}$ since $\area(\items) \leq H_{LB} \D$ and each item $i \in \cL$ satisfies $h(i) \geq \mu H_{LB}$ and $ w(i) \geq \delta \D$. 
Next, we round the height of tall items before arguing about the width of horizontal items.
 
\begin{claim}\label{claim:rounding-tall-items}
   By rounding the height of each tall item in $\cT$ to the minimal greater integer multiple of  $\eps' H_{LB}$, we increase the height of any packing by at most an additive $\eps' H_{LB}$.
\end{claim}

Since the horizontal items $\cH$ have height at most $\mu H_{LB}$ and there might be many on top of each other in a feasible packing, we cannot afford to round their height as we did for the tall items~$\cT$. 
However, we can round their widths to constantly many widths by adding an additional $4\eps' H_{LB}$ to the height of any feasible packing. 
For the sake of completeness, we describe the rounding strategy for horizontal items in detail. 
We use so-called geometric grouping, a standard rounding technique that was first introduced by Karmarkar and Karp~\cite{DBLP:conf/focs/KarmarkarK82} in the context of Bin Packing.

To this end, we partition $\cH$ into $\log\frac1\delta$ sets
$\cH_k := \{i \in \cH \mid \frac{D}{2^k} < w(i) \leq \frac{D}{2^{k-1}}\}$ for $k \in \big\{1,\dots, \log\frac1{\delta}\big\}$.
For each set $\cH_k$, we use geometric grouping (see Theorem 2 in \cite{DBLP:conf/focs/KarmarkarK82}) to round the widths of the horizontal items: 
We place the items in $\cH_k$ on top of each other in order of non-increasing width. 
The total height of this stack is given by $h(\cH_k)$. 
For each $l \in \big\{0,\dots,  \frac1{\eps'}\big\}$, we now use the width of the item~$i$ that is intersected by a horizontal line at $l \eps' h(\cH_k)$ as width of an auxiliary item $i_{k,l}$ with width $w(i)$ and height $\eps' h(\cH_k)$. 
(If $l \eps' h(\cH_k)$ coincides with the border of two items, we let $i$ be the one that is placed on top of the other in the stack.)
Formally, we fix an order of $\cH_k$ such that $w(i_1) \geq \ldots \geq w(i_{|\cH_k|})$ and set $w_{k,l} = \min \big\{ w(i_j): i_j \in \cH_k: \sum_{j' = 1 }^j h(i_{j'}) > l \eps' h(\cH_k) \big\}$ for $l \in \big\{0,\dots,  \frac1{\eps'}\big\}$. 
(Hence, the auxiliary item $i_{k,l}$ has width $w_{k,l}$.) 
We store all auxiliary items in $\rpara{\cH_k}$ and collect their widths in $\cW_k$. 
In general, we use $\rpara{\mathcal X}$ to denote items from the set $\mathcal X$ where the horizontal items are replaced by the auxiliary items $\rpara{\cH_k}$. 

For being able to later map the packing of item $i_{k,l}$ to real items, we also define $j_{k,l} = \arg \min \big\{ w(i_j): i_j \in \cH_k: \sum_{j' = 1 }^j h(i_{j'}) > l \eps' h(\cH_k) \big\}$. 
Then, we create two items $i_{k,l}^1$ and $i_{k,l}^2$ of width $w(i_{j_{k,l}})$ and set the height of these two items as follows: $h(i_{k,l}^1) =  l \eps' h(\cH_k) - \sum_{j = 1}^{j_{k,l}-1} h(i_j)$ and $h(i_{k,l}^2) = h(i_{j_{k,l}}) - h(i_{l,k}^1)$. (Note that $h(i_{k,l}^1) = 0$ is possible; in particular $h(i_{k,0}^1) = 0$.)
We now set $\cH_{k,l} = \{ i_{k,l}^2, i_{j_{k,l}+1}, \ldots, i_{j_{k,l+1}-1}, i_{k,l+1}^1\}$, that is, $\cH_{k,l}$ contains items of total height $\eps' h(\cH_k)$ such that the widths of these items is between $w_{k,l}$ and $w_{k,l+1}$. 
We summarize this rounding procedure in \cref{alg:rounding-horizontal-items}. 
Using an $\bigO(n)$ selection algorithm~\cite{BlumFPRT73} for finding the items $i_{j_{k,l}}$, the algorithm can be implemented to run in time $\bigO\big(n\cdot\min\{\log(n),\frac{1}{\eps'}\log\big(\frac1{\delta}\big)\}\big)$.
\begin{claim}\label{claim:rounding-horizontal-items_running-time}
    \cref{alg:rounding-horizontal-items} runs in time $\bigO\big(n\cdot\min\{\log(n),\frac{1}{\eps'}\log\big(\frac1{\delta}\big)\}\big)$.
\end{claim}

\begin{algorithm}
\DontPrintSemicolon
\caption{Rounding the width of horizontal items}
\label{alg:rounding-horizontal-items}
\textbf{Input:}  $(\cH, \eps', \delta)$ \tcp*[r]{$\cH$ indexed in non-increasing order of width}
\smallskip

\ForEach{
        $k \in \big\{ 1, \ldots, \log\frac1\delta \big\}$
    }{
    $\cH_k \leftarrow \{ i \in \cH: \frac{\D}{2^k} < w(i) \leq \frac{\D}{2^{k-1}} \}$ \tcp*[r]{assume $ \cH_k = \{ i_1, \ldots, i_{|\cH_k|} \}$} 
    \ForEach{
        $l \in \big\{0, \ldots,  \frac1{\eps'} -1 \big\}$ 
    }{
        $w_{k,l} \leftarrow  \min \big\{ w(i_j): i_j \in \cH_k: \sum_{j' = 1 }^j h(i_{j'}) > l \eps' h(\cH_k) \big\} $ \; 
        $j_{k,l} \leftarrow \arg \min \big\{ w(i_j): i_j \in \cH_k: \sum_{j' = 1 }^j h(i_{j'}) > l \eps' h(\cH_k) \big\}$ \tcp*[r]{find ``sliced'' item $i_{j_{k,l}}$}
        $w(i_{k,l}) \leftarrow w_{k,l}$ and $h(i_{k,l}) \leftarrow \eps' h(\cH_k)$ \tcp*[r]{define auxiliary item}
        $w(i_{k,l}^1) \leftarrow w(i_{j_{k,l}})$ and $h(i_{k,l}^1) \leftarrow l \eps' h(\cH_k) - \sum_{j = 1}^{j_{k,l}-1} h(i_j)$ \tcp*[r]{slice $i_{j_{k,l}}$ } 
        $w(i_{k,l}^2) \leftarrow w(i_{j_{k,l}})$ and $h(i_{k,l}^2) \leftarrow h(i_{j_{k,l}}) - h(i_{k,l}^1)$ \;
    }\ForEach{
        $l \in \big\{0, \ldots,  \frac1{\eps'}  \big\}$ 
    }{
        $\cH_{k,l}\leftarrow \{ i_{k,l}^2, i_{j_{k,l}+1}, \ldots, i_{j_{k,l+1}-1}, i_{k,l+1}^1\} $  \tcp*[r]{partition $\cH_k$, including sliced item part}
    }
    $\rpara{\cH_k} \leftarrow \bigcup_{l = 0}^{1/{\eps'}-1} \{ i_{k,l}\}$ \tcp*[r]{collect auxiliary items}
    $\cW_k \leftarrow \bigcup_{l = 0}^{1/{\eps'}-1} \{ w_{k,l}\}$ \tcp*[r]{store their widths}    
}
$\rpara{\cH} \leftarrow \bigcup_{k = 1}^{\log1/\delta} \cH_k $ \; 
$\cW \leftarrow \bigcup_{k = 1}^{\log1/\delta} \cW_k$ \; 
\smallskip 
\textbf{return} $\big(\rpara{\cH},(\rpara{\cH_k})_k, (\cH_{k,l})_{k,l}, \cW, (\cW_k)_k\big)$ \; 
\end{algorithm}

Since we have potentially increased the width of each item, we need to argue that we can turn any feasible packing for $\items$ into a feasible packing for the set of items where the width of the horizontal items is rounded as described above. To this end, we first introduce \emph{fractional} packings. 
A \emph{fractional} packing~$\fstart$ is characterized by a set $\fStart$ of vectors $(s,x,i) \in [0,\D]\times(0,1]\times\items$ where $(s,x,i)$ denotes the fraction~$x \in (0,1]$ of~$i \in \items$ starting at $s\in [0,\D]$.
A \emph{feasible} packing satisfies $s \leq \D - w(i)$ for all $(s,x,i) \in \fStart$ and $\sum_{(s,x,i) \in \fStart} x = 1$ for all $i \in \items$. 
If and only if $(s,1,i) \in \fStart$ for some item~$i \in \items$, we overload (and simplify) notation and use $\fstart(i) = s$. 
The \emph{height} $h(\fstart(t))$ of a fractional packing at time~$t$ is the natural generalization of the height of integral packings, i.e., $h(\fstart(t)) := \sum_{(s,x,i) \in \fStart : s \leq t < s + w(i)} x \cdot h(i)$ and the height~$h(\fstart)$ is the maximum over all~$t$, i.e., $h(\fstart) := \max_{t \in [0,\D]} h(\fstart(t))$.

Based on some packing~$\start$ for~$\items$, we construct a fractional packing~$\fstart$ of $\rpara{\items}$. 
For each item $i \in \items\setminus \cH$, we set $\fstart(i) = \start(i)$. 
Then, for each item $i \in \cH_{k,l}$, we start a fraction $x$ of the item $i_{k,l+1}$ at $\start(i)$ such that $x \cdot h(i_{k,l+1}) = h(i)$. 
The not yet packed widest items $i_{k,0}$ will be packed into an area of width~$\D$ and height~$4 \eps' H_{LB}$ using Steinberg's algorithm. 
We formalize this in \Cref{alg:integral_to_fractional_of_rounded} and analyze the packing height in \Cref{claim:integral-original-items_to_fractional-rounded-items}.

\begin{algorithm}
\DontPrintSemicolon
\caption{Packing $\rpara{\cH}$ instead of $\cH$}
\label{alg:integral_to_fractional_of_rounded}
\textbf{Input:}  $(\start, \items, (\cH_{k,l})_{k,l}, (\rpara{\cH_k})_k)$ \; 
\smallskip
\lForEach(\tcp*[f]{initialize for $\items \setminus \cH$}){$i \in \items \setminus \cH$}{
    $\fstart(i) \leftarrow \start(i)$
} 
$\fStart \leftarrow \emptyset$ \;
\ForEach{$k \in \big\{ 1, \ldots, \log\frac1\delta \big\}$}{
    \ForEach{$l \in \big\{1, \ldots,  \frac1{\eps'} -2 \big\}$}{
        \ForEach{$i \in \cH_{k,l}$}{
            $x \leftarrow \frac{h(i)}{h(i_{k,l+1})}$ \; 
            $s \leftarrow \start(i)$ \; 
            \uIf{$\exists y \in (0,1]: (s, y, i_{k,l+1}) \in \fStart$}{
                $\fStart \leftarrow \fStart \setminus \{(s,y, i_{k,l+1})\} \cup \{(s,y+x,i_{k,l+1})\}$\; 
            }
            \Else{
                $\fStart \leftarrow \fStart \cup \{(s,x,i_{k,l+1})\}$ \;
            }            
        }
    }
    $\start_S \leftarrow $ output of Steinberg$\big(\bigcup_{k=1}^{\log1/\delta} \{ i_{k,0} \}, 4 \eps' H_{LB} \big)$ \tcp*[r]{Steinberg for not yet packed items}
    \lForEach{$k \in \big\{ 1, \ldots, \log\frac1\delta \big\}$}{
        $\fstart(i_{k,0}) \leftarrow \start_S(i_{k,0}) $ 
    }
}
\textbf{return} $\fstart$ \; 
\end{algorithm}

\begin{claim}\label{claim:integral-original-items_to_fractional-rounded-items}
    If there is a feasible packing~$\start$ of~$\items$ of height $h(\start)$, then \Cref{alg:integral_to_fractional_of_rounded} returns a feasible fractional packing~$\fstart$ of $\rpara{\items}$ with height at most $h(\start) + 4\eps' H_{LB}$, and $\fstart$ packs all items in $\items\setminus \cH$ integrally.  
\end{claim}

\begin{proof}
    We start by showing that $\start_S$ is feasible and its height is at most~$4\eps' H_{LB}$. 
    By construction, item $i_{k,0}$ has width at most $\frac{\D}{2^{k-1}}$ and height exactly $\eps'h(\cH_k)$. 
    Hence, the total area of the items  $\bigcup_{k=1}^{\log(1/\delta)} \{ i_{k,0} \}$ is at most 
    \[
        \sum_{k = 1}^{\log(1/\delta)}\eps' h(\cH_k)\frac{D}{2^{k-1}} = 2 \eps' \sum_{k = 1}^{\log(1/\delta)} h(\cH_k)\frac{D}{2^{k}} \leq 2 \eps' \sum_{k = 1}^{\log(1/\delta)} \area(\cH_k) \leq 2 \eps' \cdot \area(\items) \, .
    \]
    Hence, by \cref{lem:steinberg}, packing $\start_{S}$ is feasible and has width~$\D$ and height $4 \eps' H_{LB}$.

    It remains to show that $\fstart$ restricted to the other items is feasible and has height at most $h(\start)$. 
    We observe that $w(i_{k,l+1}) \leq w(i)$ for each~$i \in \cH_{k,l}$ and that $h(\cH_{k,l}) = \eps' h(\cH_k) = h(i_{k,l+1})$ by construction. 
    Hence, each $(s,x,i_{k,l+1}) \in \fStart$ satisfies $s \leq \D - w(i_{k,l+1})$, and
    $\sum_{(s,i_{k,l+1}) \in \fStart} x = \sum_{i \in \cH_{k,l}} \frac{h(i)}{h(i_{k,l+1})} = \sum_{i \in \cH_{k,l}} \frac{h(i)}{\eps' h(\cH_k)} = 1$. 
    Therefore, $\fstart$ is indeed feasible. 
    Further, in $\fstart$, each $i \in \cH_{k,l}$ is replaced by a part of $i_{k,l+1}$ of the same height compared to~$\start$. 
    Since $\start$ and $\fstart$ coincide for~$\items \setminus \cH$, this implies that $h(\fstart) \leq h(\start) + 4 \eps' H_{LB}$ as required.      
\end{proof}

In order to only consider packings involving $\rpara{\cH}$ instead of $\cH$, we next design an algorithm that transforms a feasible packing~$\fstart$ of~$\rpara{\items}$ where only $\rpara{\cH}$ is packed fractionally into a feasible \emph{integral} packing~$\start$ of~$\items$ with height at most $h(\fstart) + 4 \eps' H_{LB}$. 
For each item $i \in \items\setminus \cH$, we set $\start(i) = \fstart(i)$. 
For each starting time $s$ of $i_{k,l}$, i.e., $(s,x,i_{k,l}) \in \fStart$ for some~$x \in (0,1]$, we place items from the set $H_{k,l}$ of height at most~$x \cdot h(i_{k,l})$ to start at~$s$. 
The not yet packed items can be packed using Steinberg's algorithm. 
We summarize this algorithm in \Cref{alg:fractional_of_rounded_to_integral} and analyze the height of the resulting packing under some additional assumptions in \Cref{claim:fractional-rounded-items_to_integral-original-items}.

\begin{algorithm}
\DontPrintSemicolon
\caption{Packing $\cH$ instead of $\rpara{\cH}$}
\label{alg:fractional_of_rounded_to_integral}
\textbf{Input:}  $(\fstart, \items,  (\cH_{k,l})_{k,l}, (\rpara{\cH_k})_k)$ \; 
\smallskip
\lForEach(\tcp*[f]{nothing changes except for $\cH$}){$i \in \items \setminus \cH$}{
    $\start(i) \leftarrow \fstart(i)$
} 
$\cH^{\start} \leftarrow \emptyset $ \tcp*[r]{items directly packed} 
\ForEach{$k \in \big\{ 1, \ldots, \log\frac1\delta \big\}$}{
    \ForEach{$l \in \big\{0, \ldots,  \frac1{\eps'} \big\}$}{
        \ForEach{$(s,x,i_{k,l}) \in \fStart$}{
            \nl\label{alg:line:frac-to-int-choose-H} greedily choose maximum $H \subseteq H_{k,l} \setminus \big(\cH^{\start} \cup \{i_{k,l}^2, i_{k,l+1}^1\} \big)$ with 
            $h(H) \leq x \cdot h(i_{k,l})$ \tcp*[r]{ subset to be packed instead of $i_{k,l}$}
            $\cH^{\start} \leftarrow \cH^{\start} \cup H $ \; 
            \nl\label{alg:line:frac-to-int-pack-H}\lForEach{$i \in H$}{$\start(i) \leftarrow s$}
        }
    }
}
$\cH^- \leftarrow \cH \setminus \cH^{\start} $ \; 
\textbf{return} $(\start, \cH^-)$ \;
\end{algorithm}

\begin{claim}\label{claim:fractional-rounded-items_to_integral-original-items}
    If~$\fstart$ is a feasible packing of $\rpara{\items}$ that integrally packs $\items \setminus \rpara{\cH}$ and uses at most $\frac{2^k-1}{\eps'}$ different starting times for items in $\rpara{\cH_k}$, then \Cref{alg:fractional_of_rounded_to_integral} returns an integral packing~$\start$ of $\items \setminus \cH^-$ of height at most $h(\fstart)$ where $\start(i) = \fstart(i)$ for all $i \notin \cH$. 
    The total area of items in $\cH^- \cap \cH_k$ is at most $\frac{2\mu}{\eps'^2 }H_{LB} \D $ and the total area of items in $\cH^-$ is at most $2 \eps' H_{LB} \D$. 
    The running time of the algorithm is at most $\bigO\big(n + \frac1{\delta(\eps')^2} \big)$.
\end{claim}

Note that the claim implies that $\cH^-$ can be packed into an area of width $\D$ and height $4\eps' H_{LB}$ using Steinberg's algorithm and \cref{lem:steinberg}. 

\begin{proof}
    Fix $k \in \big\{ 1, \ldots, \log\frac1\delta \big\}$ and $l \in \big\{0, \ldots,  \frac1{\eps'} -1 \big\}$. 
    Recall that $w(i) \leq w_{k,l} = w(i_{k,l})$ for $i \in \cH_{k,l}$ by definition. 
    Further, instead of packing an $x$-fraction of $i_{k,l}$ of height $x \cdot h(i_{k,l})$, we pack items from $\cH_{k,l}$ of total height at most $x \cdot h(i_{k,l})$ by Line~\ref{alg:line:frac-to-int-choose-H}. 
    Hence, 
    the height of~$\start$  is at most~$h(\fstart)$ and $\start$ is feasible.

    It remains to show that the total area of items in $\cH^-$ is bounded by $2 \eps' H_{LB} \D$.
    The total height of items in the set $H$ found in Line~\ref{alg:line:frac-to-int-choose-H} is close to $x \cdot h(i_{k,l})$ for each $(x,s,i_{k,l}) \in \fStart$: 
    Since $h(i) \leq \mu H_{LB}$ for $i \in \cH$, we know that $h(H) \geq x \cdot h(i_{k,l}) - \mu H_{LB}$. 
    Further, by assumption, there are at most $\frac{2^k-1}{\eps'}$ different starting positions for items in $\rpara{\cH_k}$ and at each starting position item fractions of at most $|\rpara{\cH_k}| \leq \frac{1}{\eps'}$ many different widths are started. 
    Hence, from the set $\cH_k$, at most $\frac{1}{\eps'} \big(\frac{2^k-1}{\eps'} + 1 \big)$ items are not packed in Line~\ref{alg:line:frac-to-int-pack-H}, where the additive $\frac{1}{\eps'}$ is due to the removal of the slices of $i_{j_{k,l}}$, i.e., the items $i_{k,l}^1$ and $i_{k,l}^2$. 
    Hence, the total area of items in $\cH_k$ not packed in Line~\ref{alg:line:frac-to-int-pack-H} is bounded from above by 
    \[ 
        \frac{2^k}{\eps'} \cdot \frac{1}{\eps'} \cdot \mu H_{LB} \cdot \frac{\D}{2^{k-1}} = \frac{2\mu}{\eps'^2 }H_{LB} \D 
        \, ,
    \]
    where we used that each item $i \in \cH_k$ satisfies $w(i) \leq \frac{\D}{2^{k-1}}$ and $h(i) \leq \mu H_{L,B}$.

    As the total number of groups is bounded by $\log\frac1\delta$, the total area of items in $\cH$ not yet packed is at most
    \[
        \log\left(\frac1\delta\right) \cdot \frac{2\mu}{\eps'^2} H_{LB} \D 
        = \log\left(\frac{1}{\delta}\right)\cdot \frac{\eps'^3}{\log(1/\delta)}  \cdot \frac{2}{\eps'^2} H_{LB} \D  \leq 2\eps' H_{LB} \D \, .
    \]  
    Note that the partition into groups $\cH_{k,l}$ can be computed in $\bigO\big(n\cdot\min\{\log(n),\frac{1}{\eps'}\log\big(\frac1{\delta}\big)\}\big)$ while finding the items $j_{k,l}$. Furthermore, given a start point $(s,x,i_{k,l})$ it is sufficient to choose items for $H$ until the first item does not fit anymore, i.e., would exceed the height $x h(i_{k,l})$. Therefore the algorithm can be implemented to run in time $\bigO\big(n + \frac1{\delta(\eps')^2} \big)$ if the sets $\cH_{k,l}$ are provided.
\end{proof}

Due to the just proven claim, we can transform any fractional packing~$\fstart$ of $\rpara{\items}$ with a bounded number of different starting times of the items in $\rpara{\cH_k}$ into an integral packing~$\start$ of $\items$ with almost the same height. 

The next step is to design an algorithm that transforms any fractional packing $\fstart$ of $\rpara{\items}\setminus\cS$ into a packing $\fstart'$ where the items in~$\rpara{\cH_k}$ have at most $\frac{2^k-1}{\eps'}$ distinct starting times. 
This algorithm reduces the number of starting points for horizontal items~$\rpara{\cH}$ in two steps, the first one also involves the large items~$\cL$. 

We start by setting $\fstart'(i) = \fstart(i)$ for $i \in \cT$ and round the heights of the $H$-tall items to integer multiples of $\eps' H_{LB}$. 
In the next step, we shift all non-$H$-tall items $\cL \cup \rpara{\cH}$ to the left as far as possible without increasing the packing height of~$\fstart'$. 

We reduce the total number of starting points for horizontal items further, similar to \cite{DBLP:journals/algorithmica/DeppertJKRT23}.
For each~$k \in \big\{1,\ldots,\log\frac1\delta\big\} $, we partition the segment $[0,\D]$ into $2^k$ segments of width $\frac{\D}{2^k}$.
For each such segment, we consider all parts of items of $\rpara{\cH_k}$ starting in this segment and sort them in order of increasing starting points. 
Let $h_{k,m}$ be the height of the (fractional) horizontal items in $\rpara{\cH_k}$ starting in the $m$th segment for $m \in \big\{1,\ldots, 2^k -1 \big\}$. 
We partition them into $\frac1{\eps'}$ layers of height $\eps' h_{k,m}$ and slice the horizontal items $\rpara{\cH_k}$ overlapping the layer borders. 
We first remove all (fractional) items in the bottom-most layer and place them on top of the packing, balancing out the additional height across the whole packing.  

Now we shift the items from each layer (except the bottom-most one) to the left such that they start at the latest original starting time from the layer below. 
We repeat this procedure for each segment.

In order to be able to find such a packing efficiently, the last step of the algorithm is to impose even more structure on the packing in exchange of not having to be feasible any more, i.e., there might be items $i \in \rpara{\cH_k}$ with $ \sum_{(s,x,i) \in \fStartprime} x < 1$. (We will show that the total area of ``missing'' item parts is bounded.) 
To this end, for each starting point~$s$ of item parts belonging to items in \rpara{\cH_k}, we remove item parts such that the remaining item parts have a total height that is an integer multiple of $\mu H_{LB}$. 
We summarize this algorithm in \Cref{alg:fractional_to_fractional-with-fewer-starting-times}.

\begin{algorithm}
\DontPrintSemicolon
\caption{Reducing the number of starting times}
\label{alg:fractional_to_fractional-with-fewer-starting-times}
\textbf{Input:}  $(\fstart, \items, \rpara{\cH_k}, H_{LB}, \eps')$ \; 
\smallskip
$\fstart' \leftarrow \fstart$ \tcp*[r]{initialize $\fstart$}
\lForEach(\tcp*[f]{round tall items}){$i \in \cT$}{$h(i) \leftarrow \eps' H_{LB} \cdot \big\lceil \frac{h(i)}{\eps' H_{LB}}\big\rceil$}
\ForEach(\tcp*[f]{parts of $\cL \cup \rpara{\cH}$ starting at $\tau$}){$\tau \in [0,\D] : \{\tau\}\times(0,1]\times(\cL \cup \rpara{\cH}) \cap \fStartprime \neq \emptyset$}{
    \ForEach{$ (\tau,x,i) \in  \{\tau\}\times(0,1]\times(\cL \cup \rpara{\cH}) \cap \fStart$}{
        $\tau' \leftarrow \min \{ t \in [0,\tau]: \forall t' \in [t, t+w(i)) ~         
        \sum_{ (s,x',i') \in  \fStartprime \setminus \{(\tau,x,i)\}:  s \leq t' < s + w(i')} x \cdot h(i')
        \leq h(\fstart') - x \cdot h(i)\}$ \; 
        $(\fStartprime \leftarrow \fStartprime \setminus \{(\tau,x,i)\}) \cup \{(\tau', x, i)\}$ \tcp*[r]{shift current part of $i$ to the left} 
    }
}
\ForEach{$k \in \big\{1, \ldots, \log \frac1\delta\big\}$}{
    $M_k \leftarrow \emptyset$ \tcp*[r]{item parts from $\rpara{\cH_k}$ not packed in the end}
    \ForEach{$m \in \{1, \ldots, 2^k -1 \} $}{
        $F \leftarrow \big\{ (s,x,i) \in \fStartprime: s \in \big[m \cdot \frac\D{2^k}, (m+1) \frac\D{2^k} \big) , i \in \rpara{\cH_k}\big\}$ \tcp*[r]{item parts starting in $m$th segment}
        $\{(s_1,x_1,i_1),\ldots,(s_{|F|}, x_{|F|}, i_{|F|})\} \leftarrow F$ with $s_1 \leq \ldots \leq s_{|F|}$ \tcp*[r]{index $F$}
        $h \leftarrow \sum_{(s,x,i) \in F} x \cdot h(i)$ \; 
        $f \leftarrow \emptyset$ \tcp*[r]{if an item belongs to two slices} 
        \ForEach{$l \in \big\{1,\ldots\frac1{\eps'}\big\}$}{
            $j_{l} \leftarrow \min\{j \in \{1, \ldots, |F|\} : \sum_{j'=1}^j x_{j'} \cdot h(i_{j'}) \geq l \cdot \eps' h \}$ \;
            \lIf(\tcp*[f]{item part belongs to two slices}){$\sum_{j=1}^{j_{l}} x_{j} \cdot h(i_{j}) > l \cdot \eps' h$}{
                $x' \leftarrow  \frac{l \cdot \eps' h - \sum_{j=1}^{j_{l} - 1} x_{j} \cdot h(i_{j})}{h(i_{j_{l}})}$ 
            }
            \lElse{
                $x' \leftarrow x_{j_{l}}$
            }
            $F \leftarrow f \cup \{ (s_{j_{l-1}+1}, x_{j_{l-1}+1}, i_{j_{l-1}+1}), \ldots (s_{j_{l}-1},x_{j_{l}-1},i_{j_{l}-1})\}$ \; 
            $F \leftarrow F \cup \{ (s_{j_{l}}, x', i_{j_{l}})\}$ \; 
            \uIf{$l = 1$}{
                \ForEach{$(s,x,i) \in F$}{
                    $\fStartprime \leftarrow (\fStartprime \setminus \{(s,x,i)\}) \cup
                    \bigcup_{r = 0}^{2^{k-1}-1} \big\{\big(r \cdot \frac{\D}{2^{k-1}},\frac{x}{2^{k-1}},i\big)\big\} $ \tcp*[r]{place item part on top} 
                }
            }
            \Else{
                \ForEach{$(s,x,i) \in F$}{
                    $\fStartprime  \leftarrow (\fStartprime \setminus \{(s,x,i)\}) \cup \{(\tau,x,i)\} $ \tcp*[r]{shift item part left}
                }
            }            
            \uIf{$x' < x_{j_{l}}$}{$f \leftarrow \{ (s_{j_{l}}, x_{j_{l}} -  x', i_{j_{l}})\}$}
            \Else{$f \leftarrow \emptyset$}
            $\tau \leftarrow s_{j_l}$ \tcp*[r]{starting time for next slice} 
        }
        \nl\label{alg:line:reduce-height-of-horizontal-items}\ForEach{$\tau \in [0,\D): \{\tau\} \times (0,1] \times \rpara{\cH_k} \cap \fStartprime \neq \emptyset$}{
        $h \leftarrow \sum_{(\tau,x,i) \in \fStartprime : i \in \rpara{\cH_k}} x \cdot h(i)$ \tcp*[r]{height at $\tau$}
        $F \leftarrow \{\tau\} \times (0,1] \times \rpara{\cH_k} \cap \fStartprime$ \tcp*[r]{fix some order} 
        $j \leftarrow \max \{ j \in \{1,\ldots, |F|: \sum_{j'=1}^j x_{j'} \cdot h(i_{j'}) \leq \mu H_{LB} \cdot \big\lfloor \frac{h}{\mu H_{LB}} \big\rfloor \} $ \; 
        $x' \leftarrow \frac{\mu H_{LB} \cdot \big\lfloor \frac{h}{\mu H_{LB}} \big\rfloor - \sum_{j'=1}^j x_{j'} \cdot h(i_{j'})}{h(i_{j+1})} $ \; 
        $M_k \cup \{(\tau, x_{j+1} - x', i_{j+1}), (\tau, x_{j+2}, i_{j+2}), \ldots, (\tau, x_ {|F|}, i_{|F|}) \}$  \; 
    }
    }
}

\textbf{return} $(\fstart',(M_k)_k)$ \; 
\end{algorithm}

\begin{claim}\label{claim:existence-solution-with-few-starting-points}
  If~$\fstart$ packs~$\rpara{\items} \setminus \cS$ with height~$h(\fstart) \geq H$, where $\cT$ are packed next to each other sorted by non-increasing height, with a tallest starting at~$0$, and the only fractionally packed items belong to $\rpara{\cH}$, then the packing~$\fstart'$ of \cref{alg:fractional_to_fractional-with-fewer-starting-times} before Line~\ref{alg:line:reduce-height-of-horizontal-items} satisfies
  \begin{enumerate}
      \item[(C1)] $\cT$ are packed next to each other sorted by non-increasing height, with a tallest starting at~$0$,
      \item[(C2)] the starting points of the items in $\rpara{\cH}\cup\cL$ belong to a set $S_{\cH\cL}$ with $ |S_{\cH\cL}| \leq  \left(\frac{2}{\delta\mu}\right)^{1/\delta}$, 
      \item[(C3)] the items in $\rpara{\cH_k}$ use at most $\frac{2^k-1}{\eps'}$ many distinct starting points and for each time point $\tau$ the height of the items in $\cH_k$ starting at $\tau$ is an integer multiple of $\mu H_{LB}$,  
      \item[(C4)] the only fractionally packed items belong to $\rpara{\cH}$,
  \end{enumerate}
  and $h(\fstart') \leq h(\fstart) + 2\eps' H_{LB}$. Further, 
  \begin{enumerate}
      \item[(C5)] the total area of item parts belonging to $(M_k)_{k=1}^{\log(1/\delta)}$ is at most $2 \frac{\mu}{\eps'^2} H_{LB} \D$ per group $\cH_k$. 
  \end{enumerate}
\end{claim}

\begin{proof} 
    By \cref{claim:rounding-tall-items}, rounding the height of $H$-tall items to integer multiples of~$\eps'H_{LB}$ increases the height of~$\fstart'$ by at most an additive~$\eps' H_{LB}$ compared to~$\fstart$. 
    Since $H \geq H_{LB} \geq \max_{i \in \cT} h(i)$, this implies that the $H$-tall items create a stair-shaped packing with at most~$\frac1{\eps'}$ downward steps. 
    Shifting the non-$H$-tall items~$\cL\cup\rpara{\cH}$ to the left does not increase the packing height by definition. 
    Simultaneously, this process implies that each item in~$\cL\cup\rpara{\cH}$ now either starts at the endpoint of another item in~$\cL\cup\rpara{\cH}$ or at a stair step created by the~$H$-tall items. 
    As each item in $\cL \cup \rpara{\cH}$ has a width of at least $\delta \D$, there can be at most $\frac{1}{\delta}$ of these items next to each other. 
    Therefore, the total number of possible start points $S_{\cH\cL}$ for large and horizontal items is bounded by 
    \[ 
        |S_{\cH \cL}| \leq \left(|\cL|+|\cW|+\frac{1}{\eps'}\right)^{1/\delta -1} \leq \left(\frac{2}{\delta\mu}\right)^{1/\delta} \, ,
    \]
    where we use the bounds $|\cL| \leq \frac1{\delta\mu}$ and $|\cW| \leq \log \frac1\delta \cdot \frac1{\eps'}$. 

    Now we consider the item parts that we placed on top of the packing. 
    For $k \in \{1, \ldots, \log\frac1\delta\}$, the total height of the removed item parts is bounded by~$\eps' h(\rpara{\cH_k}) = \eps' h(\cH_k)$. 
    Since any such item part has width at most~$\frac{\D}{2^{k-1}}$, we divide it into~$2^{k-1}$ parts of equal height and place them next to each other without exceeding a total width of~$\D$. 
    Hence, when placing these items on top of the packing, we add at most $\frac{\eps' h(\cH_k)}{2^{k-1}}$ to the packing height. 
    Since all the items in the set~$\cH_k$ have a width of at least $\frac{\D}{2^k}$, it holds that $\sum_{k = 1}^{\log\left(1/\delta\right)} \frac{h(\cH_k)}{2^k} \leq H_{LB}$.
    Therefore, in total, placing these items on top adds at most $\sum_{k = 1}^{\log\left(1/\delta\right)}\eps' \frac{h(\cH_k)}{2^{k-1}} \leq 2\eps' H_{LB}$ to the packing height.

    Next we consider the item parts of items~$\rpara{\cH_k}$ that are shifted to the left.
    Clearly, in each segment, the total number of starting positions of shifted item parts in~$\fstart'$ is at most $\frac1{\eps'}-1$ while the items placed on top potentially add another starting time (the left end of the current segment). 
    Since there are $2^k - 1$ segments, we reduce the total number of item-part starting positions of items in $\rpara{\cH_k}$ to $\frac{2^k - 1}{\eps'}$ as required. 
    
    It remains to argue that this shift did not increase the packing height. 
    To this end, we fix~$k$ and some~$m \in \{1, \ldots, 2^k - 1\}$ such that item parts of items in $\rpara{\cH_k}$ start in the $m$th segment.   
    Since each item in $\rpara{\cH_k}$ has width greater than $\frac\D{2^k}$, all item parts belonging to items in $\rpara{\cH_k}$ that start in the $m$th segment finish in the $(m+1)$st segment.     
    Since we removed the bottom-most layer and start the new layer at the latest original starting time~$\tau$ of the layer below, we know that the height between $\tau$ and $(m+1) \frac{\D}{2^k}$ in $\fstart'$ is still bounded by $h(\fstart) + 2 \eps' H_{LB}$.
    In the $(m+1)$st segment, we observe that each item part does not end later in $\fstart'$ than in $\fstart$. 
    Since in $\fstart$ each item part completely covers the time from $(m+1) \frac{\D}{2^k}$ to its end point, the height of $\fstart'$ is still bounded by $h(\fstart) + 2\eps' H_{LB}$. 

    For the last condition, we upper-bound the total area of removed items per group $\rpara{\cH_k}$ as follows: 
    There are $\frac{2^k-1}{\eps'}$ starting times. 
    The number of items with different widths starting at some time point is bounded by $|\mathcal{W}_k| \leq \frac1{\eps'}$.
    Since items in group $\rpara{\cH}$ have a width of at most $\frac{\D}{2^{k-1}}$, the total area that needs to be removed is bounded by 
    \[ 
        \frac{2^k-1}{\eps'} \cdot \frac{1}{\eps'} \cdot \mu H_{LB} \cdot \frac{\D}{2^{k-1}} = \frac{(2^k-1)\mu}{\eps'^2 2^{k-1}}H_{LB} \D \leq 2 \frac{\mu}{\eps'^2} H_{LB} \D \, .
    \]
\end{proof}

Observe that the packing $\fstart'$ returned by \cref{alg:fractional_to_fractional-with-fewer-starting-times} has a lot of structure. 
In fact, for each group $\cH_k$, the state at each of the $\big\{1,\dots,\frac{2^k-1}{\eps'}\big\}$ reduced starting points can be specified by a \emph{starting vector} $f_{k} = (s, h_{1},\dots,h_{|\mathcal{W}_k|}) \in S_{HL} \times \big\{1, \ldots, \frac1\mu \big\}^{|\mathcal{W}_k|}$ where $s \in S_{HL}$ denotes the starting point and the total height of the items with width $w_{k,l} \in \mathcal{W}_k$ starting at $s$ is given by $h_{l} \cdot \mu H_{LB}$ for $h_{l} \in \{0,\dots,  \frac1\mu \}$. 
Let $\cF_k$ denote the set of possible starting vectors for group $\cH_k$. 
Using the bound on $S_{\cH\cL}$ from \cref{claim:existence-solution-with-few-starting-points} and that $\cW_k \leq \frac1{\eps'}$, we can bound the total number of different vectors in $\cF_k$ by $\frac{|S_{\cH\cL}|} {\mu^{|\mathcal{W}_k|}} \leq \left(\frac{2}{\delta\mu^2}\right)^{1/\eps'}$.  
The next algorithm relies on this observation to efficiently find a packing $\fstart'$ satisfying (C1) to (C5) or asserts that such a packing is not possible. 

Formally, this algorithm starts by removing the items $\cS$; they will be packed at the end using the Squeezing Algorithm~\ref{alg:iterated-squeezing}.  
Then, the algorithm rounds the widths of the horizontal items using \cref{alg:rounding-horizontal-items}. 
The $H$-tall items~$\cT$ are packed next to each other in order of non-increasing height with a tallest one starting at time~$0$ and the height of these items is rounded up to the next multiple of $\eps' H_{LB}$, where $H_{LB} = \max\{h_{\max}(\items), \area(\items)/\D\}$.
Having the staircase shape created by $\cT$, the different widths of items in $\cL$, and the widths $\cW$ of the horizontal items at hand, the algorithm now computes the set $S_{\cH\cL}$ of distinct starting times for items in $\cH \cup \cL$. 
The algorithm checks the feasibility of each possible combination of starting large items $\cL$ and choosing at most $\frac1{\eps'}$ vectors from $\cF_k$ for group $\cH_k$.  
To this end, it first checks if the resulting packing has height at most $\frac32 H + 7\eps' H_{LB}$ at every point in time. 
If this is the case, it greedily fills the placeholders created by the starting vectors as follows: suppose that $(s, h_{1}, \ldots, h_{|\cW_k|}) \in \cF_k$ is considered in the current combination. 
Then, for $w_l \in \cW_k$, the algorithm starts horizontal items from $\cH_k$ with $w_{k,l} \geq w(i) \geq w_{k,l+1}$ and total height at most $h_l \cdot \mu H_{LB}$ at $s$ for each $l \in \{1,\ldots, |\cW_k|\}$. 
Due to the integral packing of the horizontal items in $\cH_k$, not all horizontal items might fit into the placeholders created by vectors in $\cF_k$. 
However, if the total area reserved for each item group is large enough, we know that the total area of unpacked horizontal items is at most $4 \eps' H_{LB} \D$.
To this end, the algorithm additionally checks whether the total height reserved for each group is larger than the group's own height.
The unpacked horizontal items are packed with Steinberg's algorithm and placed on top of the packing. 
Finally, the items in $\cS$ are packed using the Squeezing Algorithm~\ref{alg:iterated-squeezing}.  
We summarize the algorithm in \cref{alg:approx-if-tall-next-to-each-other}.

\begin{algorithm}
\DontPrintSemicolon
\caption{Extended Squeezing (add squeezable items)}
\label{alg:squeezingAdd}
\textbf{input:} $(\start, H, \eps',\items_{\mathrm{add}})$ \; 
\smallskip 
$\sigma' \leftarrow$ result of \cref{alg:squeezing} with $(\start,H,\eps')$ \tcp*[r]{shift items to the left using the squeezing algorithm}  
\While{$\items_{\mathrm{add}} \not = \emptyset$}{
    $\tau \leftarrow \min\{t \geq \tau:  h(\itemsT{\start}(t)) \leq (1+\eps')H \} $\;
    $i\leftarrow \mathrm{pop}(\items_{\mathrm{add}})$\;
    $\start(i) \leftarrow \tau$\;
}
\textbf{return} $\start$ \;
\end{algorithm}

\begin{algorithm}
\DontPrintSemicolon
\caption{Algorithm that packs tall items next to each other}
\label{alg:approx-if-tall-next-to-each-other}
\textbf{Input:}  $(\cT, \cL, \cS, \cH, H, H_{LB}, \eps', \delta)$ \; 
\smallskip
$\fstart \leftarrow $ output of \Cref{alg_items_next_to_each_other} with $(\cT,0)$ \tcp*[r]{sort tall by height}  
\lForEach(\tcp*[f]{round tall items}){$i \in \cT$}{$h(i) \leftarrow \eps' H_{LB} \cdot \big\lceil \frac{h(i)}{\eps' H_{LB}}\big\rceil$}
$\big(\rpara{\cH},(\rpara{\cH_k})_k, (\cH_{k,l})_{k,l}, \cW, (\cW_k)_k\big) \leftarrow$ output of \Cref{alg:rounding-horizontal-items} with $(\cH, \eps', \delta)$ \tcp*[r]{round width of $\cH$}  
compute $S_{\cH\cL}$ \; 
\lForEach(\tcp*[f]{compute set of starting vectors}){$k \in \big\{1, \ldots \log\frac1\delta\big\}$}{compute $\cF_k$}
\ForEach{feasible $(\fstart(i))_{i \in \cL} \in (S_{\cH\cL})^{\cL}$}{
    \ForEach{feasible $(F_k)_{k=1}^{\log 1/\delta} \subset (\cF_k)_{k=1}^{\log 1/\delta}$ with $|F_k| \leq \frac1{\eps'}$}{
    \ForEach{$k \in \big\{1, \ldots \log\frac1\delta\big\}$}{
        \ForEach(\tcp*[f]{transform $F_k$ into fractional packing $\fstart$}){$f = (s, h_1,\ldots,h_{|\cW_k|}) \in F_k$}{
            \lForEach{$l \in  \{1,\ldots, |\cW_k|\}$}{
                $F^{\fstart} \leftarrow F^{\fstart} \cup \big\{\big(s, \frac{\mu H_{LB} \cdot h_l}{\eps' h(\cH_k)}    ,i_{k,l}\big)\big\} $ 
            }
        }
    }
    \nl\label{alg:line:approx:check-height}\lIf(\tcp*[f]{packing too high}){$h(\fstart) > \big(\frac32 + 7 \eps'\big) H$}{
        {break} 
    }
    \nl\label{alg:line:approx:check-area}\If{ $\sum_{f \in F_k}h_l \geq h(H_{k,l})$ for each $k,l$}
    {
        $(\start, \cH^-) \leftarrow$ output of \Cref{alg:fractional_of_rounded_to_integral} with $\big(\fstart, \items \setminus \cS, (\cH_{k,l})_{k,l}, (\rpara{\cH_k})_{k}\big)$ \tcp*[r]{pack $\cH \setminus \cH^-$}
        $\start_S \leftarrow$ Steinberg$(\cH^-, 8\eps' H_{LB})$ \;
        \lForEach{$i \in \cH^-$}{$\start(i) \leftarrow \start_S(i)$} 
        $\start \leftarrow $ output of \cref{alg:squeezingAdd} with $(\start, H, \eps,\cS) $ \tcp*[r]{squeeze squeezable items in} 
        \textbf{return} $\start$ \; 
    }
    }
}
\textbf{return} $\start$ does not exist \; 
\end{algorithm}

\cref{lem:algorithm_if_tall_sorted} follows immediately from the next lemma.

\begin{lemma}
    Let $\eps > 0$ and $\eps' \leq \frac\eps{15}$. If the set of items $(\items,\D)$ admits an $(\eps',H)$-neat packing for $H \geq H_{LB}$, 
    then \cref{alg:approx-if-tall-next-to-each-other} returns an $(\eps,H)$-neat packing $\start$. 
    Otherwise, the algorithm correctly decides that no such packing~$\sopt$ exists. 
    \cref{alg:approx-if-tall-next-to-each-other} runs in time $\mathcal{O}(n (\log n+\frac{1}{\eps^3})) + \frac{1}{\eps^{\mathcal{O}(1/\eps^5)}}$.     
\end{lemma}

\begin{proof}
    Suppose that $(\items,\D)$ admits $(\eps',H)$-neat packing for $H \geq H_{LB}$. 
    We want to show that \cref{alg:approx-if-tall-next-to-each-other} returns a packing $\start$ with the claimed height bound. 
    We observe that by \cref{claim:integral-original-items_to_fractional-rounded-items}, there exists a fractional packing of $\rpara{\items}$ of height at most $h(\sopt) + 4 \eps' H_{LB}$, where all items in $\items \setminus \cH$ are packed integrally.
    With \cref{claim:existence-solution-with-few-starting-points} applied to this packing, we can guarantee the existence of a fractional packing $\fstart$ of the items $\items \setminus \bigcup_{k=1}^{\log 1/\delta} M_k$ of height at most $h(\sopt) + 6 \eps' H_{LB}$. 
    The same claim also guarantees that the total area of items belonging to $\bigcup_{k=1}^{\log 1/\delta} M_k$ is bounded by $2 \frac{\mu}{\eps'^2} H_{LB}\D$ per set $M_k$. 
    Packing $\fstart$ packs the $H$-tall items~$\cT$ next to each other sorted in non-increasing order of height with a tallest starting at $0$ and the starting points of items in $\rpara{\cH} \cup \cL$ belong to the set $S_{\cH\cL}$. 
    Further, the height of a fraction of $\rpara{\cH_k}$ starting at some point in $\fstart$ is an integral multiple of $\mu H_{LB}$. 
    As argued above, this allows us to represent the packing for the items $\rpara{\cH_k}$ by at most $\frac1{\eps'}$ vectors from the set $\cF_k$.  
    Since $S_{\cH\cL}$ also encompasses the starting points of the large items~$\cL$, \cref{alg:approx-if-tall-next-to-each-other} indeed considers packing $\fstart$. 
    Since $h(\fstart) \leq h(\sopt) + 6 \eps' H_{LB} \leq \big(\frac32 + 7 \eps'\big) H$ by assumption on $\sopt$, the packing $\fstart$ also survives the check in Line~\ref{alg:line:approx:check-height}. 
    Note that $\fstart$ does not pack horizontal items from $\cH_k$ of total area at most $2 \frac{\mu}{\eps'^2} H_{LB}\D$. 
    Hence, turning $\fstart$ into an integral packing of $\items \setminus \cS$ may lead to items from $\cH_k$ of total area at most $4 \frac{\mu}{\eps'^2} H_{LB}\D$ not packed by \Cref{claim:fractional-rounded-items_to_integral-original-items}. 
    In total, the area of not yet packed items in $\cH$ is at most 
    \[
        \log\left(\frac1\delta\right) \cdot \frac{4\mu}{\eps'^2} H_{LB} \D 
        = \log\left(\frac{1}{\delta}\right)\cdot \frac{\eps'^3}{\log(1/\delta)}  \cdot \frac{4}{\eps'^2} H_{LB} \D  \leq 4\eps' H_{LB} \D \, .
    \]
    Hence, $\start$ also survives the check in Line~\ref{alg:line:approx:check-area}.
    Since any item in $\cH^-$ has height at most $\mu H_{LB} \ll \eps' H_{LB}$, Steinberg's Lemma~\ref{lem:steinberg} implies that $\start_S$ is a feasible packing of $\cH^-$ of height at most $ 8 \eps' H_{LB}$ and width at most $\D$. 
    Combining with our previous height bound, we conclude that the height of $\start$ before the squeezing step is at most $ \big(\frac32 + 7 \eps'\big) H + 8 \eps' H_{LB} \leq \big(\frac32 + \eps\big) H$ by our choice of \alertBound{$\eps' = \frac\eps{16}$}. 
    Hence, by \cref{lem:squeezing}, after squeezing every squeezable item into the packing using \Cref{alg:squeezing}, $\start$ is a packing of $\items$ of height at most $\big(\frac32 + \eps\big) H$ as required.

    It remains to bound the running time of \cref{alg:approx-if-tall-next-to-each-other}. 
    To this end, we first calculate the number of iterations of the outer two for-loops. Iterating all possible start points for large and start vectors for horizontal items can be done in 
    \[ |S_{\cH\cL}|^{|\cL|}\cdot \left(\frac{2}{\delta\mu^2}\right)^{4/\delta\eps'^2} \leq \left(\frac{2}{\delta\mu^2}\right)^{4/\delta\eps'^2 + 1/\delta\mu} \leq \frac{1}{\eps^{\mathcal{O}(1/\eps^5)}}\]
    steps. 

    Once a suitable choice of starting vectors $(F_k)_{k=1}^{\log 1/\delta}$ is found, the algorithm packs the horizontal items integrally.
    Packing the horizontal items belonging to $\cH_{k,l}$ needs $\bigO\big(n + \frac1{\delta(\eps')^2} \big)$ operations as claimed in \Cref{claim:fractional-rounded-items_to_integral-original-items}.
    As rounding the horizontal and tall items needs to be done at most once, this adds at most $\bigO\big(n \min\big\{\log(n),\frac{1}{\eps'}\log\big(\frac1{\delta}\big)\big\}\big)$ to the algorithm's running time.
    Further, the running time of the Squeezing algorithm \ref{alg:squeezing} is at most $\mathcal O(n \log n)$.
    Hence using the extended Squeezing algorithm \ref{alg:squeezingAdd} leads to an additional term of $\mathcal O(n \log n)$ using the same argument as in the proof of \cref{lem:squeezing}.
    Therefore, the total running time of the algorithm is bounded by 
\[
\bigO\bigg(n \cdot \min\bigg\{\log(n),\frac{1}{\eps}\log\frac1{\eps}\bigg\}\bigg) + \frac{1}{\eps^{\mathcal{O}(1/\eps^5)}}.
\] 
\end{proof}

\bibliographystyle{plain} 
\bibliography{main_arxiv}

\end{document}